%% file: KN13.tex
\documentclass[showpacs,aps,nofootinbib,superscriptaddress]{revtex4}
\usepackage[pdftex]{graphicx}
\usepackage{bm}
\usepackage{amssymb}
\usepackage{dcolumn}
\usepackage{bm}
\usepackage{latexsym}
\usepackage{ifthen}
\usepackage{color}
\usepackage{tabularx}
\usepackage{amssymb}
\usepackage{hyperref}

\begin{document}
\newcommand{\ti}[1]{\mbox{\tiny{#1}}}
\newcommand{\im}{\mathop{\mathrm{Im}}}
\def\be{\begin{equation}}
\def\ee{\end{equation}}
\def\bea{\begin{eqnarray}}
\def\eea{\end{eqnarray}}
\newcommand{\il}{~}
\newcommand{\tb}[1]{\textbf{\texttt{#1}}}
\newcommand{\rtb}[1]{\textcolor[rgb]{0.98,0.00,0.00}{\textbf{\texttt{#1}}}}
\newcommand{\ptb}[1]{\textcolor[rgb]{0.50,0.00,0.80}{\textbf{\texttt{#1}}}}
\newcommand{\gtb}[1]{\textcolor[rgb]{0.17,0.69,0.25}{\textbf{\texttt{#1}}}}
\newcommand{\ytb}[1]{\textcolor[rgb]{0.87,0.89,0.46}{\textbf{\texttt{#1}}}}
\newcommand{\btb}[1]{\textcolor[rgb]{0.0,0.0,1.00}{\textbf{\texttt{#1}}}}
\def\bes{\begin{equation*}}
\def\ees{\end{equation*}}
\title{ Equatorial circular orbits of neutral test particles
in the Kerr--Newman spacetime}
\author{Daniela Pugliese}
\affiliation{ School of Mathematical Sciences, Queen Mary, University of London,\\
      Mile End Road, London E1 4NS, United Kingdom}
\author{Hernando Quevedo$^{2,3,4}$, and Remo Ruffini$^{2}$}
\affiliation{
      $^2$Dipartimento di Fisica and ICRA, Universit\`a di Roma ``La Sapienza",
                 Piazzale Aldo Moro 5, I-00185 Roma, Italy\\
                 ICRANet, Piazzale della Repubblica 10, I-65122 Pescara, Italy \\
$^3$Instituto de Ciencias Nucleares, Universidad Nacional Aut\'onoma de M\'exico,\\
AP 70543, M\'exico, DF 04510, Mexico\\
$^4$Instituto de Cosmologia, Relatividade e Astrofisica and ICRANet - CBPF\\
Rua Dr. Xavier Sigaud, 150, CEP 22290-180, Rio de Janeiro, Brazil
}
\email{d.pugliese.physics@gmail.com, quevedo@nucleares.unam.mx, ruffini@icra.it}
\date{\today}
\begin{abstract}
We present a detailed analysis of the orbital circular motion of electrically neutral test particles on the equatorial plane of the Kerr-Newman spacetime. Many details of the motion in the cases of  black hole and naked singularity  sources  are pointed out.
We identify four different types of orbital regions, which depend on the properties of the orbital angular momentum,  and define four different kinds of naked singularities, according to the values of the charge-to-mass ratio of the source. It is shown that the presence of a particular type of counter-rotating test particles is sufficient to uniquely identify naked singularities. It is pointed out that the structure of the stability regions can be used to differentiate between black holes and naked singularities.
 \end{abstract}
\maketitle
\section{Introduction}\label{KNint}
The Kerr-Newman (KN) spacetime is an exact solution of the Einstein-Maxwell equations that describes the exterior gravitational and electromagnetic
fields of a rotating charged source with mass $M$, angular momentum $J$ and electric charge $Q$. In Boyer-Lindquist coordinates, the KN line element can be written as
\bea\nonumber
 ds^2=-\frac{(\Delta -a^2\sin^2\theta)}{\Sigma
}dt^2-\frac{2a\sin^2\theta (r^2+a^2-\Delta)}{\Sigma}dtd\phi+
\eea
\be\label{knnetric}+\left[\frac{(r^2+a^2)^2-\Delta
a^2\sin^2\theta}{\Sigma}\right]\sin^2\theta
d\phi^2+\frac{\Sigma}{\Delta}dr^2+\Sigma d\theta^2, \ee
where we used geometrized units with $G = 1 = c$, and
\be
\Delta = r^2-2M r+a^2+Q^2,
\ee
and
\be
\Sigma =r^2+a^2\cos^2\theta.
\ee
The parameter $a$ stands for the angular momentum per unit mass, as measured
by a distant observer.
The corresponding electromagnetic vector potential
\be
A_\alpha=-\frac{Q r}{\Sigma}\left[(dt)_\alpha-a\sin^2\theta (d\phi)_\alpha\right],
\ee
depends on the charge
$Q$ and the specific angular momentum $a$. It then follows that the magnetic field is generated by the rotation of the charge distribution.

The limiting cases of the KN metric are the Kerr metric for $Q=0$, the Schwarzschild metric which is recovered for $a=Q=0$, the Reissner-Nordstr\"om (RN) spacetime for $a=0$, and the Minkowski metric of special relativity for $a=Q=M=0$. The KN spacetime is asymptotically flat and free of curvature singularities outside a region situated very close to the origin of coordinates.

Several critical points characterize the geometric and physical properties of this spacetime. In particular,
the function $\Delta$ vanishes at the radii
\be
r_{\pm} \equiv M \pm \sqrt{M^2 - a^ 2 - Q^2}
\ee
which are real only if the condition $M^2\geq Q^2 + a^2$ is satisfied. In this case, $r_+$ and $r_-$ represent the radius of the outer and
inner horizon, respectively, and the KN solution is interpreted as describing the exterior field of a rotating charged black hole (BH).
In the case $M^2<a^2+ Q^2$,  no zeros of $\Delta$ exist and the gravitational field corresponds to that of a naked ring singularity situated at
\be
r^2 + a^2 \cos^2\theta =0.
\ee

The norm of  the timelike
Killing vector field $\xi^\alpha_t=\delta^\alpha_t$ for the KN metric in Boyer-Lindquist coordinates (\ref{knnetric}) reads
\be
g_{tt}=-\frac{\Delta -a^2\sin^2\theta}{\Sigma},
\ee
Then, the norm is positive in the region where
\be
r^2+a^2\cos^2\theta+ Q^2-2Mr<0,
\ee
or
\be
r^-_{\epsilon}<r<r^+_{\epsilon}
\quad\mbox{with\quad}r^{\pm}_{\epsilon}\equiv M\pm\sqrt{M^2-Q^2-a^2\cos^2\theta}
\ee
where $M^2-Q^2-a^2\cos^2\theta\geq0$ is satisfied. In particular, for a black hole it is $r_+<r<r^+_{\epsilon}$
a region which is known as the ergosphere. In the regions where $\xi_t$ becomes spacelike, the coordinate $t$ cannot
be used as a time coordinate and the analysis of the physical properties of this spacetime requires a different approach.

The extrema of $g_{tt}$ are located in
\be \label{soa}r^*_{\pm}=\frac{Q^2\pm\sqrt{2 a^2 M^2+Q^4+2 a^2 M^2
\cos(2 \theta)}}{2 M},
\ee
which is a function of the parameters $a$ and $Q$, and the coordinate $\theta$. At these radii we find that
\be -g_{tt}(r^*_{\pm})=\frac{2 a^2+\left(Q^2\mp\sqrt{2 a^2 M^2+Q^4+2
a^2 M^2 \cos(2\theta)}\right) \sec^2\theta}{2 a^2},
\ee
so that $g_{tt}$ has a maximum at $r=r^*_{+}$. The limiting cases of these radii are of interest. When $a=0$
(Reissner-Nordstr\"om spacetime) we have
\be
r^*_+=\frac{Q^2}{M},\quad r^*_{-}=0,\quad
g_{tt}(r^*_{+})=-1+\frac{M^2}{Q^2}.
\ee
For a Kerr spacetime, $Q=0$, we have that
\be
r^*_{+}=\sqrt{a^2 \cos^2\theta},\quad
r^*_{-}=-\sqrt{a^2 \cos^2\theta}, \quad g_{tt}
(r^*_{+})=-\left(1-\frac{M^2}{\sqrt{a^2 M^2 \cos^2\theta}}\right).
\ee

In this work, we will perform a detailed analysis of the motion of test particles along circular
orbits on the equatorial plane of the KN spacetime. Since test particles moving along circular orbits are particularly appropriate to measure the effects generated
by naked singularities, we will focus on the study of the physical differences between a black hole and a naked
singularity; see also \cite{BiHoe85,Saa:2011wq,Lopez83} and for further applications \cite{Bambi:2013eb,Wunsch:2013s}.
{
Although most known astrophysical compact objects possess only a small net charge or no charge at all, the study of the KN spacetime is important from the conceptual  and theoretical points of view. The KN is probably the most important exact solution of the Einstein-Maxwell equations and, therefore, represents an ideal framework to study the interaction between the electromagnetic field and the gravitoelectric and gravitomagnetic components of gravity. Moreover, the detailed study of an exact solution usually sheds light on the extent of applicability of the solution and the theory itself. In this work, we study the KN solution from the point of view of the neutral circular motion, and compare our results with those obtained in the RN and Kerr spacetimes. To this end, the spin and charge are gradually changed between the limiting  cases  of $a/M=0$, $Q/M=0$  and  the  KN case. We first fix the charge-to-mass ratio and let the spin change from zero
to the black hole and the naked singularity regime. Then, we fix the spin and let the charge change, comparing    the  naked singularity and black hole spacetime properties.}

This work follows a series of papers \cite{Pu:Neutral,Pu:Kerr,Pu:Charged} in which we showed the presence of a typical band structure, i.e., disconnected regions
of stable orbits for the spacetimes generated by  Kerr and RN naked singularities. This structure is absent in the case of black holes.
It was also found that there exist two types of singularities that affect the properties of the orbits and their stability.
We generalize here those  works considering a field source  with charge  and intrinsic angular momentum.

The plan of this  paper is the  following:
In Sec.\il\ref{Sec:qua_te}, we introduce  the effective potential for a neutral test particle moving along a circular orbit on the equatorial plane, and we set the main notations to be used in this work. In Sec.\il\ref{Sec:BH_KN}, we explore the motion  in the spacetime of a KN black hole while the case of a naked singularity is explored in Sec.\il\ref{Sec:NS_Case}. Section\il\ref{Sec:BH-vs-NS} summarizes the main results and compare the two cases. Finally, the conclusions in Sec.\il\ref{Sec:Conclusions} close this article.
\input{sign_V13.tex}

%
\input{BH_KN13.tex}
%
\input{NS_L0_13.tex}
\input{NS_QmaggM13.tex}
\input{NS_QminM13.tex}

\input{CFRT_BH_NS_13.tex}
%
\section{Conclusions}\label{Sec:Conclusions}
In this work, we performed a detailed analysis of the behavior of test particles moving along circular orbits on the equatorial plane of the KN spacetime. We use the method based on the investigation of the effective potential from which one can derive the location and the stability
properties of the circular orbits. We also analyzed the energy and angular momentum of all possible orbits, and used this information to identify different orbital regions which allow us to find out the differences between black holes and naked singularities.

We have identified four orbital regions, namely:  region \textbf{I}  characterized by $L=L_-$, region \textbf{II} with $L=(L_-,-L_+)$, region \textbf{III} with $L=-L_{\pm}$, and a fourth orbital region \textbf{IV} with $L=-L_-$. The region \textbf{I} contains only co-rotating particles whereas regions \textbf{III} and \textbf{IV} are for counter-rotating particles. It is interesting that region \textbf{II} contains both co-rotating and counter-rotating particles with different orbital angular momentum.

\begin{figure}[h!]
\begin{center}
\begin{tabular}{c}
\includegraphics[width=0.71\hsize,clip]{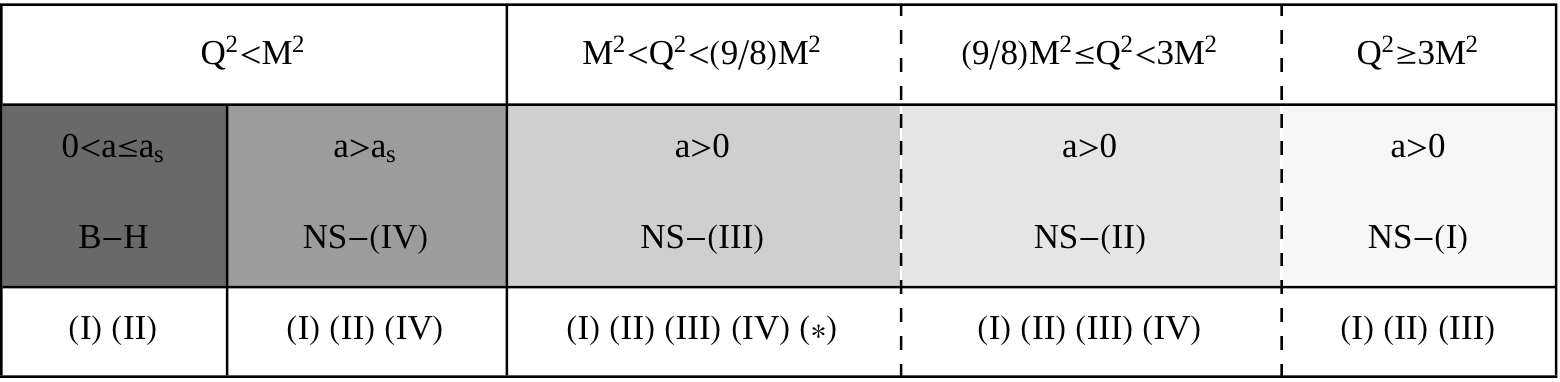}
\end{tabular}
\caption[font={footnotesize,it}]{\footnotesize{Classification of KN compact object according to the charge-to-mass ratio  $Q/M$ and the spin $a/M$ of the source. Black holes can have regions of type \textbf{I} and \textbf{II} and only naked singularities can accommodate regions of all four types.
The type  NS-III (with $*$) has two forbidden
orbital  regions while NS-II has only one  forbidden orbital region.}}
 \label{Table:Everxcf}
\end{center}
\end{figure}

We found that naked singularities can be divided into four classes that depend on the value of the
 charge-to-mass ratio:    \textbf{NS-I} for  ($Q/M\geq\sqrt{3})$ with orbital regions of the type \textbf{I-II-III},
  \textbf{NS-II} for $(3/(2\sqrt{2})\leq Q/M<\sqrt{3})$  with orbital regions of the type \textbf{I-II-III-IV} and one forbidden region,  \textbf{NS-III}  for $(1< Q/M< 3/(2\sqrt{2}))$ with orbital regions of the type \textbf{I-II-III-IV} and two forbidden regions and
  \textbf{NS-IV} for  $(Q/M<1)$  with orbital regions of the type \textbf{I-II-IV}. Figure\il\ref{Table:Everxcf} illustrates this classification also for the case of black holes.

We found that from the point of view of the circular motion, there exists always a forbidden area ($r<r_*)$ which is completely inaccessible for test particles. In the case of black holes, the radius $r_*$ is situated inside the outer horizon. In the case of naked singularities, it can be considered as a surface that ``covers'' the singularity itself and prevents it from being studied with orbiting test particles. The value of the radius  $r_*=Q^2/M$ does not depend on the source angular momentum so that it is a property of the gravitational field generated only  by the electric charge only . It is interesting that the expression for $r_*$ coincides with the classical radius of an electric charge $Q$ with mass $M$ which is obtained by applying a completely different approach. In the limiting case of a RN naked singularity, this radius has a very particular significance because it corresponds to particles with vanishing angular momentum $(L=0)$ \cite{Pu:Neutral}.

Another interesting result is that only naked singularities allow orbiting test particles of the type \textbf{III} and \textbf{IV}, i.e., counter-rotating particles with $L=-L_\pm$. Since black holes can have counter-rotating particles only with $L=-L_+$, it follows that the presence of counter-rotating particles with $L=-L_-$ would necessarily imply the existence of KN naked singularities.

In the case of black holes, we have shown that there exists only one stability region for circular orbits whereas naked singularities present two stability regions separated by one zone of instability. This peculiar structure suggests the possibility of distinguishing black holes from naked singularities by investigating the structure of accretion disks around the central source. Indeed, if we
imagine an idealized accretion disk made of test particles which are moving along circular orbits on the equatorial plane of a KN spacetime,
then in the case of a black hole the accretion disk is continuous whereas in the case of a naked singularity it is discontinuous. It would be
interesting to find out if this difference will remain in the case of more realistic accretion disks \cite{Babichev:2008dy,
StuBicBal01,
StuJu96,
StuHle98,
KoStuKara0607,
EsteMarr90,JaAbraPac90},
and if it could lead to observable effects. This question was also recently analyzed in \cite{Kovacs:2010xm}.

During the past few years, much effort has been made to understand the question about the formation of naked singularities \cite{Pat2010,Joshi:2012mk,Patil:2011uf}, and about the properties of solutions that describe the gravitational field of such
hypothetical  compact objects \cite{Gautrou,DiCriscienzo:2010gh,Liang74,Dotti:2008ta,Casadio:2003iv,Toth:2011ab,Virbhadra:2007kw,Virbhadra:2002ju}. The present
work contributes to this discussion in the sense that it shows the possibility of identifying KN naked singularities by investigating the
properties of their accretion disks.

However, there are indications that the rigid structure of the KN solution is not adequate to correctly describe naked singularities. In fact, several studies show that KN singularities are highly unstable configurations \cite{dotti,pani}  that, therefore, cannot exist in Nature.
{Nevertheless, it is important to emphasize that the  stability problem of the KN solution is still a subject under debate. For instance, in \cite{Pani:2013ija} the slow-rotation limit of the KN black holes has been addressed recently
by using gravito-electromagnetic perturbations. Using a self-consistent calculation of scalar, electromagnetic and
gravitational quasi normal modes  up to linear
order in the spin and arbitrary values of the charge, a fully-consistent stability analysis of the KN is provided. Since none
of these modes is unstable, the authors suggest that their  calculation provides solid
(numerical) evidence for the stability of the  KN metric in the non extremal regime, an analysis that, in principle, could be generalized to the naked singularity regime. We conclude that it is important to continue the study of the stability of the KN spacetime. It is also possible
to consider more general solutions that generalize the KN spacetime and could describe the gravitational field of compact objects.
The first possibility is to consider solutions with quadrupole and higher moments \cite{quev90}. Since already in the case of a static
field with only quadrupole moment, there are many possible solutions \cite{weyl,chaz,cur,erro59,gm,quev87,solutions} and most of them are
quite difficult to handle, we propose to use the recently proposed reinterpretation of the Zipoy-Voorhees \cite{zip66,voor70,quev11} solution which  from the mathematical point of view is the simplest generalization of the Schwarzschild black hole solution with a quadrupole moment. Nevertheless,  to understand properly the main features of the  solutions with quadrupole  it is important to address  the limiting  cases of the exact KN metric.
We expect to investigate
this problem in the near future.
\section*{Acknowledgments}
One of us (DP) gratefully acknowledges financial support from the Blanceflor Boncompagni-Ludovisi n\'ee Bildt Foundation. This work was  also supported by  A. Della Riccia Foundation.
This work was supported by CONACyT-Mexico, Grant No. 166391, by DGAPA-UNAM,  and by CNPq-Brazil.
\input{Appendix_spin_radius13}

\input{BiB_KN13}
\end{document}

%% file: sign_V13.tex
\section{Circular orbits}\label{Sec:qua_te}
We consider the circular motion of a test particle of mass $\mu$ in the
background represented by the KN metric (\ref{knnetric}). We limit ourselves
to the case of orbits situated on the equatorial plane.
We adopt the effective potential approach to the study of the test particle dynamics.
The test particle  motion  is therefore described as
the one--dimensional motion of a classical particle in the effective potential $V(r)$.
In the case of equatorial geodesics ($\theta=\pi/2$), one obtains
the effective potential \cite{MTW}
\begin{equation}\label{Eq:tra_0}
V=-\frac{B}{2A}+\frac{\sqrt{B^2-4 A C}}{2A},
\end{equation}
where
\bea
A&\equiv&\left(r^2+a^2\right)^2-a^2\Delta,\\
B&\equiv&-2aL\left(r^2+a^2-\Delta
\right),\\
C&\equiv& a^2L^2-\left(M^2r^2+L^2\right)\Delta\ .
\eea
Here $L=\mu g_{\alpha\beta} \xi^\alpha_{\phi} u^\beta$ is a constant of motion associated with the
angular momentum of the test particle with mass $\mu$ and 4-velocity $u^\beta$.  The Killing vector $\xi_\phi =\partial_\phi$
represents the axial symmetry of the rotating source. The additional Killing vector $\xi_t = \partial_t$ is timelike and represents
the stationarity of the field configuration. It also generates a constant of motion $E=-\mu g_{\alpha\beta} \xi^\alpha_t u^\beta$
which is associated with the total energy of the test particle, as measured by a static observer at infinity.

Notice that the effective potential (\ref{Eq:tra_0}) was obtained originally as the solution of a quadratic algebraic equation so that the solution
\begin{equation}\label{Eq:tra_01}
V^-=-\frac{B}{2A}-\frac{\sqrt{B^2-4 A C}}{2A}
\end{equation}
is also allowed. Nevertheless, this solution does not need to be analyzed separately, because its properties can be obtained from $V$ by using the symmetry $V^-(-L) = - V(L)$.

Circular orbits are given by the simultaneous solutions of the equations
\be\label{E:proprio}
V= E/\mu\quad\mbox{and}\quad \frac{dV}{dr}=0.
\ee
The potential (\ref{Eq:tra_0}) is quadratic in the charge $Q$. We therefore will limit ourselves to the study of the case $Q>0$. Moreover, the effective
potential  is invariant under a simultaneous  change of the sign of $a$ and $L$. Consequently, without lost of generality, we can restrict the analysis to the case $a>0$, and will differentiate between corotating particles with $L>0$ and counter-rotating particles with $L<0$.

To perform a detailed analysis of the effective potential of a naked singularity  with
\(
a^2+ Q^2>M^2
\),
we will consider the cases
\be
a>M, \quad\mbox{or}\quad Q>M,
\ee
otherwise
\be\label{Eq:as_intro}
M>a>a_s, \quad\mbox{and}\quad Q<M,
\quad a_s\equiv\sqrt{M^2-Q^2}\ ,
\ee
or
\be
M>Q>Q_s, \quad\mbox{and}\quad a<M,
\quad Q_s\equiv\sqrt{M^2-a^2} \ .
\ee
For a discussion of the motion in a KN spacetime see
\cite{Stu80, BiStuBa89,BaBiStu89,Kovar:2010ty,Kovar:2008ta,StuHle2000,Cin89,Stuc81,BekDenTre74}
and also \cite{RaVi96,StuHleJu00,Kovar:2013pf}.

We introduce also the radii $r^{\ti{RN}}_{\pm}\equiv r_{\pm}(a=0)=M\pm\sqrt{M^2-Q^2}$ for the horizons of the Reissner--Nordstr\"om black hole, and $r^{\ti{K}}_{\pm}\equiv r_{\pm}(Q=0)=M\pm\sqrt{M^2-a^2}$ for the horizons of the Kerr black hole, respectively.
Notice that $r^{\ti{RN}}_{\pm}$ coincides with the ergosphere boundaries  $r_{\epsilon}^{\pm}$ on the equatorial plane $\theta=\pi/2$. On this plane the ergosphere is completely independent from the source spin. In particular, for a KN naked singularity there is a change of sign in the time-time component of the metric tensor (and in the norm of the timelike Killing vector) inside  the region $[r_{\epsilon}^-,r_{\epsilon}^+]$, for $a\leq M$ and $Q_s<Q<M$ and $a>M$ with $Q<M$.
%
%
\subsection{Circular motion around black holes and naked singularities}\label{Sec:circular_motion_BH+NS}
In this section, we introduce the basic equations of the  circular motion in the KN spacetime,  we  fix the main notations
and introduce  the most important quantities that are relevant for the characterization of the motion both in the black hole and  in the naked singularity case.

\begin{figure}[h!]
\begin{tabular}{ccc}
\includegraphics[width=0.3\hsize,clip]{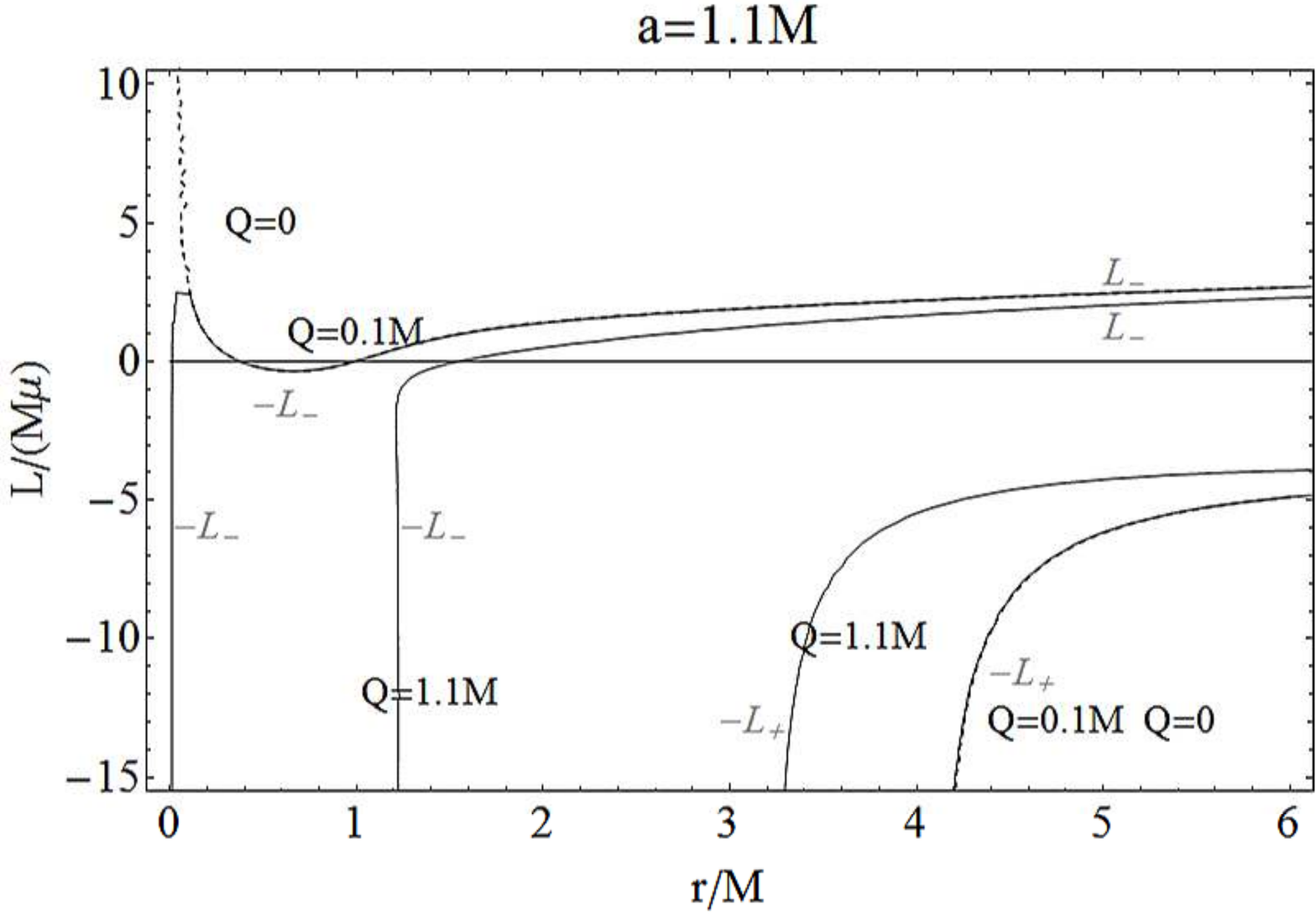}&
\includegraphics[width=0.3\hsize,clip]{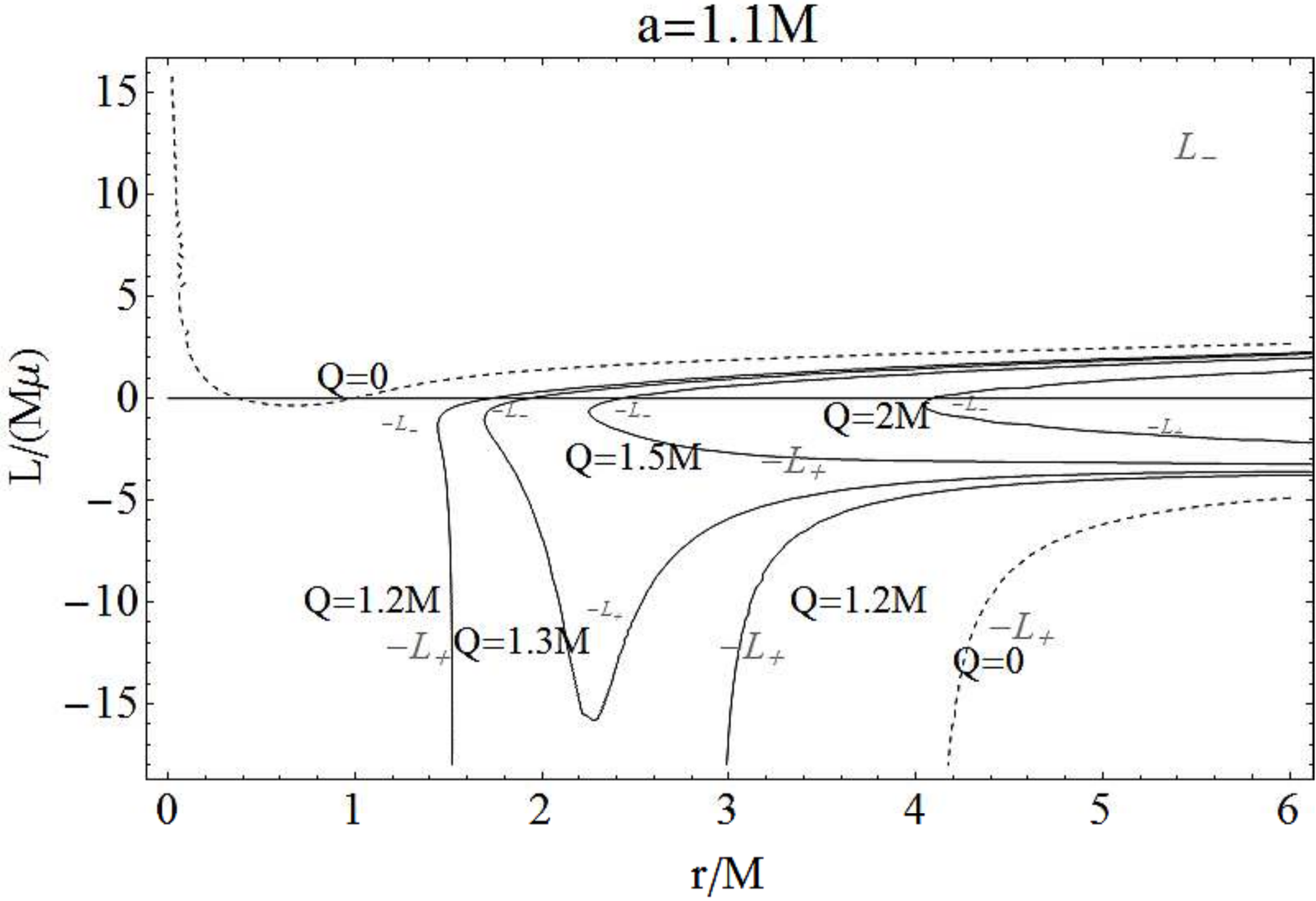}&
\includegraphics[width=0.3\hsize,clip]{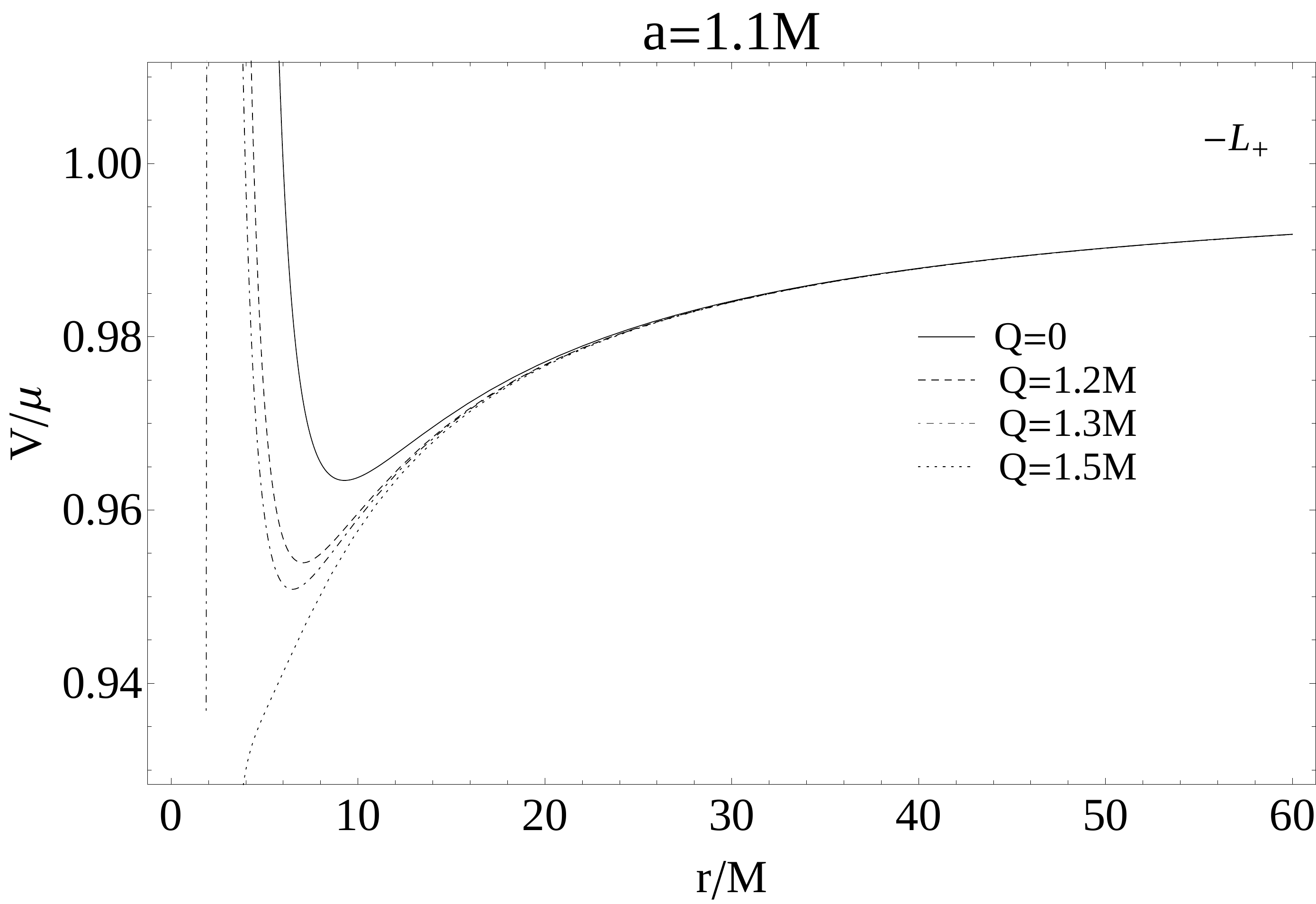}
\end{tabular}
\caption[font={footnotesize,it}]{ The angular momentum of circular orbits is plotted
as a function of the orbital radius for different  source charge--mass ratio $Q/M$. Circular orbits in the
KN spacetime with $a=1.1M$ are explored. Right: The energy $V/\mu$ of circular orbits as a function of $r/M$ for selected values of $Q/M$.
 } \label{PlotSPOIU2uno}
\end{figure}

\begin{figure}[h!]
\begin{tabular}{ccc}
\includegraphics[width=0.3\hsize,clip]{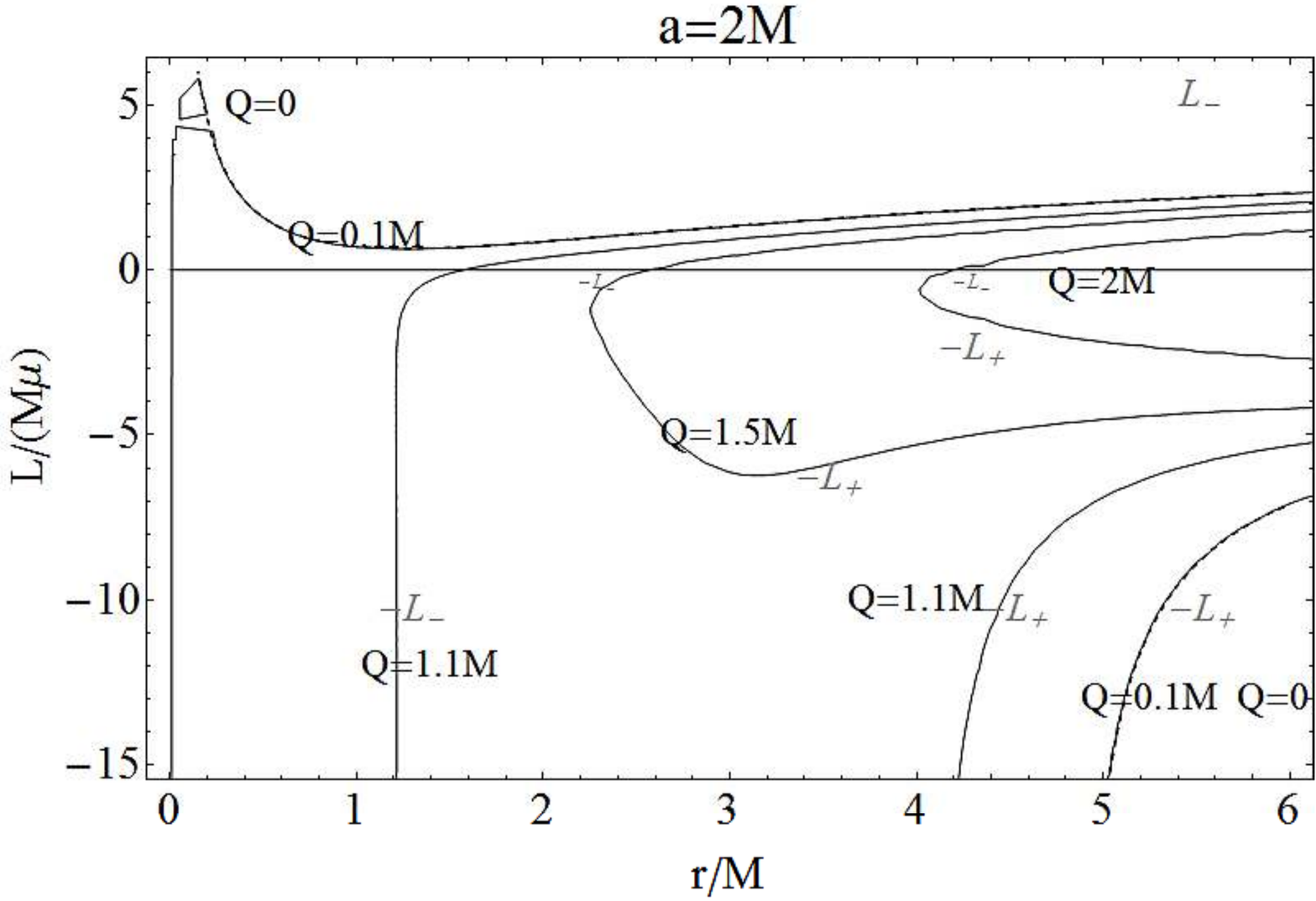}&
\includegraphics[width=0.3\hsize,clip]{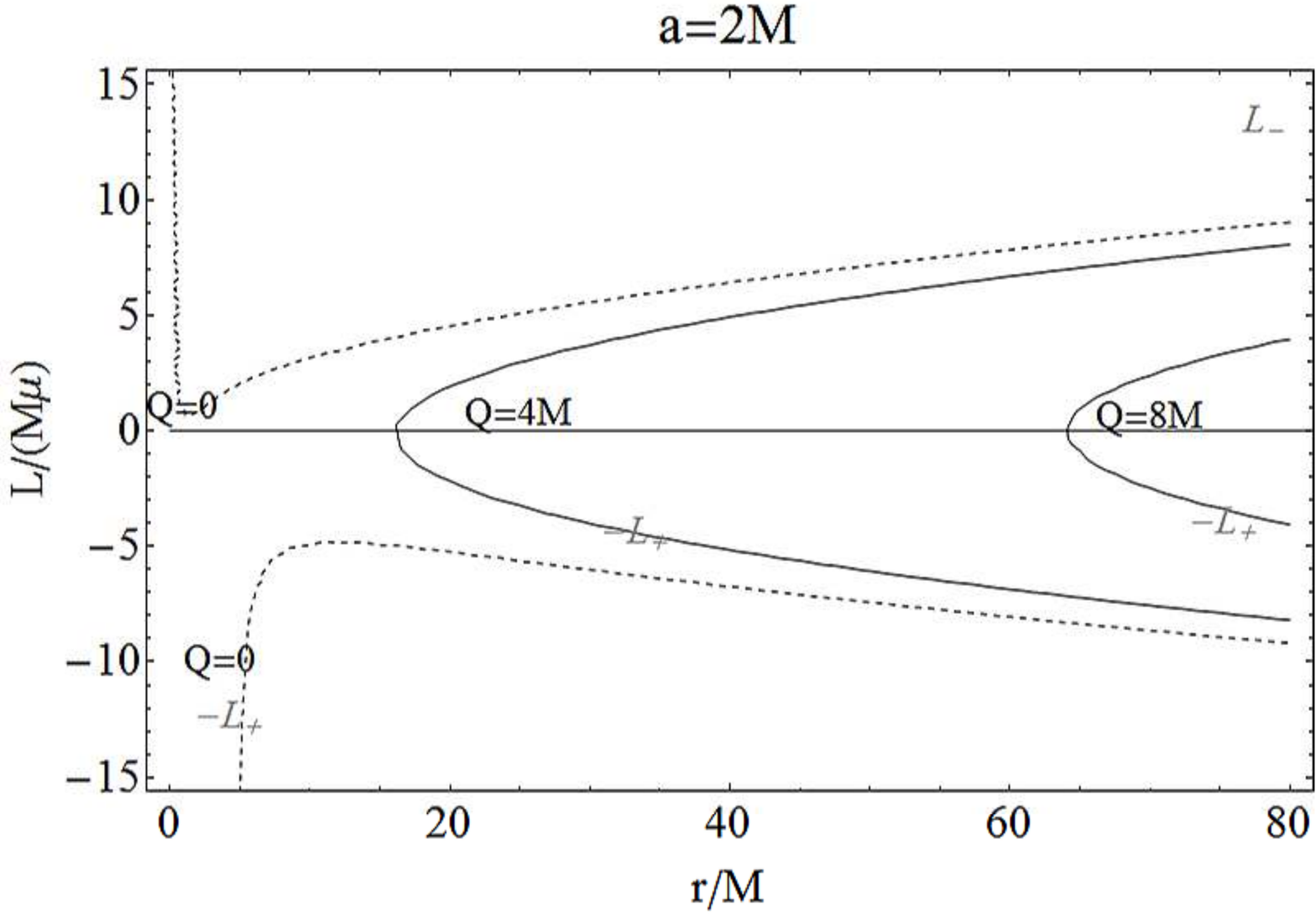}&
\includegraphics[width=0.3\hsize,clip]{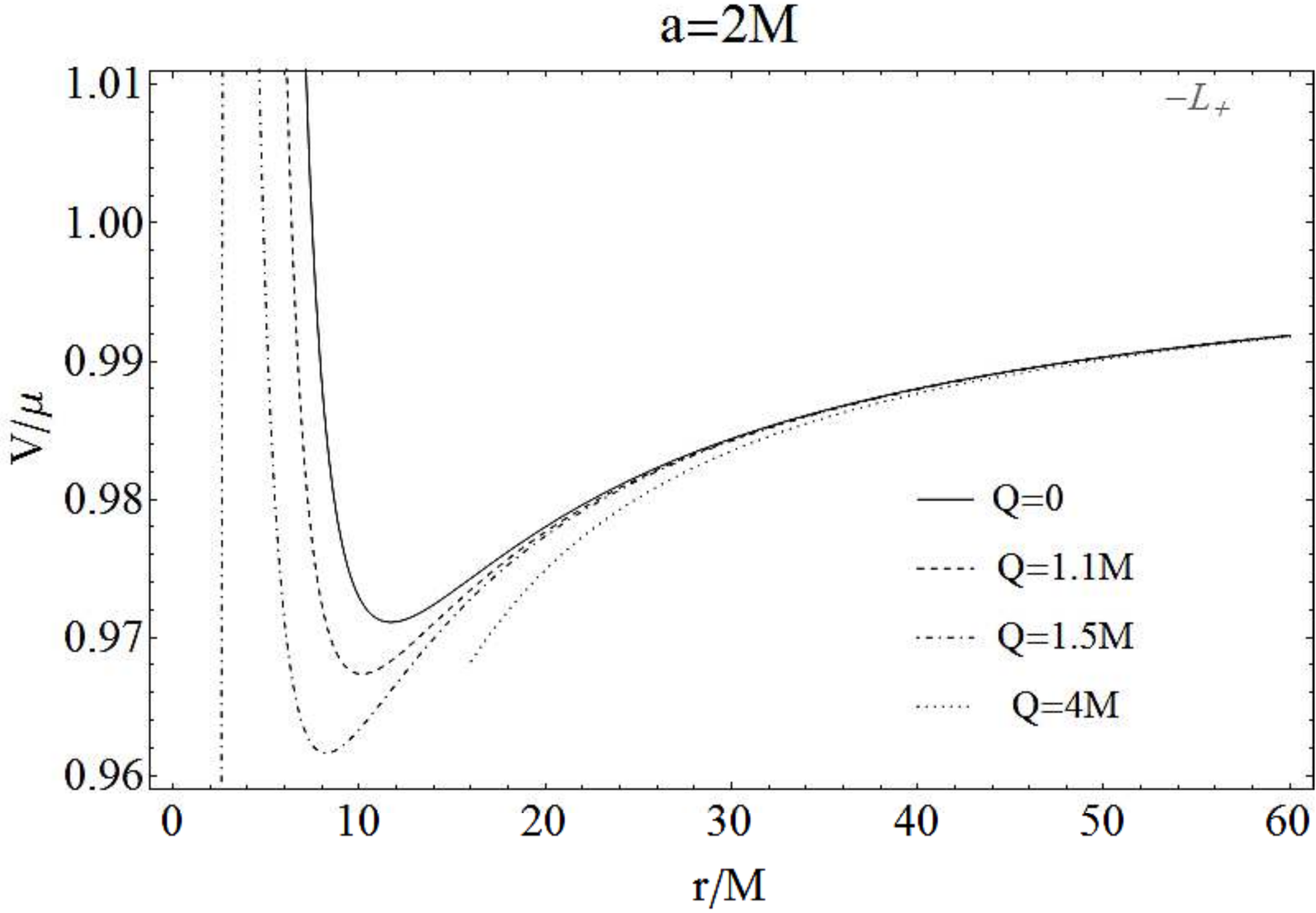}
\end{tabular}
\caption[font={footnotesize,it}]{\footnotesize{Circular orbits in the
KN spacetime with $a=2M$ are explored. The angular momentum of circular orbits is plotted
as a function of the orbital radius for different  source charge--mass ratii $Q/M$. Right: The energy $V/\mu$ of circular orbits as a function of $r/M$ for selected values of $Q/M$.
} } \label{PlotSPOIUA2uno}
\end{figure}

In order to explore the circular motion around black holes and naked singularities, we first study the condition $V'=0$. Thus,
we  analyze  the equations for the circular motion
\be\label{E:bicchi}
V'(r,L,a,Q)=0,\quad V=E/\mu,
\ee
and solve them with respect to  the angular momentum of the test particle.
We obtain that the general solution corresponds
to $L=\pm L_{\pm}$, where
\be\label{lpspr}
\frac{L_{\pm}}{\mu }\equiv \frac{1}{r^2}\sqrt{\frac{\Sigma\pm2M^2 \sqrt{\Pi}}{\Psi }}
\ee
with
\bea
\Psi&\equiv&4 a^2 \left(Q^2-rM\right)+\left[2 Q^2+(r-3M) r\right]^2
\\\label{E:pidef}
\Pi&\equiv&-a^2 \left(Q^2-rM\right) r^4 \left[a^2+Q^2+(r-2M) r\right]^2 \left[a^2 \left(Q^2-Mr\right)+\left(2 Q^2-3Mr\right) r^2\right]^2,
\\
\nonumber\Sigma&\equiv& r^2 \left\{-\left(Q^2-Mr\right) r^4 \left[2 Q^2+(r-3M) r\right]+a^4 \left(Q^2-Mr\right) \left[2 Q^2-(5M+r) r \right]+\right.\\
&&\left. a^2 \left[2 Q^6+Q^4 (r-11M) r-2 Q^2 (r-2M) r^2 (5M+r)+2M r^3 [r (3M+r)-6M^2]\right]\right\}.
\eea
The evaluation of the corresponding energies leads to
\be
E({\pm L_{\mp}})=\frac{- (\pm L_{\mp}) \left(Q^2-2 M r\right)+\sqrt{r^2 \left[a^2+Q^2+(r-2M) r\right] \left(r^2 \left(L_{\mp}^2+r^2\right)+a^2 \left[r (2M+r)-Q^2\right]\right)}}{r^4+a^2 \left[r (2M+r)-Q^2\right]}\ .
\ee
Now, the main point is to find the regions in the space of the radial coordinate and parameters $\{r/M, Q/M, a/M\}$ where the solutions for the angular momentum and the energy are well defined. For instance,
in Fig.\il\ref{PlotSPOIU2uno}, we show the behavior of the particle angular momentum and energy in the case
$a=1.1M$ for different values of the naked singularity charge $Q$.
In Fig.\il\ref{PlotSPOIUA2uno}, we present the numerical results for the case
$a=2M$ and different values of the naked singularity charge $Q$.
First, we study the  orbital regions in which circular motion occurs, that is, where  solutions of Eq.\il(\ref{E:bicchi}) for $L$ exist. These regions have as boundaries the solutions of $\Psi=0$ or explicitly
\be\label{anaozoress}
r^4 -6Mr^3+\left(4Q^2+9M^2\right) r^2-4\left(3Q^2+
a^2\right)M r+4 Q^4+ 4a^2Q^2=0.
\ee
The solutions of Eq.\il(\ref{anaozoress}) are  the radii
\bea\label{Eq:defrprm}
r^{+}_{ \pm}&\equiv& \frac{1}{6} \left[9M+ \sqrt{3\Upsilon}\pm\sqrt{3}
\sqrt{27M^2-24 Q^2+\frac{24 \sqrt{3}M a^2}{\sqrt{\Upsilon}}-\Upsilon}\right],
\\\nonumber
r^{-}_{\pm}&\equiv&\frac{1}{6} \left[9M- \sqrt{3\Upsilon}\pm\sqrt{3}
\sqrt{27M^2-24 Q^2-\frac{24 \sqrt{3}M a^2}{\sqrt{\Upsilon}}-\Upsilon}\right],
\eea
where
\bea\nonumber
\Upsilon&\equiv&9M^2-8 Q^2+\frac{\left(9M^2-8 Q^2\right)^2+24 a^2 \left(2 Q^2-3M^2\right)}{\varsigma ^{1/3}}+\varsigma ^{1/3},
\\
\nonumber
\varsigma&\equiv&
216 M^2a^4-\left(8 Q^2-9M^2\right)^3-36 a^2 \left(27M^4-42M^2 Q^2+16 Q^4\right)+
\\
&&24 \sqrt{3} a^2\sqrt{\left(Q^2+a^2-M^2\right) \left[27 M^4a^2-Q^2 \left(9M^2-8 Q^2\right)^2\right]}\ .
\eea
The study of the regions confined by these radii  will be the key point  to analyze the circular orbits of neutral particles in a KN naked singularity and to distinguish this case  from the black hole one.

\begin{figure}[h!]
\begin{tabular}{ccc}
\includegraphics[width=0.3\hsize,clip]{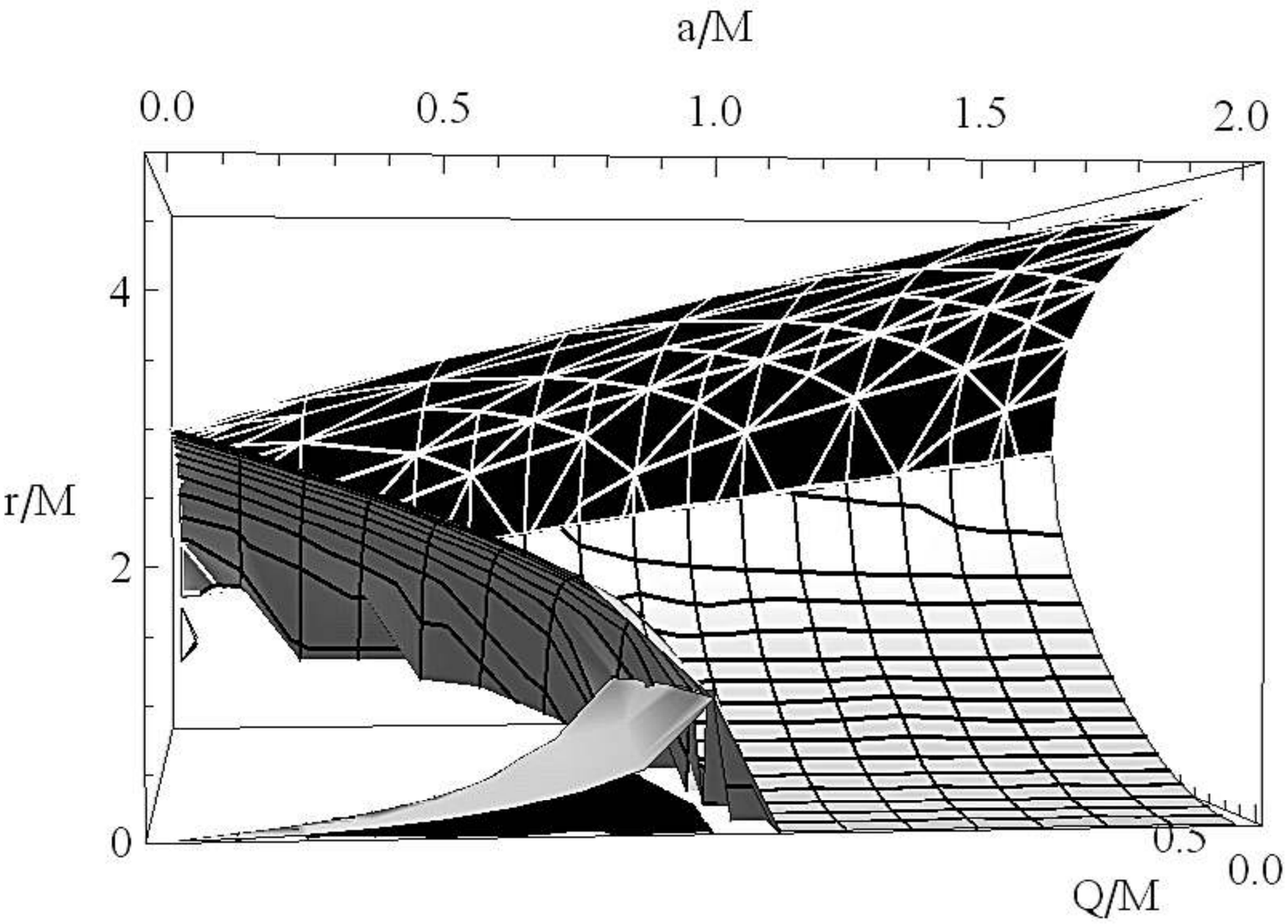}&
\includegraphics[width=0.3\hsize,clip]{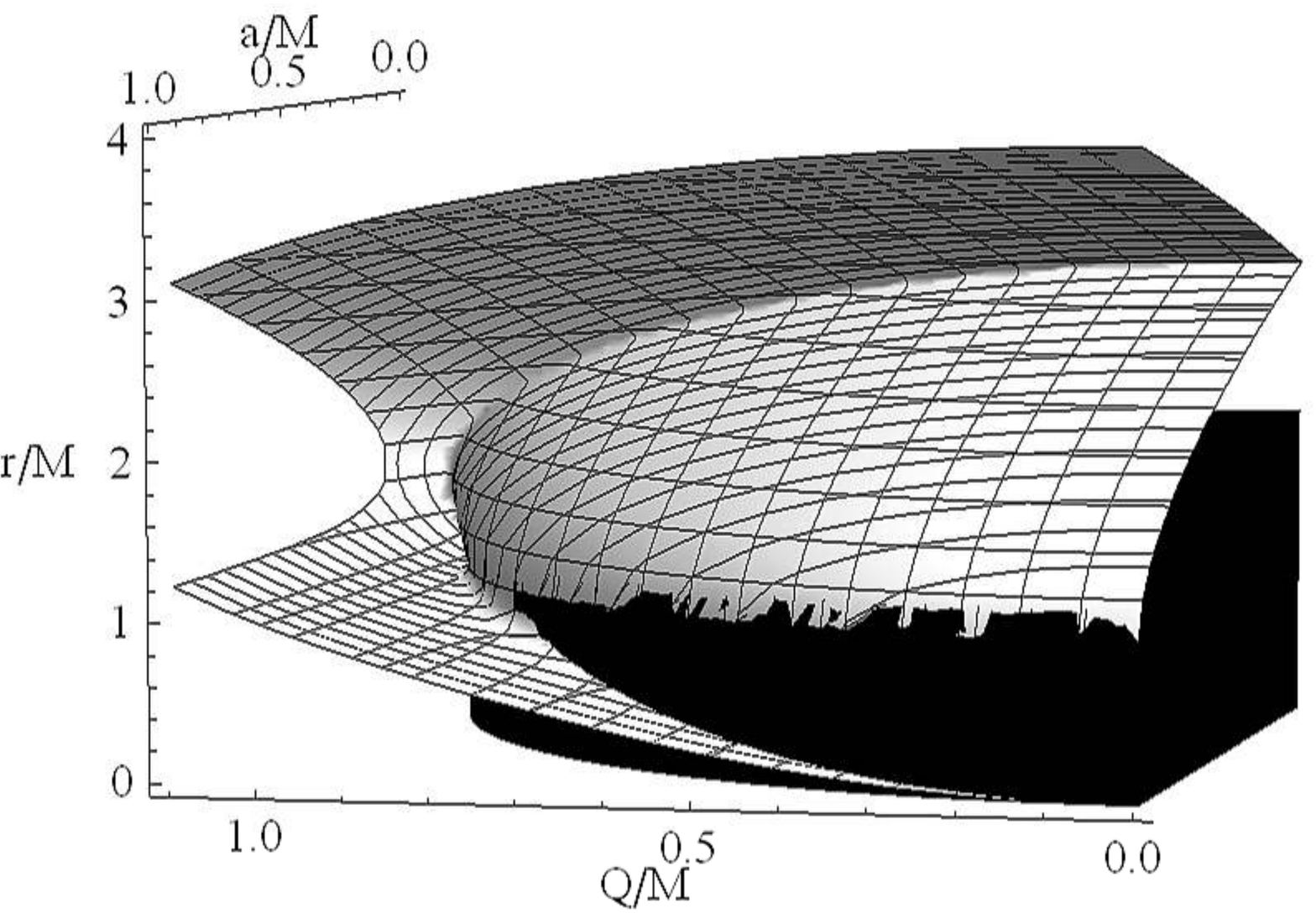}&
\includegraphics[width=0.3\hsize,clip]{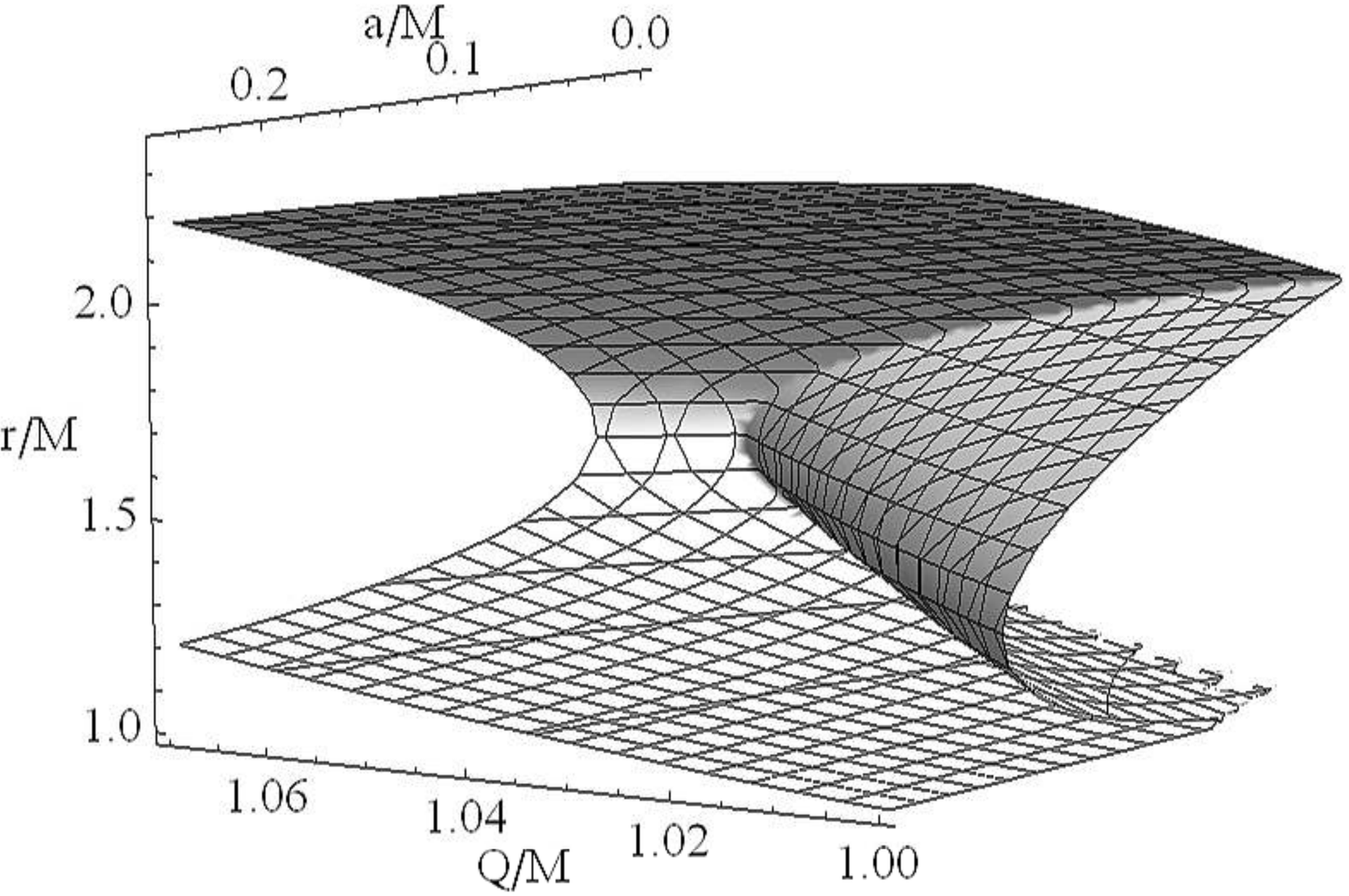}
\end{tabular}
\caption{{Left: The radii $r_1$ (black surface), $r_2$ (white surface), $r_3$ (gray, black-shaded, surface), and $r_4$ (black, white-shaded surface) plotted as functions of $Q/M\in(0,2)$ and $a/M\in(0,2)$; it is $r_1<r_2<r_3<r_4$.
Center: Surfaces $(r_1,r_2,r_3,r_4)$ in the the region $Q/M\in\{0, 1.1\}$, $a/M\in\{0, 1\}$, $r/M\in\{0, 4\}$. The black region represents $r<r_+$, where $r_+$ is the outer horizon of the black hole. Right: The region $Q/M\in\{1, 1.07\}$, $a/M\in\{0, .251\}$, $r/M\in\{1, 2.35683\}$ corresponds to the case of a naked singularity.}}
 \label{R1R2R3R4T}
\end{figure}

\begin{figure}[h!]
\begin{tabular}{cc}
\includegraphics[width=0.4\hsize,clip]{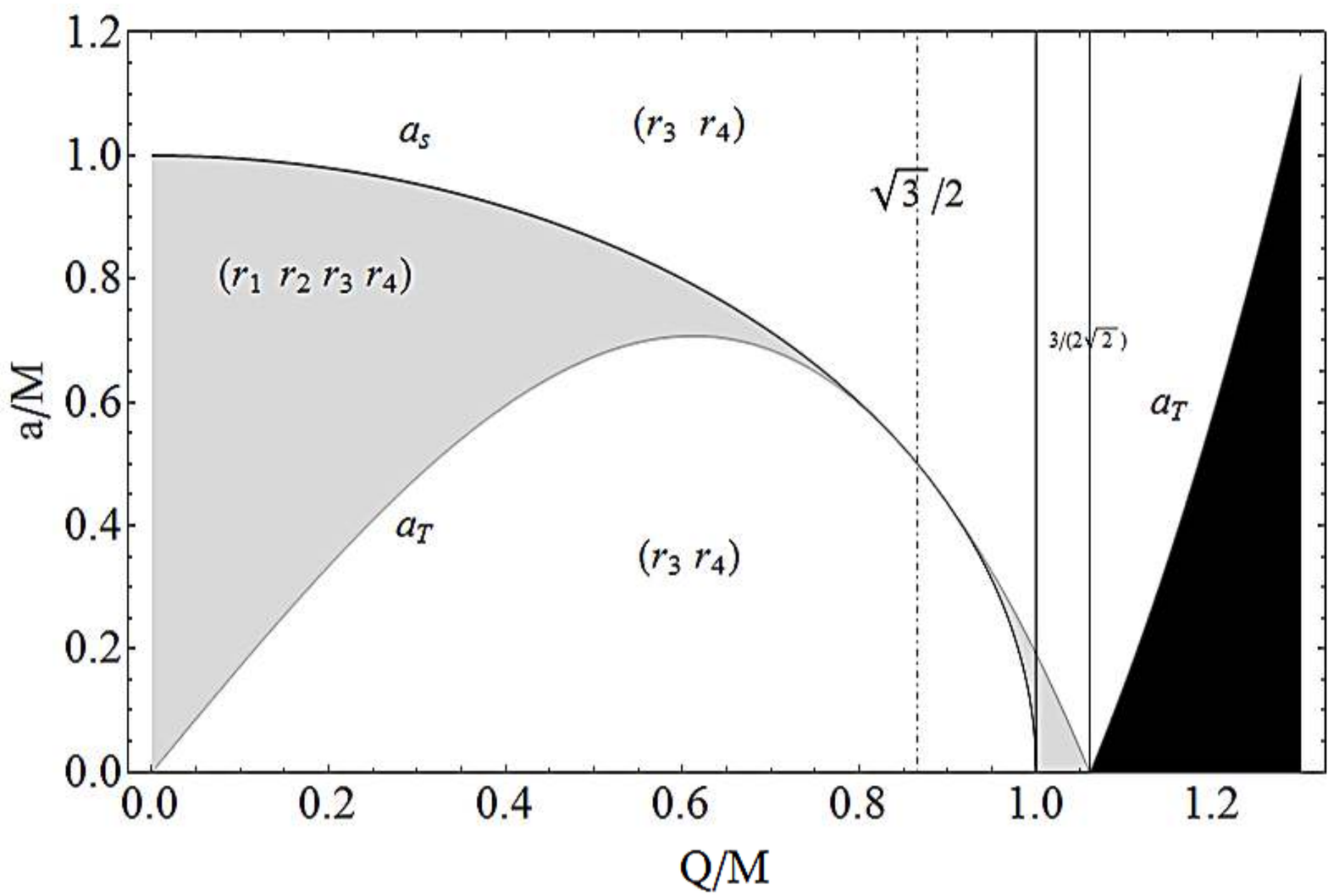}&
\includegraphics[width=0.4\hsize,clip]{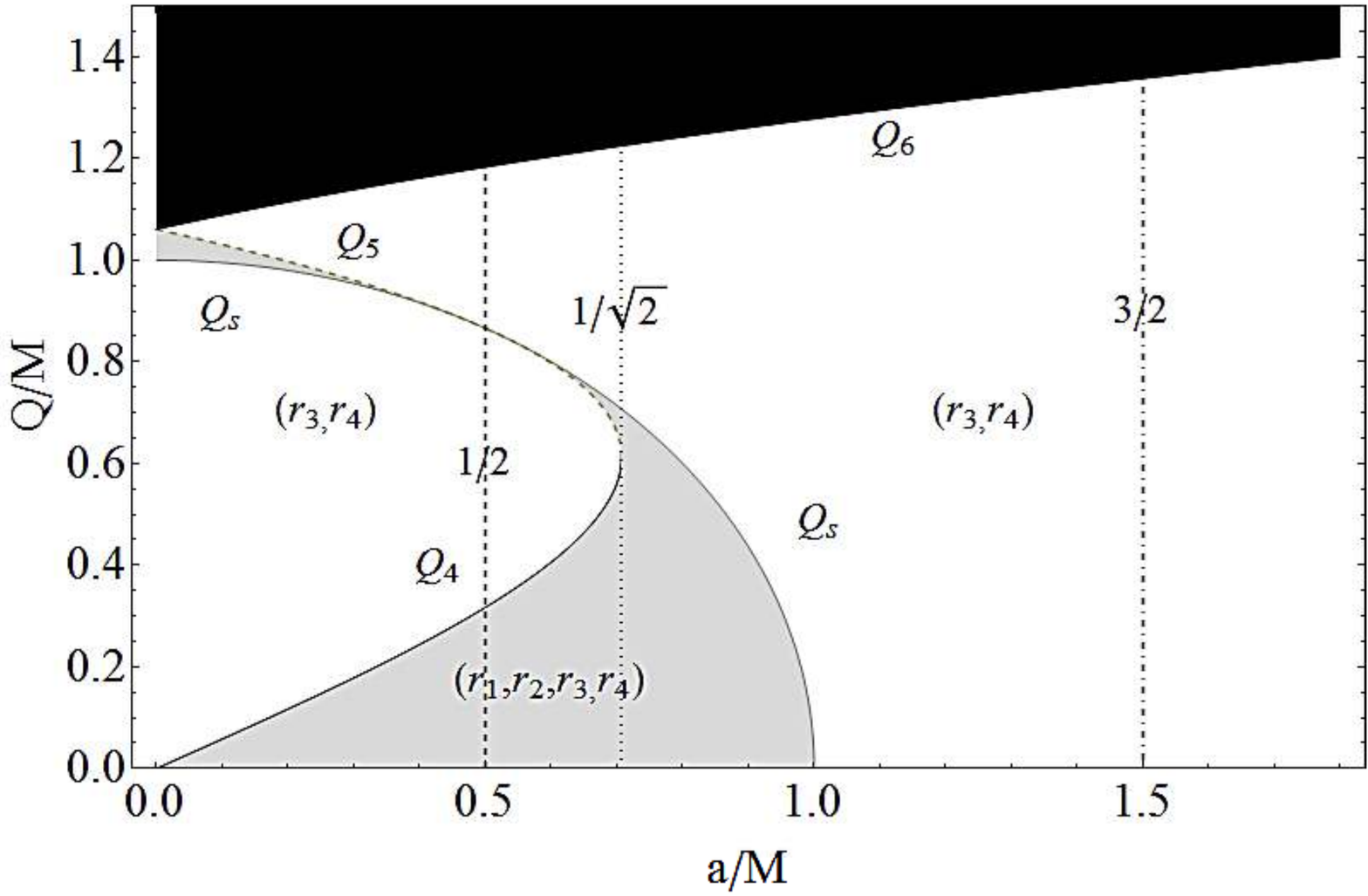}
\end{tabular}
\caption[font={footnotesize,it}]{\footnotesize{ Left: Plot of $a_T$ and $a_s$ as a function of  $Q/M$. Right: Plot of ${Q_4}$  (black curve), $Q_s$ (gray curve), ${Q_5}$ (dashed curve), ${Q_6}$ (black thick curve) as  functions of $a/M$.
Black regions are forbidden: no solution of Eq.\il(\ref{anaozoress}) exits. In the white regions, Eq.\il(\ref{anaozoress}) is satisfied for  $r=r_3$ and $r=r_4$, and in the light gray regions, it is satisfied for $r=r_1$, $r=r_2$, $r=r_3$ and $r=r_4$. In particular, for $Q=0$ and $a=0$ the solution is $r=3M$, for  $0<a<M$ it is $r=r_2$, $r=r_3$ and $r=r_4$. For $a=M$ it is $r=M$ and $r=4M$, finally for $a>M$ it is $r=r_3=r_4$.
}}
\label{Fig:ma_n:a}
\end{figure}

We see that in general there are four possible different radii and all of them depend
on the value of the parameters $M$, $Q$ and $a$ that characterize the rotating source. For a given value of the source parameters not all the solutions lead to real and positive radii so that many different cases can arise.
For a better presentation of the results it is convenient to introduce the notation
\bea\label{r3r4label}
r_{1}&\equiv&r^{-}_{-},\quad r_{2}\equiv r^{-}_{+},\quad
r_{3}\equiv r^{+}_{-},\quad
r_{4}\equiv r^{+}_{+}.
\eea
The behavior of these radii is illustrated in Fig.\il(\ref{R1R2R3R4T}). Notice that in general $r_1<r_2<r_3<r_4$.
The existence of the surfaces is studied in detail in Fig.\il\ref{Fig:ma_n:a}. %
The shaded regions contain all the points at which  Eq.\il(\ref{anaozoress}) does not possess real positive solutions. In the light gray regions of Fig.\il\ref{Fig:ma_n:a}, four solutions exist, and in the white regions only  certain radii, as indicated in the plots.
Thus, to analyze these regions in detail, it is convenient to introduce
the following charges
\bea
Q_4/M&\equiv&\sqrt{\frac{3}{2} \text{Sin}\left[\frac{1}{6} \arccos\left[1-4\frac{ a^2}{M^2}\right]\right]^2},
\quad
Q_5/M\equiv\sqrt{\frac{3}{4} \left(1+\text{Sin}\left[\frac{1}{3} \arcsin\left[1-4 \frac{a^2}{M^2}\right]\right]\right)},\\
Q_6/M&\equiv&\sqrt{\frac{3}{2} \cos\left[\frac{1}{6} \arccos\left[-1+4 \frac{a^2}{M^2}\right]\right]^2}.
\eea
 and the following spin parameter
\bea
a_{\ti{T}}\equiv\frac{\sqrt{Q^2 \left(8 Q^2-9M^2\right)^2}}{3 \sqrt{3}M}
\eea
which guarantees that $\varsigma$ in Eq.\il(\ref{Eq:defrprm}) is real.

\begin{figure}[h!]
\begin{tabular}{cc}
\includegraphics[width=0.4\hsize,clip]{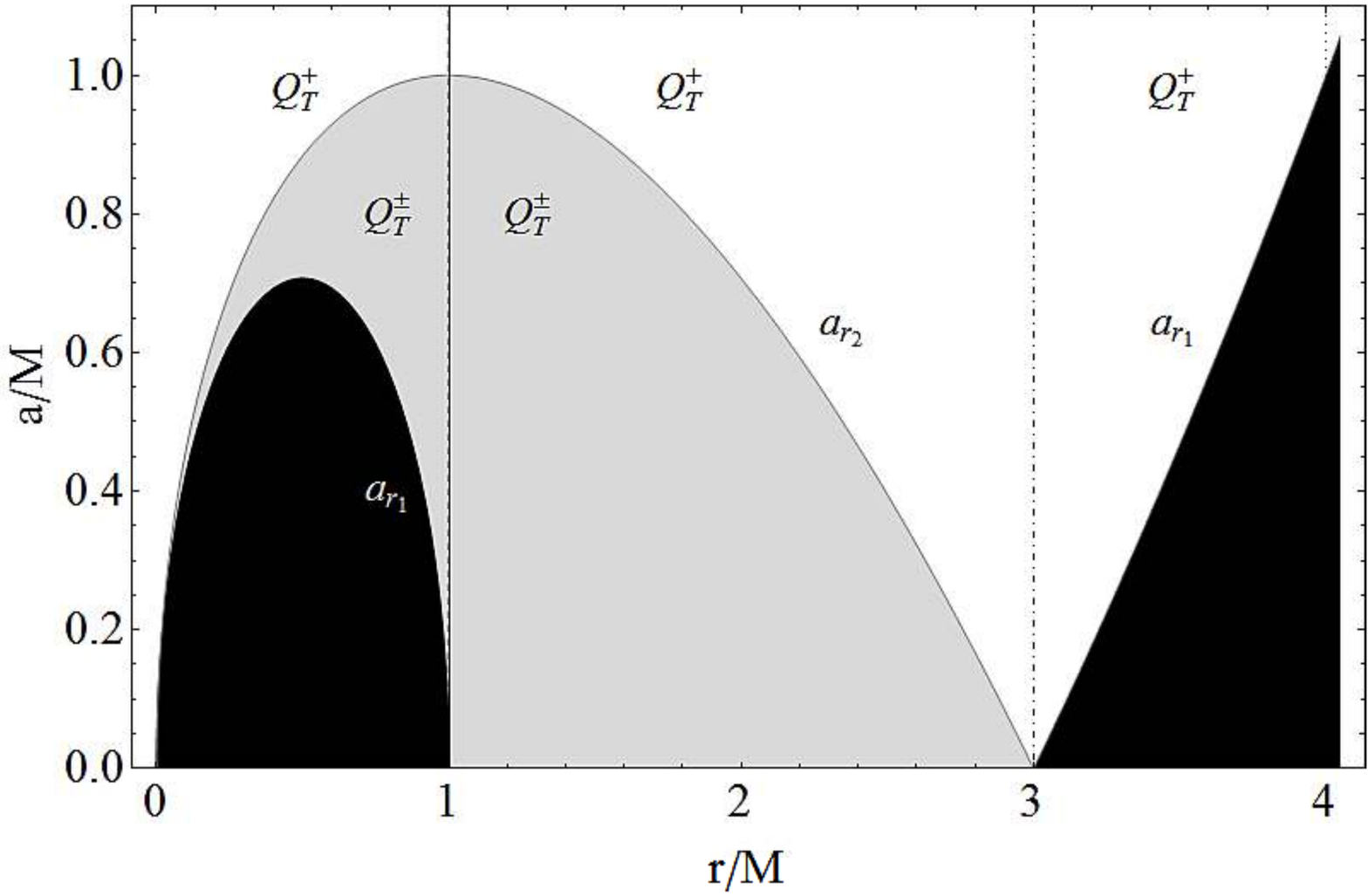}&
\includegraphics[width=0.4\hsize,clip]{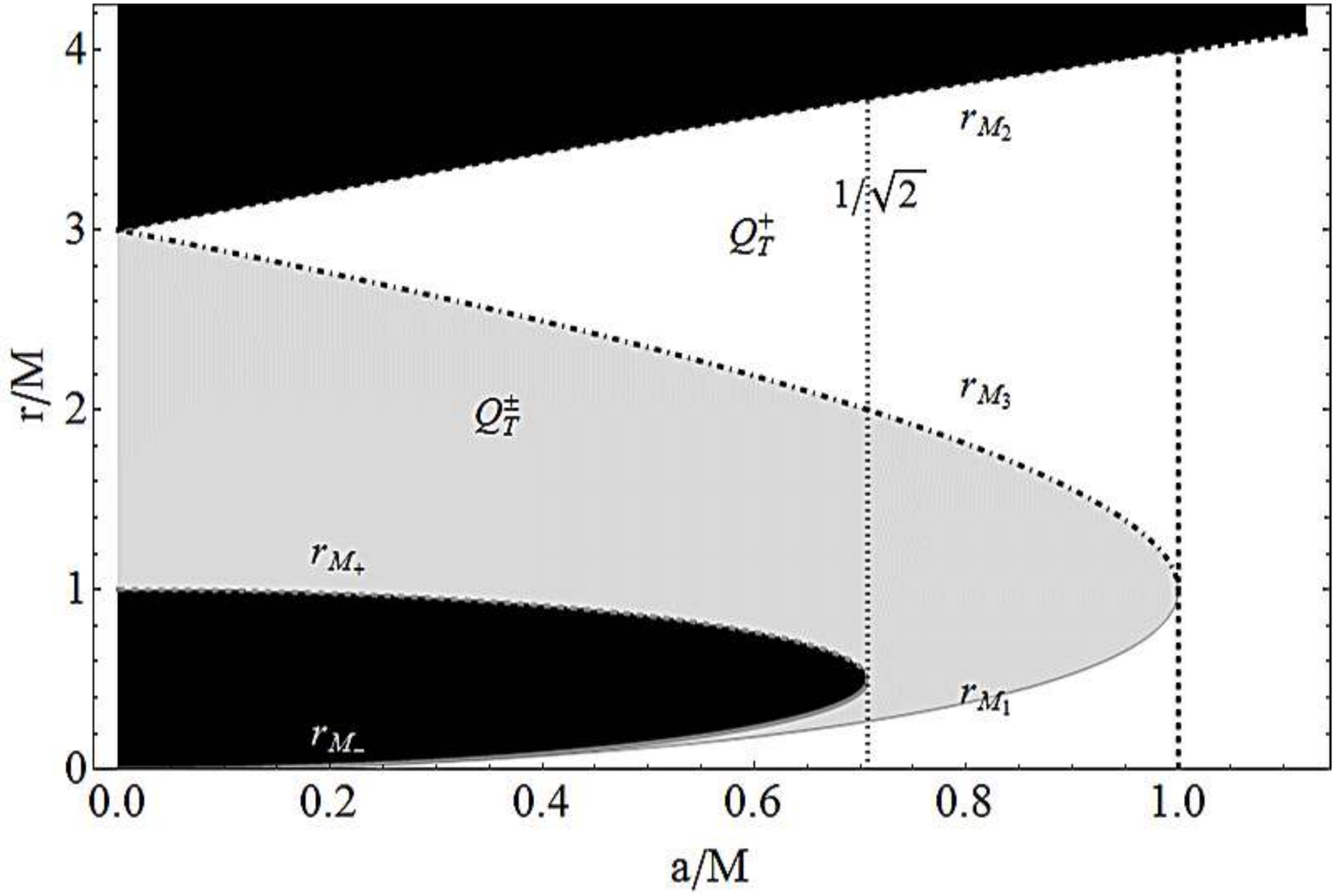}
\end{tabular}
\caption[font={footnotesize,it}]{\footnotesize{Black regions are forbidden: no solution of Eq.\il(\ref{anaozoress}) exits. Left: Plot of $a_{r_1}$ and $a_{r_2}$ as  functions of $r/M$. The solutions $Q_{\ti{T}}^{\pm}$ are shown in the light-gray region  and $Q_{\ti{T}}^{+}$ in the white regions.  Right: Plot of $r_{{\ti{M}_1}}$ (gray curve), $r_{{\ti{M}_2}}$  (dashed black curve), $r_{{\ti{M}_3}}$  (dotted-dashed black curve), $r_{\ti{M}_-}$ (gray curve), $r_{\ti{M}_+}$  (gray dashed curve) as functions of $a/M$. Black regions are forbidden: no solution of Eq.\il(\ref{anaozoress}) exists. In the white regions, the solution is $Q=Q_{\ti{T}}^+$; in the light gray regions the solutions are $Q=Q_{\ti{T}}^{\pm}$}}
\label{Fig:d_me_ra}
\end{figure}

Alternatively, it is also possible to investigate the solutions of the equation $ \Psi = 0 $ in terms of the source charge  as a function of $ (a, r) $. In this way we will have a different view of the regions of existence of circular orbits. Thus, the solution of $ \Psi = 0 $ is
\bea
Q_{\ti{T}}^{\pm}\equiv\sqrt{\frac{1}{2} \left[(3M-r)M-a^2 \frac{r}{M}\pm\sqrt{a^2 \left(a^2+2 (r-M) r\right)}\right]},
\eea
The corresponding surfaces are studied in detail the left plot of Figs.\il\ref{Fig:d_me_ra}, which also includes the regions where these charges exist.
It is convenient to introduce the following spin parameter
\bea
a_{r_1}\equiv\sqrt{2 r(M-r)},\quad a_{r_2}\equiv \frac{1}{2} \sqrt{(r-3M)^2 r/M},
\eea
and the radii
\bea
r_{\ti{M}_{\pm}}&\equiv&\frac{1}{2}\left(M\pm\frac{1}{2} \sqrt{M^2-2 a^2}\right) ,
\eea
which define the region $r\in]r_{M_-},r_{M_+}[$ where no solution of Eq.\il(\ref{anaozoress}) exists (see the right panel of
Fig.\il\ref{Fig:d_me_ra}-right).
The radii
\bea
r_{{\ti{M}_1}}/M&\equiv& 4 \sin\left[\frac{1}{6}  \arccos\left(1-2 a^2/M^2\right)\right]^2,
\quad
r_{{\ti{M}_2}}/M\equiv4 \cos\left[\frac{1}{6} \arccos\left(2 a^2/M^2-1\right)\right]^2,
\\
r_{{\ti{M}_3}}/M&\equiv& 2 \left(1+\sin\left[\frac{1}{3}  \arcsin\left(1-2 a^2/M^2\right)\right]\right),
\eea
were introduced in \cite{Pu:Kerr}, with a different notation, in order to study the existence of circular orbits in the Kerr spacetime.

\begin{figure}[h!]
\begin{tabular}{ccc}
\includegraphics[width=0.35\hsize,clip]{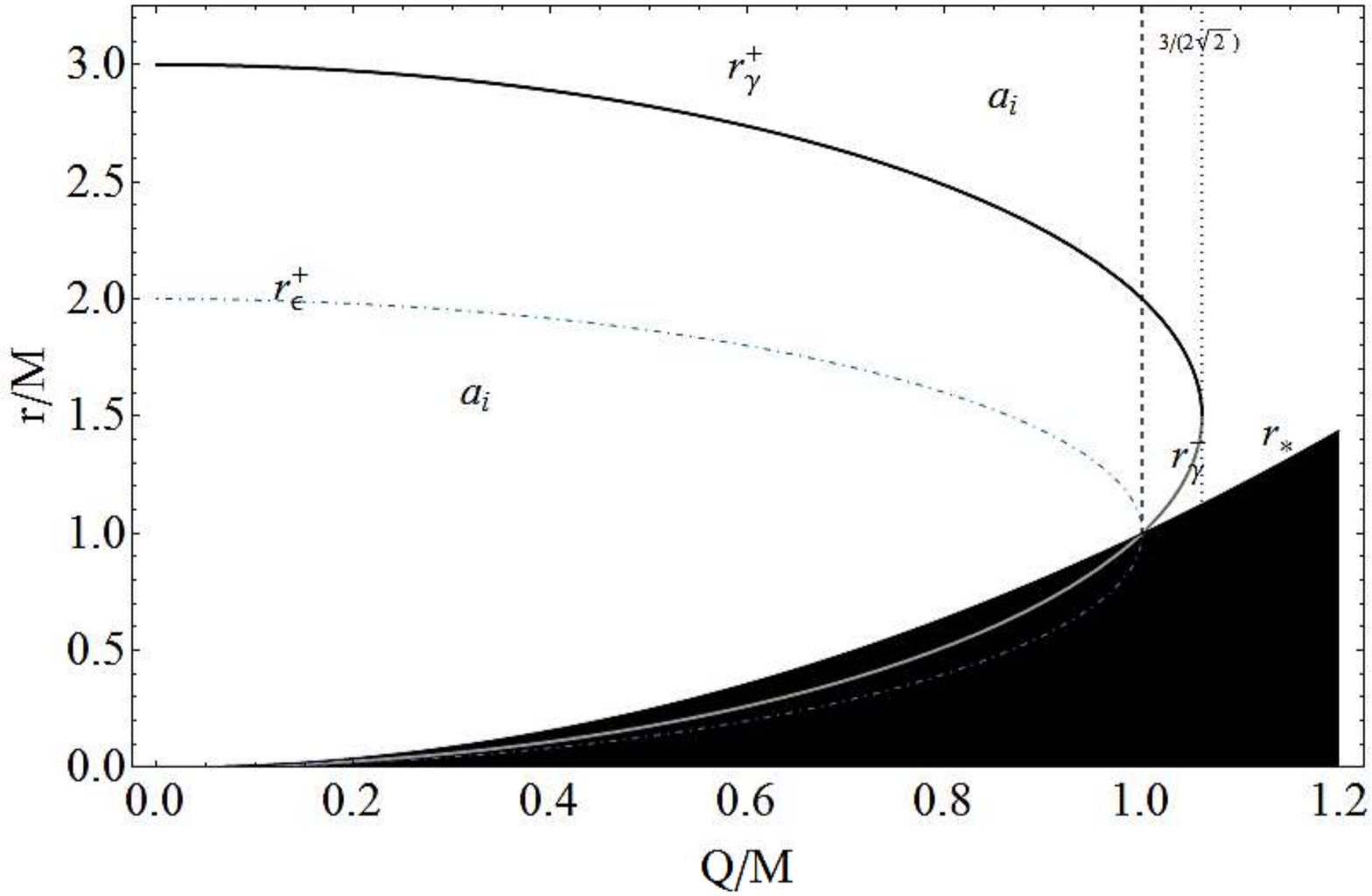}&
\includegraphics[width=0.35\hsize,clip]{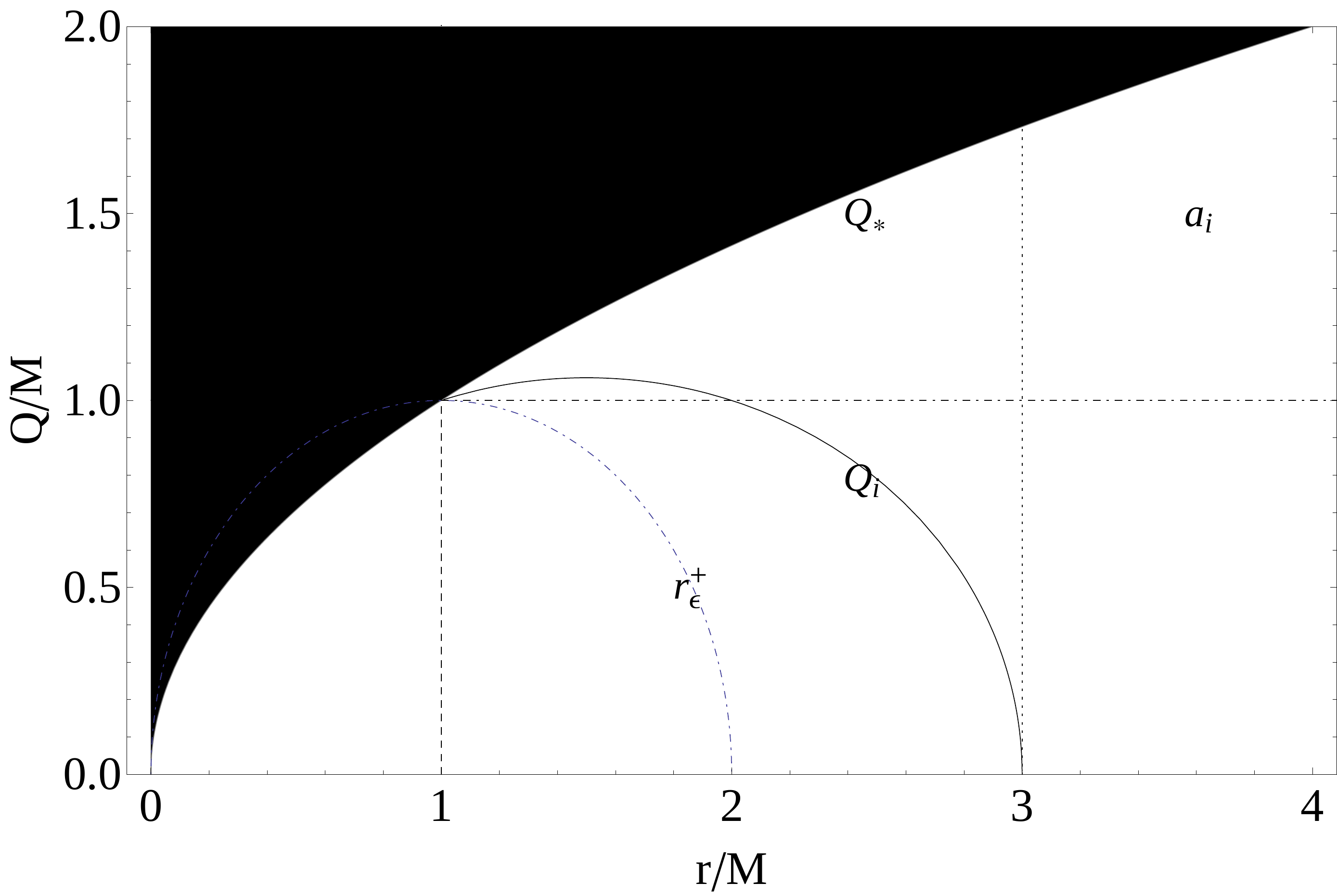}&
\includegraphics[width=0.3\hsize,clip]{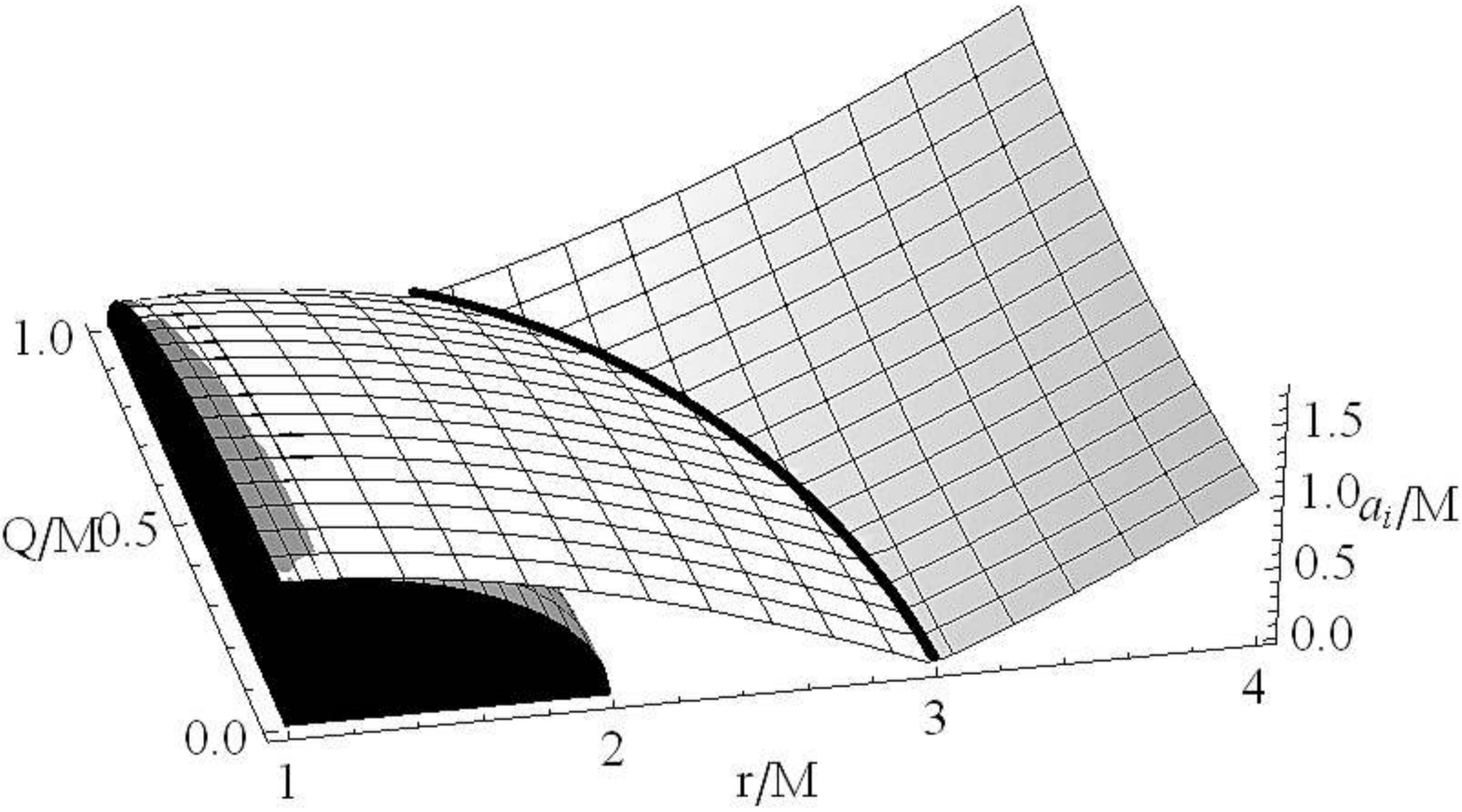}
\end{tabular}
\caption[font={footnotesize,it}]{\footnotesize{
Left: Plot of $r_{\gamma}^+$ (thick black curve), $r_{\gamma}^-$ (thick gray curve), $r_*$ (black curve),$r_{\epsilon}$ (dotted-dashed curve) as functions of $Q/M$. The solution of Eq.\il(\ref{anaozoress}) is $a_i$ in the white regions. Center: Plot of $Q_i$ and $Q_*/M=\sqrt{r/M}$ as functions of the radius $r/M$.  Dotted-dashed curve signs $r_{\epsilon}$.
In the white regions, Eq.\il(\ref{anaozoress}) is satisfied for $a=a_i$. The plot is therefore divided  in the regions: $0<r<M$, $M<r\leq3M$, $r>3M$.  In particular, at $r=M$ and $Q=1$, a solution exists for $a\geq0$. The dotted-dashed line is $Q=M$, the dashed line is $r=M$ and the dotted line is $r=3M$. Right:  The black hole angular momentum $a_i$ is  plotted as a function of $r/M$, and the charge-to-mass ratio $Q/M$ of the black hole. The outer horizon plane $r_+$ (gray) is also plotted. The black region is $r<r_+$. The curve  $r=r_{\gamma}^+$ (black thick) is
also plotted.}}
\label{Fig:no_men}
\end{figure}

In Figs.\il\ref{Fig:no_men},
we show the solution of the equation $\Psi=0$ in terms of the spin parameter $a_i$
\bea\label{Eq:Def:a_i}
a_i\equiv\frac{1}{2} \sqrt{\frac{\left[2 Q^2+(r-3M) r\right]^2}{Mr-Q^2}},
\eea
where we used the  definitions
\bea
r_{\gamma}^{\pm}&\equiv&\frac{1}{2}\left(3M\pm \sqrt{9M^2-8 Q^2}\right),\quad r_*\equiv Q^2/M,
\eea
and $ Q_i/M\equiv\sqrt{\frac{(3M-r) r}{2M}}$ and
with  $Q_*/M \equiv\sqrt{r/M}$ (that corresponds to the definition $r_*=Q^2/M$). The
radii ($r_{\gamma}^{\pm}, r_*$), introduced in \cite{Pu:Neutral}, characterize the dynamics around a Reissner-Nordstr\"om  naked singularity $(Q/M > 1)$ where   the following inequality holds
$r_*< r_{\gamma}^{-}<r_{\gamma}^{+}$, and $r_{\gamma}^{\pm}$ are the radii at which the value of the angular momentum
and the energy of the test particle diverge. At the classical radius $r_*$ circular orbits exist with ``zero''
angular momentum. This means that a static observer situated at infinity would interpret
this situation as a test particle that remains motionless  as
time passes. This phenomena can take place only in the case of a naked singularity and is
interpreted as a consequence of the ``repulsive'' force generated by the charge distribution. It is interesting to note that these radii have an important role even in the presence of a  non zero source spin. In particular, it turns out that from Eq.\il(\ref{lpspr})  the orbital angular momentum $L$ is defined when $\Pi\geq0$, and from  Eq.\il(\ref{E:pidef}) we recover $r\geq r_*$, at $r=r_*$ it is
$L_{\pm}=\mu M\sqrt{\frac{M^2a^2}{Q^2 \left(Q^2-M^2\right)}}$, which is in fact zero in the Reissner-Nordstr\"om case.
In Fig.\il\ref{Fig:no_men}, we plot
the behavior of these radii as functions of the ratio $Q/M$ and $a/M$.

\begin{figure}[h!]
\begin{tabular}{ccc}
\includegraphics[width=0.3\hsize,clip]{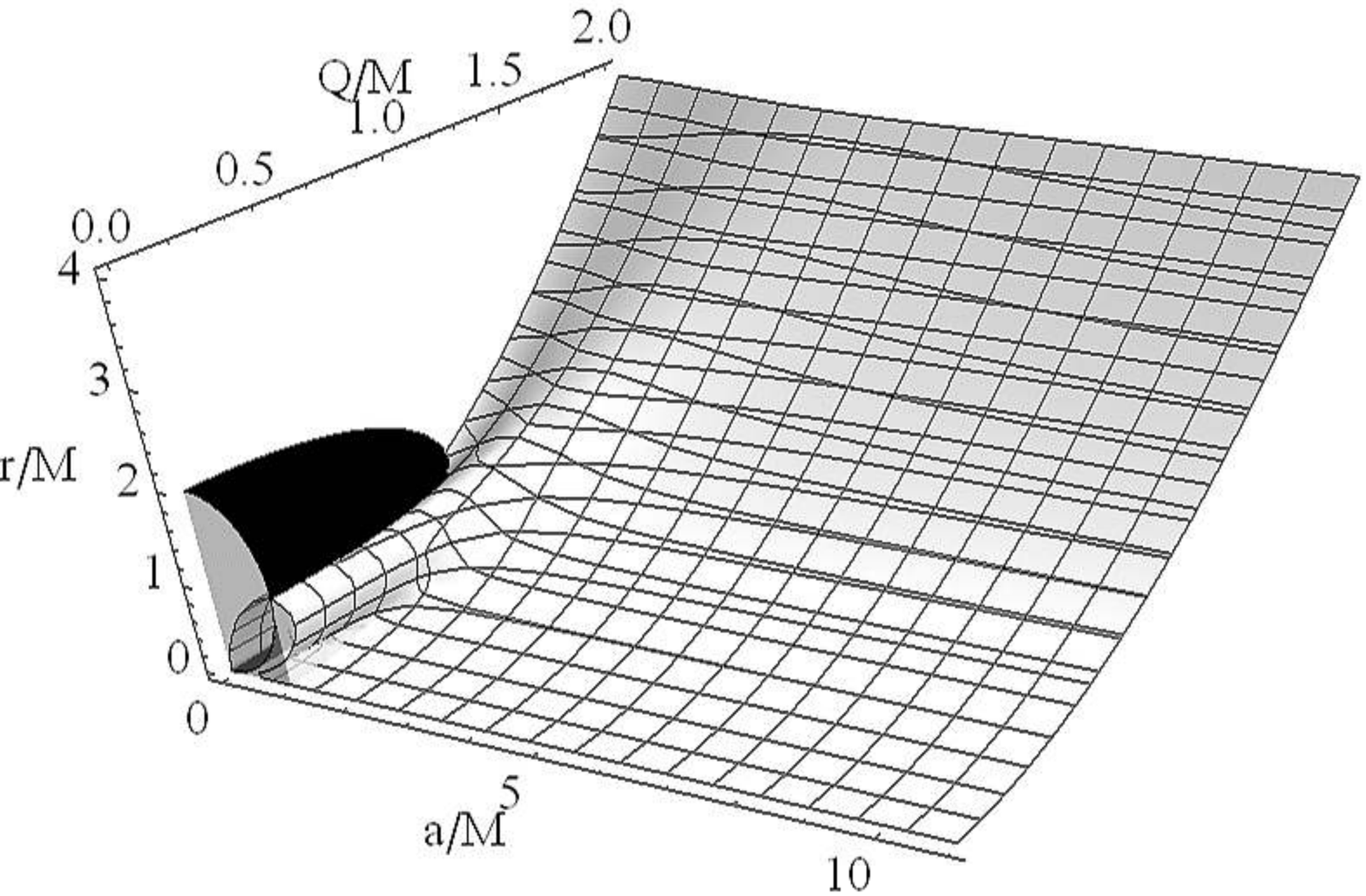}&
\includegraphics[width=0.3\hsize,clip]{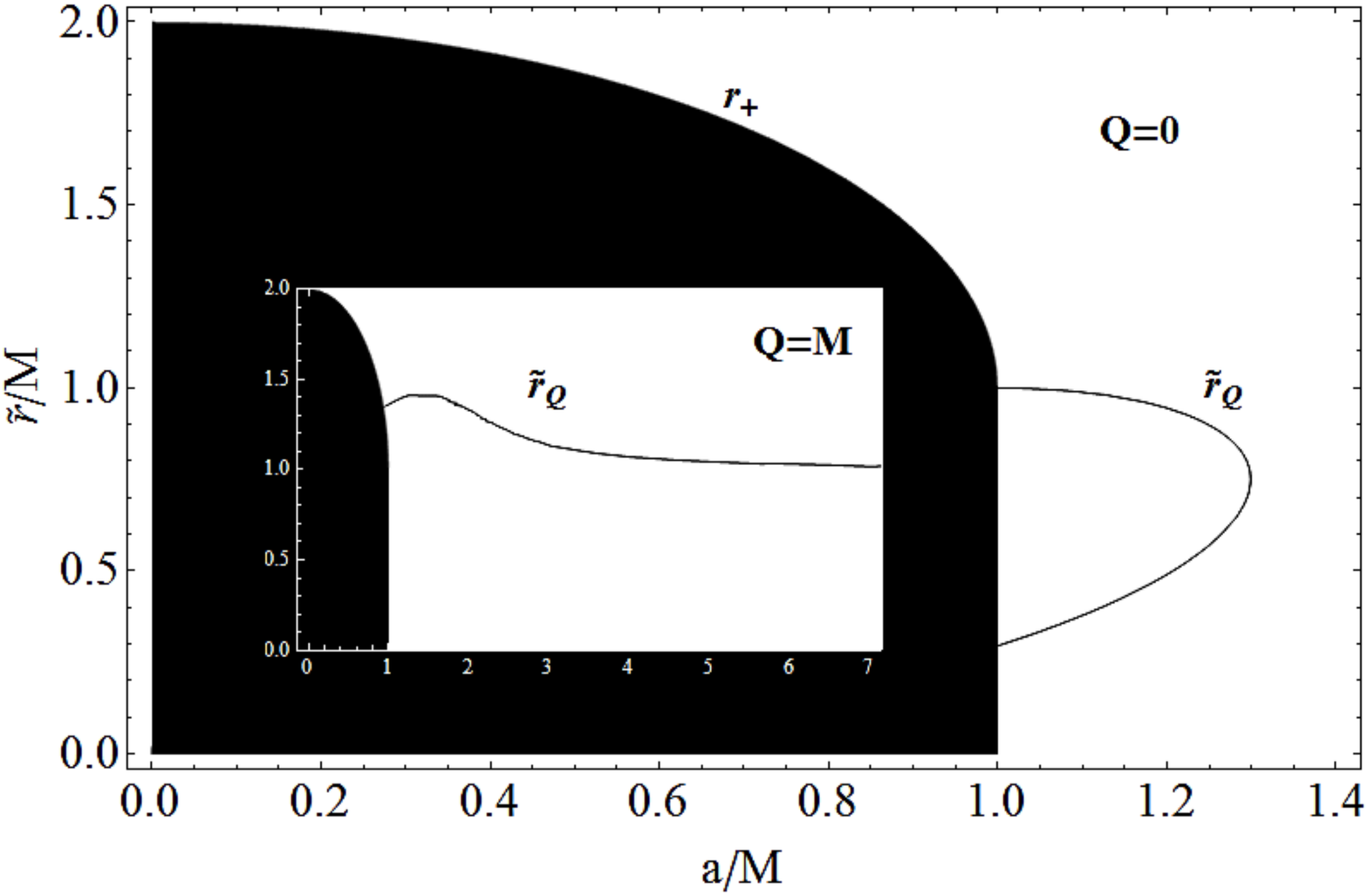}&
\includegraphics[width=0.3\hsize,clip]{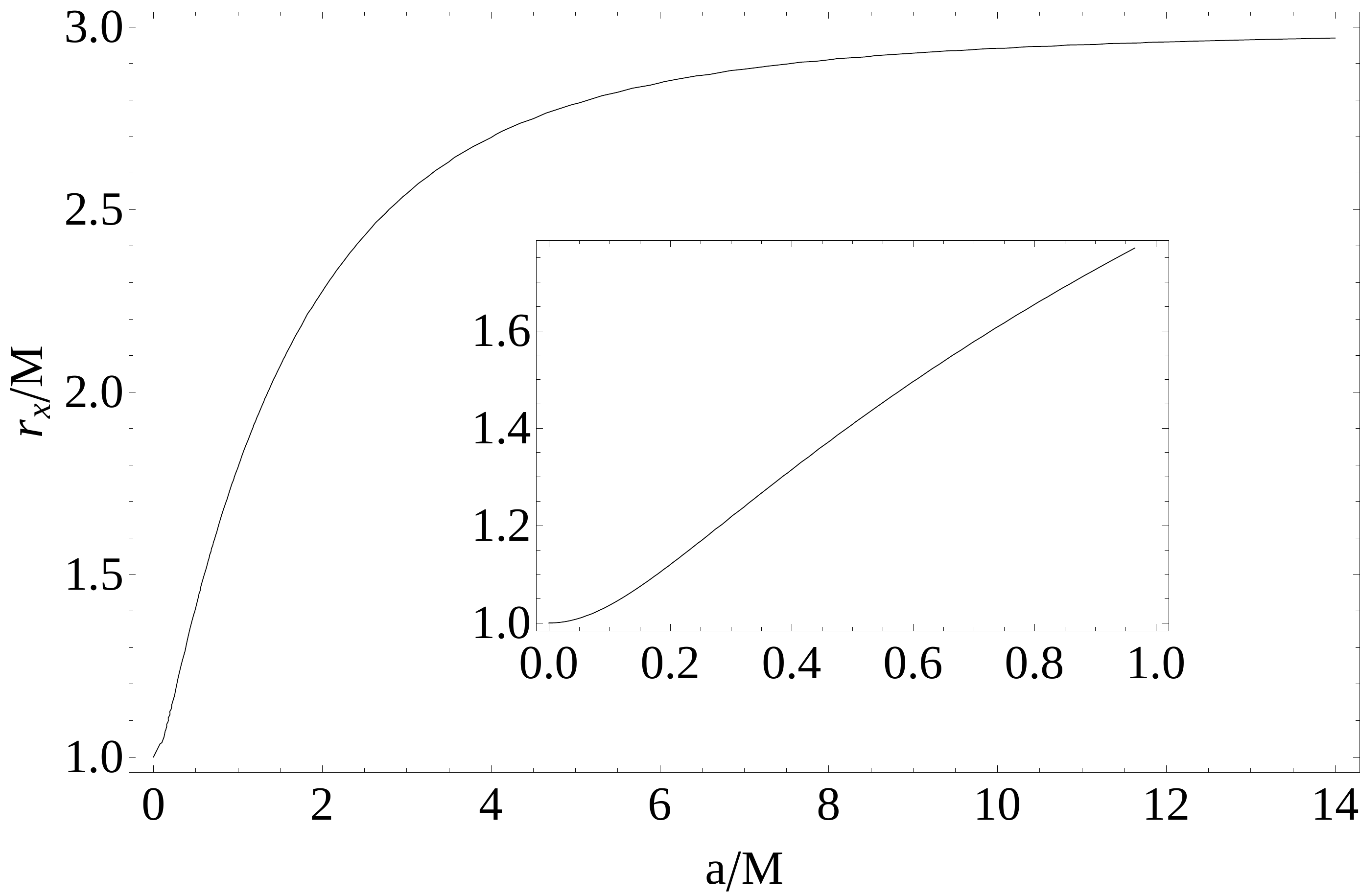}
\end{tabular}
\caption[font={footnotesize,it}]{\footnotesize{
Left: Plots of $\tilde{r}$ as functions of $Q/M$ and $a/M$.
The  surfaces $r=r_+$ and $r=r_-$, delimited by the outer and inner horizon respectively are also plotted. The region $r<r_+$ is shaded. Center: Plot of $\tilde{r}_{Q}$, {that is $\tilde{r}$} in the case $Q=0$ and $Q=M$ (inset plot). Right: Plot of $r_x$ in the range $[0,3M)$ as a function of $a/M$. The inset plot shows a zoom for $a\leq M$. }} \label{Fig:c_olin}
\end{figure}

Now,  for the  existence of solutions of the Eq.\il(\ref{E:bicchi}) it
 is however necessary to demand that  the solution (\ref{lpspr}) be real. Therefore, we study the solutions of the equation $\Sigma^2-4M^4 \Pi=0$, that is   $L=0$, or
\be\label{Eq:quintarx}
-Q^4a^2-Q^2 a^4+\left(4Q^2 a^2+a^4\right)Mr-2\left(2M^2+Q^2\right)a^2 r^2+2M a^2 r^3-Q^2r^4+ M r^5=0,
\ee
which
will be denoted by $\tilde{r}$ (see Fig.\il\ref{Fig:c_olin}).  But in general we can also find the regions of circular motion by fixing the value of the source angular momentum $a$ and plotting the ratio $Q/M$ in terms of the radial coordinate. Thus, we can also express the solution of Eq.\il(\ref{Eq:quintarx}) in terms of the charge $Q_c$
\bea
Q_{c}&\equiv&\sqrt{\frac{-a^4+a^2 \left(-2 (r-2M) r+\sqrt{a^4+2 a^2 (r-2M) r+r^4}\right)+r^2 \left(-r^2+\sqrt{a^4+2 a^2 (r-2M) r+r^4}\right)}{2a^2}}\ .
\eea
In this case, it is convenient to introduce the charge
\bea
 Q_r&\equiv&\sqrt{-a^2-(r-2M) r}\ ,
\eea
which is the solution of the equation  $\Delta=0$,
and the radius $r_x$ which represents the real solutions of Eq.\il(\ref{Eq:quintarx}) when the source charge is $Q_{\ti{T}}^{\pm}$, or
\be\label{Eq:quinta}
-12M a^4+a^2\left(4 a^2-M^2\right) r-12M a^2 r^2+5 a^2 r^3-2M r^4+2 r^5=0\ .
\ee
This radius is  plotted in Fig.\il\ref{Fig:c_olin} for different values of the ratio $a/M$. Notice in particular that $r_x$ does not depend
on the source charge but only on its angular momentum.

%% file: BH_KN13.tex
\section{Black holes}\label{Sec:BH_KN}
In this section, we study the Kerr--Newman  black hole case, considering the constraint $\delta\geq0$, where $\delta\equiv M^2-Q^2-a^2$. In particular it is $\delta=0$ for  $Q^2+a^2=M^2$, and  $Q\in[0,M]$ and $a\in[0,M]$. For $Q=a=0$, Schwarzschild black hole, it is $\delta=M$.
\begin{figure}[h!]
\begin{tabular}{ccc}
\includegraphics[width=0.3\hsize,clip]{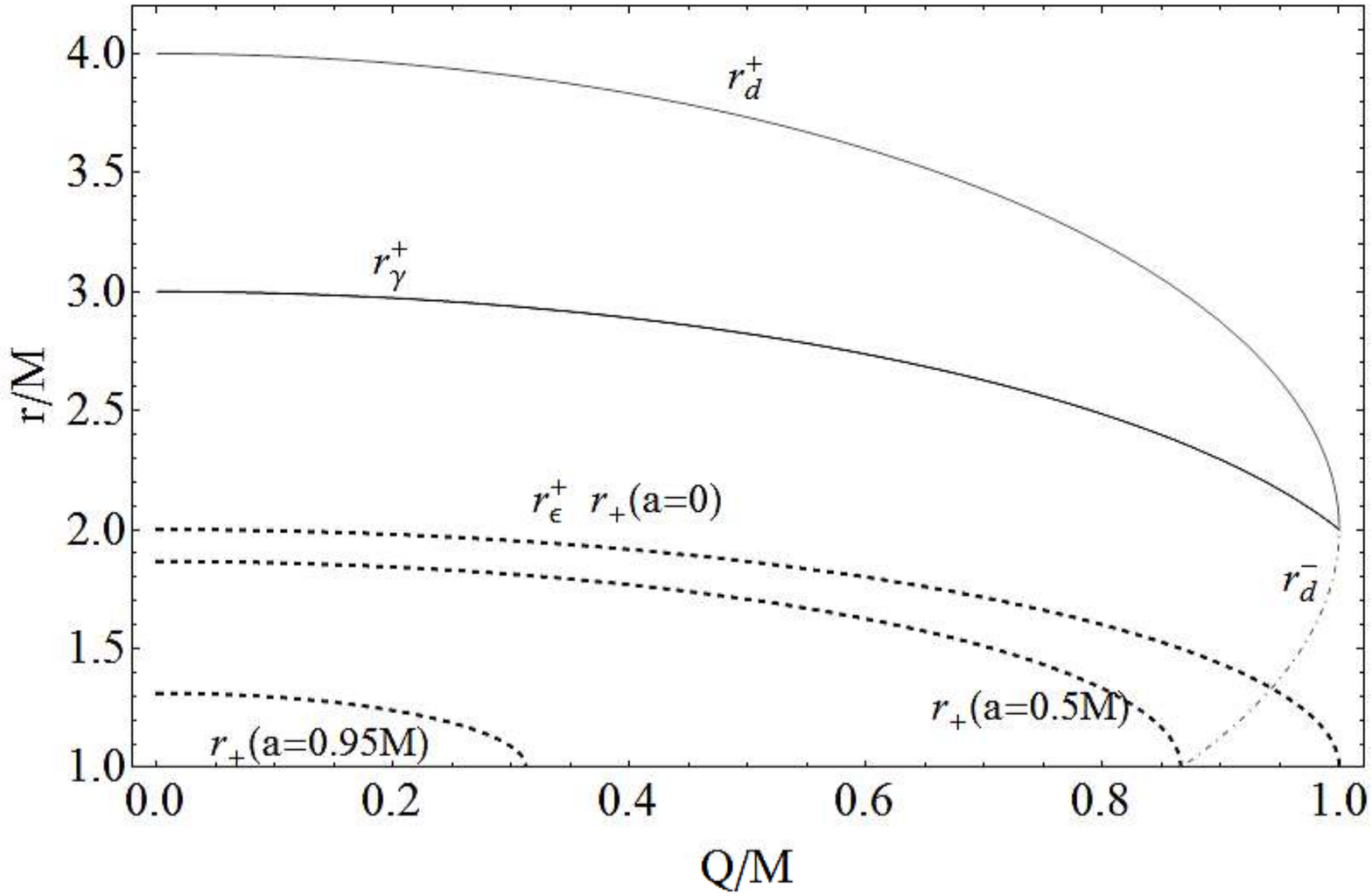}&
\includegraphics[width=0.3\hsize,clip]{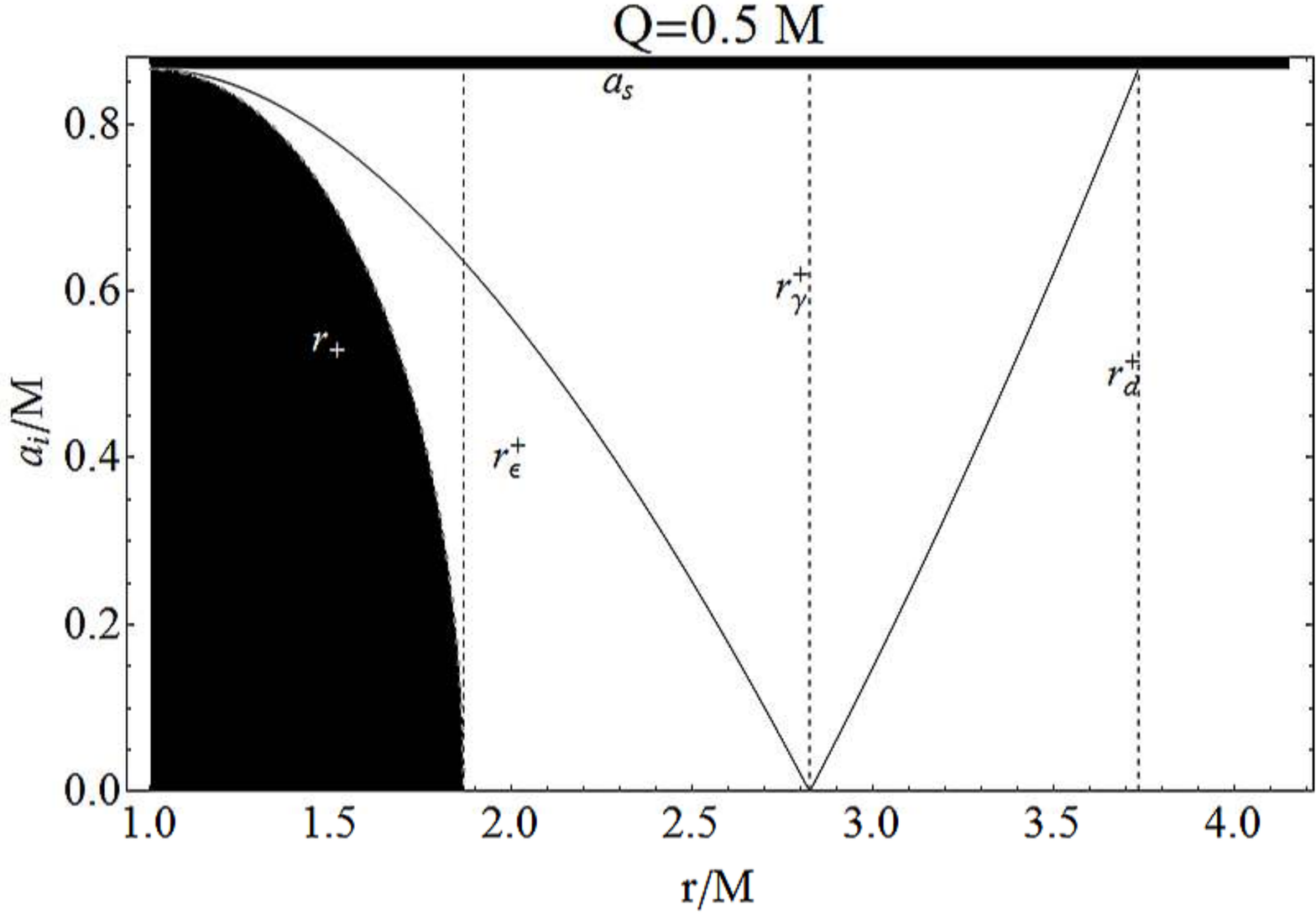}&
\includegraphics[width=0.3\hsize,clip]{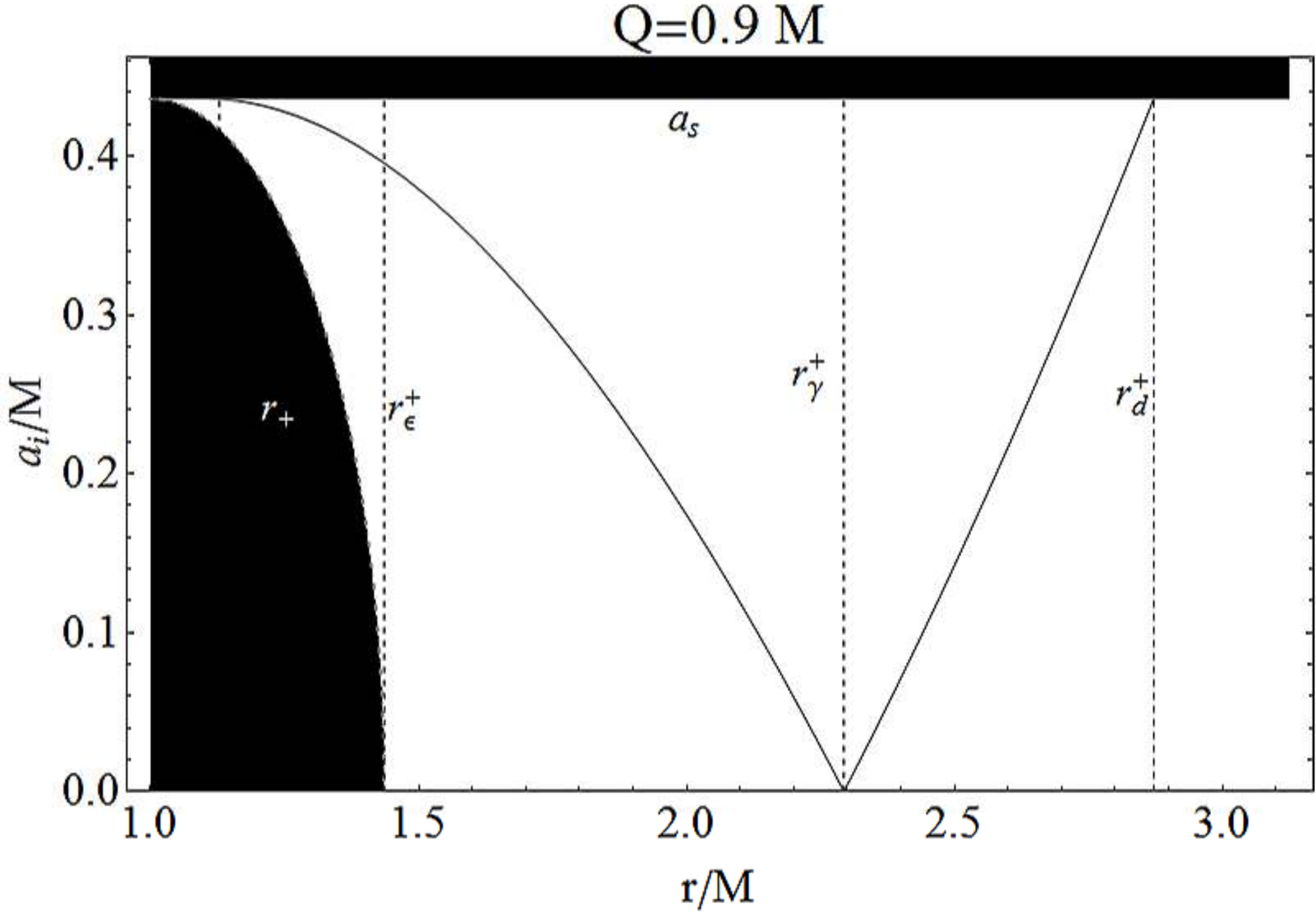}
\end{tabular}
\caption[font={footnotesize,it}]{\footnotesize{
Left: The outer horizon $r_+=M + \sqrt{M^2 - a^ 2 - Q^2}$, the outer ergosphere radius on the equatorial plane $r_{\epsilon}^+=M + \sqrt{M^2  - Q^2}$ and the radius $r_d^+=2r_{\epsilon}^+$, are plotted as functions of the black hole charge-to-mass ratio $Q/M$, for different values of the intrinsic angular momentum  $a/M$. Center and Right: The outer horizon
$r_+=M + \sqrt{M^2 -a^2 - Q^2}$, the outer ergosphere radius on the equatorial plane $r_{\epsilon}^+=M + \sqrt{M^2 - Q^2}$ and the radius $r_d^{\pm}=2r_{\epsilon}^{\pm}$ are plotted as functions of  $r/M$ and the black hole charge-to-mass ratio $Q/M$. The angular momentum $a_s=\sqrt{M^2-Q^2}$ is plotted (gray curve); for the black hole case it is $a<a_s$.
Center: $Q=0.5M$ , $a_s=0.866M$, $r_{\epsilon}^+=1.866M$, $r_d^{+}=3.732M$, $r_d^{-}=0.268M$.
Right: $Q=0.9M$,  $a_s=0.436M$, $r_{\epsilon}^+=1.436M$, $r_d^{+}=2.872M$, $r_d^{-}=1.1282M$.
Shaded regions are forbidden: the horizon constraint of the KN black hole implies $0\leq a\leq a_s\leq 1$ and $r>r_+$.
 }}\label{Fig:BH-Hori}
\end{figure}
In Fig.\il\ref{Fig:BH-Hori},  the outer $r_+$ and inner $r_-$ horizon surfaces are plotted as function of $a/M$ and $Q/M$.
We are interested in the solutions of   Eq.\il(\ref{E:bicchi})  in the region $r>r_+$.
Solving  Eq.\il(\ref{E:bicchi}) for the angular momentum $L$,  we obtain the solutions given in  Eq.\il(\ref{lpspr}).
There are  zones in which two kinds of orbits, characterized by two different angular momentum, but same orbital radius, are possible.
In particular, it is possible to show that there are no circular orbits in the KN black hole spacetime with zero angular momentum, i. e., there are no solutions of the following equations
\be
V'(r,Q,a,L)=0,\quad L=0 .
\ee
However, one can see that $V'(r,Q,a,L=0)>0$ in the region $r>r_+$; this means that for  particle with an ``angular momentum'' $L=0$   the effective potential monotonically increases  as the  outer regions of spacetime are approached.

But before we deal with the general case  $a\neq0$ and $Q\neq0$, it is useful to  summarize briefly the known outcomes for the two limiting cases of Kerr  $(Q=0)$ and RN $(a=0)$ spacetimes.  We refer for details to the extensive literature on the subject and  to the aforementioned works \cite{Pu:Neutral,Pu:Kerr,Pu:Charged}.
In fact,  we will see that the source spin  and  the charge   deform the regions of circular motion and the stability properties of a RN and Kerr spacetime, respectively.
To discuss the dynamics of a black hole spacetime, it is convenient to introduce here the following radius %
\be
r_{d}^{\pm}=2(M\pm\sqrt{M^2-Q^2}).
\ee
We notice that   $r_{d}^{\pm}=2r^{\ti{RN}}_{\pm}$ and, for  circular motion on the equatorial plane  ($\theta=\pi/2)$, it is  $r_{d}^{\pm}=2r^{\ti{RN}}_{\pm}=2 r_{\epsilon}^{\pm}$,
see Figs.\il\ref{Fig:BH-Hori}.
Furthermore,
it is possible to prove that  the following limits are satisfied:
\bea
r_{\ti{M}_3}&=&r_3,\quad \mbox{for}\quad Q=0\quad  \mbox{(Kerr metric)},
\\
r_{\gamma}^+&=&r_3=r_4,\quad \mbox{for}\quad a=0\quad \mbox{(Reissner--Nordstr\"om metric)}.
\eea
Moreover, following the  notation of \cite{Pu:Kerr},  we define the radius  $r_{a}\equiv r_4\ (Q=0)$, i. e.,
\be
r_a\equiv 4 M\cos\left[\frac{1}{6} \arccos\left[2 \frac{a^2}{M^2}-1\right]\right]^2.
\ee

Tables \ref{Tab:M_corrected} and \ref{Tab:Man:ure} show the case $Q = 0 $  from two different points of view. This differentiation in the  presentation of the orbital regions is important for the study of the stability problem also in the general case of a KN spacetime.
In Table\il\ref{Tab:M_corrected}, we consider different values for the source angular momentum $a$, ranging from the Schwarzschild black hole,
$a=0$, to the extreme Kerr black hole with $a=M$. Each range is then divided in various spatial regions, each of which is characterized by a particular value of the orbital angular momentum $L$ of the test particle.
%
\begin{table}[h!]
\caption{\label{Tab:M_corrected}\footnotesize{Angular momentum of circular orbits $L$ in the region $r>M$ of a Kerr spacetime for specific values of the source angular momentum $a$.
}}
\begin{ruledtabular}
\begin{tabular}{ll|ll|llll}
Case: $Q=0$ (Kerr) \\
$a=0$& (Schwarzschild) &$0<a< M$  &  (Kerr)&$ a= M$  &  (extreme Kerr)
\\
\hline
Region&$L$& Region&$L$&Region&$L$
\\
$r>3M $ &$\pm L_-$ & $]r_{\ti{M}_3},r_a]$  &$L_{-}$& $]M,4M]$  &$L_-$
\\
  & & $]r_{a},\infty[$ &$(L_-,-L_+)$ & $]4M,\infty[$ &$(L_-,-L_+)$\\
\end{tabular}
\end{ruledtabular}
\end{table}

In Table\il\ref{Tab:Man:ure}, we present an alternative classification of the circular orbits. The radial coordinate is split into various
regions that range from $r=M$ to $r\rightarrow\infty$. It turns out that the radii $r=3M$ and $r=4M$ are of particular interest because they determine spatial boundaries in which the value of the  orbital angular momentum $L$ drastically depends on the source angular momentum $a$.
\begin{table}[h!]
\caption{\label{Tab:Man:ure}\footnotesize{Angular momentum $L$ of  circular orbits in the region $r>M$ of a Kerr spacetime for specific
values of the radial coordinate $r$.}}
\begin{ruledtabular}
\begin{tabular}{ll|ll|ll|ll}
Case: $Q=0$ (Kerr) \\
\hline
$M<r\leq3M$& &$3M<r<4M$ &  &$r=4M$&  &$r>4M$    &
\\
\hline
$a$&$L$&$a$&$L$&$a$&$L$&$a$&$L$
\\
$]a_i,M] $ &$L_-$ & $[0,a_i[$  &$(L_-,-L_+)$&$[0,M[$ &$(L_-,-L_+)$ & $[0,M]$  &$(L_-,-L_+)$
\\
$$  &$$& $[a_i,M]$ &$L_-$ & $M$ &$13/(4\sqrt{2})M\mu$ & $$  &$$
\\
\end{tabular}
\end{ruledtabular}
\end{table}
In each spatial region, we present the type of allowed circular orbits (rotating and counter-rotating) with the corresponding value of the
angular momentum of the test particle $(L_\pm)$. More details about this classification can be found in \cite{Pu:Neutral,Pu:Kerr}.
Finally, in the limiting case of a RN black hole ($a=0$), one can see that circular orbits with $L=\pm L_-$ are allowed only in the region outside the radius  $r_{\gamma}^+$.

The main question now is to determine how the electric charge of the gravitational source will affect this classification. Since the charge generates its own gravitational field, it is expected that its presence will modify the spatial regions as well as the ranges of the
source angular momentum in which circular motion can take place.

%
\subsection{Orbital regions around a  KN black hole}\label{Sec:orbitre}
We give here a description of  circular orbits  with $L\neq0$ around a  KN black hole  from the outer horizon $r_+$ to infinity.
Since  it is always $r_+>M$, in order to  simplify the exploration of the dynamics in the region $r>r_+$, we approach the more general problem of finding the extremes of the effective potential in the larger region $r>M$ with the constraints $0\leq a\leq a_s$ and $0\leq Q\leq M$.
The results of this analysis are summarized in Tables\il\ref{Tab:M_g_a} and \ref{Tab:Lamp_corr}.
Notice that the spin parameter $a=a_s$ plays an important role for the definition of the different ranges of Table.\il\ref{Tab:M_g_a},
because for  $Q>M$ a black hole case occurs if the source spin satisfies the condition $a\leq a_s$.
We point out that there are
 two different orbital regions, namely, the region  \textbf{I},  with $L=L_-$, and the  region \textbf{II}, with $L=(L_-, -L_+)$.
Table\il\ref{Tab:M_g_a} shows  the allowed angular momenta of the test particle $(L)$, depending on the the values of
source spin $(a)$, for fixed ranges of the charge $Q/M$ and different orbital regions defined by $r/M$.
We also include in this table the particular case of an extreme RN  black hole,  $Q=M$ and $a=0$.
This is a generalization of the Table\il\ref{Tab:Man:ure} for the case $Q\neq0$.
We can see that the regions  in which circular motion is allowed  do not differ qualitatively from the
configuration shown in Tab.\il\ref{Tab:Man:ure} for the case $Q = 0$ in which the limits $r_{\gamma}^+= 3M$ and $r^+_d = 4M$ hold.
We conclude that the introduction of a
source charge  does not change qualitatively   the
 structure of the orbital regions, but only their extensions, i.e., the boundaries become modified as shown
in Table\il\ref{Tab:M_g_a}.

The extensions of  the orbital regions \textbf{I} and \textbf{II} will depend depend on $a/M $ and $ Q/M $ but, at least for the characteristic values of black hole spacetimes,  we do not see the appearance of an additional type of orbit. For example, there are no
counter-rotating orbits with  $L =-L_-$ in $r>r_+$ as in the limiting case $a = 0$.

\begin{figure}[h!]
\begin{tabular}{ccc}
\includegraphics[width=0.3\hsize,clip]{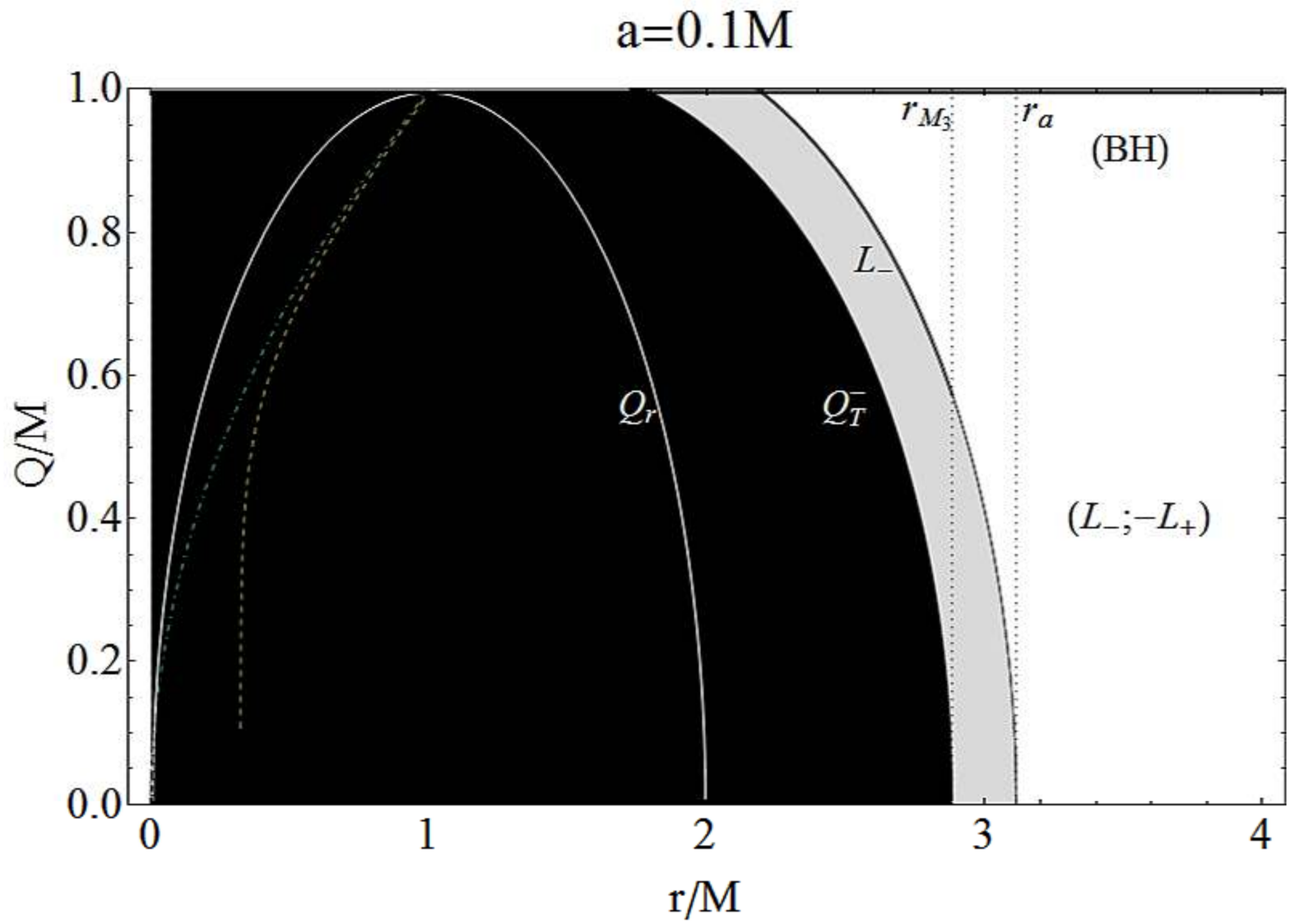}&
\includegraphics[width=0.3\hsize,clip]{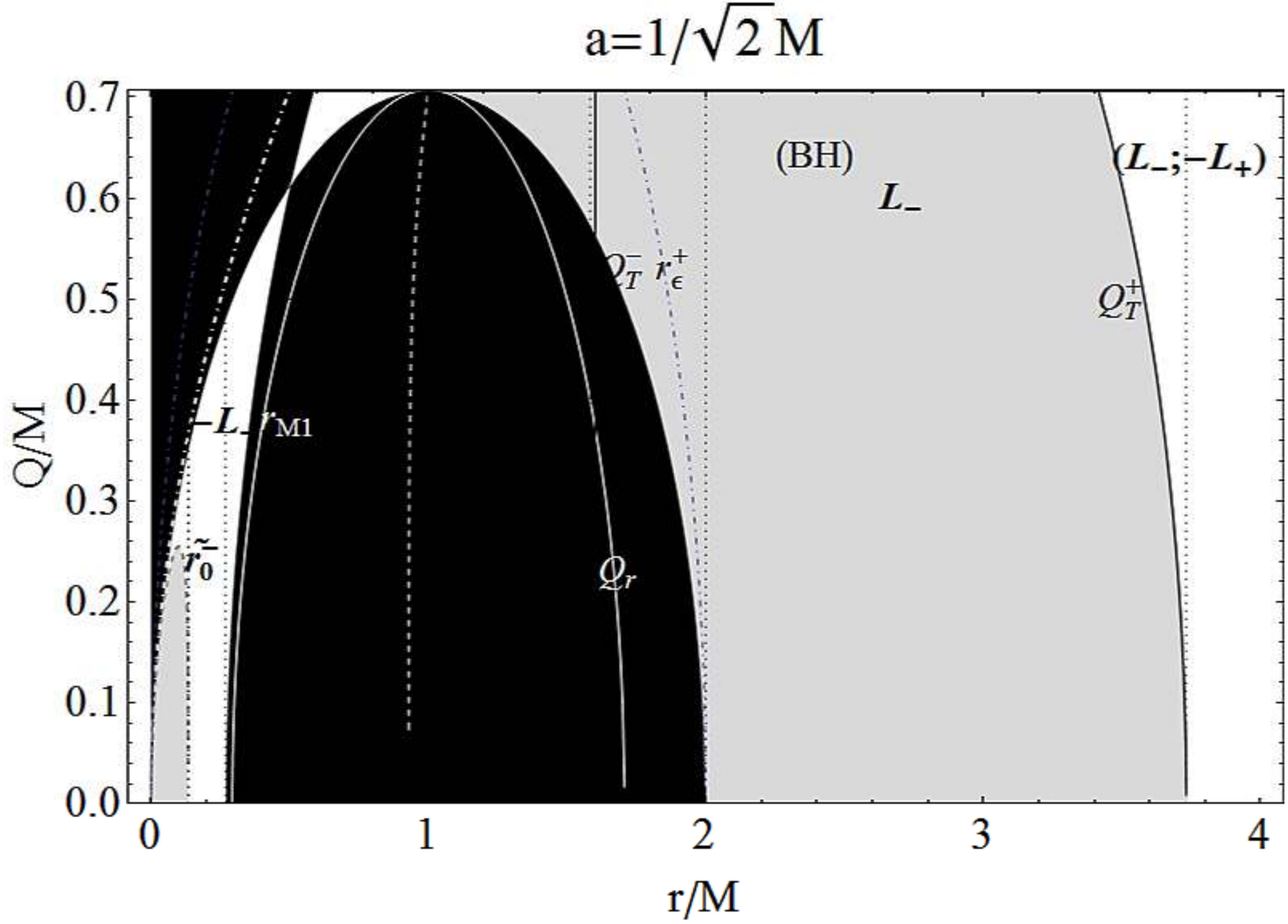}&
\includegraphics[width=0.3\hsize,clip]{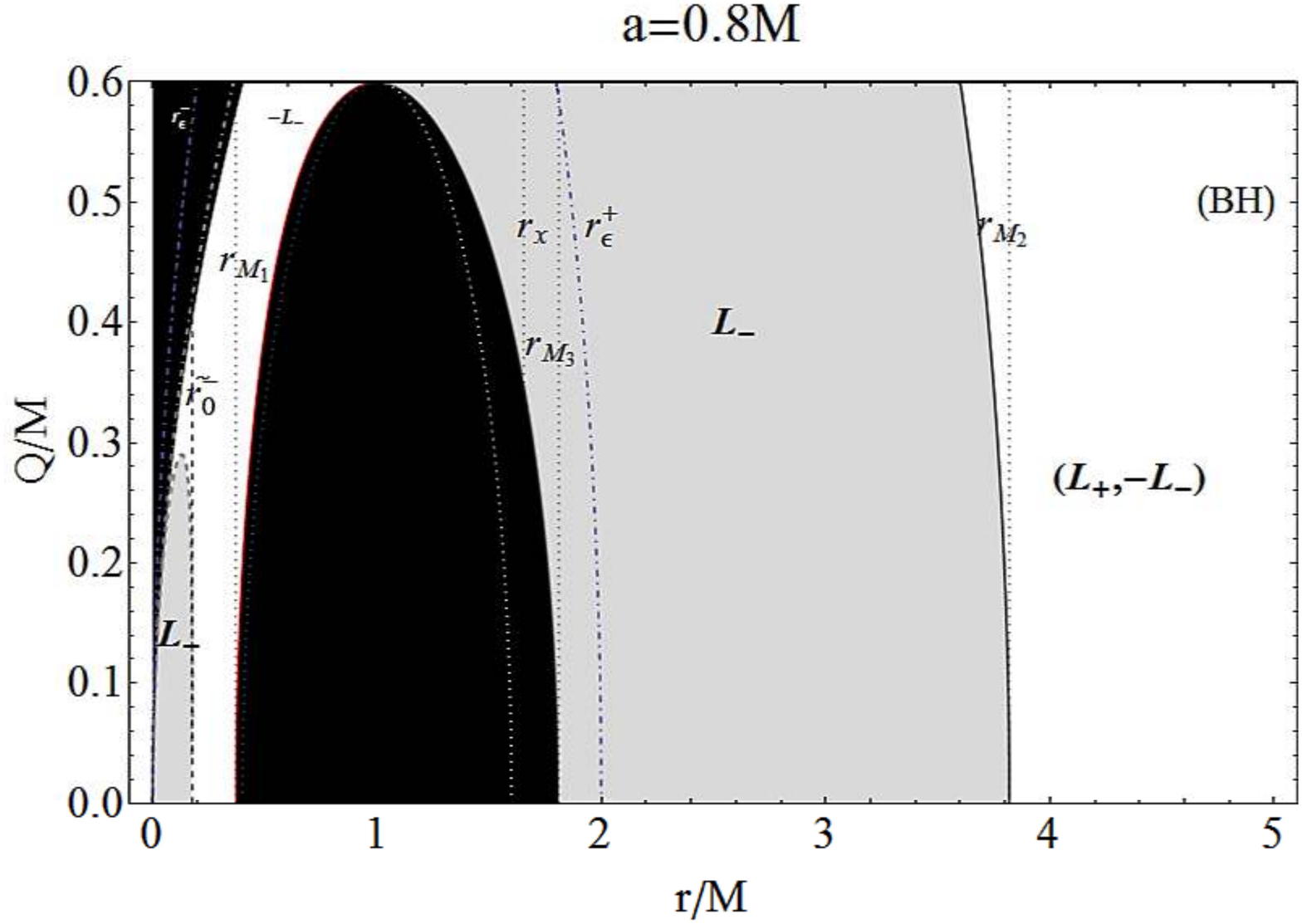}
\end{tabular}
\caption[font={footnotesize,it}]{\footnotesize{Black hole case: circular orbital regions in the plane $(Q/M, r/M)$ for fixed values of $a/M$. The angular momentum of the orbiting  test particle is shown in each spatial region. The curves   $Q_{\ti{T}}^{\pm}/M$, $Q_*/M$  are also plotted.  See also Table\il\ref{Tab:M_g_a}. The horizons $r_{\pm}$ are represented as white curves. The dotted-dashed curve is the outer ergosphere  $r_{\epsilon}^+$.
}}\label{Fig:bicchic1_BH}
\end{figure}

\begin{figure}[h!]
\begin{tabular}{ccc}
\includegraphics[width=0.3\hsize,clip]{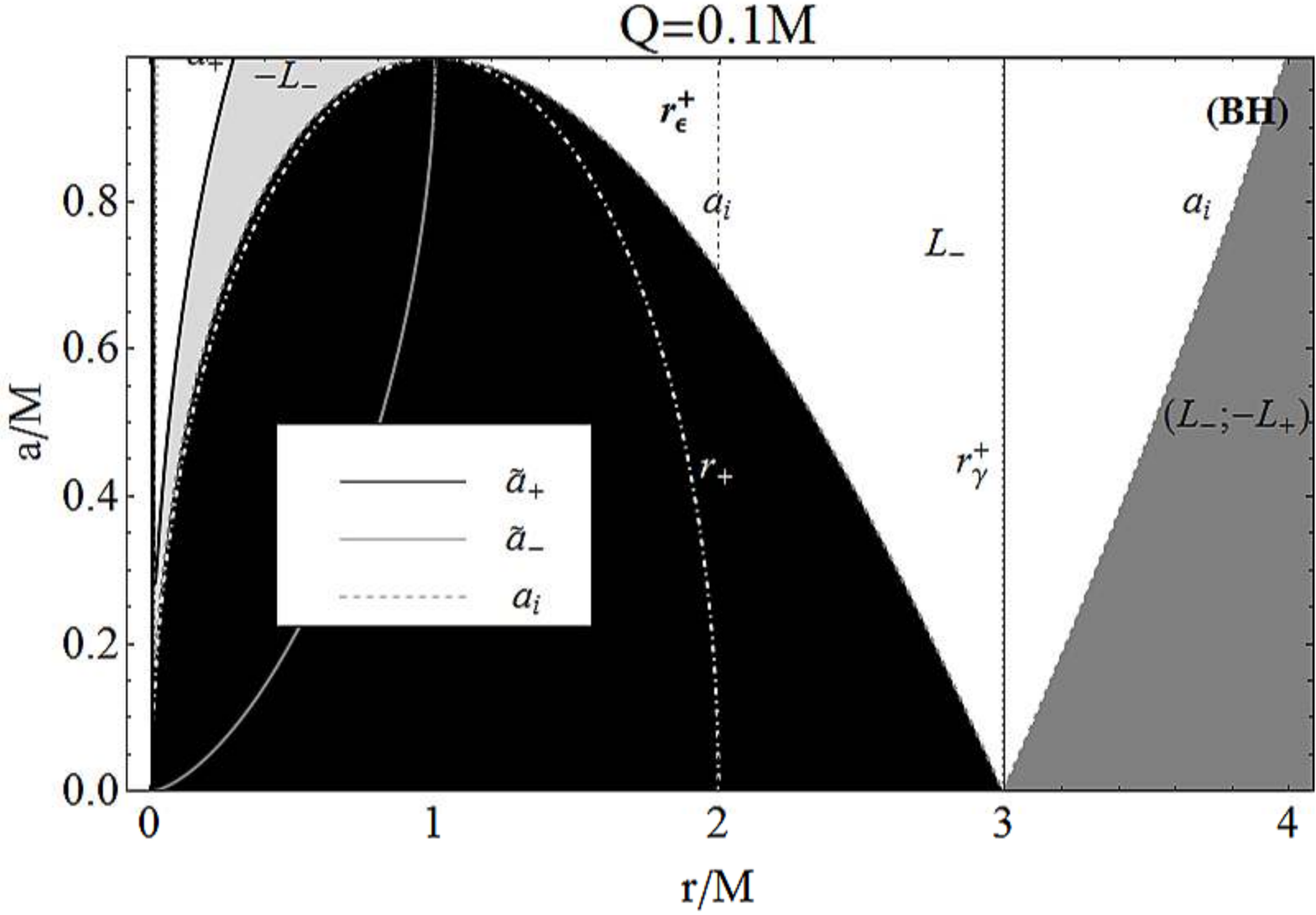}&
\includegraphics[width=0.3\hsize,clip]{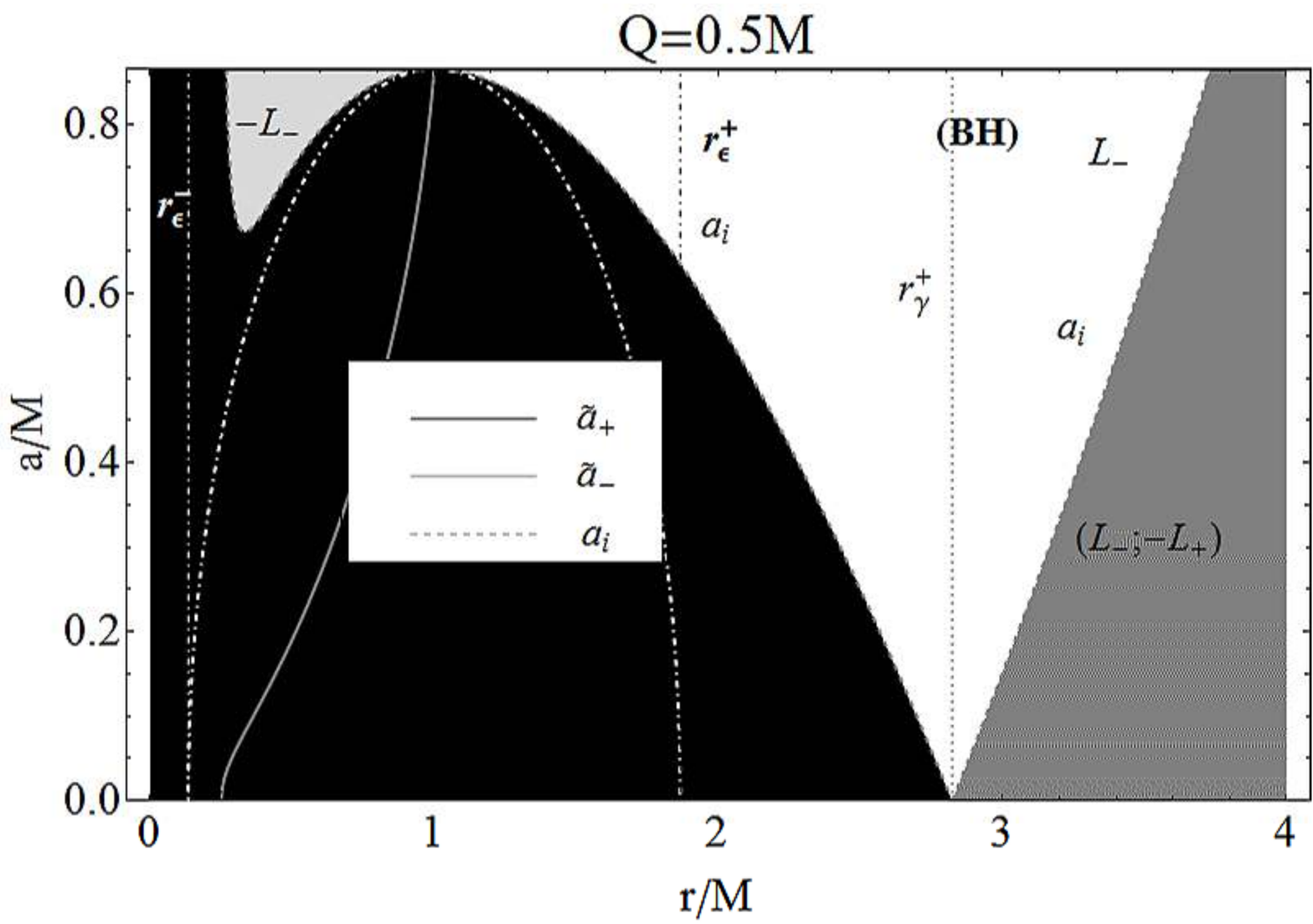}&
\includegraphics[width=0.3\hsize,clip]{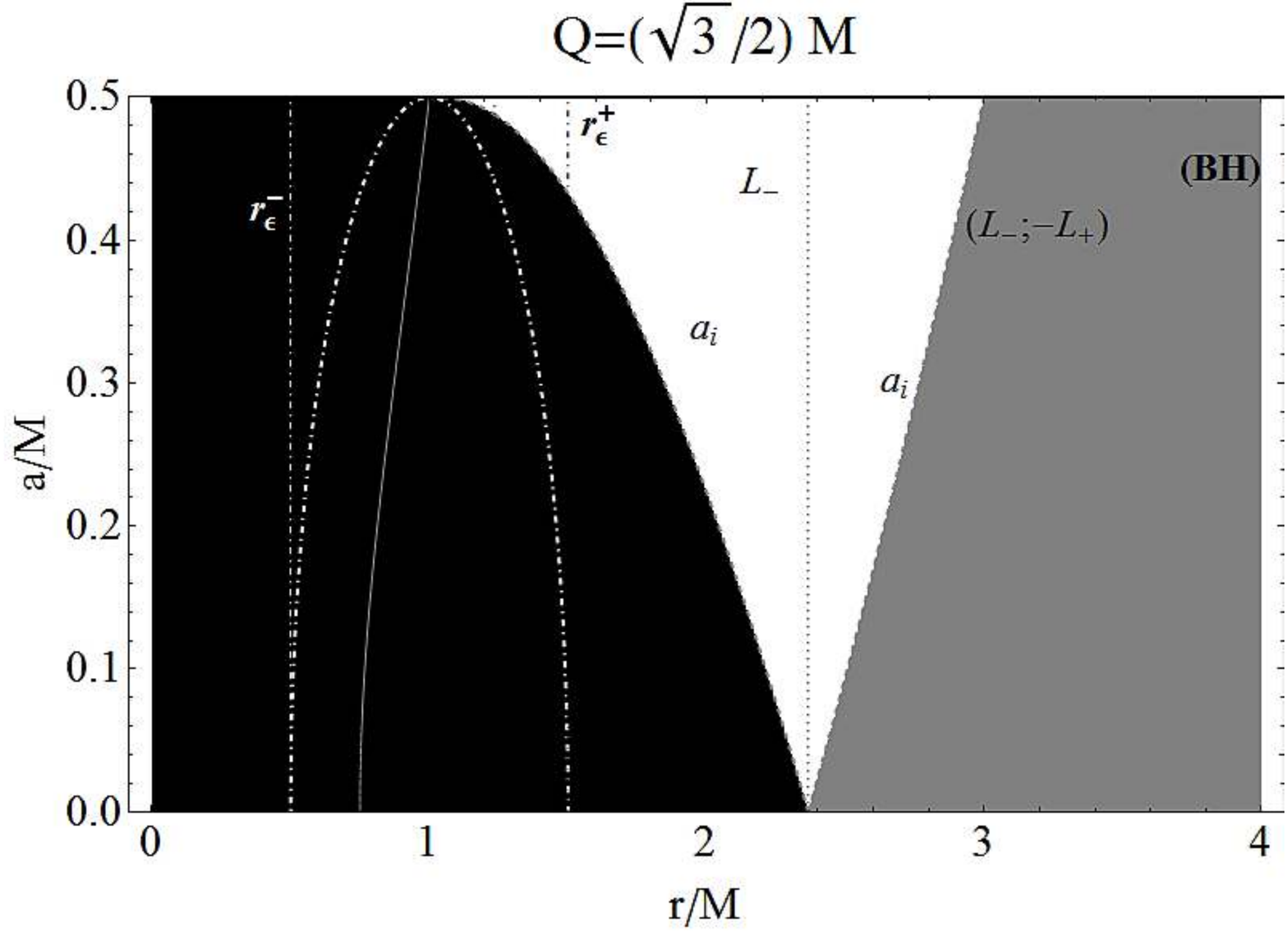}
\end{tabular}
\caption[font={footnotesize,it}]{Black hole case: circular orbital regions in the plane $(a/M, r/M)$ for fixed $Q/M$. The angular momentum of the orbiting  test particle is shown in each spatial region. The intrinsic KN angular momentum $\tilde{a}_-/M$ (gray curve),  $\tilde{a}_+/M$ (black curve)  and  $a_i/M$ (dashed gray  curve) are also plotted as functions of $r/M$.  The horizons $r_{\pm}$ are represented as white curves. See also Table\il\ref{Tab:Lamp_corr}. The dotted-dashed line is the outer ergosphere $r_{\epsilon}^+$.}
\label{Fig:bicchic1due_BH}
\end{figure}

\begin{table}[h!]
\caption{\label{Tab:M_g_a}\footnotesize{Description of the circular orbits for a test particle in a Kerr--Newman black hole  with $0<Q\leq M$, in the region $r>M$. The orbital angular momentum is given for each region. This is a generalization of Table\il\ref{Tab:Man:ure} to the case $Q\neq0$.
}}
\begin{ruledtabular}
\begin{tabular}{ll|ll|ll|ll}
Case: $0<Q\leq \sqrt{3}M/2$ & $(\sqrt{3}M/2<Q<M)$ & $$\\
\hline
$(r_d^-)M<r\leq r_{\gamma}^+$& &$r_{\gamma}^+<r<r_d^+$&  &$r=r_d^+$&  &$r>r_d^+$    &
\\
\hline
$a$&$L$&$a$&$L$&$a$&$L$&$a$&$L$
\\
$]a_i,a_s] $ &$L_- $ & $[0,a_i[$  &$(L_-,-L_+)$&$[0,a_s[$ &$(L_-
,-L_+)$ & $[0,a_s]$  &$(L_-,-L_+)$
\\
 $$ &$$& $[a_i,a_s]$ &$L_-
 $ &$a_s$ &$L_-$   &$$
\\
\hline
Case: $Q=M$&
\\
\cline{1-2}
$r>2M$&
\\
\cline{1-2}
$a$&$L$&
\\
$a=0$ &$\pm L_{-}$
\\
\end{tabular}
\end{ruledtabular}
\end{table}
The results presented in Table\il\ref{Tab:M_g_a} are illustrated in Figs.\il\ref{Fig:bicchic1_BH} and \ref{Fig:bicchic1due_BH}.
\begin{table}
\caption{\label{Tab:Lamp_corr}\footnotesize{Description of the circular orbits for a test particle in a Kerr--Newman black hole  with $0\leq Q\leq M$, in the region $r>M$. The orbital angular momentum is given for different values of the black hole spin $a/M$.
This is a generalization of Table\il\ref{Tab:M_corrected} to the case $Q\neq0$.
}}
\begin{ruledtabular}
\begin{tabular}{ll|ll|lllll}
Case: $0<Q<M$\\
\hline
$a=0$& (Reissner Norstr\"om)&$0<a< a_s$  &  &$a=a_s$  &
\\
\hline
Region&$L$&Region&$L$&Region&$L$
\\
$r>r_{\gamma}^+ $ &$(L_-,-L_+)$ & $]r_3,r_4]$  &$L_-$&$]r_3,r_4]$  &$L_-$
\\
  & & $]r_4,\infty[$ &$(L_-,-L_+)$  &$]r_4,\infty[$ &$(L_-,-L_+)$
\\
\hline
Case: $Q=M$
\\
\hline
$a=0$& (Reissner Nordstr\"om)&
\\
\cline{1-2}
Region&$L$
\\
$r>2M $ &$(\pm L_-)$
\\
\end{tabular}
\end{ruledtabular}
\end{table}

On the other hand,
Table\il\ref{Tab:Lamp_corr} shows, for different intervals of the specific charge $Q/M$, the regions of existence of
circular orbits for different ranges of the source spin $a$.
For this analysis, it is convenient to define the following angular momentum parameters
\bea
\tilde{a}_{\pm}&\equiv&\sqrt{\frac{Q^4-2 r (r-2M)(rM\mp Q^2)+(2rM\mp Q^2)\sqrt{Q^4+4 Q^2M (r-M) r-4 (r-M)M r^2}}{2(rM-Q^2)}},
\eea
which determine the boundaries of the different regions. In the limiting case $\tilde{a}_{+}=\tilde{a}_{-}$, these regions reduce and very particular orbits appear {which are studied in detail in  Appendix\il\ref{Sec:appendix_spin_radius}}.

The orbital region boundaries  are determined  by the radii $r_{\gamma}^+$, $r_3$ and $r_4$, which for fixed $Q/M$ are in general functions of $a/M$. The results of Table\il\ref{Tab:Lamp_corr} are illustrated in Figs.\il\ref{Fig:bicchic1due_BH}.

Finally,  we notice that in Figs.\il\ref{Fig:bicchic1_BH} and \ref{Fig:bicchic1due_BH}  we have represented the entire orbital range  $r> 0$, although for the study of circular orbits we are restricting ourselves to the physical region outside the outer horizon $r>r_+$. Nevertheless, it is interesting to mention that, according to Figs.\il\ref{Fig:bicchic1_BH} and \ref{Fig:bicchic1due_BH}, there exist circular orbits with
$L=-L_-$  in the region $r<r_-$. The study of this motion inside the inner horizon is outside the scope of the present work.
\subsection{Stability of circular orbits}\label{SubSec:BH-sta}
\begin{figure}[h!]
\begin{tabular}{ccc}
\includegraphics[width=0.3\hsize,clip]{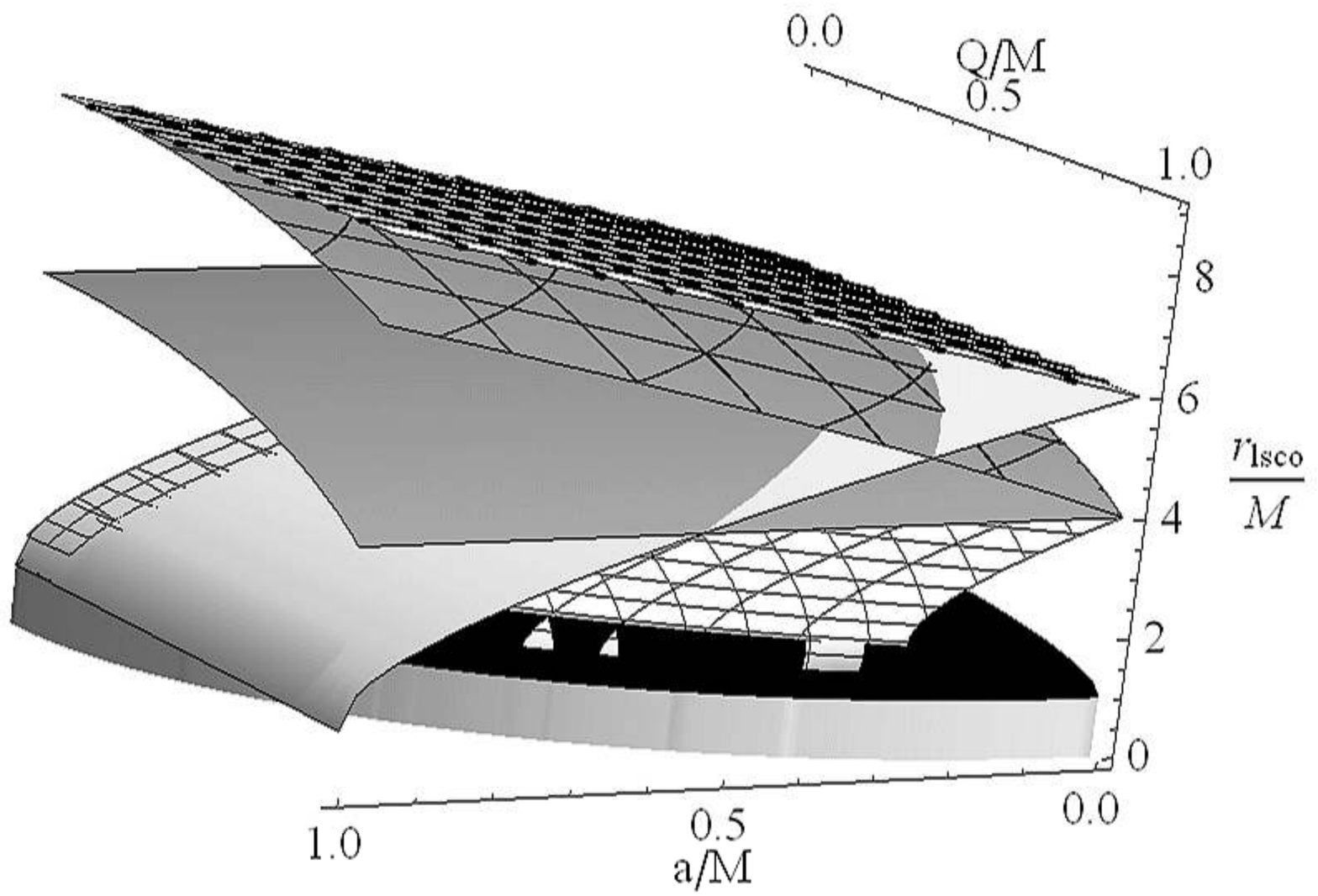}&
\includegraphics[width=0.3\hsize,clip]{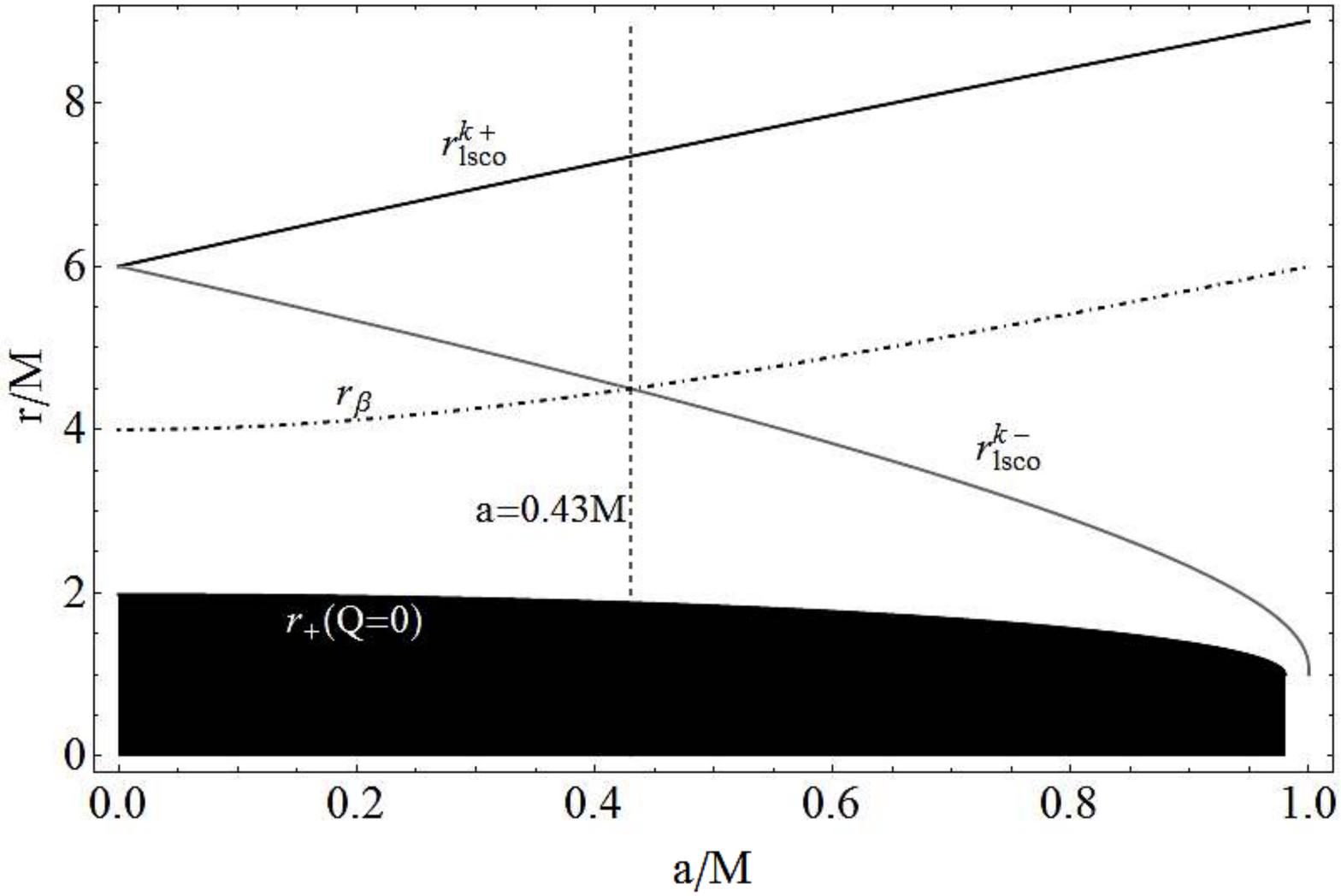}&
\includegraphics[width=0.3\hsize,clip]{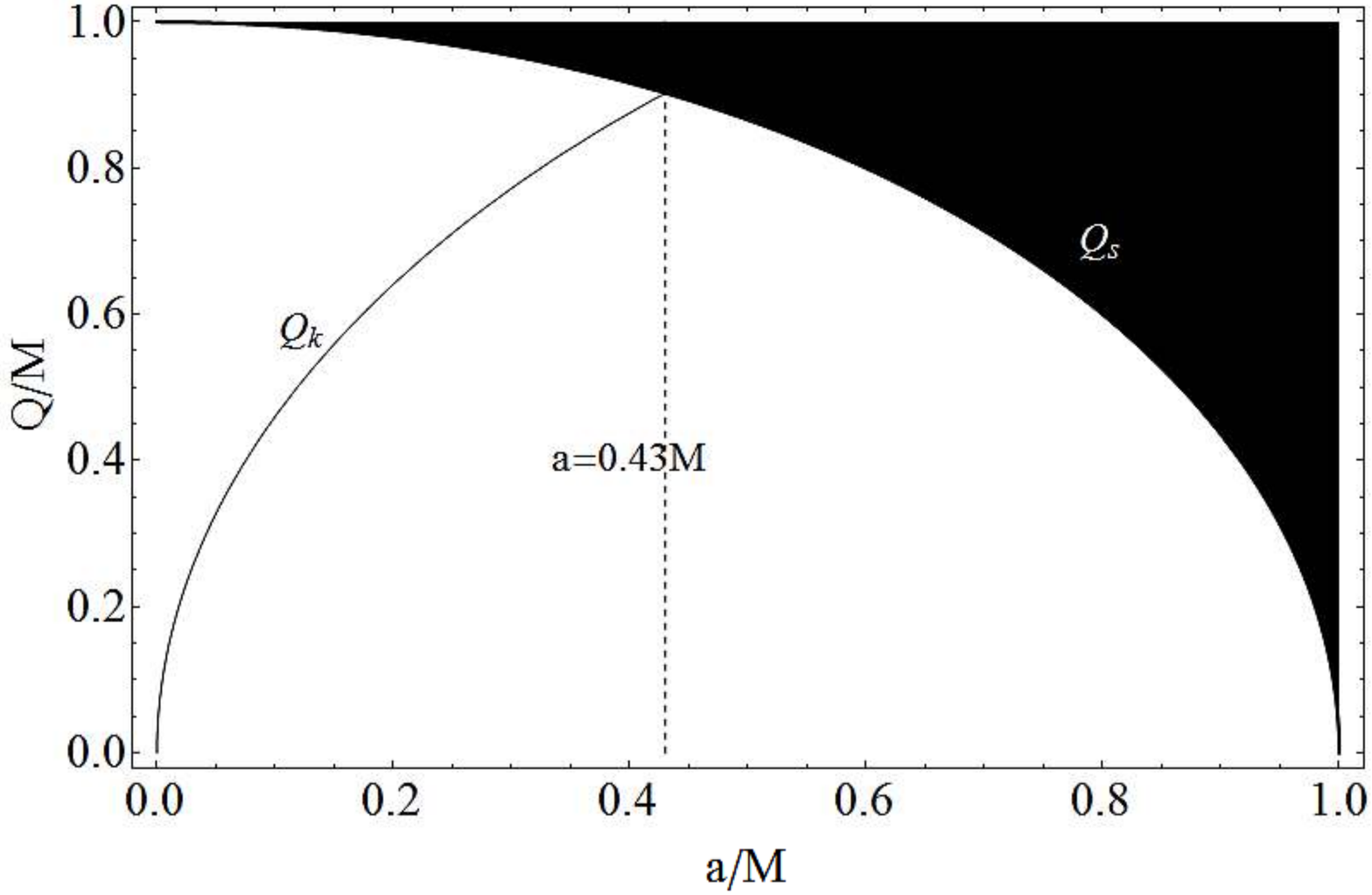}
\end{tabular}
\caption{
Left:  The outer horizon  (black surface), $r^{RN}_{lsco}$ (gray surface),  $r^{K-}_{lsco}$ (light-gray surface),  $r^{K+}_{lsco}$
 (white- black-thick-mashed surface), the surface $r_{lsco}(L_-)$ (light-gray, meshed surface) and  $r_{lsco}(-L_+)$   (gray surface with black thick mash). Note that
$r^{K+}_{lsco}=r^{K-}_{lsco}$ at $r= 6M $ and $r^{K-}_{lsco}=r^{RN}_{lsco}$ along the curve $Q=Q_k$ (see text) and $r_{lsco}(-L_+)<r_{lsco}(L_-)$.
Center: Last stable circular orbit radius $r_{lsco}$ for a Kerr  black hole $(Q=0)$ -- $r^{\ti{K-}}_{lsco}$ (gray curve) and $r^{\ti{K}+}_{lsco}$ (black curve) --, the outer horizon $r_+$, and the radius $r_{\beta}$ (dotted-dashed curve)  are plotted as  functions of the KN black hole spin $a/M$.
The dashed line represents $a=0.430075M$.
Right: The charge $Q_{\ti{K}}$ (black curve) of a KN black hole is plotted as a function of the momentum $a/M$, where $r^{\ti{RN}}_{lsco}=r^{\ti{K}-}_{lsco}$ (see text). A naked singularity exists in the region $Q>Q_s$.} \label{Fig:Das_cri_1}
\end{figure}

The radius of the last stable circular orbit  for a neutral particle orbiting on the equatorial plane of a KN black hole is defined as the solution of the following equations:
\be
V(r_{lsco})=E,\quad  V'(r_{lsco})=0,\quad V''(r_{lsco})=0,\quad \mbox{with}\quad Q^2+a^2\leq M^2.
\ee
Stable (unstable) orbits are characterized by the condition  $V''(r_{lsco})>0$ ($V''(r_{lsco})<0$).
In Fig.\il\ref{Fig:Das_cri_1}, we plot $r_{lsco}$ as a function of the intrinsic angular momentum $a$ and the charge $Q$.
We note that the radius $r_{lsco}$ is given as a surface that has as a limiting case the planes identified by the radii $r^{\ti{RN}}_{lsco}$, the last stable orbit for  $a=0$,   and $r^{\ti{K}\mp}_{lsco}$, the last stable orbit  for  $Q=0$.  The exact expressions for these radii
were given in \cite{Pu:Kerr} and \cite{Pu:Neutral}, respectively. We can see from Fig.\il\ref{Fig:Das_cri_1} that the surface contained between the two limiting cases  is reduced to a curve if the condition
\(
r^{\ti{RN}}_{lsco}=r^{\ti{K}-}_{lsco}
\)
is satisfied. This happens when $Q$ and $a$ are linked by the following relation
\be
Q=Q_{\ti{K}}\equiv\frac{r^{\ti{K}-}_{lsco} }{8M}\left[9M- \sqrt{16 r^{\ti{K}-}_{lsco} -15M }\right].
\ee
In particular, if $a=0$, we obtain that $4M\leq r_{lsco}\leq 6M$ and
if $0<a<M$, then  the last stable orbit in the spacetime of a KN black hole is located within the bounded region
$r_{\beta}\leq r_{lsco}\leq6M$ where
\be
r_{\beta}/M\equiv\frac{3 a^2+\left(M^2+\left[M^6+a^2 M^2\left(5M^2+2 a^2\right)+\sqrt{a^2 \left(a^2-M^2\right)^2 \left(M^2+4 a^2\right)}\right]^{1/3}\right)^2}{\left[M^6+a^2M^2 \left(5M^2+2 a^2\right)+\sqrt{a^2 \left(a^2-M^2\right)^2 \left(M^2+4 a^2\right)}\right]^{1/3}}.
\ee
This behavior is depicted in Fig.\il\ref{Fig:Das_cri_1}. It is interesting to mention the identity
\(
r_{\beta}={r^{\ti{RN}}}_{lsco}(Q=a_s),
\)
which means that at the boundary $r_\beta$ the effective intrinsic angular momentum $\sqrt{1-(a/M)^2}$ plays the role of the specific charge
$Q/M$. Moreover, for the particular value $a=0.430075M$ we obtain that
$r_\beta= r^{\ti{K}-}_{lsco}$. See Fig.\il\ref{Fig:Das_cri_1}.
Finally, for $a=0$ and  $Q=0$ we have that $r_{lsco}=6M$.

In particular, we note that $E(-L_+)>E(L_-)$ in the regions where both of them are allowed.
As a function of the radial coordinate, the energy of circular orbits shows  points of minimum values, i.e., stable orbits. There are also radii ($r_3$ and $r_4$)
at which the energy necessary for a test particle  to remain in orbit around the source becomes infinite.
This kind of circular orbits is also found in the RN and Kerr spacetimes.

\begin{figure}[h!]
\begin{tabular}{ccc}
\includegraphics[width=0.3\hsize,clip]{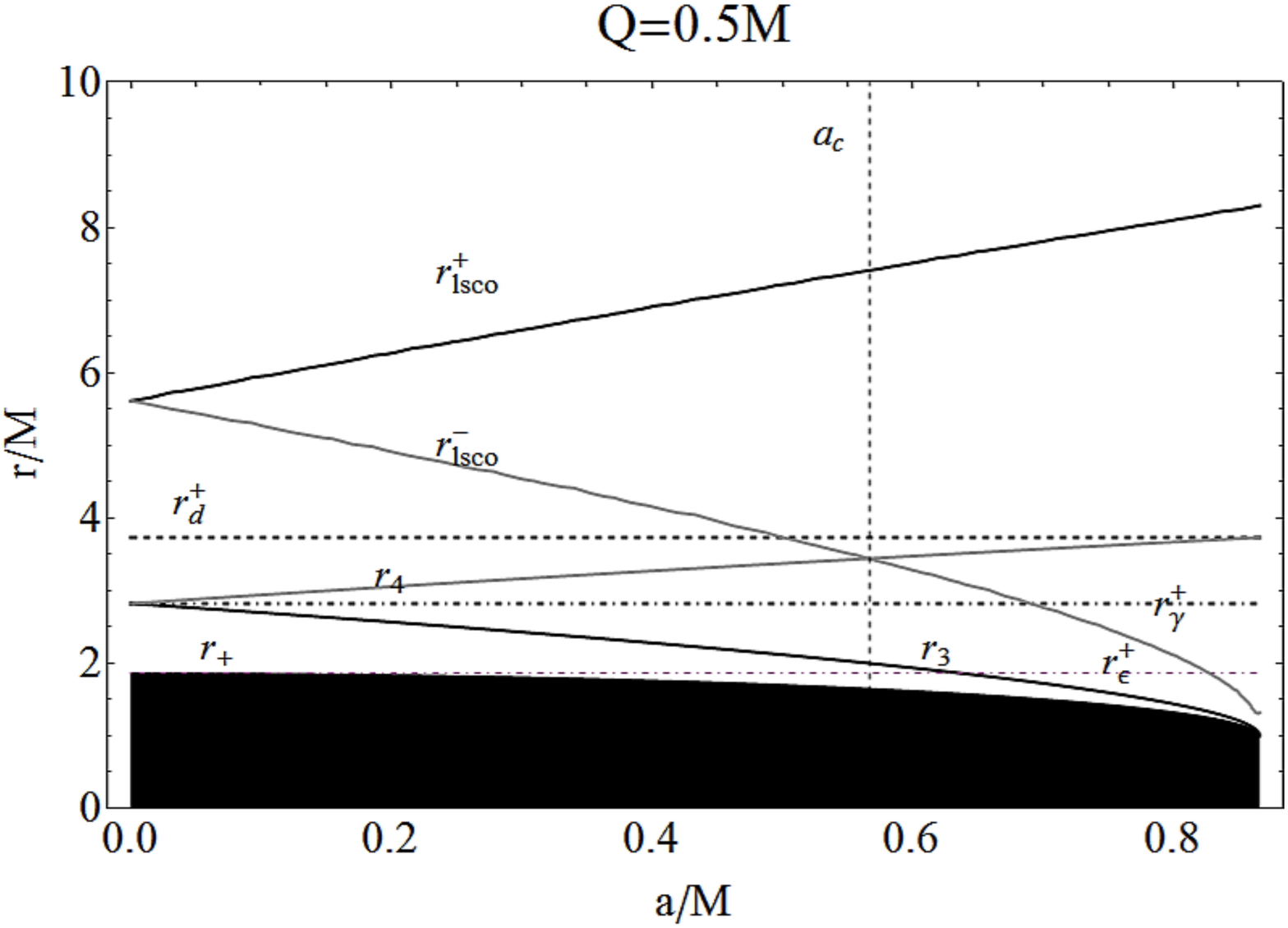}&
\includegraphics[width=0.3\hsize,clip]{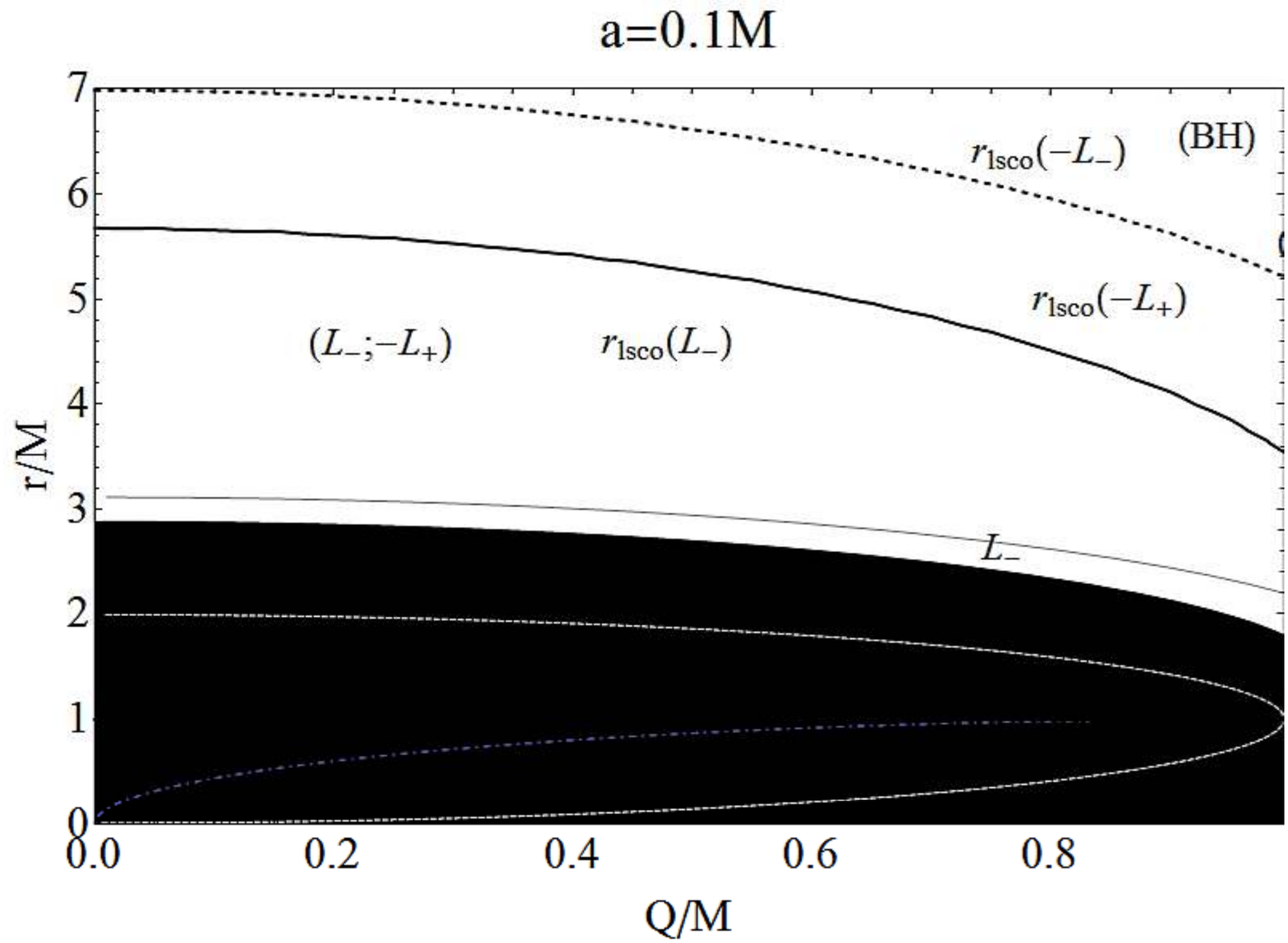}&
\includegraphics[width=0.3\hsize,clip]{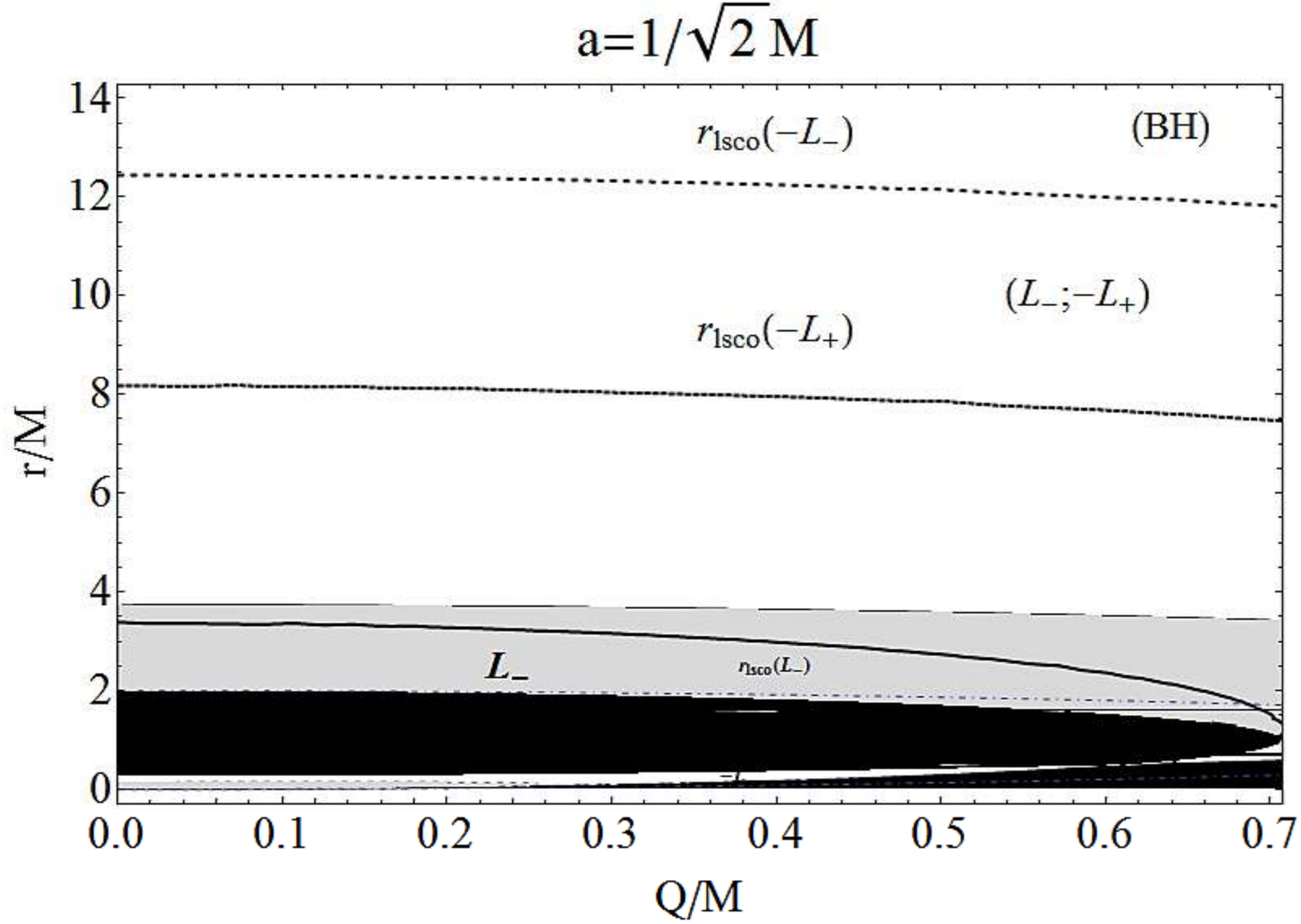}
\end{tabular}
\caption[font={footnotesize,it}]{\footnotesize{Left: The radius $r^{+}_{lsco}\equiv r_{lsco}(-L_+)$ and $r^{-}_{lsco}\equiv r_{lsco}(L_-)$  of the last stable circular orbits for a KN black hole, with  fixed charge-to-mass ratio $Q/M$, as a function of the black hole spin $0\leq a\leq a_s$. The horizon  $r+$, the radius $r_{\gamma}^+$ (dotted line), $r_{d}^+$ (dashed line), $r_{3}$ (black curve), $r_{4}$ (gray curve) are also plotted. The black region $(r<r_+)$ is forbidden. The value of $a_c$, the crossing point between $r_3$ and $r^{-}_{lsco}$, is represented by a dashed line. The dotted-dashed line is the outer ergosphere  $r_{\epsilon}^+$.  Center and Right: Last stable circular orbits $r_{lsco}(L_-)$ (thick black curve),   $r_{lsco}(-L_-)$ (thick black dashed curve) and  $r_{lsco}(-L_+)$ (thick black  dotted curve) as functions of $Q/M$, for increasing values of the source spin. The angular momentum of an orbiting test particle is shown  in each region. Black regions are forbidden. In the region $r<r_{lsco}$ ($r>r_{lsco}$), circular orbits are unstable (stable). The curves $L=-L_-$ exist in a region located in $r<r_-$.} } \label{Fig:xyz}
\end{figure}

The stability of these orbits can be studied by comparison with the Kerr black hole case.
In Fig.\il\ref{Fig:xyz}, the results are shown for black holes with different charges.
Following \cite{Pu:Kerr}, we define two different radii for the last stable circular orbits as follows
\be
r^{+}_{lsco}\equiv r_{lsco}(-L_+),\quad r^{-}_{lsco}\equiv r_{lsco}(L_-)
\ee
for the orbits with angular momentum $L=-L_+$ and $L=L_-$, respectively.
We now compare the general configuration for  different values of $Q/M\neq0$ with the configuration in the case of a Kerr black hole $(Q=0)$.
We show in Fig.\il\ref{Fig:xyz} that for a KN black hole  the orbital   regions of stability  are qualitatively similar to those of a Kerr black hole \cite{Pu:Kerr}. The main feature is that if the charge-to-mass ratio increases, for fixed black hole spin $a/M$, the radius of the last stable circular orbit $r^{\pm}_{lsco}$ decreases.
In particular, the plots of  Fig.\il\ref{Fig:xyz}  show that, for fixed  $a/M$ and for different values of $Q/M$, the stability regions  bounded by  $r_{lsco}(-L_+)$ and  $r_{lsco}(L_-)$ are connected regions. This is the main result of this section and it will be important when comparing with the case of naked singularities.
We can clarify the arrangement of the stability regions in a schematic way: In the region $r_+<r<r_{lsco}(L_-)$ all orbits are unstable, in the region  $r_{lsco}(L_-)<r<r_{lsco}(-L_+)$ the orbits with  $L_-$ are stable while the orbits with   $L=-L_+$ are unstable, and finally in the region $r>r_{lsco}(-L_+)$ all orbits are stable.
This scheme does not change significantly when the parameters $Q/M$ and $a/M$ are changed; however, we note that the radius $r_{lsco}(L_-)$ can lie on the right of the region  \textbf{I} so that all the orbits with $L=L_-$ are unstable. This occurs for values of $a/M$ sufficiently small. As $a/M$ increases, the radius $r_{lsco}(L_-)$ becomes smaller and approaches the horizon, reducing in this way the regions of instability for the orbits $L=L_-$ which belong to both the \textbf{I} and \textbf{II} regions (see Fig.\il\ref{Fig:xyz}-left).
Furthermore, the radius $r_{lsco}(L_-)$ splits the region \textbf{I} into two different stability zones: Orbits with  $r>r_{lsco}(L_-)$ are stable whereas orbits with   $r<r_{lsco}(L_-)$ are instable (see Fig.\il\ref{Fig:xyz}-Right).


%% file: NS_L0_13.tex
\section{Naked singularities}\label{Sec:NS_Case}
\subsection{Orbits with zero angular momentum}\label{Sec:NS_L0}
In this section, we start  the study of the motion around a KN naked singularity. We will
focus first on the ``orbits'' that are solutions of the following equations
\be\label{E:cro_cee_dl}
V'(r,Q,a,L)=0,\quad L=0.
\ee
Note that real solutions can exist only in the case of naked singularities. This particular type of circular
motion can occur in many axially symmetric spacetimes, and has been investigated in detail for neutral and charged test particles
in the RN and Kerr spacetimes in \cite{Pu:Neutral,Pu:Kerr,Pu:Charged}. In the RN spacetime, this  particular orbit was interpreted as corresponding to a particle  located at rest at $r=r_* =Q^2/M$  with respect to static observers located at infinity.
The radius $r_*\equiv Q^2/M$
coincides with the value of the classical radius of an
electric charge, which is usually obtained by using a completely different approach (see also \cite{AbraMillStuc93,Burinskii:2005mm,Arcos:2002ip}).
 In general,  the motion defined by the conditions
\ref{E:cro_cee_dl}, is governed only by the gravitational component of the effective potential  (and the  electromagnetic interaction in the case of charged particles), with a  zero   centrifugal force. Therefore, if a  configuration with $L= 0$    is  possible, it  should be the result of a balance of forces.
Thus, this fact  can be interpreted in terms of  repulsive gravity  effects. It is then important  to note that these limiting conditions  can exist  only in the case of naked  singularities, even for $a=0$ or  $Q=0$.
This means that the ``repulsive component''  of the effective mass in the case of a Kerr naked singularity is guaranteed by the so-called  ``spin-charge'' of the source. The presence of both charges in the KN spacetime, in Eq.\il\ref{Eq:tra_0}, makes this effect to become more complex  due to the combination of these two components. For more details about  the repulsive gravitation effects in  naked singularities, see also
\cite{deFelice74}.

The solutions of Eq.\il\ref{E:cro_cee_dl} are listed in Table\il\ref{Tab:di_gno} and plotted in Fig.\il\ref{Fig:c_olin}-Left
\begin{table}[h!]
\caption{\label{Tab:di_gno}\footnotesize{Description of the circular orbits for a test particle in a Kerr--Newman naked singularity   with $L=0$.}}
\begin{ruledtabular}\begin{tabular}{ll|ll|ll|llll|lll}
Case: $Q=0$ (Kerr metric)&&$0<Q\leq0.344263M$&&$0.344263M<Q<\sqrt{5}M/4$
\\
\hline
$a/M$&Radius& $a/M$&Radius&  $a/M$&Radius
\\
\hline
$M<a<3\sqrt{3}M/4$&$(\tilde{r}_2,\tilde{r}_3)$ & $a_s<a<a_2$&$(\tilde{r}_1,\tilde{r}_2,\tilde{r}_3)$
& $a_s<a<a_1$&$\tilde{r}_1$
\\
$a=3\sqrt{3}M/4$  &$\tilde{r}_2$& $a=a_2$ &$(\tilde{r}_1,\tilde{r}_2)$
& $a=a_1$ &$(\tilde{r}_1,\tilde{r}_3)$
\\
$$  &$$& $a>a_2$ &$\tilde{r}_1$  &$a_1<a<a_2$  &$(\tilde{r}_1,\tilde{r}_2,\tilde{r}_3)$
\\
&&&&$a>a_2$&$\tilde{r}_1$
\\
\hline
Case: $\sqrt{5}M/4\leq Q\leq M$&&$Q>M$&
\\
\cline{1-4}
$a/M$&Radius& $a/M$&Radius
\\
\cline{1-4}
$a>a_s$ & $\tilde{r}_1$&$a=0$&$Q^2/M$\\
& &$a>0$&$\tilde{r}_1$
\\
\end{tabular}
\end{ruledtabular}
\end{table}
from which it is possible  to see that for  $Q>M$ there always exists at least  one ``circular orbit'' radius characterized by the conditions given in Eq.\il\ref{E:cro_cee_dl} for $a>a_s$, and  in the case of   a RN naked singularity; thus, when $a=0$ and $Q>M$,  it is $r=r_*\equiv Q^2/M$.
For $Q=0$, ``circular orbits'' defined by Eq.\il\ref{E:cro_cee_dl}  exist only in a closed interval $a/M\in[1,3\sqrt{3}/4]$.
A numerical analysis shows that the  orbital regions in the plane $(a,r)$  are bounded by the  intrinsic spins $a_1, a_2$ whose
exact form can be found in Appendix\il\ref{App:QmM}.

\begin{figure}[h!]
\begin{tabular}{ccc}
\includegraphics[width=0.3\hsize,clip]{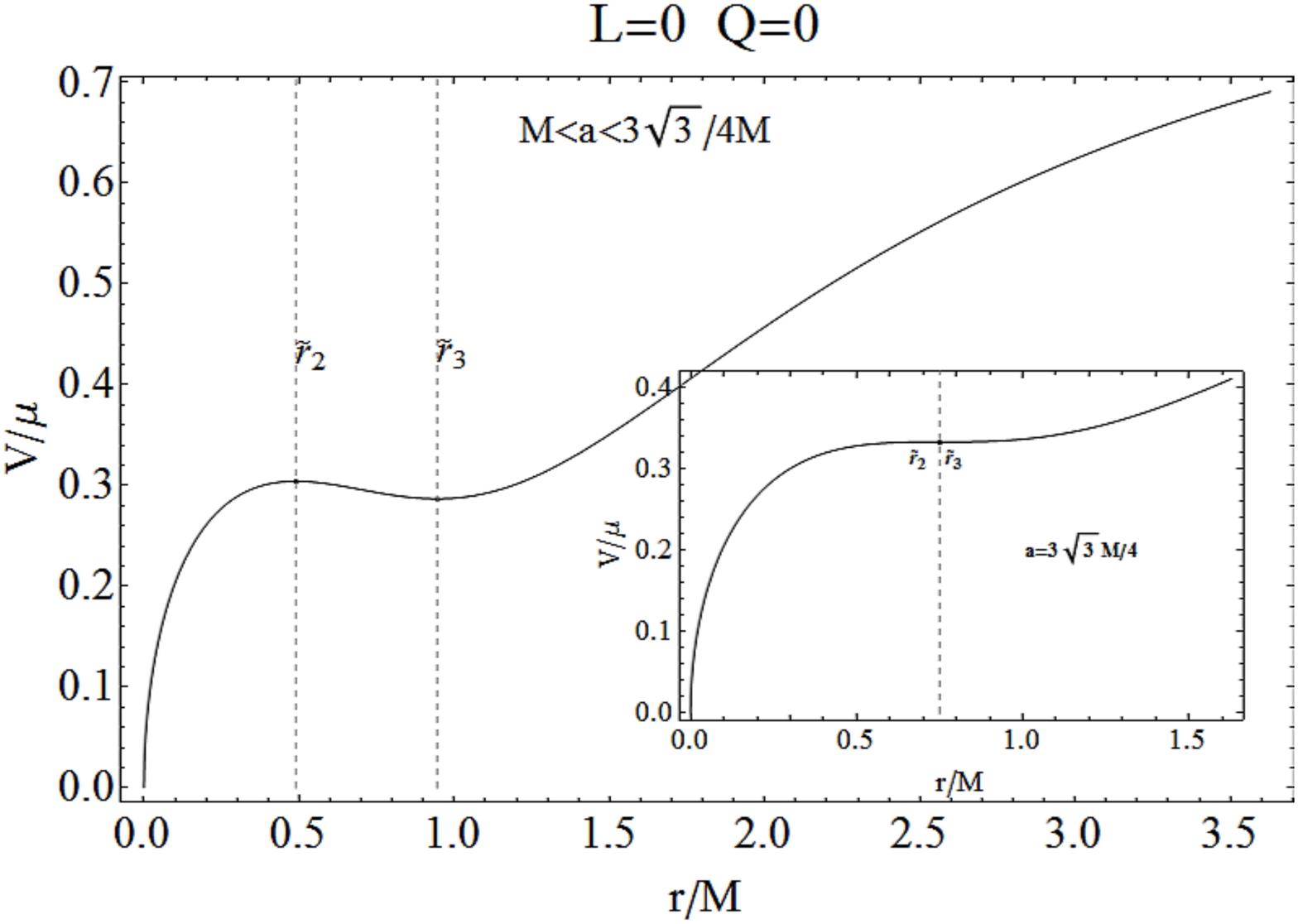}
\includegraphics[width=0.3\hsize,clip]{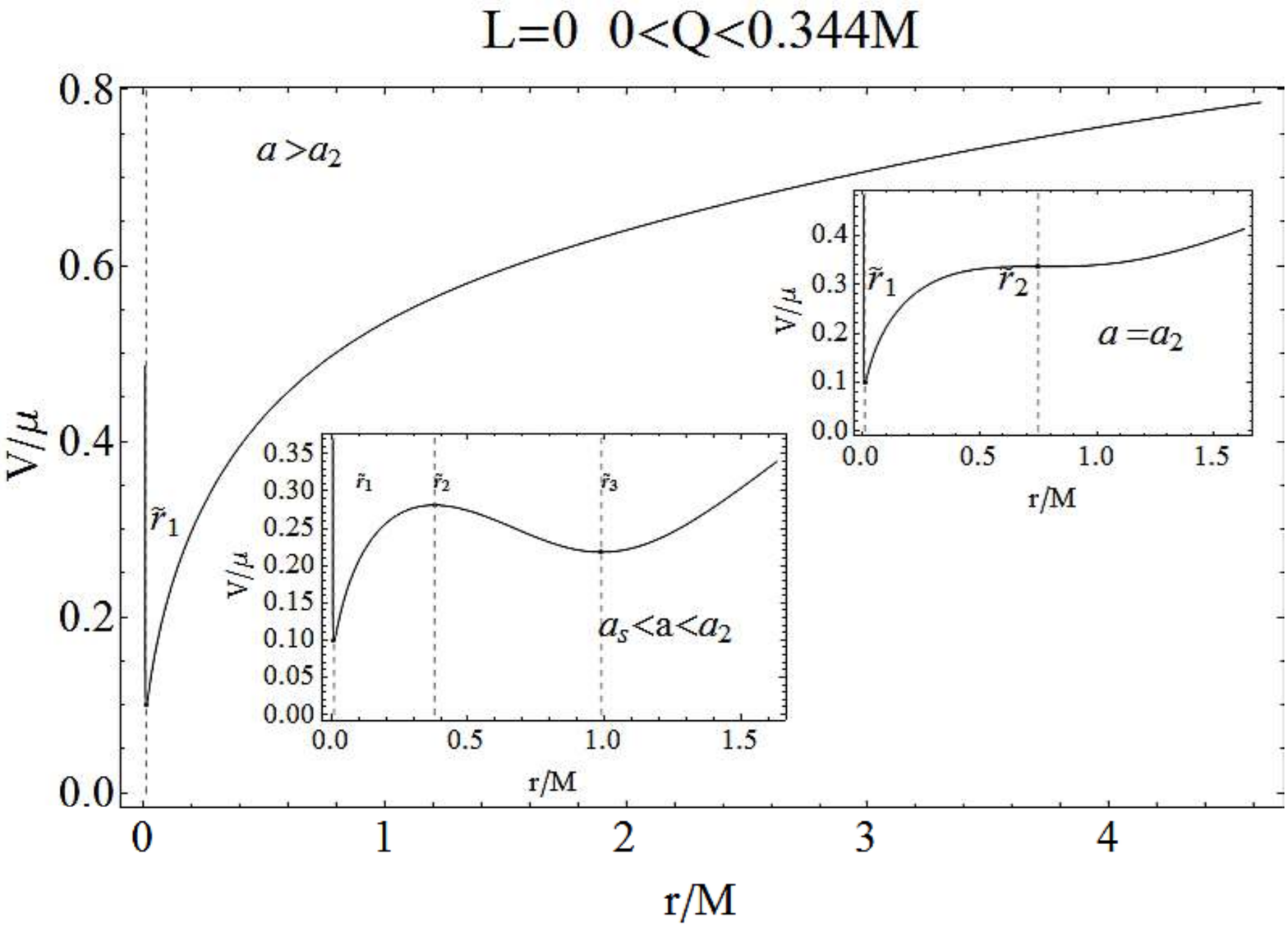}
\includegraphics[width=0.3\hsize,clip]{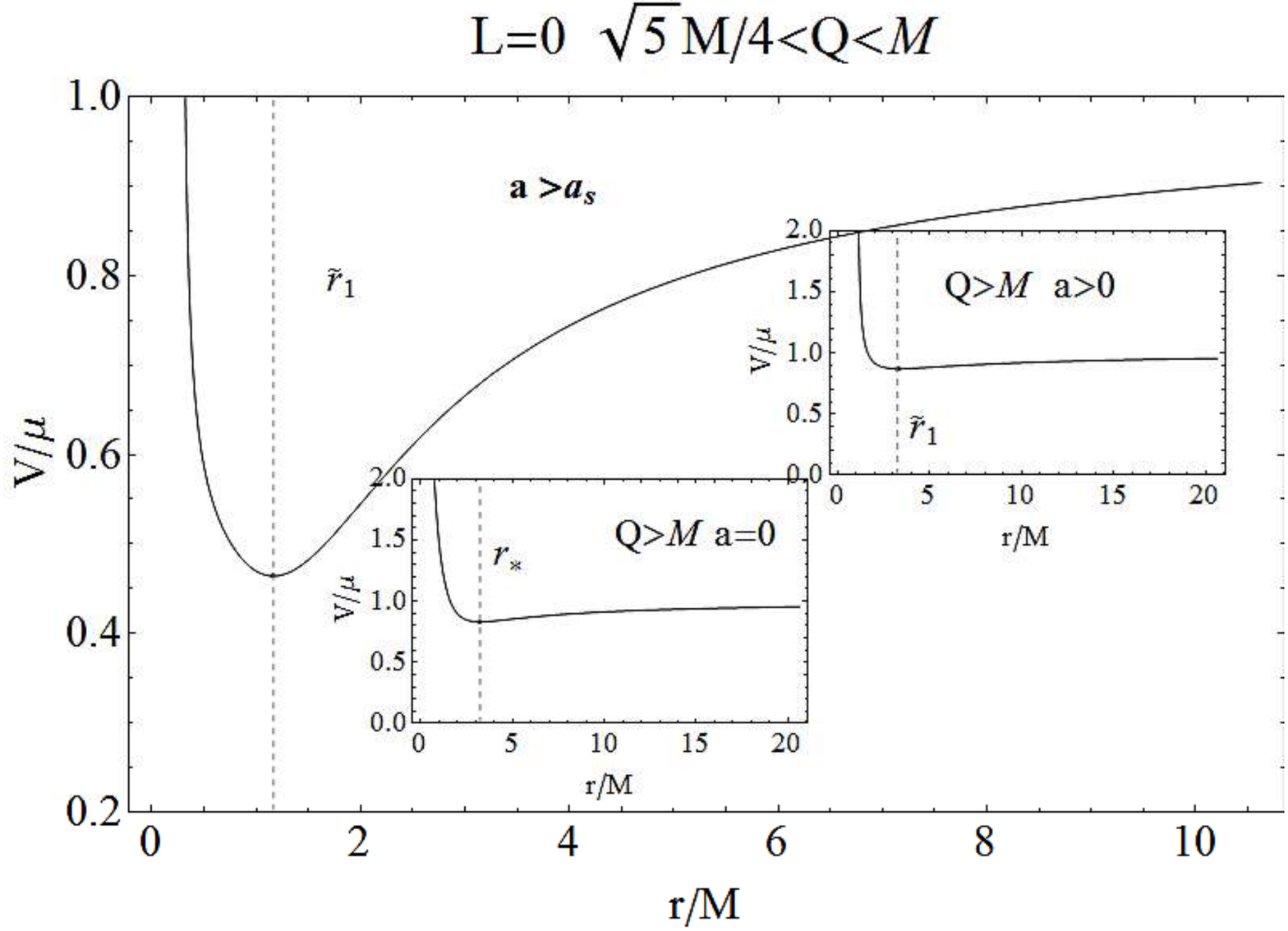}
\\
\includegraphics[width=0.3\hsize,clip]{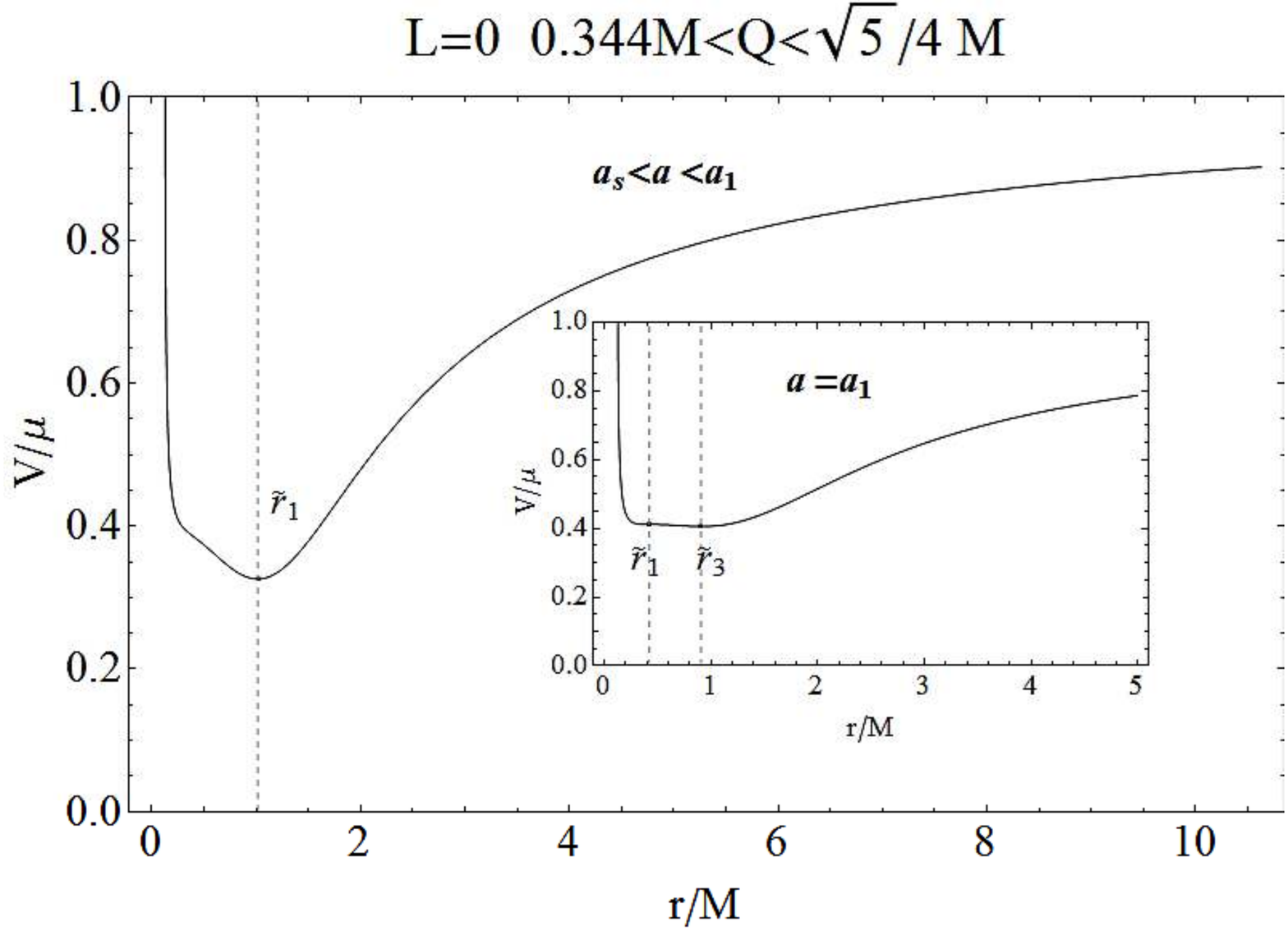}
\includegraphics[width=0.3\hsize,clip]{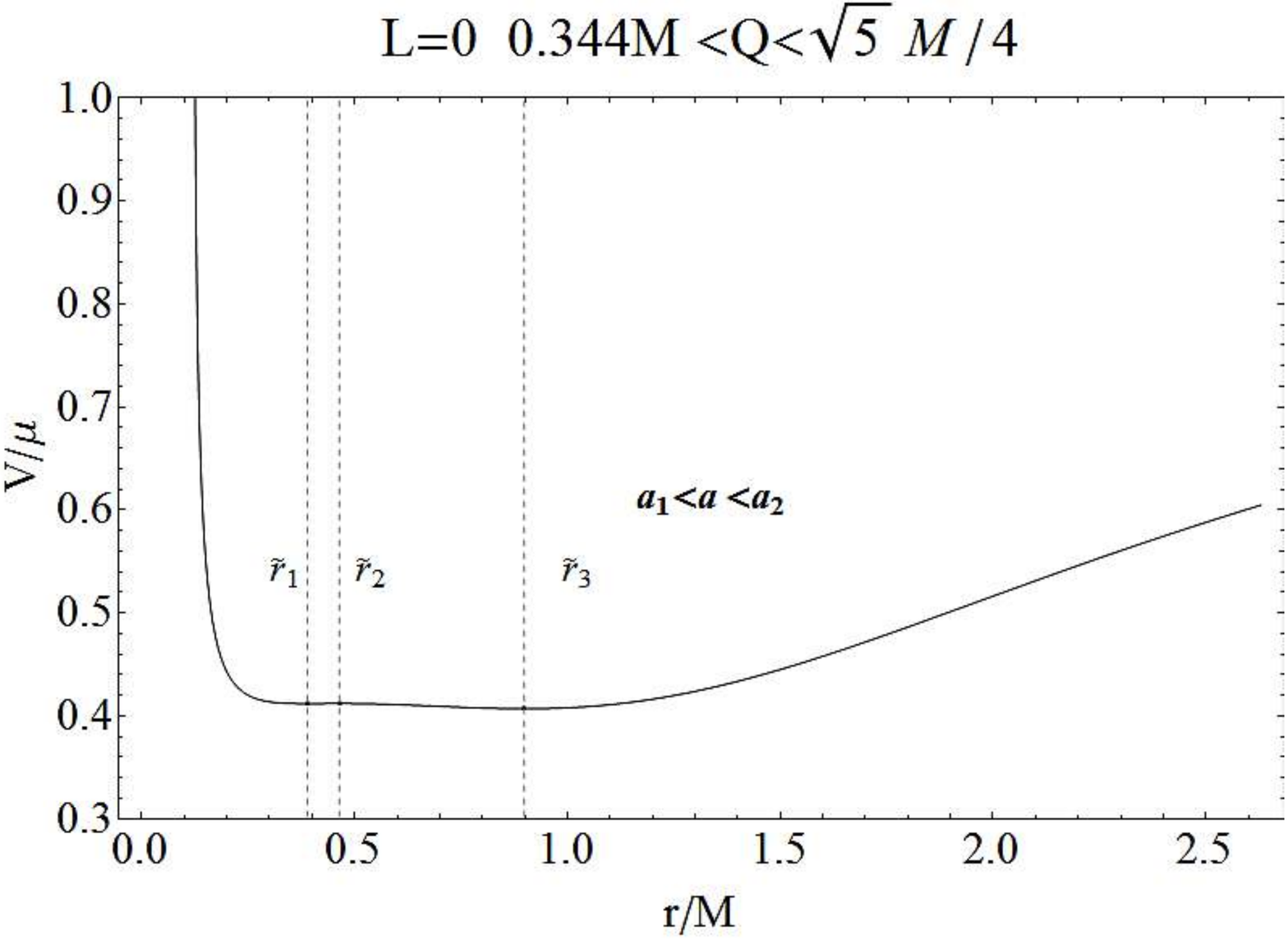}
\includegraphics[width=0.3\hsize,clip]{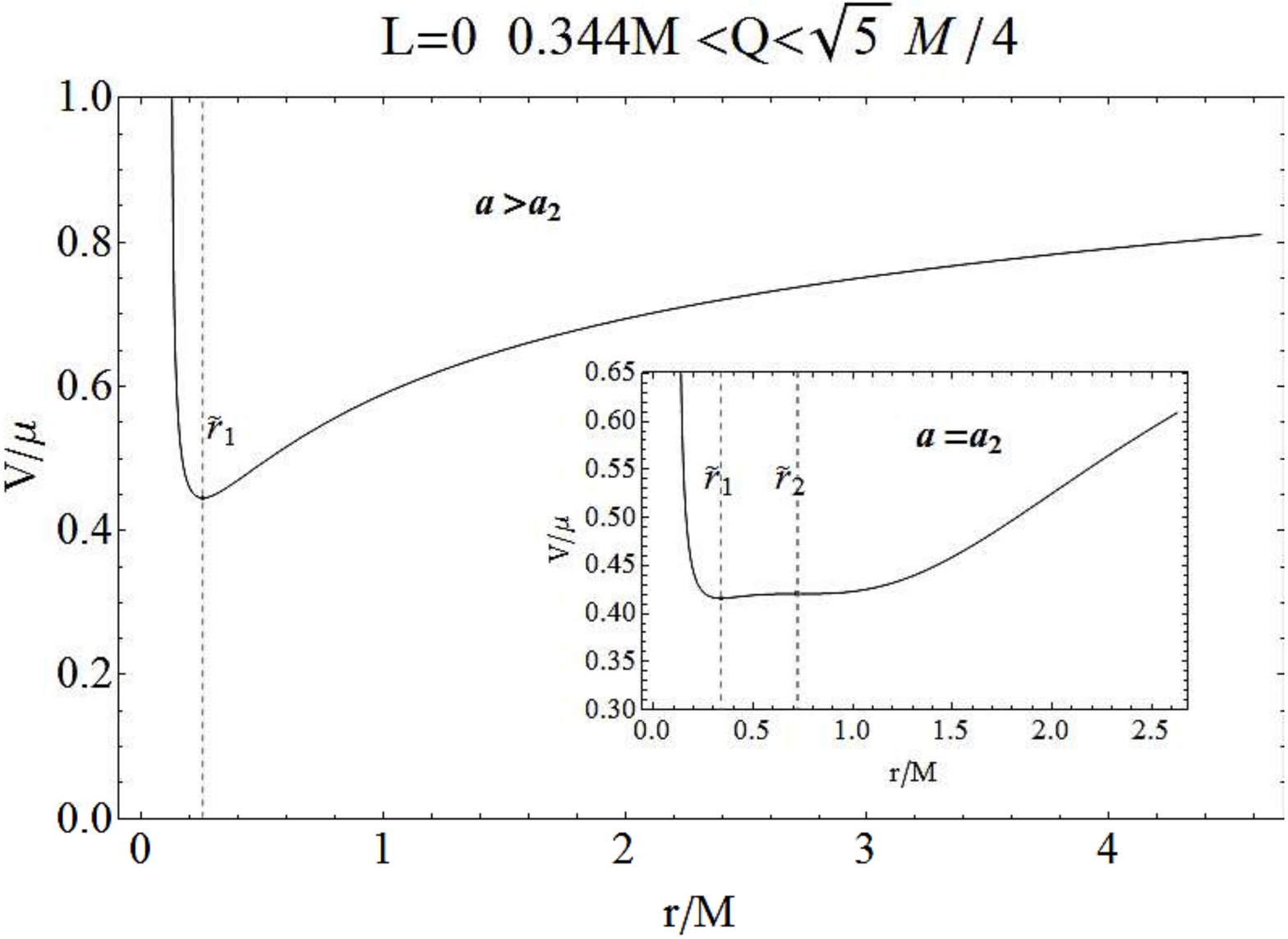}
\end{tabular}
\caption[font={footnotesize,it}]{\footnotesize{{The effective potential $V/\mu$ for a neutral test particle  with angular momentum $L=0$ in the field of a  Kerr--Newman naked singularity with different values of $Q/M$ and $a/M$. See also Table\il\ref{Tab:di_gno}. We also plot the radii $\tilde{r}_1$, $\tilde{r}_2$ and $\tilde{r}_3$ (dashed gray line).
}}} \label{Fig:P0aqrfj2}
\end{figure}

Moreover, for $Q/M<\sqrt{5}/4$ there are several radii  satisfying the conditions \il\ref{E:cro_cee_dl}.
This  is  a feature of the Kerr naked singularity, $Q=0$, as well as of the KN singularity.
However, in the KN case the situation is much more complex, and  the possibility appears to have three radii $(\tilde{r}_1, \tilde{r}_2, \tilde{r}_3)$, but only in the range $a<a_2$.
In Figs.\il\ref{Fig:P0aqrfj2} we show some specific examples of these cases. This analysis is also useful to show that eventually a  minimum appears, at which these peculiar conditions are satisfied,
 that corresponds to the interesting case of particle at rest $(L = 0)$ on a stability point.

%% file: NS_QmaggM13.tex
\subsection{Circular motion and stability problem}\label{SubSec:NS-stab}
\begin{figure}[h!]
\begin{tabular}{ccc}
\includegraphics[width=0.3\hsize,clip]{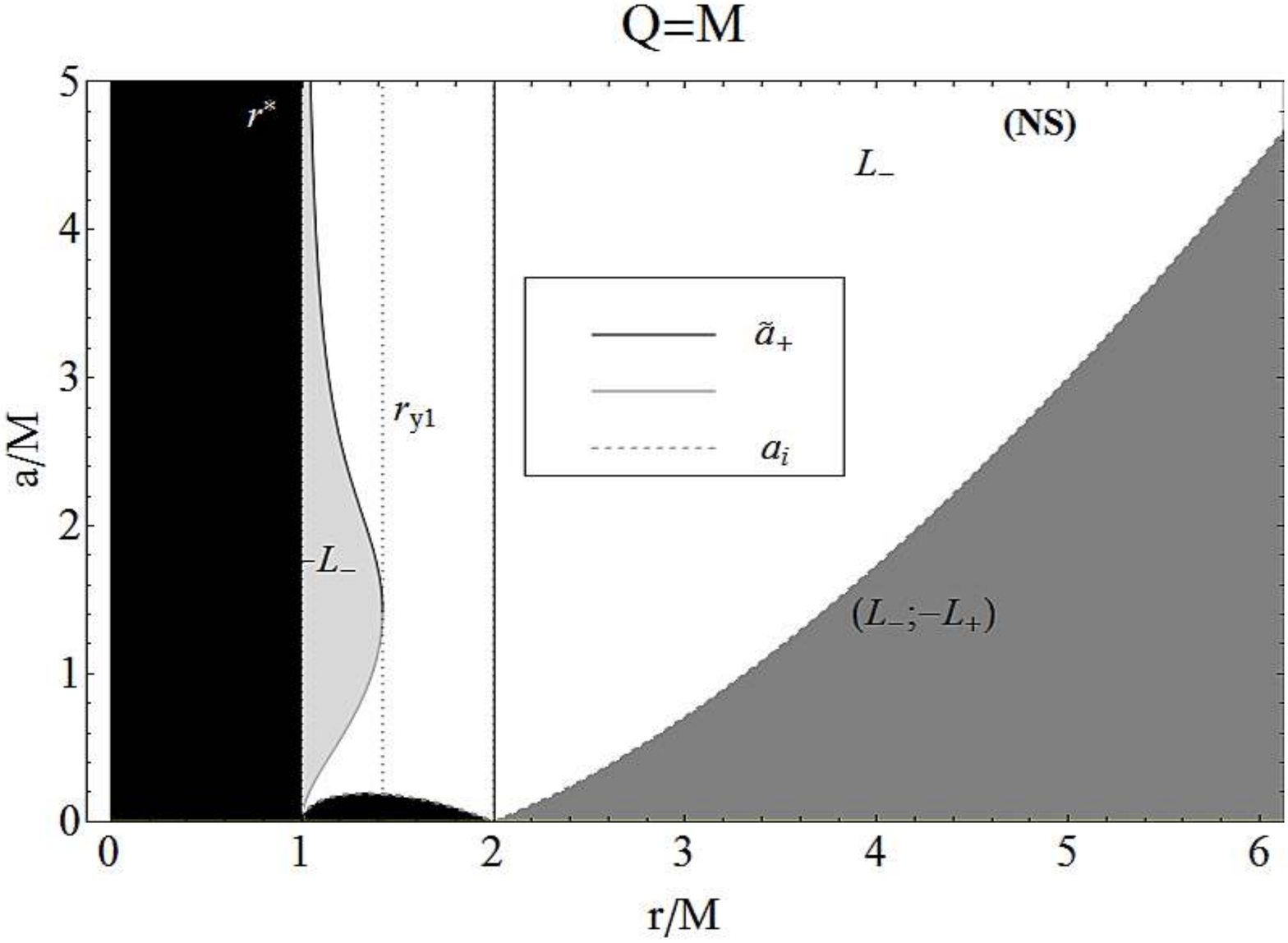}&
\includegraphics[width=0.3\hsize,clip]{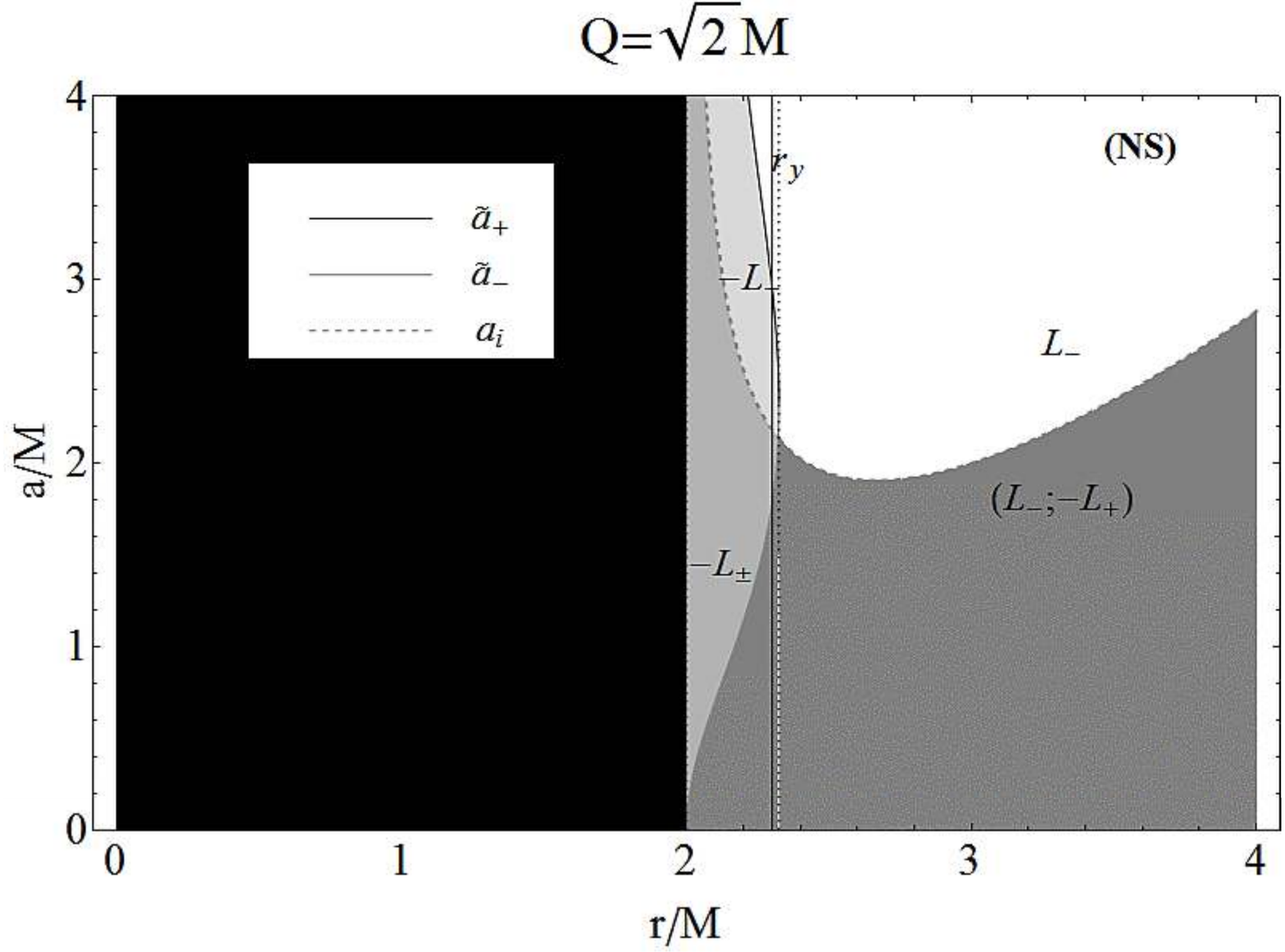}&
\includegraphics[width=0.3\hsize,clip]{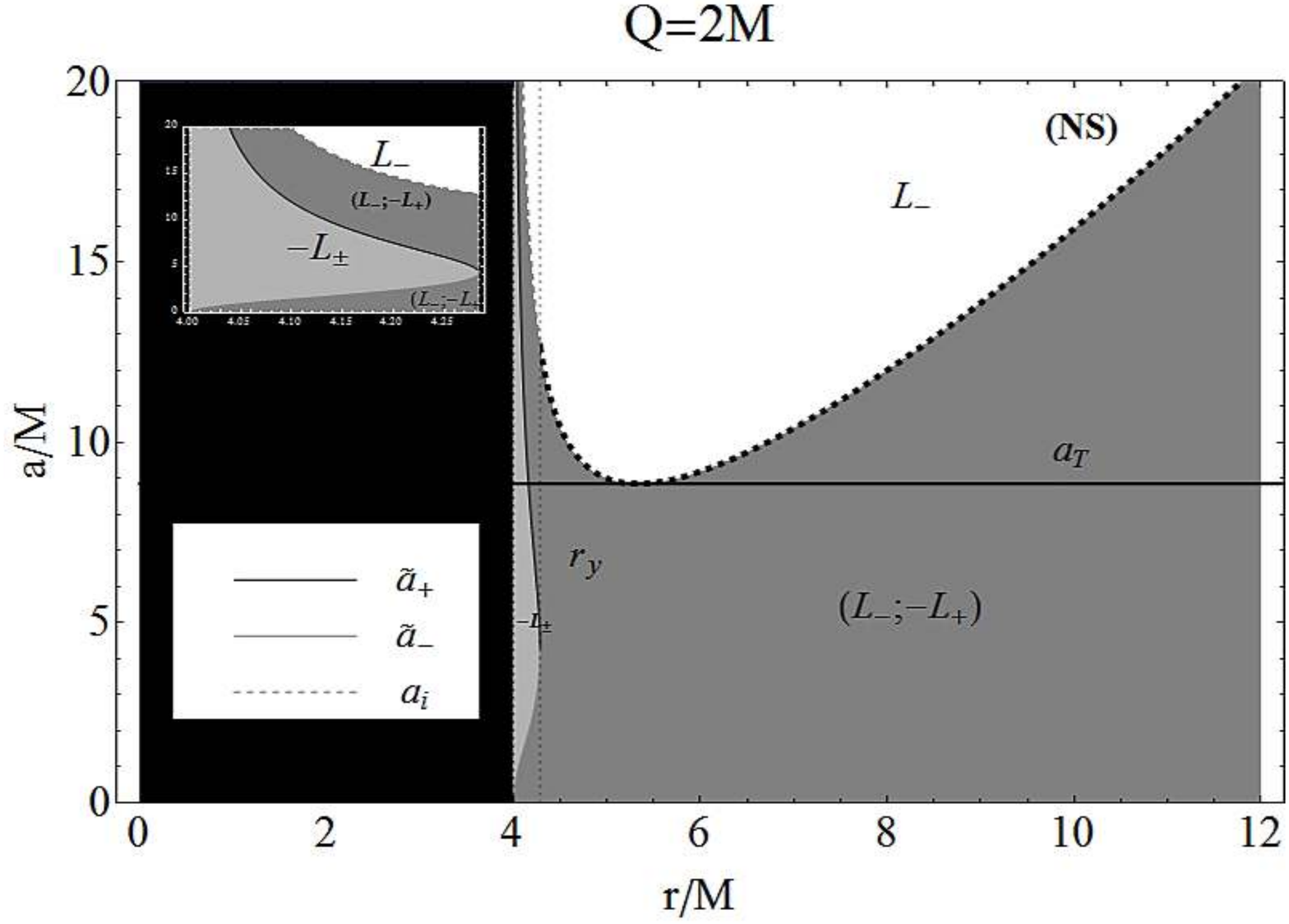}
\end{tabular}
\caption[font={footnotesize,it}]{\footnotesize{
 Plots of the intrinsic KN angular momentum $\tilde{a}_-/M$ (gray curve),  $\tilde{a}_+/M$ (black curve)  and  $a_i/M$ (dashed gray  curve)  as functions of $r/M$, for increasing values of the charge-to-mass ratio $Q/M$. The angular momentum of the orbiting test particle is shown in each region. Black regions are forbidden: circular motion is not possible there. Dotted lines represent the the radii  $r_*$, $r_{\gamma}^{\pm}$ and $r_{y}$}}\label{Fig:3_NS_circ}
\end{figure}

In this section, we focus on  general  circular motion   with $L\neq0$.
First, it turns out to be convenient to analyze separately the case
 $Q>M$, in which the  spacetime metric  has  a singularity not covered by the  horizon for any value of the spin $a/M$,
 and the case $Q\leq M$, in which  the source  spin  must comply with the naked singularity constraint $a>a_s$.
  In order to simplify the discussion it is useful to analyze first the case $\mathbf{Q>\sqrt{3}M}$;
in fact from Figs.\il\ref{Fig:3_NS_circ} and Figs.\il\ref{bicchic1due} -- which will be discussed below --we expect for $Q\geq\sqrt{3}M$ a forbidden region only for $r <r_*$,
 and  only three orbital regions namely (we use a classification similar to that introduced in Sec.\il\ref{Sec:BH_KN}):
 region \textbf{I}  characterized by $L=L_-$, region \textbf{II} with $L=(L_-,-L_+)$, and region  \textbf{III} with $L=-L_{\pm}$.
 A  fourth orbital region \textbf{IV} with $L=-L_-$  appears for $M\leq Q<\sqrt{3}M$. The characterization of circular motion  for these values of charge-to-mass ratios is considerably more complex and, therefore, requires a different approach.

Thus for $Q>\sqrt{3}M$ and  ${0<a<a_{\ti{T}}}$ circular orbits exist within the interval  $r_*<r\leq\tilde{r}$.
The explicit value of $\tilde r $ depends on the ratios $a/M$ and $Q/M$ (see Fig.\il\ref{Fig:3_NS_circ}).
It can be shown that  in this
interval, the orbital angular momentum is $L=- L_{\pm}$. On the other hand, for $r>\tilde{r}$ there exist circular orbits with both
$L=L_{-}$ and $L=-L_+$. For sufficiently large values of $Q/M$,  all these circular orbits are stable. The last stable circular orbit ($r_{lsco}$) is, therefore, located at $r=r_*$, where the particle angular momentum and the energy are
\be
L_*/(M\mu)=-\sqrt{\frac{a^2M^2}{Q^2 \left(Q^2-M^2\right)}},
\quad
E_*/\mu=\sqrt{\frac{Q^2-M^2}{Q^2}},
\label{lsco}
\ee
respectively.
Note that this was exactly the case for $a=0$, in which for  $Q/M>\sqrt{5}/2$  all the orbits with $r>r_*$ were stable, while for  lower values of $Q/M$ we had a closed region of instability.
The energy is always positive and independent of the value of the intrinsic angular momentum of the gravitational source;  it approaches the value $E_*\rightarrow \mu$ for large values of $Q$ (see Figs.\il\ref{Fig:3_NS_circ}).

For $\mathbf{a>a_{\ti{T}}}$ circular orbits exist at $r=r_*$ with angular momentum and energy as given in Eq.(\ref{lsco}).
For the remaining
radial distances there are several possibilities. In fact, in the interval
$r_*<r\leq \tilde{r}$, the orbits have $L=-L_{\pm}$, in $\tilde{r}<r<r_3$ the  orbital angular momentum can be either $L=-L_{+}$ or $L=L_-$,
in $r_3\leq r\leq r_4$ it is $L=L_{-}$, and, finally, for $r>r_4$ there are circular orbits with $L=L_-$ and $L=-L_+$.
The situation for $Q>\sqrt{3}M$ is summarized in Table\il\ref{Tab:dff_kn}.
\begin{table}[h!]
\caption{\footnotesize{\label{Tab:dff_kn}Description of the circular orbits for a test particle in a KN naked singularity with $Q>\sqrt{3}M$.
}}
\begin{ruledtabular}
\resizebox{.8\textwidth}{!}{%
\begin{tabular}{ccc}
&Case: $Q>\sqrt{3}M$& \\
&Region&Angular momentum
\\
\hline
&$a>a_{\ti{T}}$  &
\\
& $]Q^2/M,\tilde{r}]$  &$-L_{\pm}$
\\
& $]\tilde{r},r_{3}[$ &($L_-$, $-L_+$)
\\
& $]r_{3},r_{4}]$ &$L_-$
\\
& $]r_{4},\infty[$  &($L_-$, $-L_+$)
\\
&$0<a<{a_{\ti{T}}}$  &
\\
& $]Q^2/M,\tilde{r}]$  &$-L_{\pm}$
\\
& $]\tilde{r},\infty[$ &$(L_-,-L_+$)
\\
\end{tabular}}
\end{ruledtabular}
\end{table}

The case  $\mathbf{Q<\sqrt{3}M}$ requires a different analysis because there are some important details in the dynamics around  naked singularities that must be emphasized. In Figs.\il\ref{Fig:3_NS_circ},
(for $Q>M$) and Figs.\il\ref{Fig:3_NS_litt} (for $\mathbf{Q<\sqrt{3}/2M}$),
we plot the orbital radius in terms of the intrinsic angular momentum $a/M$ for different values of $Q/M$.
For a better presentation, we have ``rotated'' the figures, plotting  the inverse of the function $a = a(r)$,
and using the definition  $a_i$ in Eq.\il(\ref{Eq:Def:a_i}).
We want to emphasize that in the plane $(r,a)$, these curves coincide with $\tilde{r}$ so that the correspondence between the results in Table\il\ref{Tab:dff_kn} and Figs.\il\ref{Fig:3_NS_circ} does not give rise to confusion.
Thus the highlighted regions in Figs.\il\ref{Fig:3_NS_circ} are the orbital regions with fixed $Q/M$ for different values of the spin.
 In all the cases,   the  region $r<r_*$ is forbidden.
 For $\mathbf{Q\geq\sqrt{3}M}$
 the situation is rather simple: As one can see in Figs.\il\ref{Fig:3_NS_circ},  and it is also shown with details in Table\il\ref{Tab:dff_kn}, there are three  allowed orbital regions which we have denoted as follows:
 \textbf{I}  characterized by $L=L_-$,
 \textbf{II} with $L=(L_-,-L_+)$, and \textbf{III} with $L=-L_{\pm}$.
Using this classification, we can  interpret the results in Table\il\ref{Tab:dff_kn}
as follows: The region \textbf{III} has a finite, bounded spatial extension that reaches a maximum $r_{max}$  at a given value of the spin  $a_{max}$, and then decreases as the spin of the source increases. The values $(r_{max},a_{max})$  depend in general on the source charge-to-mass ratio  $Q/M$. The size of this region increases with  increasing  $Q/M$.
 The regions \textbf{I} and  \textbf{II}, instead, have an infinite extension.

The situation for $\mathbf{M<Q<\sqrt{3}M}$ is much more complex and  in addition to the three orbital regions, which characterize the motion in  $Q\geq\sqrt{3}M$,  there is now a fourth orbital region \textbf{IV} with $ L=-L_-$.
 The size of the orbital region \textbf{IV}  increases as  $Q/M$ decreases. Instead, the region \textbf{III}  decreases with decreasing $Q/M$ until it disappears in the limit $Q=M$ (see Fig.\il\ref{Fig:3_NS_circ}).
The orbital region \textbf{IV}  increases  until it reaches a maximum and then decreases. The figure highlights the presence of a further type of  naked singularity for   $Q^2<(9/8)M^2$; in fact, for $Q^2=(9/8)M^2$, an additional banned orbital zone  appears  for  $a<a_i$ and $r<r_{\gamma}^+$. This zone increases as  $Q/M$ increases, and for  fixed values of $a/M$ and $Q/M$ it is localized in a finite spatial region. Notice that at  $r>r_{\gamma}^+$  there are orbits for spin $a<a_i$  with $(L_-,-L_+)$, that is, there exists  a  region of type \textbf{II}.
In Table\il\ref{Tab:tuTt_nAs}, we include the different cases.

\begin{table}[h!]
\caption{\label{Tab:tuTt_nAs}\footnotesize{Description of the circular orbits for a test particle in a Kerr--Newman spacetime. At $r=r_{y}$  it is $\tilde{a}_+=\tilde{a}_-$.  In the naked singularity case for $Q<M$, it is always $a>a_s$.  And $a_i<a_s$ in $r\in]r_d^-,r_d^+[$, whereas $a_i=a_s$ for $r_d^{\pm}$ and $r=M$. See also Appendix\il\ref{Sec:appendix_spin_radius} and Figs.\il\ref{bicchic1due}.}}
\begin{ruledtabular}
\begin{tabular}{|lr|lr|lr|lr|}
\multicolumn{2}{c}{$0<Q<M$ }\vline& \multicolumn{2}{c}{\textbf{{NS-III: }} $M<Q<\sqrt{9/8}M$}\vline& \multicolumn{2}{c}{\textbf{{NS-II: }}  $\sqrt{9/8}M<Q<\sqrt{3}M$}\vline& \multicolumn{2}{c}{\textbf{{NS-I: }}  $Q>\sqrt{3}M$ }
\\
\hline\hline
\multicolumn{2}{c}{$r_*<r<r_y$ }\vline& \multicolumn{2}{c}{$r_*<r<r_{\rho}$ }\vline&  \multicolumn{4}{c}{$r_*<r<r_y$ }
\\
 \hline\hline
$a_i<a<\tilde{a}_-$&$L_-$  & $0<a<\tilde{a}_-$&$(L_-,-L_+)$& $0<a<\tilde{a}_-$&$(L_-,-L_+)$
& $0<a<\tilde{a}_-$&$(L_-,-L_+)$
\\
$\tilde{a}_-<a<\tilde{a}_+$  &$-L_-$& $\tilde{a}_-<a<a_i$ &$-L_{\pm}$& $a_i<a<\tilde{a}_-$ &$-L_{\pm}$
&$\tilde{a}_-<a<\tilde{a}_+$ &$-L_{\pm}$
\\
$a>\tilde{a}_+$  &$L_-$&$a_i<a<\tilde{a}_+$ &$-L_-$ &  $\tilde{a}_+<a<a_i$ &$-L_-$  &$\tilde{a}_+<a<a_i$ &$(L_-,-L_+)$
\\
&&$a>\tilde{a}_+$&$L_-$&$a>\tilde{a}_+$&$L_-$&$a>a_i$&$L_-$
\\
\hline\hline
  \multicolumn{2}{c}{$r_y<r<r_{\gamma}^+$ }\vline& \multicolumn{2}{c}{$r_{\rho}<r<r_{\gamma}^-$ } \vline& \multicolumn{4}{c}{$r>r_y$ }
  \\
 \hline\hline
$a>a_i$ & $L_-$&$0<a<a_i$&$(L_-,-L_+)$&$0<a<a_i$&$(L_-,-L_+)$&$0<a<a_i$&$(L_-,-L_+)$\\
& &$a_i<a<\tilde{a}_-$&$L_-$&$a>a_i$ & $L_-$&$a>a_i$ & $L_-$
\\
& &$\tilde{a}_-<a<\tilde{a}_+$&$-L_-$&$ $ & $ $&$ $ & $ $
\\
& &$a>\tilde{a}^+$&$L_-$&$ $ & $ $&$ $ & $ $
\\
\hline\hline
 \multicolumn{2}{c}{$r>r_{\gamma}^+$ }\vline& \multicolumn{2}{c}{$r_{\gamma}^-<r<r_y$ }\vline
 \\
 \cline{1-4}\cline{1-4}
 $0<a<a_i$ & $(L_-,-L_+)$&$a_i<a<\tilde{a}_-$ & $L_-$ \\
  $a>a_i$ & $L_-$&$\tilde{a}_-<a<\tilde{a}_+$&$-L_-$
  \\
& $ $ & $a>\tilde{a}_+$&$L_-$
\\
 \cline{1-4}\cline{1-4}
 &&\multicolumn{2}{c}{$r>r_{\gamma}^+$ }
 \\
\cline{3-4}
 &  & $0<a<a_i$ & $(L_-,-L_+)$
 \\
  &&$a>a_i$ & $L_-$
 \\
\end{tabular}
\end{ruledtabular}
\end{table}

%% file: NS_QminM13.tex
Finally for $\mathbf{{Q<M}}$, and $a>a_{s}$ (condition of existence of a naked singularity with $Q <M$, Eq.\il(\ref{Eq:as_intro}),  the analysis  is schematically summarized   in Figs.\il\ref{Fig:3_NS_litt}. Some examples are discussed  in Appendix.\il\ref{App:QmM}.
\begin{figure}[h!]
\begin{tabular}{ccc}
\includegraphics[width=0.3\hsize,clip]{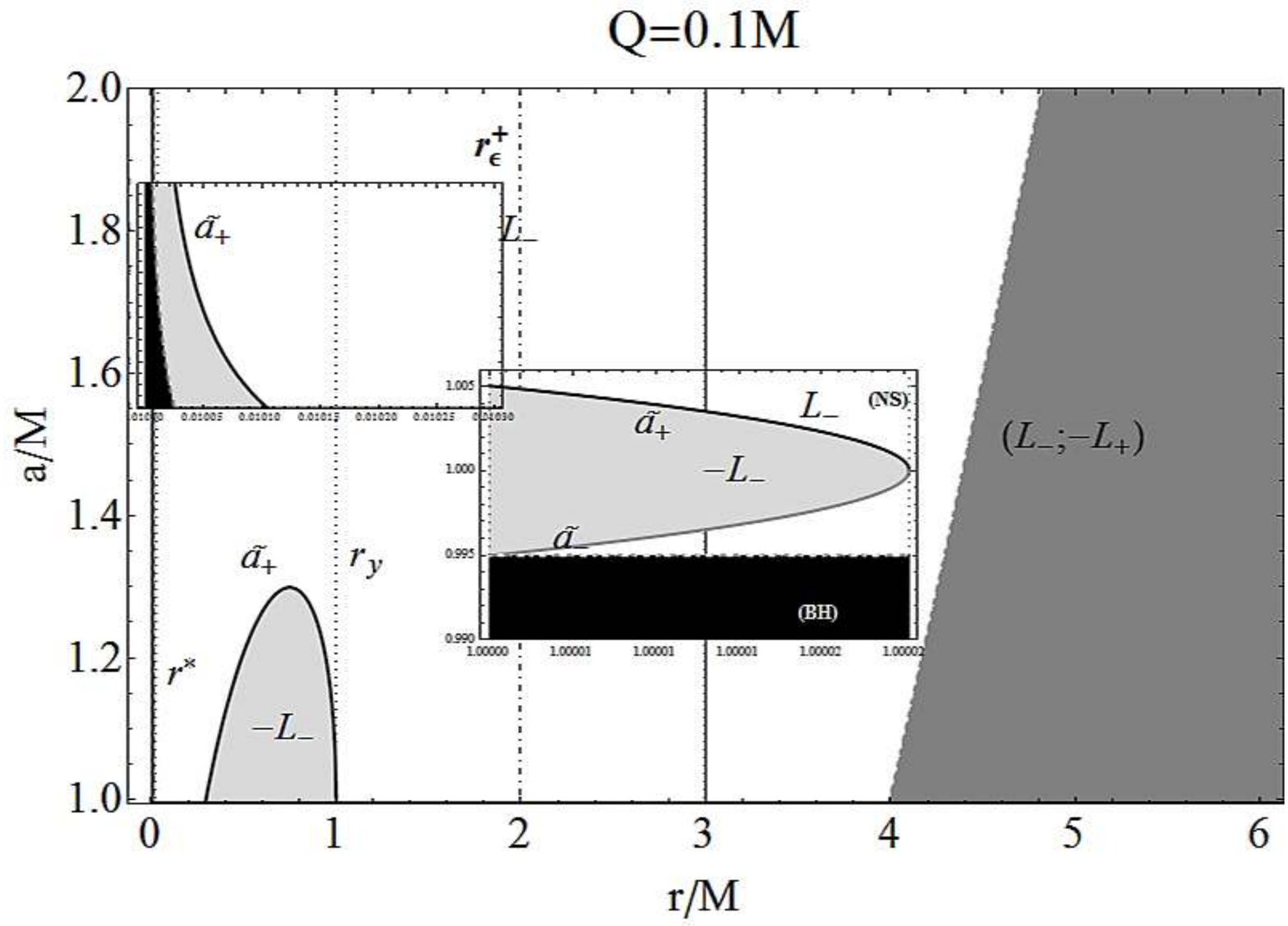}
\includegraphics[width=0.3\hsize,clip]{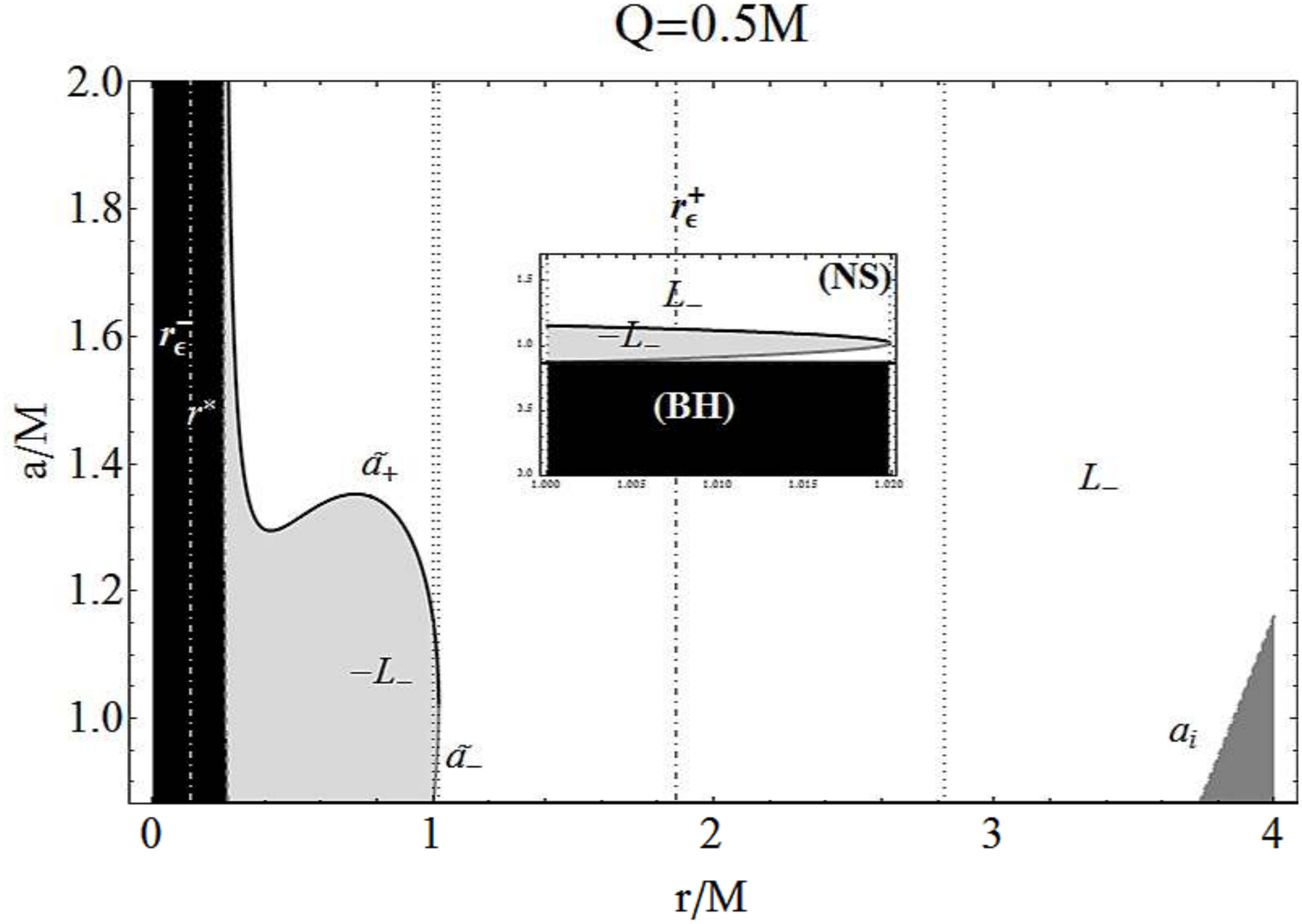}
\includegraphics[width=0.3\hsize,clip]{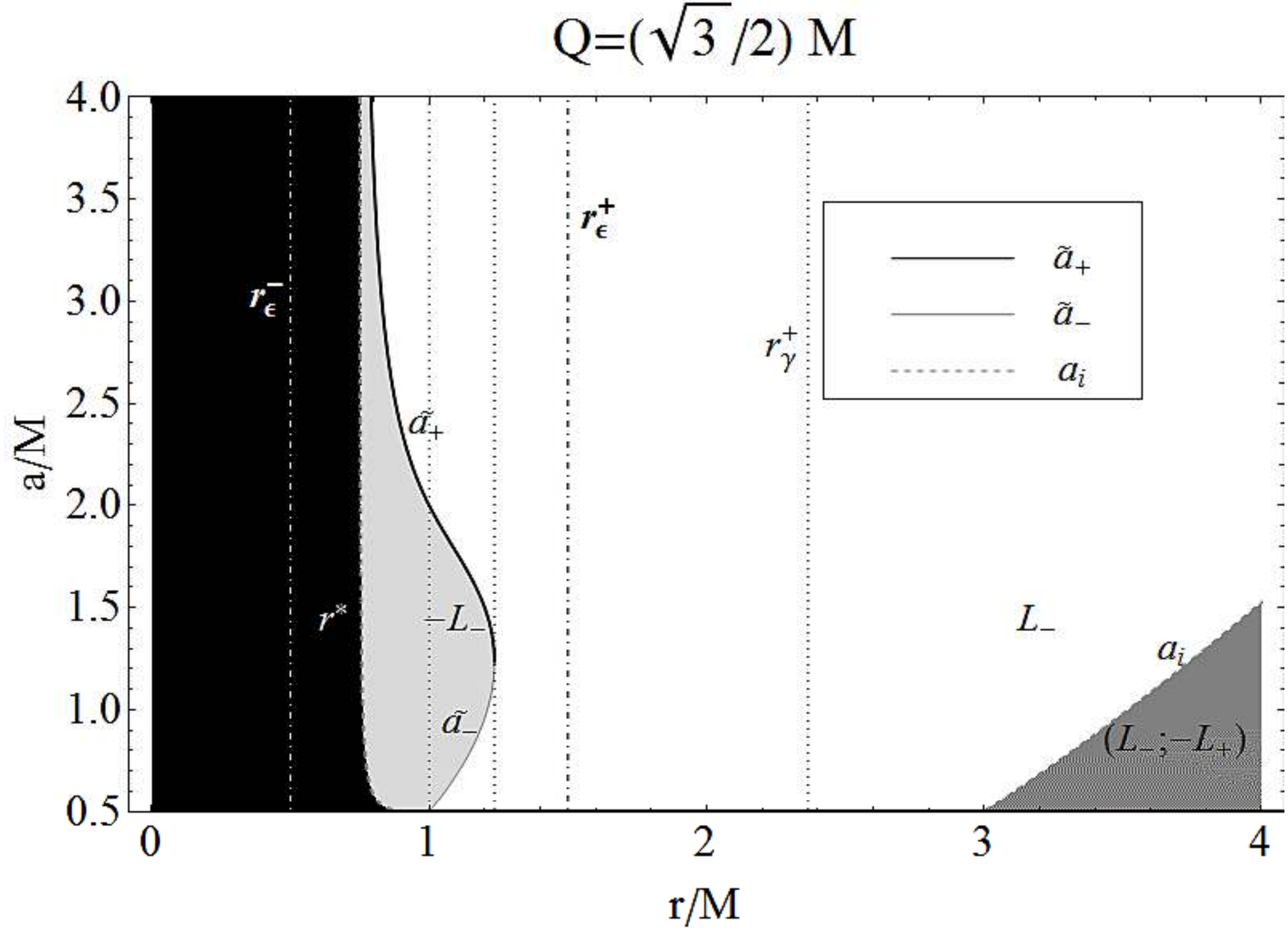}
\end{tabular}
\caption[font={footnotesize,it}]{\footnotesize{
Plots of the intrinsic KN angular momentum $\tilde{a}_-/M$ (gray curve),  $\tilde{a}_+/M$ (black curve)  and  $a_i/M$ (dashed gray  curve)  as functions of $r/M$, for increasing values of the charge-to-mass ratio $Q/M=0.1,0.5,0.866025$.  The region  $a>a_s$ is explored:   a naked singularity occurs for $Q<M$ and $a>a_s$. The angular momentum of the orbiting test particle is shown in each region. Black region $r<r_*$ is forbidden; no circular motion is possible there. The dotted-dashed line is the outer ergosphere  $r_{\epsilon}^+$.
} }\label{Fig:3_NS_litt}
\end{figure}
We note that the region \textbf{IV}, with $L= -L_-$, has a very complex structure   and, in contrast to  the case.
$M\leq Q<\sqrt{3}M$, it not only possesses a maximum  at $r_{max}$, but also a minimum $r_{min}$  that appears as the charge decreases.
This situation is illustrated for $Q=1/2M$ in the central plot of   Fig.\il\ref{Fig:3_NS_litt}. We see that, for a suitable range of values of the source angular momentum, as the radial distance increases there appear two disconnected regions of type  \textbf{IV}. This situation becomes extreme with decreasing values  of $Q/M$; for instance, for $Q=1/10M$
the two regions are   completely separated. The  outer region, farthest from the singularity, has a
finite extension
and the inner region, closer to the singularity,  has an  infinite extension.

The radius at which the region \textbf{II}, with $(L_-,-L_+)$, starts, increases as the source charge-to-mass ratio $Q/M$ decreases.
This can be interpreted as a consequence of the fact that the effects due to the spin become increasingly predominant as the charge decreases, affecting the properties of the  counter-rotating orbits with $L=-L_+$. As a result, circular motion is no more allowed at very close to the central source where the effects of the gravitational stress due to the singularity are  expected to be very strong. This effect is corroborated by the fact that a similar behavior is found when the charge remains fixed and the spin of the source is increased.

It is important to notice that once the two parameters $(a/M, Q/M)$ are fixed within  in a suitable  interval, the  regions where circular orbits are allowed become disconnected. In particular, for $ Q <M $ and orbits
 sufficiently close to the singularity,  the regions
affected by this phenomenon are of the types  \textbf{IV} and \textbf{I}.

As a result of this analysis, we can identify four types of naked singularities according to the charge-to-mass ratio:

\begin{description}

\item[\textbf{NS-I}]: For $Q/M\geq\sqrt{3}$ with orbital regions of the type \textbf{I}, \textbf{II} and \textbf{III}

\item[\textbf{NS-II}]: For $3/2\sqrt{2}\leq Q/M<\sqrt{3}$ with orbital regions of the type \textbf{I}, \textbf{II}, \textbf{III},
\textbf{IV} and one forbidden region

\item[\textbf{NS-III}]: For $1< Q/M< 3/(2\sqrt{2}$ with orbital regions of the type \textbf{I}, \textbf{II}, \textbf{III},
\textbf{IV} and two forbidden regions

\item[\textbf{NS-IV}]: For $Q/M<1$ with orbital regions of the type \textbf{I}, \textbf{II} and \textbf{IV}

\end{description}

Notice that in naked singularities of the type \textbf{NS-IV}, it is necessary to consider the position of the ergosphere region $[r_{\epsilon}^-,r_{\epsilon}^+]$, because, as shown in Figs.\il\ref{Fig:3_NS_litt}, in this case $r<r_{\epsilon}^+$.

\begin{figure}[h!]
\begin{tabular}{ccc}
\includegraphics[width=0.3\hsize,clip]{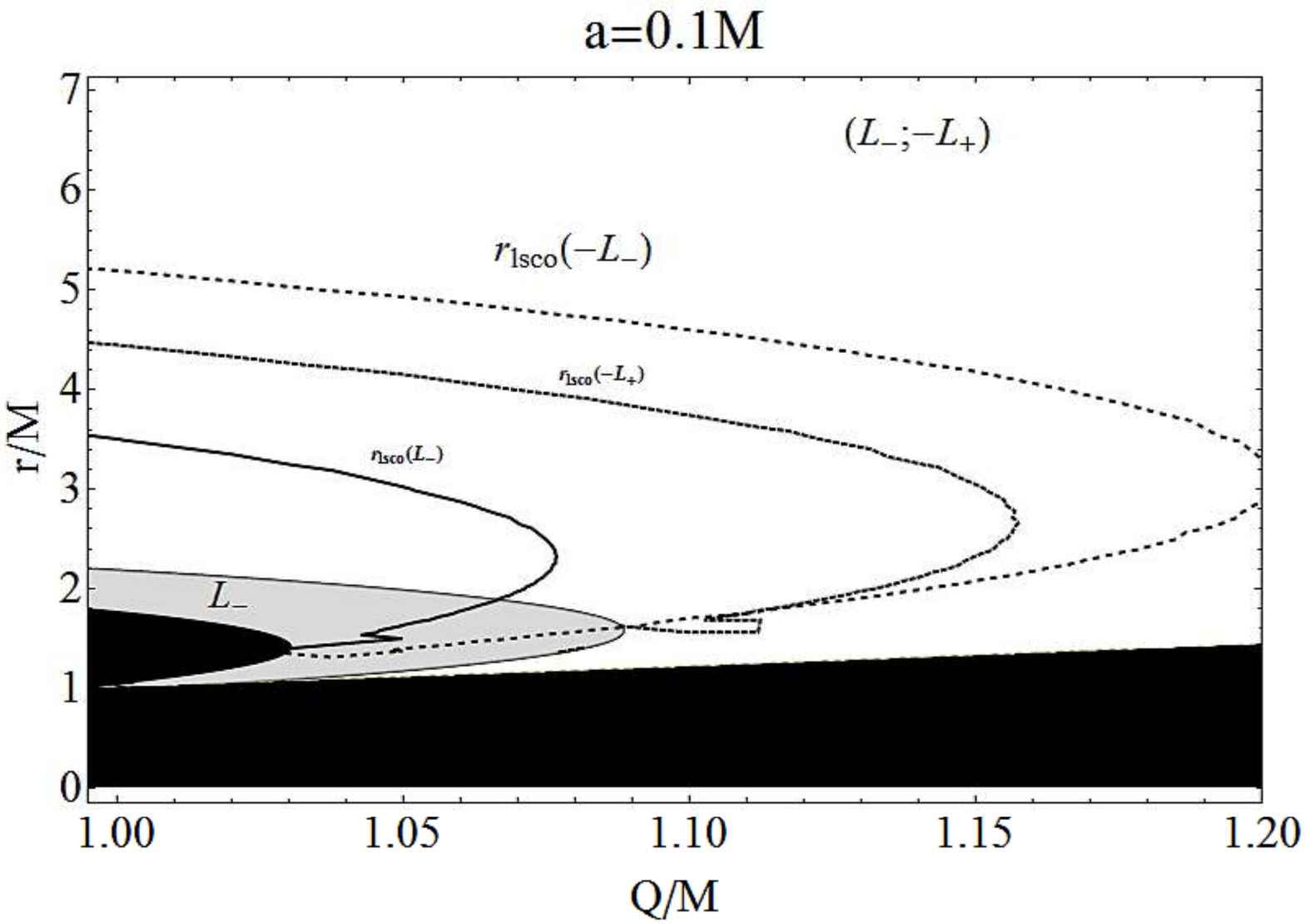}
\includegraphics[width=0.3\hsize,clip]{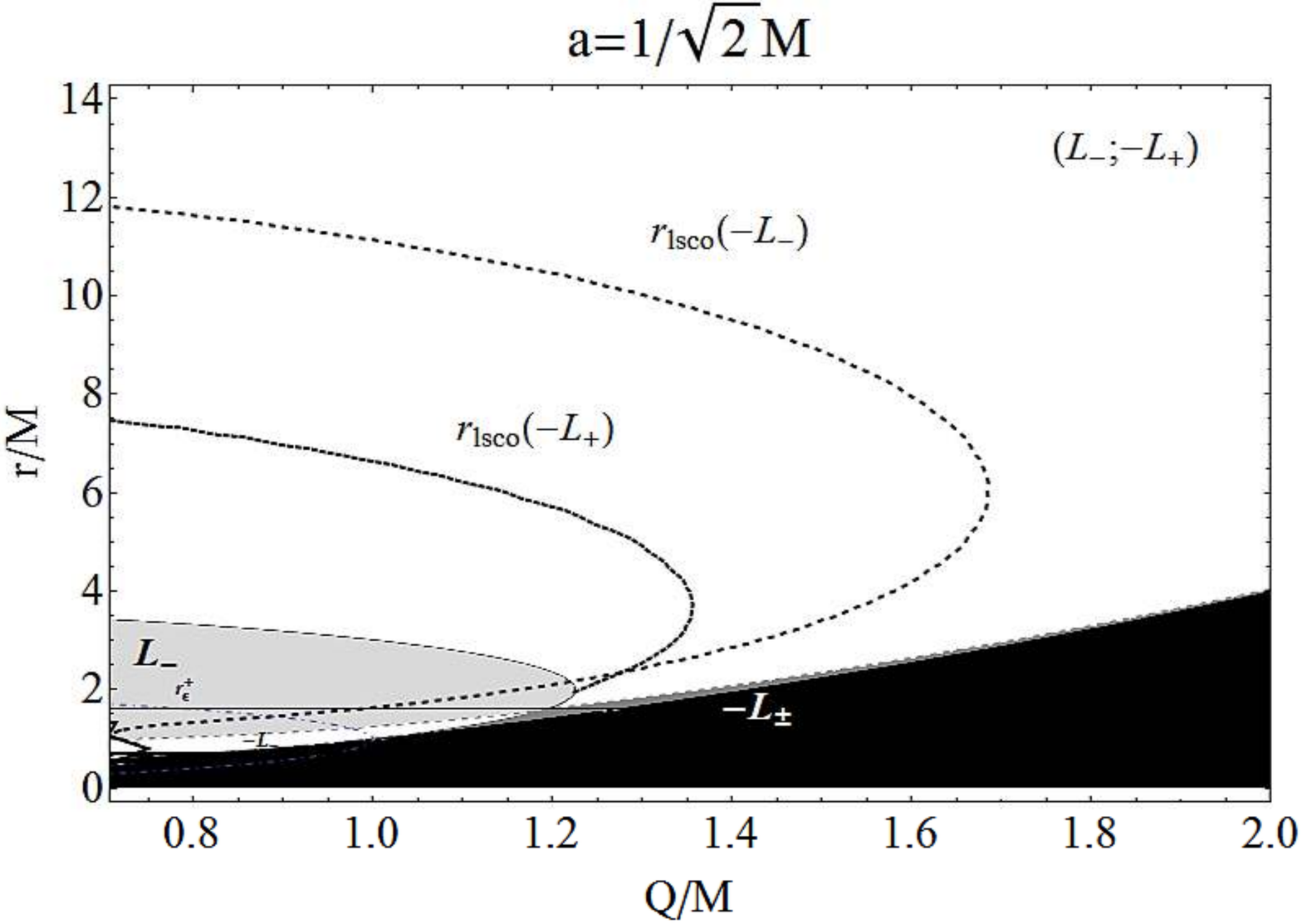}
\includegraphics[width=0.3\hsize,clip]{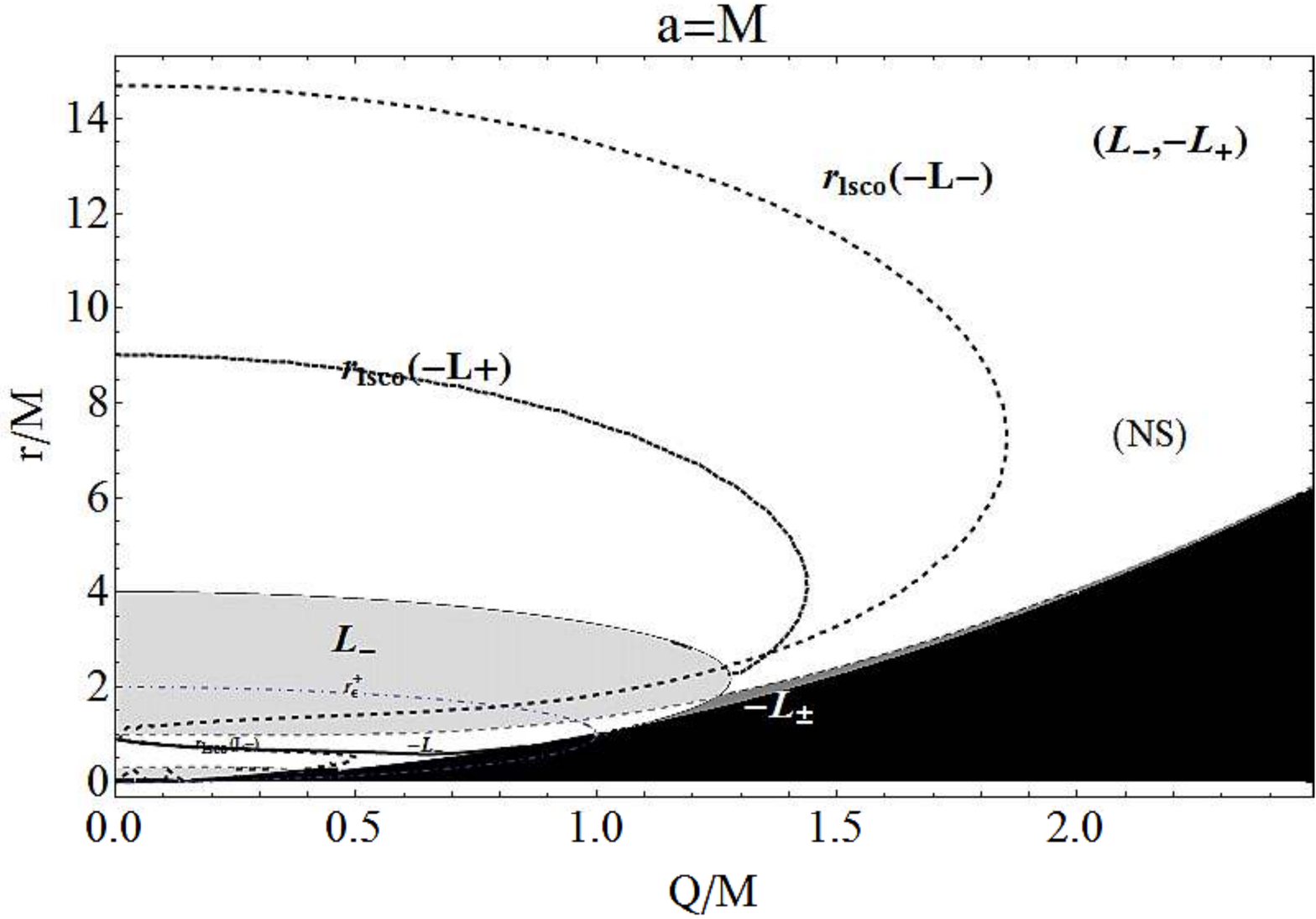}\\
\hspace{2cm}\includegraphics[width=0.35\hsize,clip]{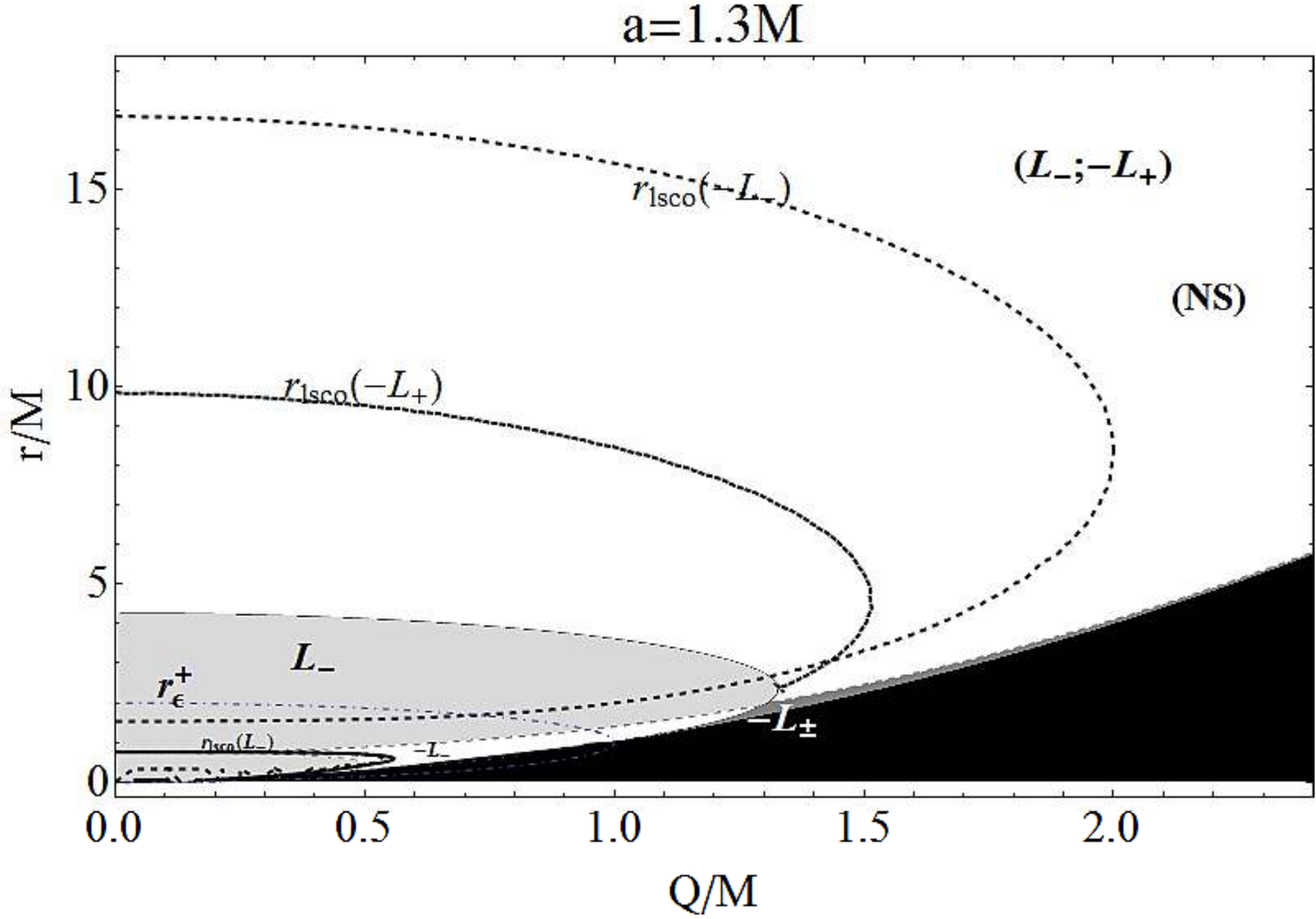}
\includegraphics[width=0.35\hsize,clip]{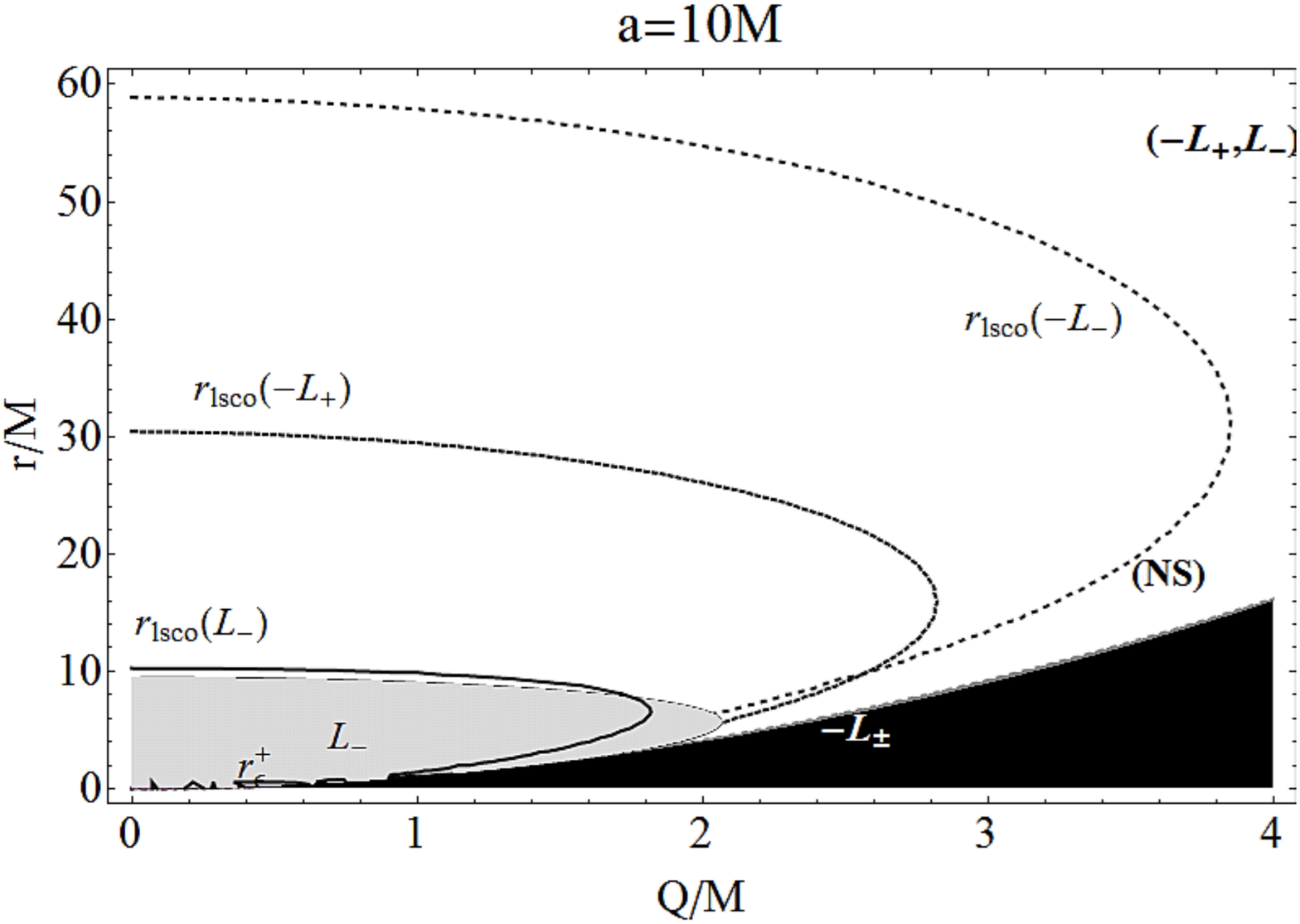}
\end{tabular}
\caption[font={footnotesize,it}]{\footnotesize{The charge-to-mass ratio   $Q_{\ti{T}}^{\pm}/M$, $Q_*/M$ and $Q_{c_2}$  as functions of $r/M$, for increasing values of the source spin. The angular momentum of the orbiting test particle is denoted in each region. Black regions are forbidden. The radius of the last stable circular motion $r_{lsco}$ is plotted: $r_{lsco}(L_-)$ (thick black curve),   $r_{lsco}(-L_-)$ (thick black dashed curve) and  $r_{lsco}(-L_+)$ (thick black  dotted curve). In the region $r<r_{lsco}$ ($r>r_{lsco}$), circular orbits are unstable (stable).
 A naked singularity occurs for $a>M$  and $Q>Q_s$; then, $Q_s=0.994987M$ for $a=0.1M$, $Q_s=0.707107M$ for $a=1/\sqrt{2}M$  and $Q_s=0$ for $a=M$. The dotted-dashed curve is the  ergosphere  boundary $r_{\epsilon}$.
}} \label{Fig:enolimNS_sopra}
\end{figure}

The results of studying the problem of stability are represented in Figs.\il\ref{Fig:enolimNS_sopra}.
The open curves shown in  Figs.\il\ref{Fig:xyz} for black holes, find a smooth continuation here in the naked singularity case, as illustrated in Figs.\ref{Fig:enolimNS_sopra}. As a result, for fixed values of $Q/M$, a finite region of instability appears that splits the plane into two regions of stability. The plots show a disconnected structure that resembles the structures found previously in the case of
RN \cite{Pu:Neutral} and Kerr \cite{Pu:Kerr} naked singularities.

Several features appear in the case of naked singularities. Firstly, there is a new region of  exclusively counter-rotating  orbits (type  \textbf{III}) that is practically always stable. Also, in general, the radii of the orbits $r_{lsco}$ intersect now at a point, thereby creating a complex structure which depends on the source spin and  charge. The region \textbf{I},  which is closer to the singularity  (say at $r\lesssim5M$),  presents the most complex structure so that by varying $a/M$,  we can have regions completely filled with either stable or unstable orbits or also partially filled with stable and unstable orbits.
In this regard, for  small enough spin (say approximately $a<10M$), the region  \textbf{I} splits   into two disconnected regions that become larger as $a/M$ decrease, until they become entirely separated by a type \textbf{IV} region ($L=-L_-$), as illustrated in
Figs.\il\ref{Fig:enolim}. This region appears to be always unstable and is bounded by the curve  $r_{lsco}(L_-)$.  Then, the region of instability, which increases with increasing $a/M$, will eventually cover at $a=10M$ the single type \textbf{I} region,
which therefore results in an almost complete island of instability.
For sufficiently large values of $a/M$,  the radius $r_{lsco}(L_-)>r_{\epsilon}^+$.

In conclusion,  the existence, extent and stability properties of the region  \textbf{I} is mainly determined by the spin of the source.

%% file: CFRT_BH_NS_13.tex
\section{Black holes vs naked singularities}\label{Sec:BH-vs-NS}
This section summarizes the main results concerning the circular motion and stability properties obtained for the black hole case  in Sec.\il\ref{Sec:BH_KN} and for  the naked singularity case in Sec.\il\ref{Sec:NS_Case}.  We show a benchmarking  and discuss   the key differences between these  two cases.
We start by considering the extension of  the stability analysis performed in Sec.\il\ref{Sec:BH_KN} and illustrated in Figs.\il\ref{Fig:xyz}   to include the naked singularity region.
\begin{figure}[h!]
\begin{tabular}{cc}
\includegraphics[width=0.3\hsize,clip]{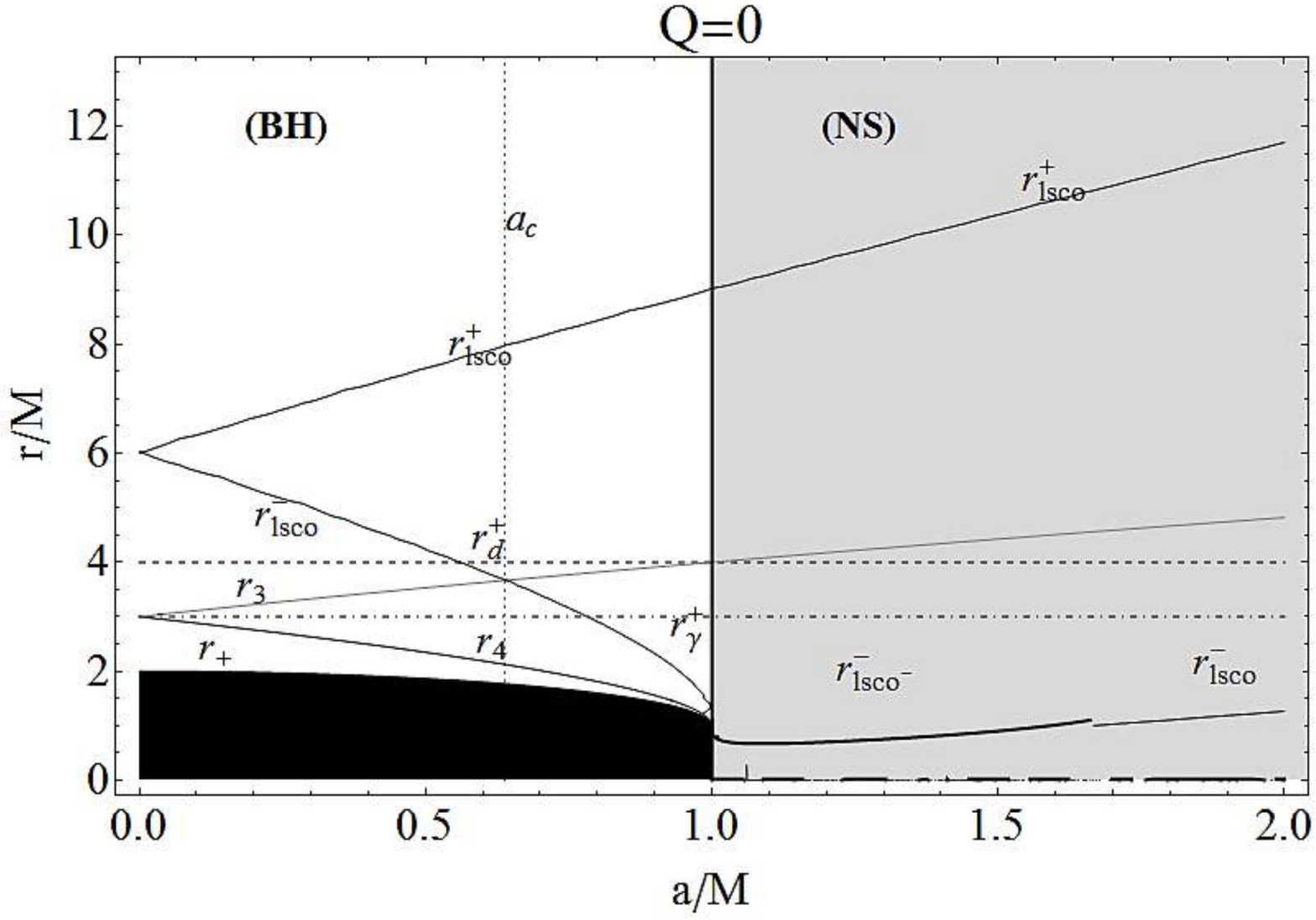}&
\includegraphics[width=0.3\hsize,clip]{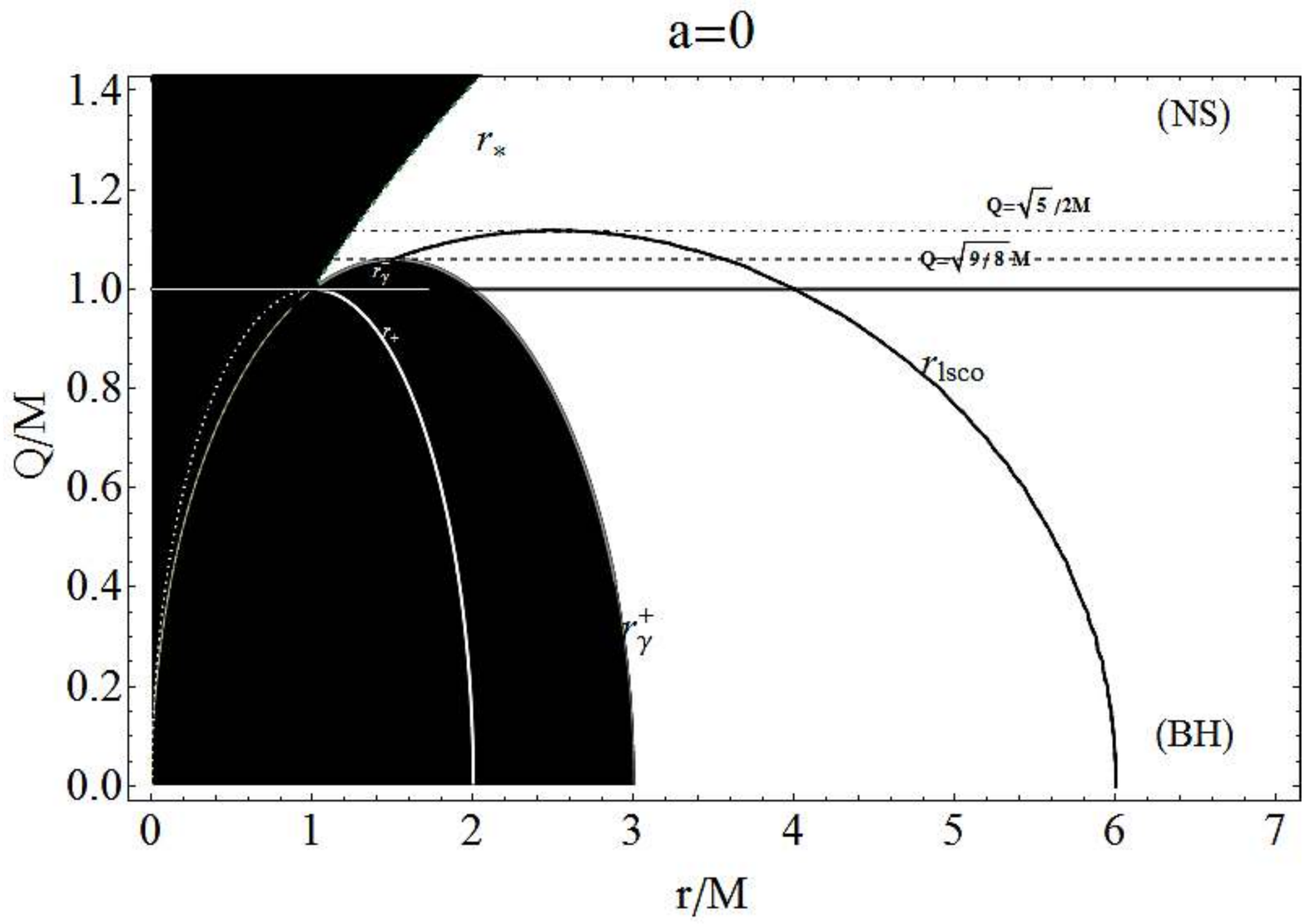}
%
\end{tabular}
\caption[font={footnotesize,it}]{\footnotesize{Left: The radii $r^{+}_{lsco}\equiv r_{lsco}(-L_+)$, $r^{-}_{lsco}\equiv r_{lsco}(L_-)$, and $r^{-}_{lsco^-}\equiv r_{lsco}(-L_-)$ of the last stable circular orbit in the Kerr spacetime as functions of the source angular momentum $a$
The horizon  $r_+$ and the radii $r_{\gamma}^+$ (dotted line), $r_{d}^+$ (dashed line), $r_{3}$ (black curve), and $r_{4}$ (gray curve) are also plotted. The black region $(r<r_+)$ is forbidden. The value $a_c$, at the crossing point between $r_3$ and $r^{-}_{lsco}$, is represented by a dashed line. The gray banned regions correspond to the naked singularity case \textbf{(NS)}, the white ones to  the black hole case \textbf{(BH)}.
Right: The last stable circular orbit radius $r_{lsco}$  in the Reissner-Nordstr\"om spacetime for a black hole
$Q\leq M$ and a naked singularity $Q >M$ with different charge-to-mass ratios $Q/M$.
} } \label{Fig:xyzns}
\end{figure}
The left plot show
the case  $Q=0$ (Kerr spacetime)  discussed  in Sec.\il\ref{Sec:BH_KN}  and   extensively studied  in \cite{Pu:Kerr}. For all values of $a/M$,  the last stable circular orbit $r_{lsco}(-L_+)$ exists,  that is, for a  naked singularity  it is the continuous extension of the radius for the black hole case. On the other hand, the radius $r_{lsco}^-$, the last stable circular orbit in the black hole case with momentum $L=L_-$, does not extend to the values $a>M$. Then, in accordance with \cite{Pu:Kerr}, two additional radii determining the stability of circular motion appear: the radius  $r_{lsco}(-L_-)$ for $1<a/M<3\sqrt{3}/4$ and  $r_{lsco}(L_-)$ for  $a/M>3\sqrt{3}/4$.
The introduction of a source charge changes qualitatively the stability pattern. Firstly, the behavior of the two radii with domain only in the naked singularity region  drastically depends on the value of $Q/M$.
A similar behavior characterized  the case $a=0$. Indeed, as shown in the naked singularity case in Sec.\il\ref{Sec:BH_KN} for sufficiently small charge source $1<Q/M<\sqrt{9/8}$, there appears only one
$r_{lsco}$, while  in the range $1<Q/M<\sqrt{5/2}$ the  curve $r_{lsco}$  changes direction so that for $Q/M>\sqrt{5/2}$ all  allowed orbits are stable.

The stability analysis of the naked singularity case must be carried out in detail for orbital regions near the singularity.
We will do this by using the analysis performed in Sec.\il\ref{Sec:NS_Case} for a naked singularity, by extending the results
of  Figs.\il\ref{Fig:enolimNS_sopra} to the black hole case, and by matching Figs.\il\ref{Fig:xyz}-Right and Figs.\il\ref{Fig:enolimNS_sopra}. We will also use all the quantities introduced in Sec.\il\ref{Sec:circular_motion_BH+NS}.
Thus, the  analysis will be carried out in parallel in two alternative ways:
 \textbf{(1)} We reinterpret the analysis in Figs.\il\ref{Fig:xyzns}  by studying the function $r = r (a)$:
 by fixing a particular value of the charge-to-mass ratio, we vary the source spin in order to cover both the black hole and the naked singularity regions. However, it turns out to be convenient to use the inverse function $a= a (r) $ Figs.\il\ref{Fig:xyzns}.
 The obtained regions  are the same as in  Figs.\il\ref{Fig:xyzns}, but with a suitable enlargement of some regions close to the singularity. This is the generalization of  the work performed in \cite{Pu:Kerr} to include a charge-to-mass ratio. This analysis turns out to be very useful to understand the main difference between the allowed regions for the circular motion around the two sources.
\textbf{(2)} In the second analysis, we fix the source spin $a/M$  and vary the charge $Q/M$ to  study the orbits $r=r(Q)$.
This is a generalization of  the case  $a\neq0$ of the left plot of  Fig.\il\ref{Fig:xyzns} for the RN spacetime. This allows us to determine  how the rotation of the source affects the results of the plot. This is the generalization of the work performed in \cite{Pu:Neutral}
to take into account the spin-to-mass ratio. It turns out also in this case to be convenient to use the inverse function $Q=Q(r)$ in order to
explore the details of the orbital regions.
This comparative analysis is very important to study in detail the stability properties of the circular motion around black holes and naked singularities.
\subsection{Comparative analysis of the circular motion}
\begin{figure}[h!]
\begin{tabular}{ccc}
\includegraphics[width=0.3\hsize,clip]{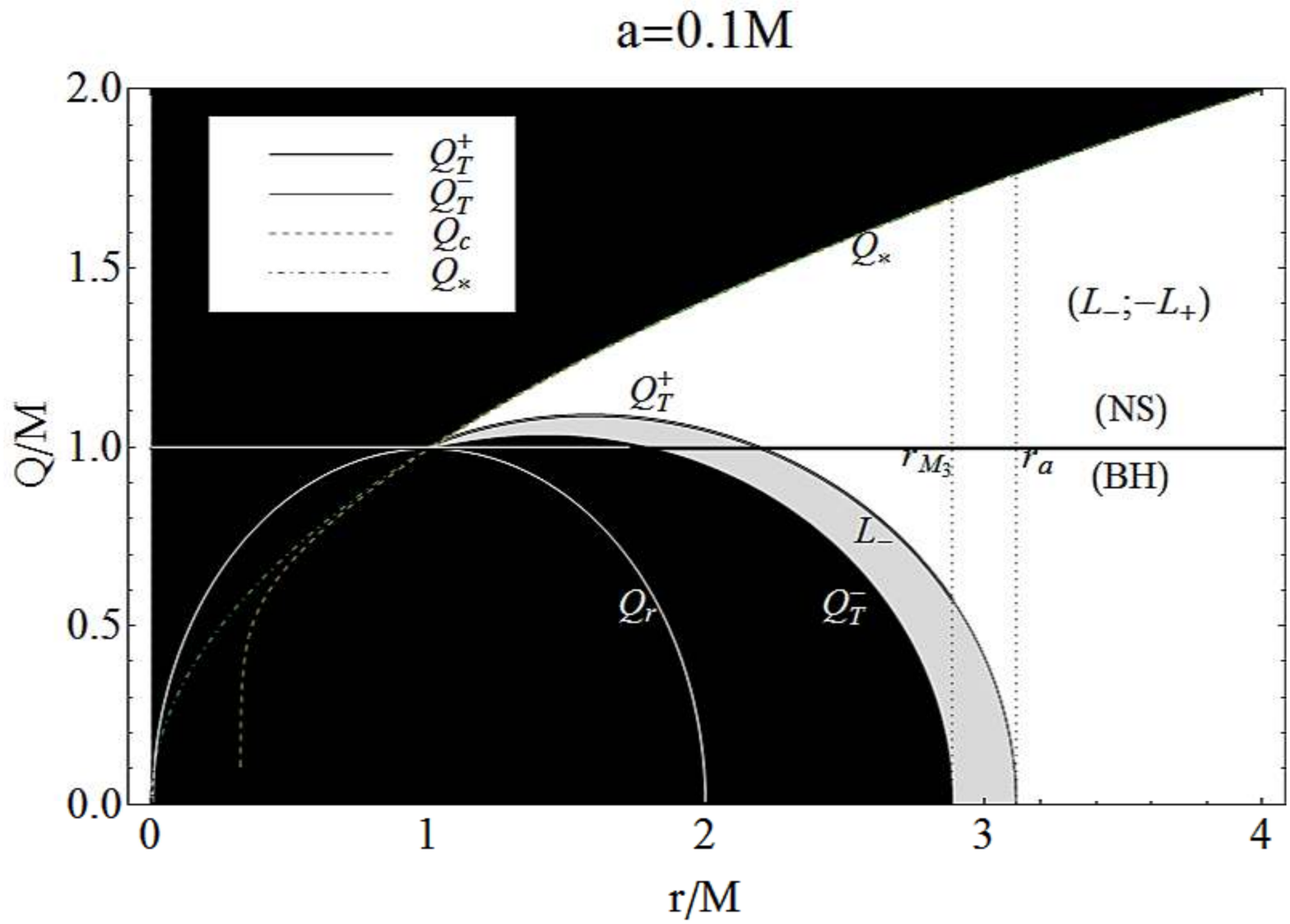}&
\includegraphics[width=0.3\hsize,clip]{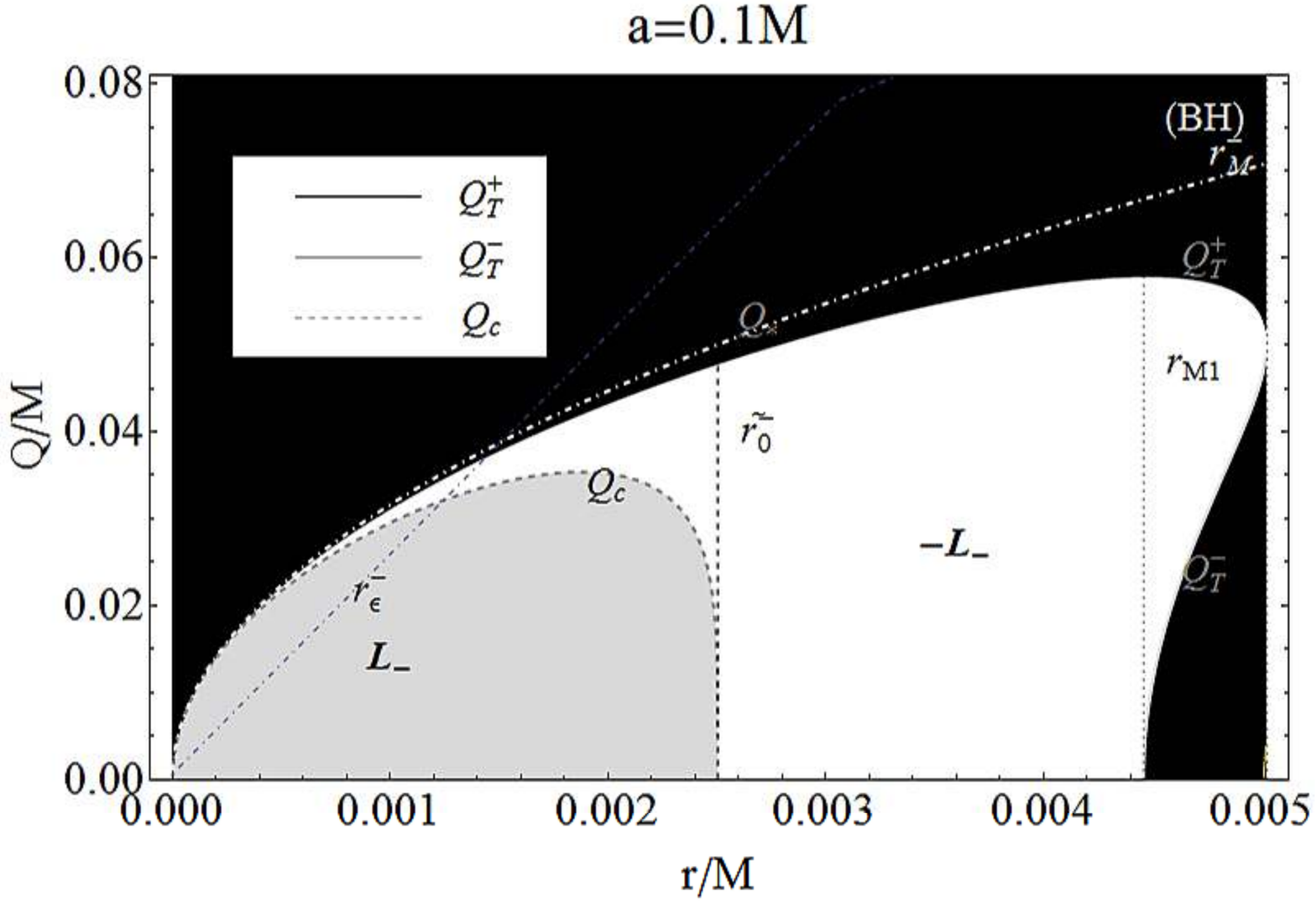}&
\includegraphics[width=0.3\hsize,clip]{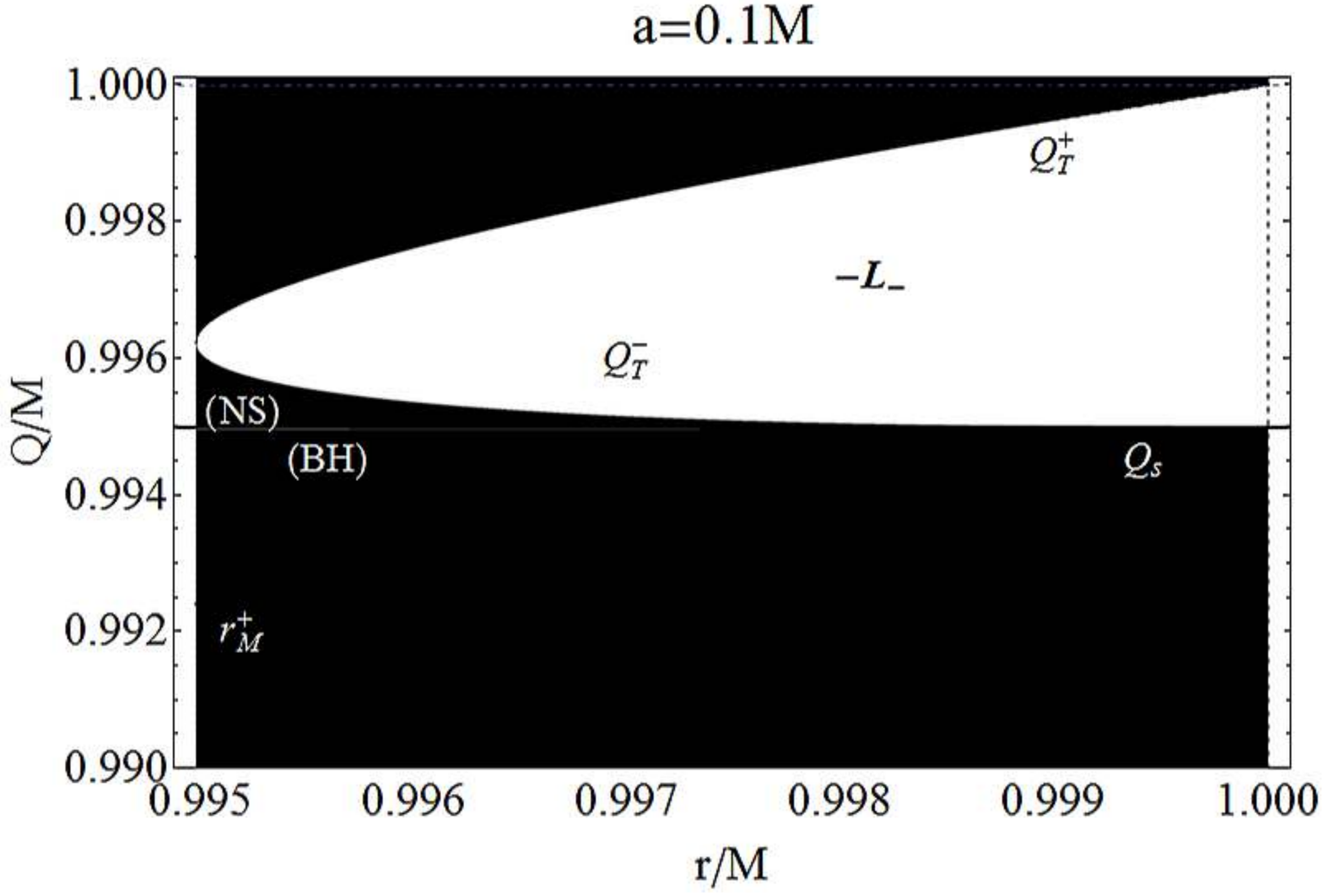}\\
\includegraphics[width=0.3\hsize,clip]{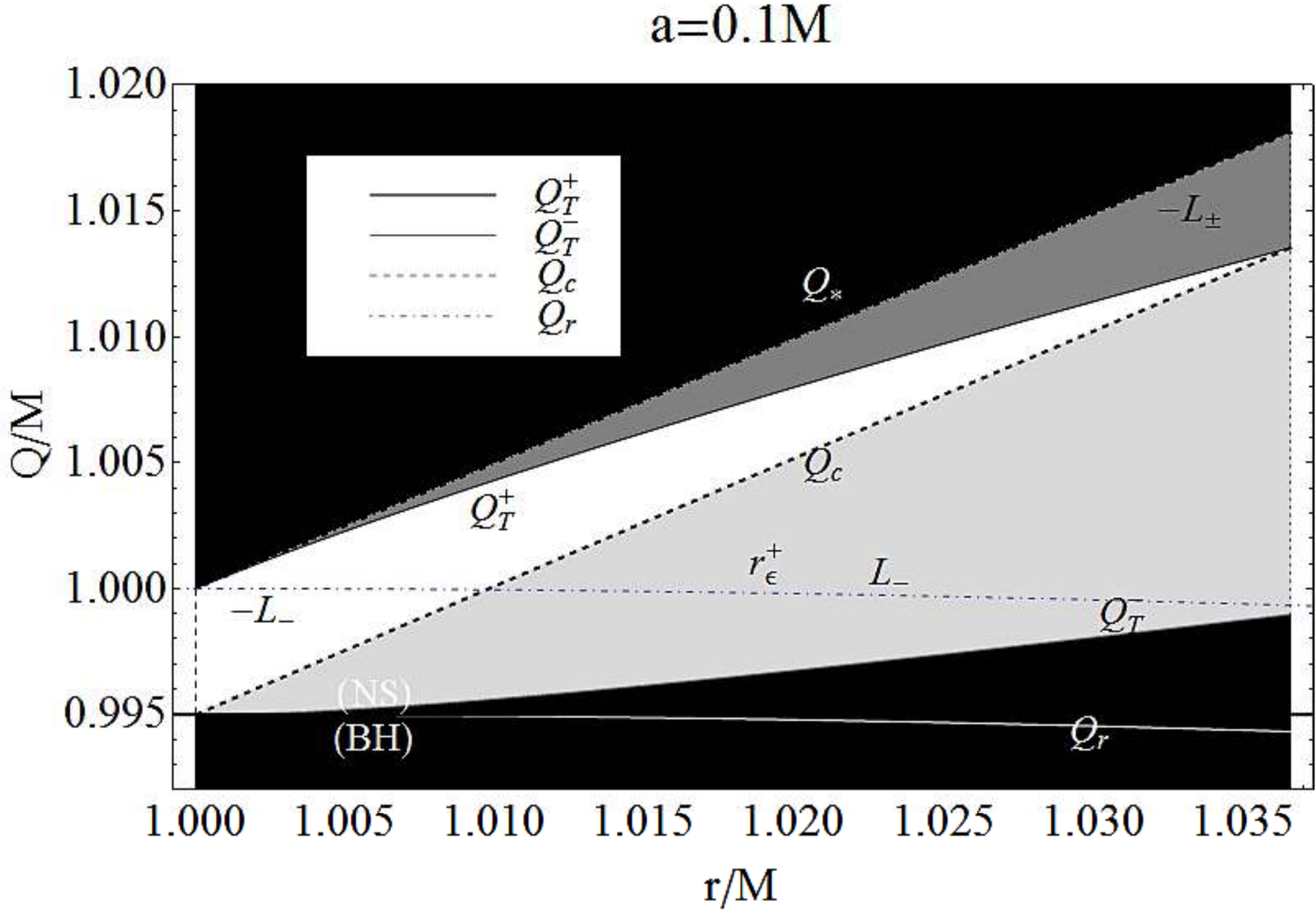}&
\includegraphics[width=0.3\hsize,clip]{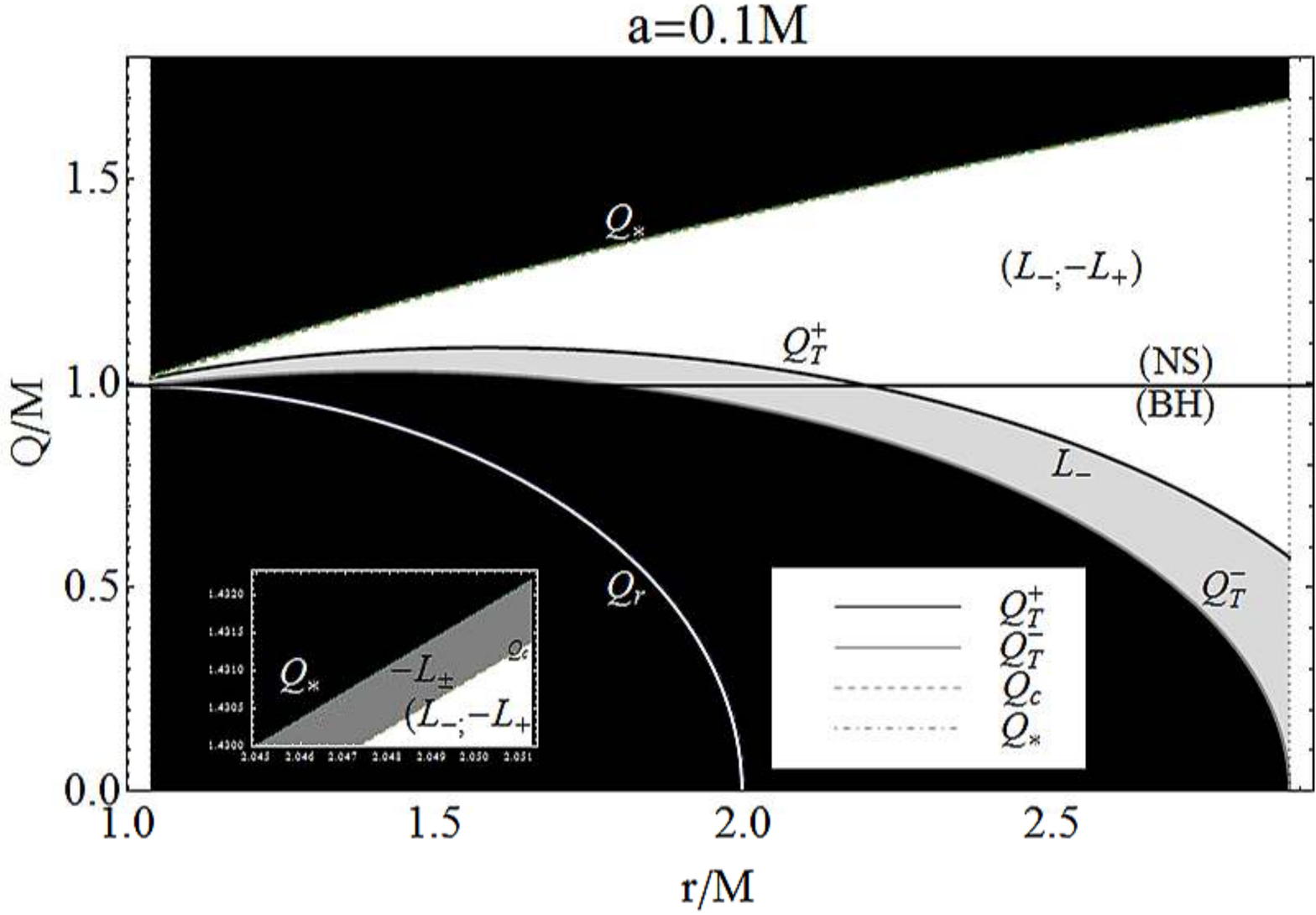}&
\includegraphics[width=0.3\hsize,clip]{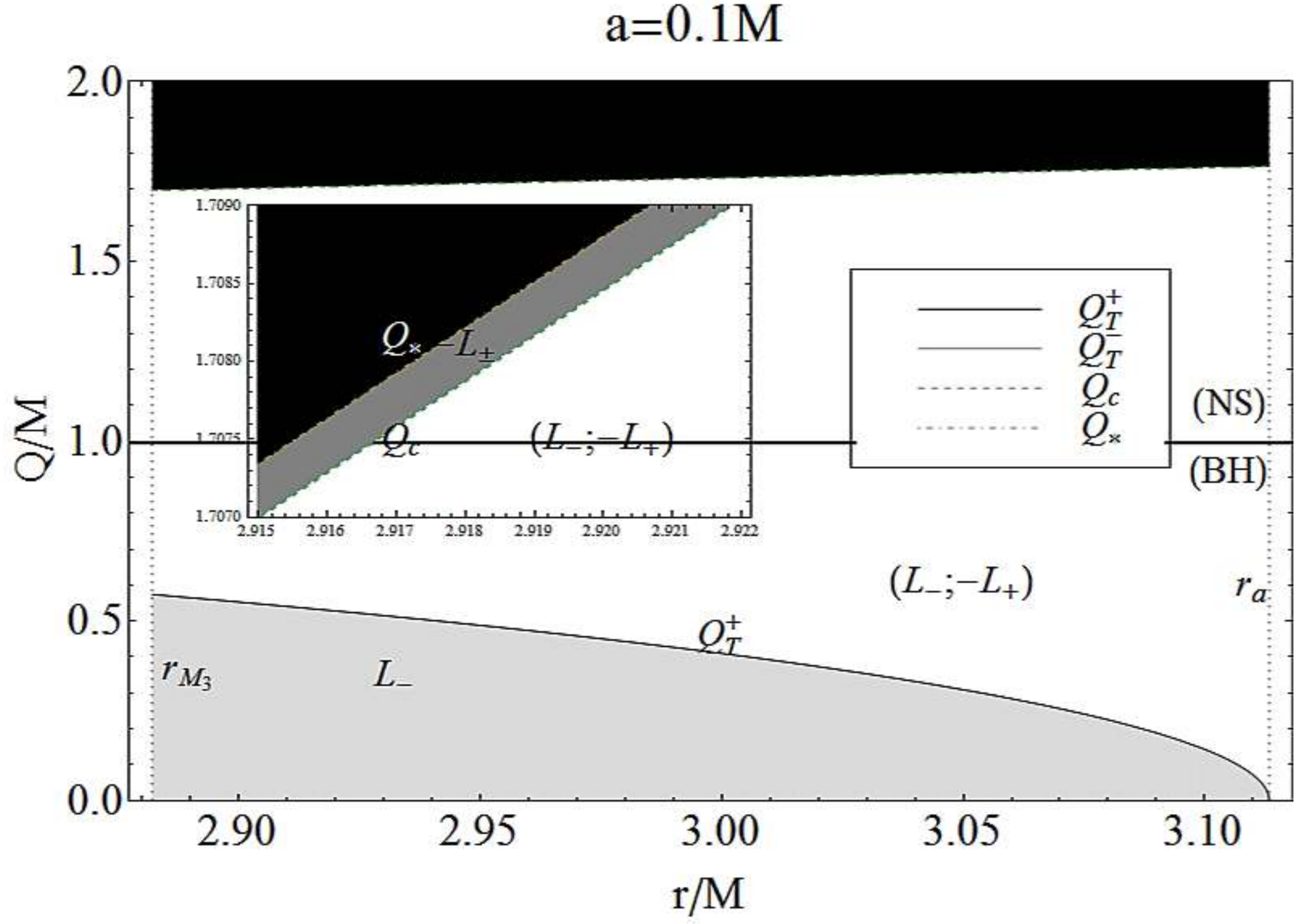}\\
&
\includegraphics[width=0.3\hsize,clip]{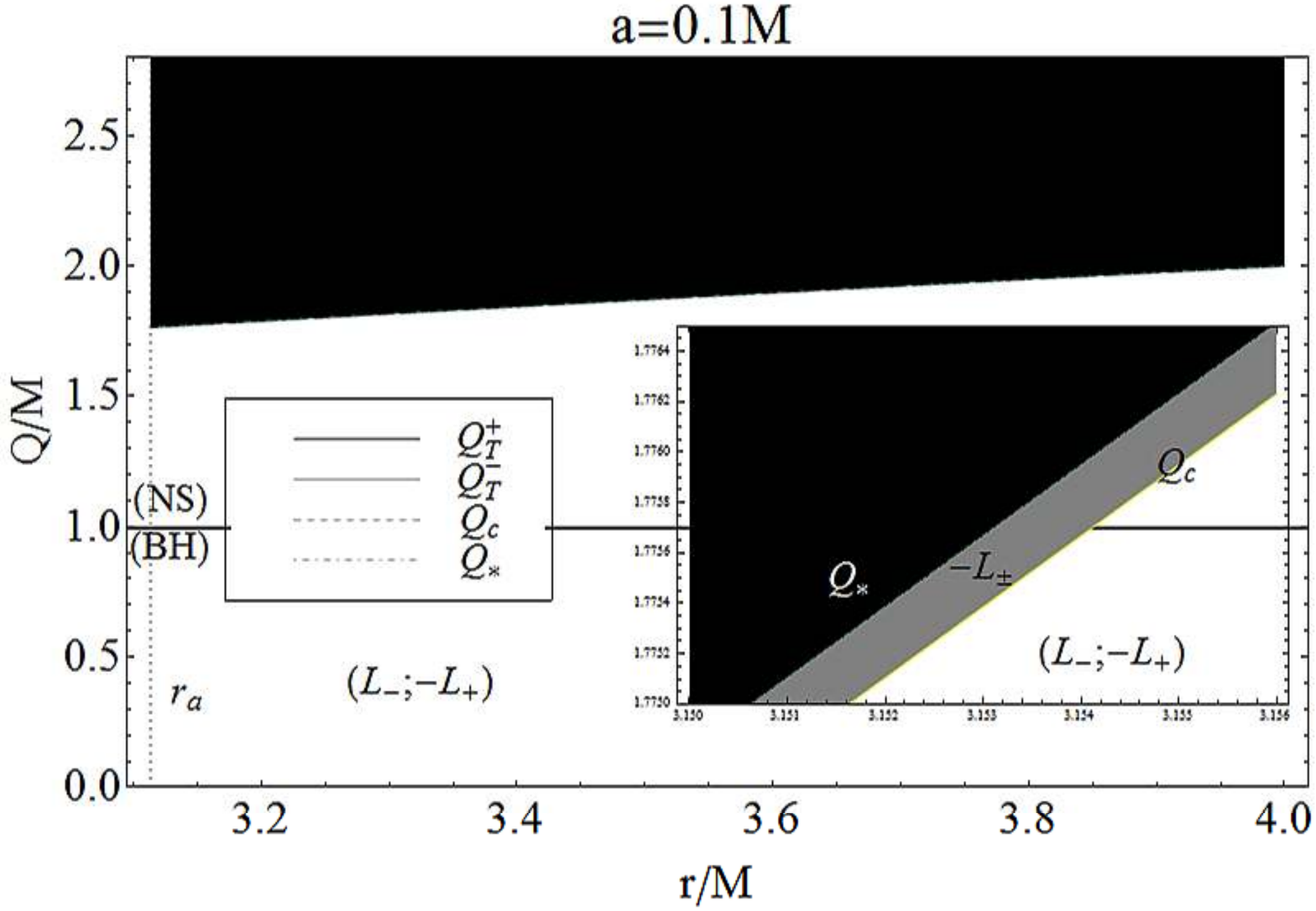}
\end{tabular}
\caption[font={footnotesize,it}]{\footnotesize{ The case: $a=0.1M$}. Properties of circular orbits in a KN spacetime, including black holes and naked singularities.  The charge-to-mass ratio $Q_{\ti{T}}^{\pm}/M$, $Q_*/M$ and $Q_{c_2}$ are plotted as functions of $r/M$. The value of the angular momentum of the orbiting  test particle is shown in each region. Black regions are forbidden; no circular motion is possible there. Dotted lines represent the radii $\tilde{r}_0$, $r_{\ti{M}_2}$, and $r_*$. Upper-left shows the range $r/M\in[0,4]$. The remaining plots magnify the intervals where  non-trivial behaviors are observed.  The solid (dashed) white curve is the outer  (inner) horizon $r_+$  ($r_-$).  We denoted with $\tilde{r}^-_0$  the lowest value of $\tilde{r}(Q=0)$. The horizontal thick black line represents $Q=Q_s$. A naked singularity occurs for $a>M$  and $Q>Q_s$, i.e., $Q_s=0.994987M$ in this case. The black hole \textbf{(BH)} and naked singularity \textbf{(NS)} regions are denoted explicitly. The dotted-dashed curve represents  the  ergosphere boundaries $r_{\epsilon}^{\pm}$.} \label{bicchi}
\end{figure}

\begin{figure}[h!]
\begin{tabular}{ccc}
\includegraphics[width=0.3\hsize,clip]{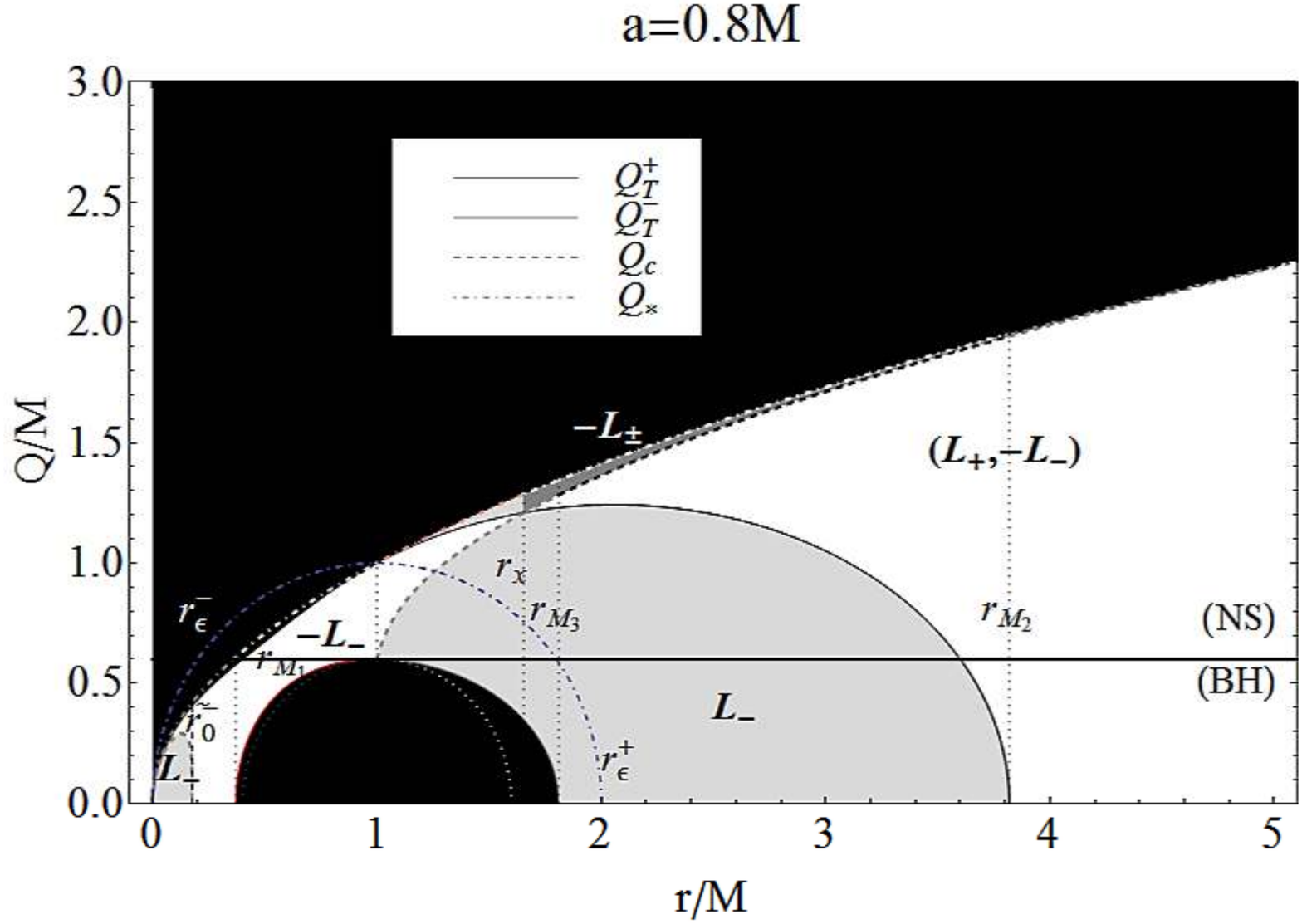}&
\includegraphics[width=0.3\hsize,clip]{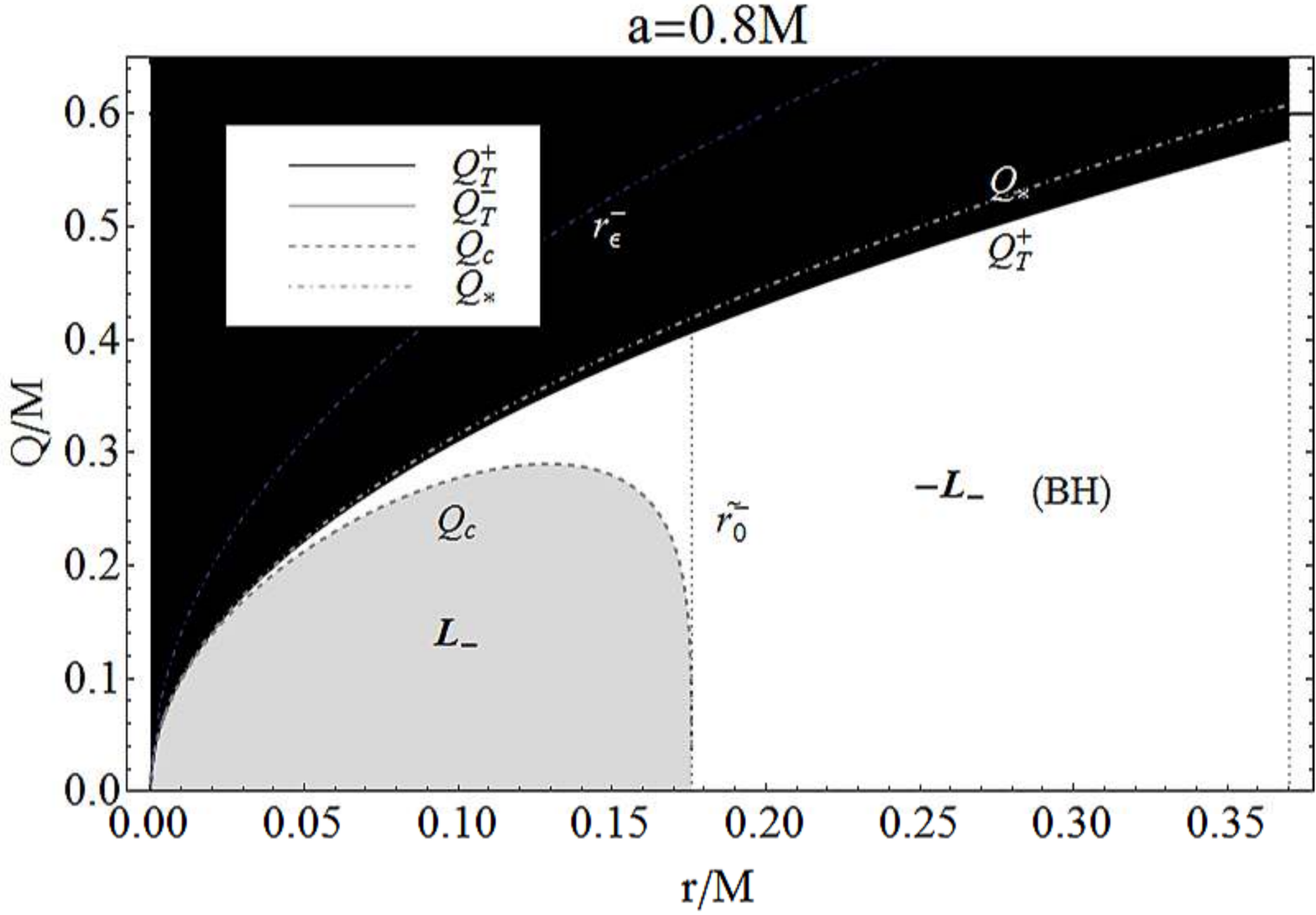}&
\includegraphics[width=0.3\hsize,clip]{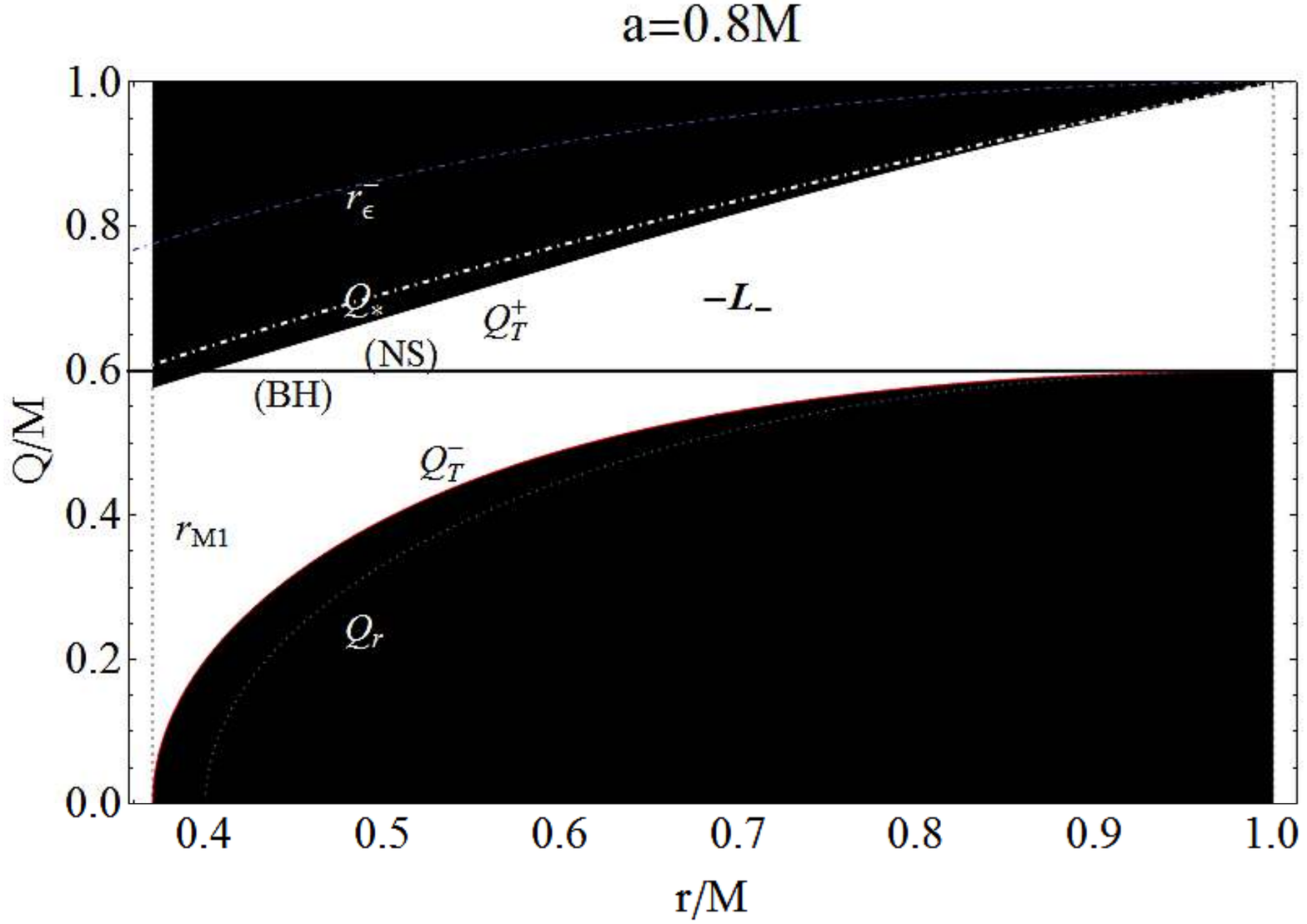}\\
\includegraphics[width=0.3\hsize,clip]{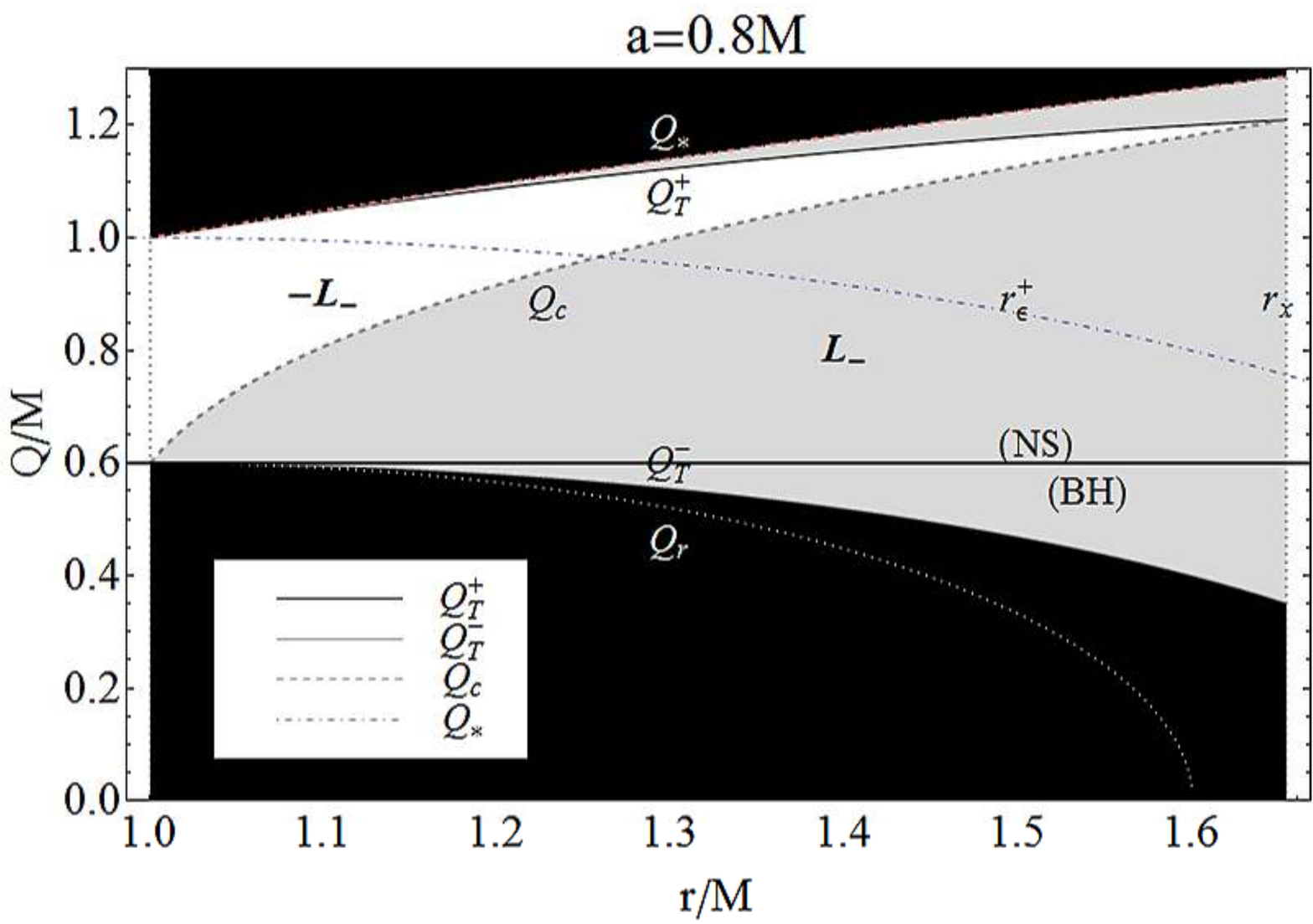}&
\includegraphics[width=0.3\hsize,clip]{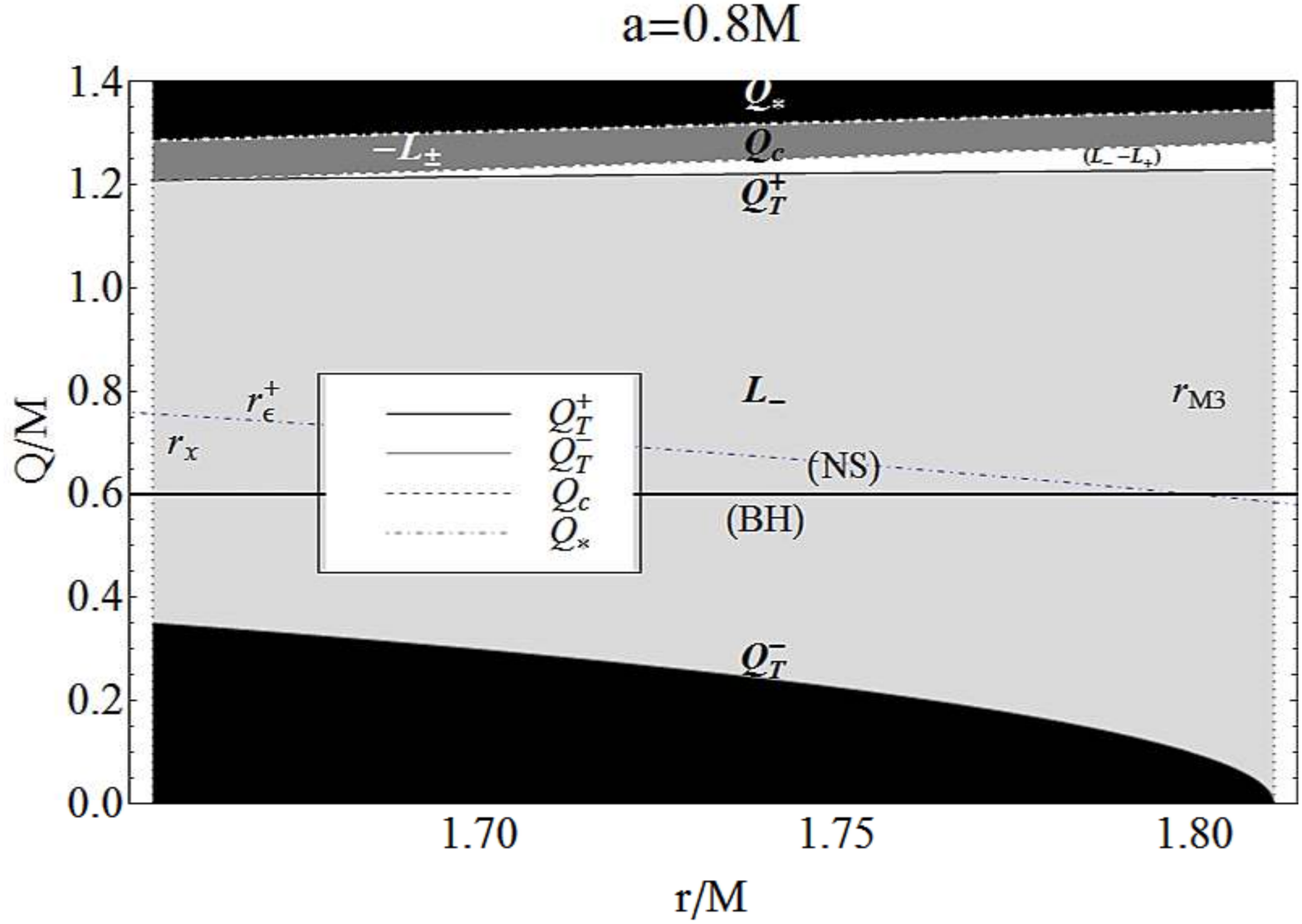}&
\includegraphics[width=0.3\hsize,clip]{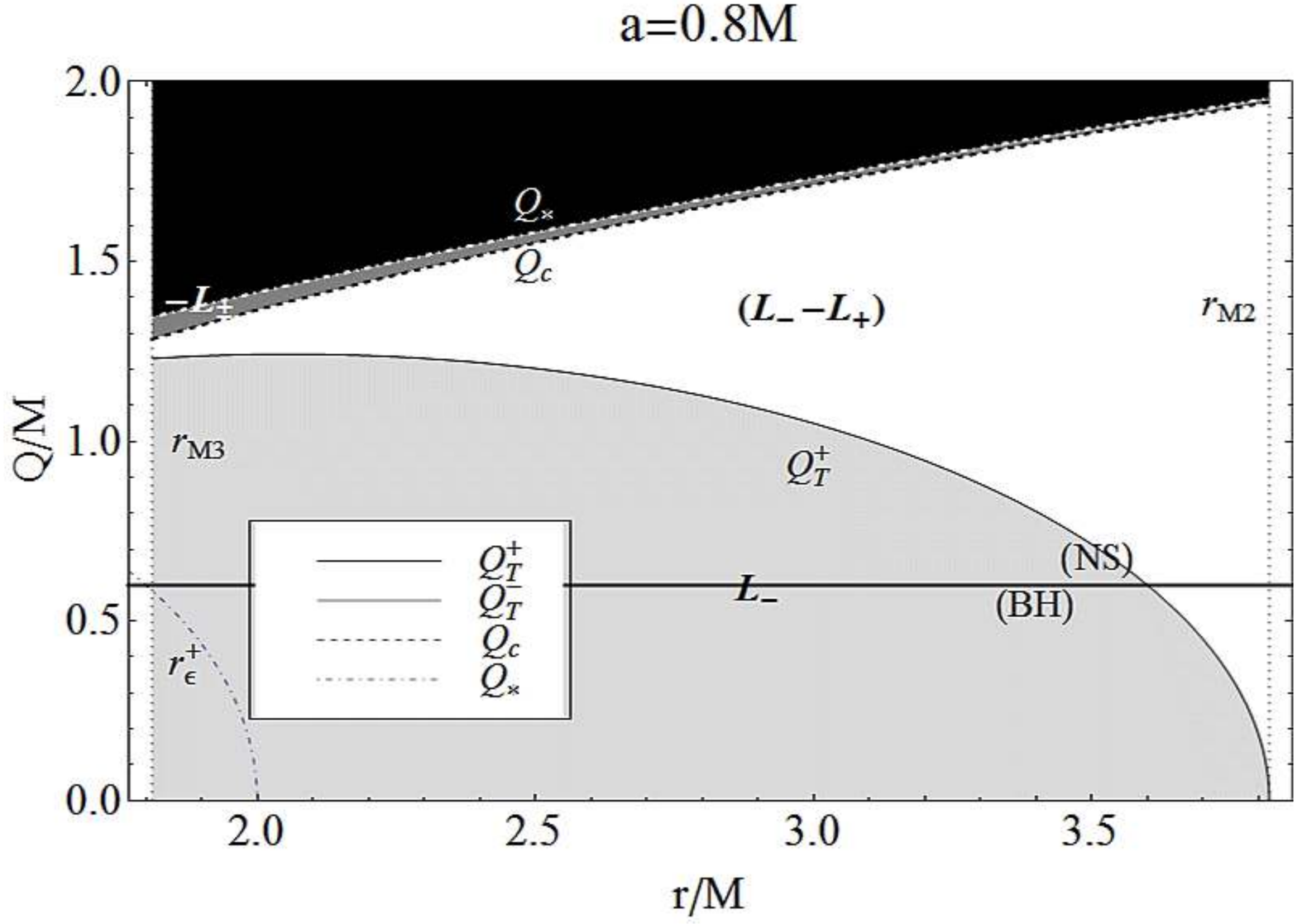}\\
\hspace{6cm}
&\includegraphics[width=0.3\hsize,clip]{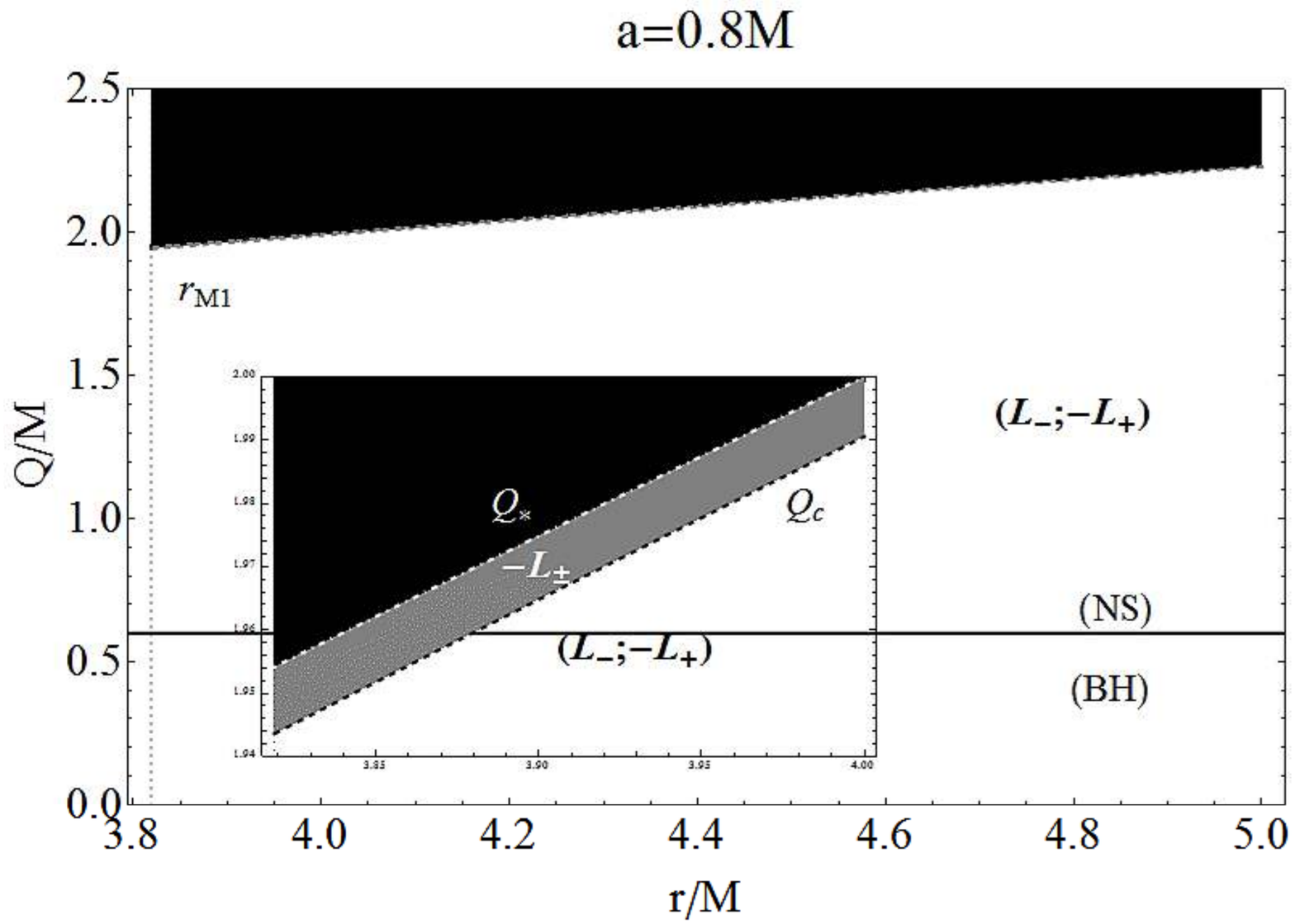}
\end{tabular}
\caption[font={footnotesize,it}]{\footnotesize{The case: $a=0.8M$. Properties of circular orbits in a KN spacetime, including black holes and naked singularities.  The charge-to-mass ratio $Q_{\ti{T}}^{\pm}/M$, $Q_*/M$ and $Q_{c_2}$ are plotted as functions of $r/M$. The angular momentum of the orbiting test particle is shown in each region. Black regions are forbidden; no circular motion is possible there. Dotted lines represent the radii $\tilde{r}_0$, $r_{\ti{M}_2}$, $r_{\ti{M}_3}$, $r_*$, $r_x$. Upper-left shows the range $a/M\in[0,5]$. The remaining plots magnify the intervals where  non-trivial behaviors are observed. The solid (dashed) white curve is the outer  (inner) horizon $r_+$  ($r_-$).   We denoted with $\tilde{r}^-_0$  the lowest value of $\tilde{r}(Q=0)$.  The horizontal thick black line represents $Q=Q_s$. A naked singularity occurs for $a>M$  and $Q>Q_s$, i.e., $Q_s=0.6M$ in this case. The black hole \textbf{(BH)} and naked singularity \textbf{(NS)} regions are denoted explicitly. The dotted-dashed curve represents  the  ergosphere boundaries $r_{\epsilon}^{\pm}$.} }\label{bicchi9}
\end{figure}

\begin{figure}[h!]
\begin{tabular}{ccc}
\includegraphics[width=0.3\hsize,clip]{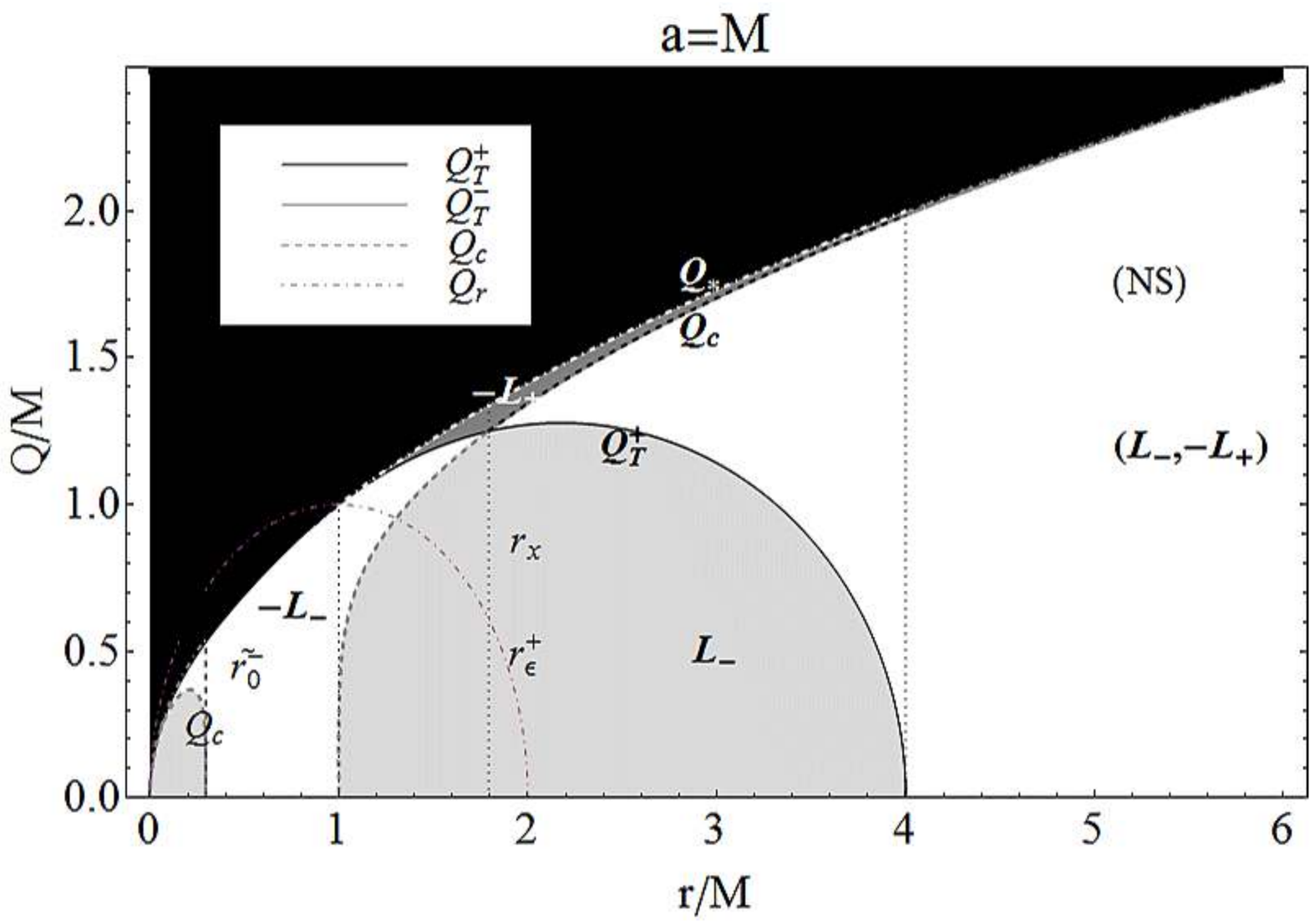}&
\includegraphics[width=0.3\hsize,clip]{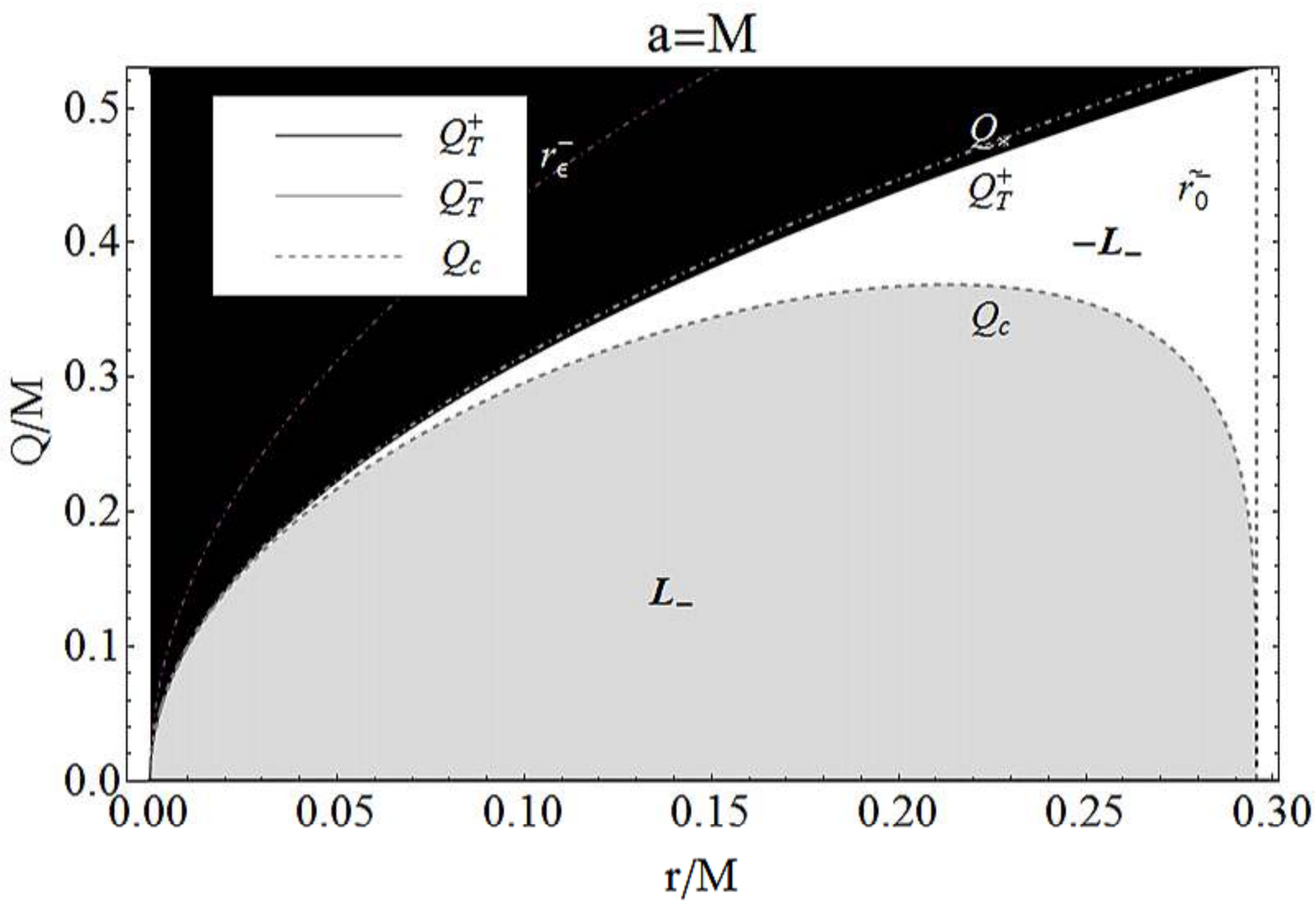}&
\includegraphics[width=0.3\hsize,clip]{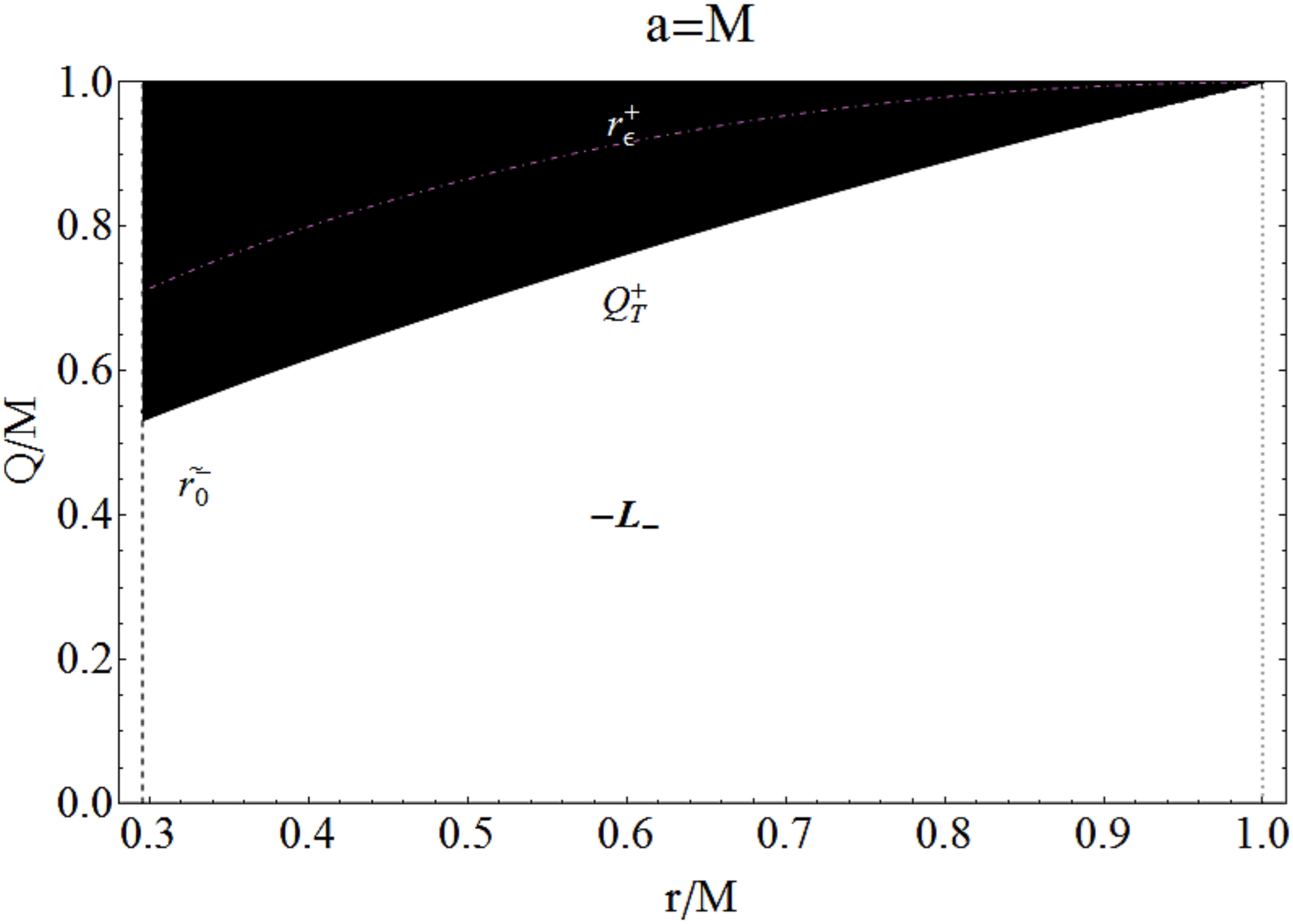}\\
\includegraphics[width=0.3\hsize,clip]{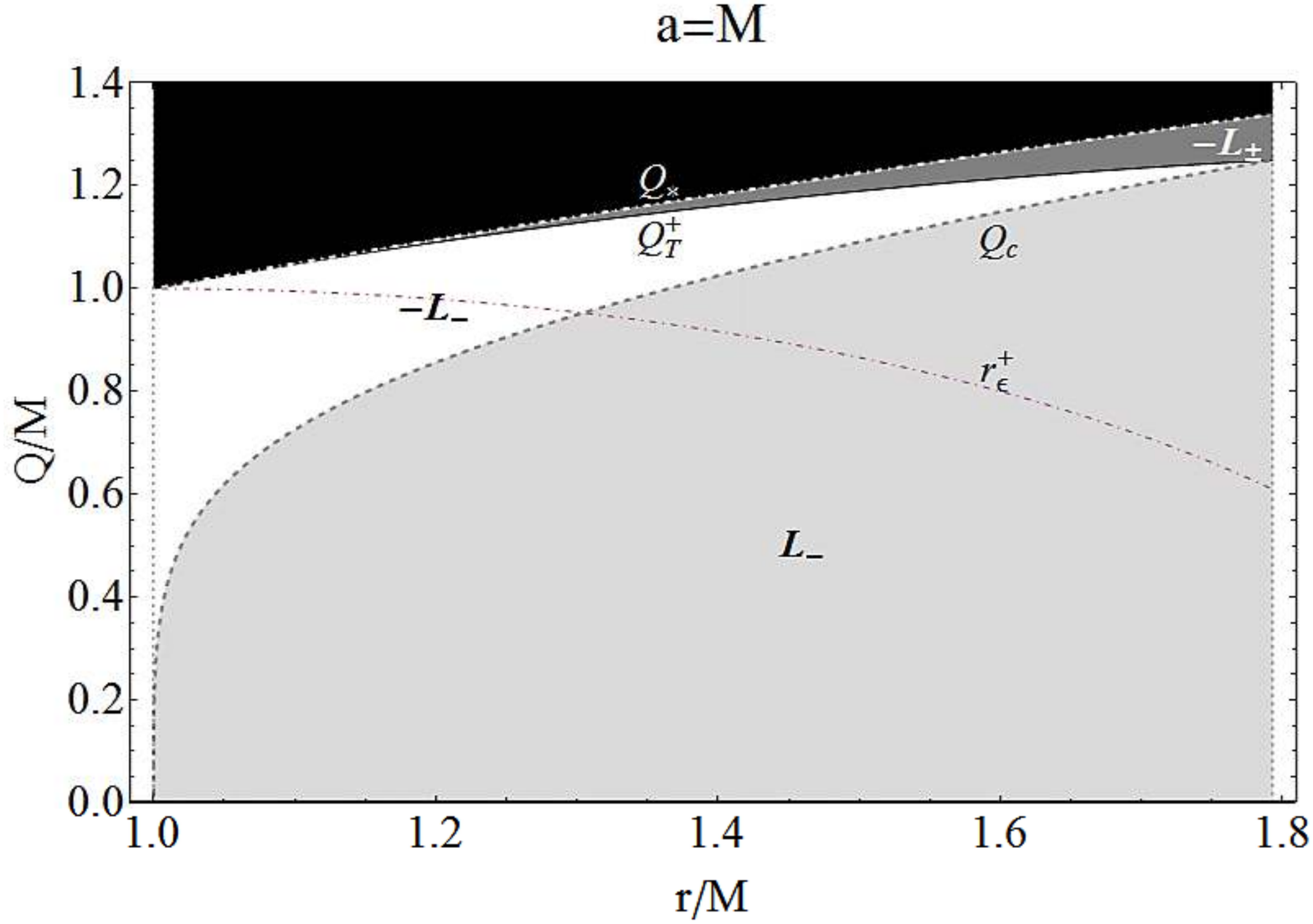}&
\includegraphics[width=0.3\hsize,clip]{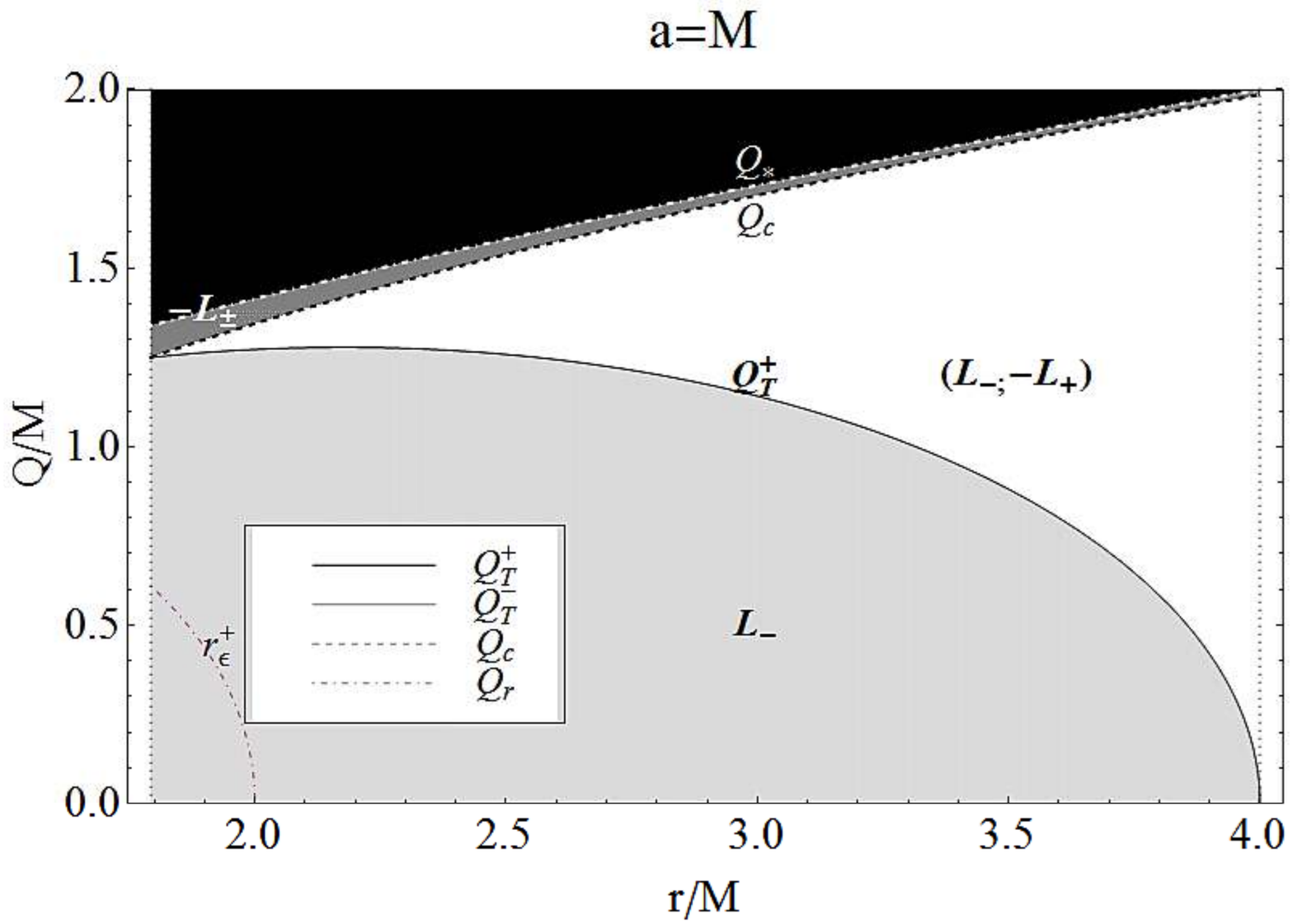}&
\includegraphics[width=0.3\hsize,clip]{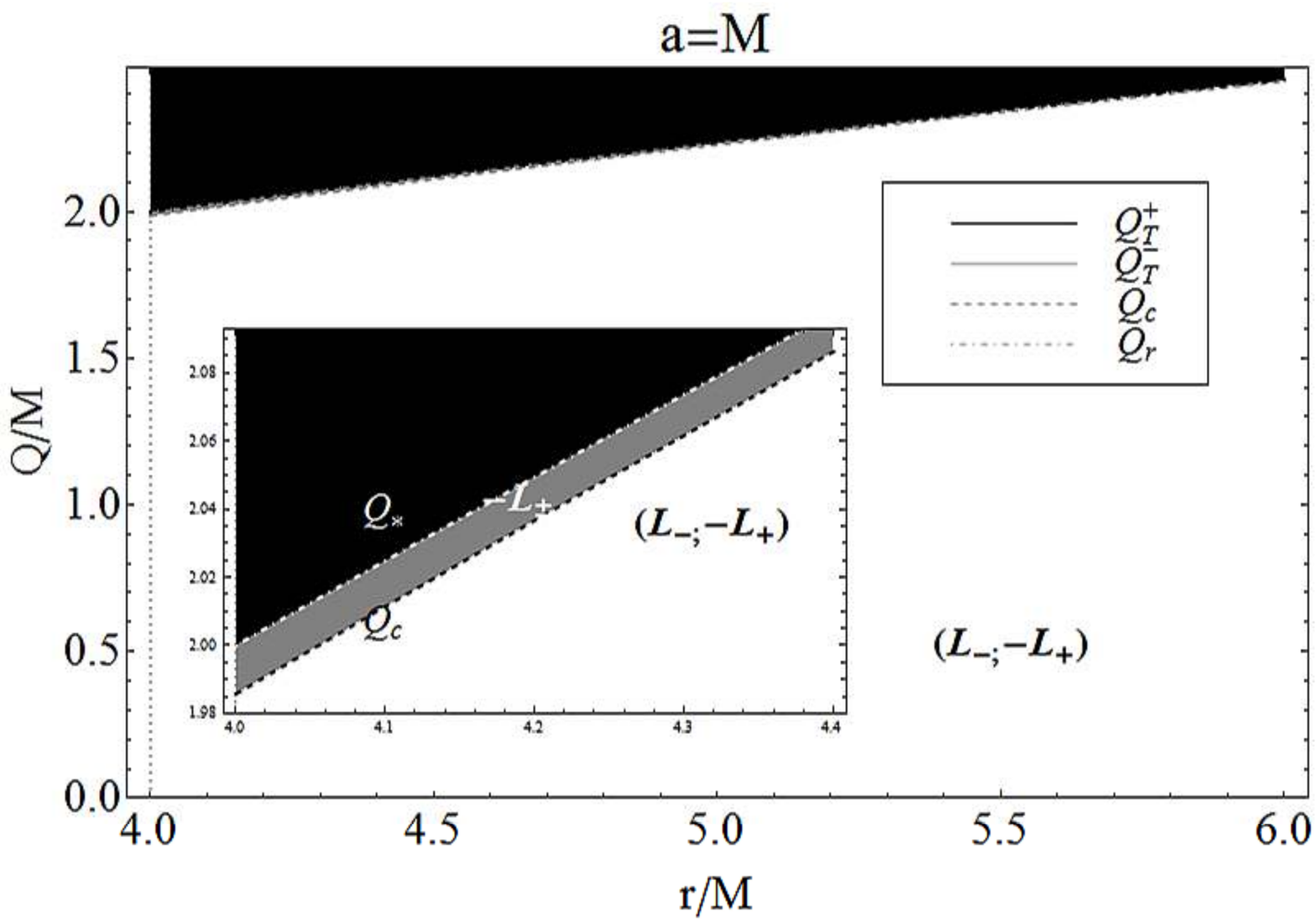}
\\
\end{tabular}
\caption[font={footnotesize,it}]{\footnotesize{The case: $a=M$.  Properties of circular orbits in a KN spacetime, including black holes and naked singularities.  The charge-to-mass ratio $Q_{\ti{T}}^{\pm}/M$, $Q_*/M$ and $Q_{c_2}$ are plotted as functions of $r/M$. The dotted-dashed curve represents  the  ergosphere boundaries $r_{\epsilon}^{\pm}$. The angular momentum of the orbiting test particle is shown in each region. Black regions are forbidden; no circular motion is possible there. Dotted lines represent the radii $\tilde{r}_0$,  $r_*$, $r_x$. Upper-left shows the range $a/M\in[0,6]$. The remaining plots magnify the intervals where  non-trivial behaviors are observed. The solid (dashed) white curve is the outer  (inner) horizon $r_+$  ($r_-$).   We denoted with $\tilde{r}^-_0$  the lowest value of $\tilde{r}(Q=0)$. A naked singularity occurs for $a>M$  and $Q>Q_s$, i.e., $Q_s=M$ in this case.} }\label{bicchi1io}
\end{figure}

\begin{figure}[h!]
\begin{tabular}{ccc}
\includegraphics[width=0.3\hsize,clip]{PLOT/Vivienne}
\includegraphics[width=0.3\hsize,clip]{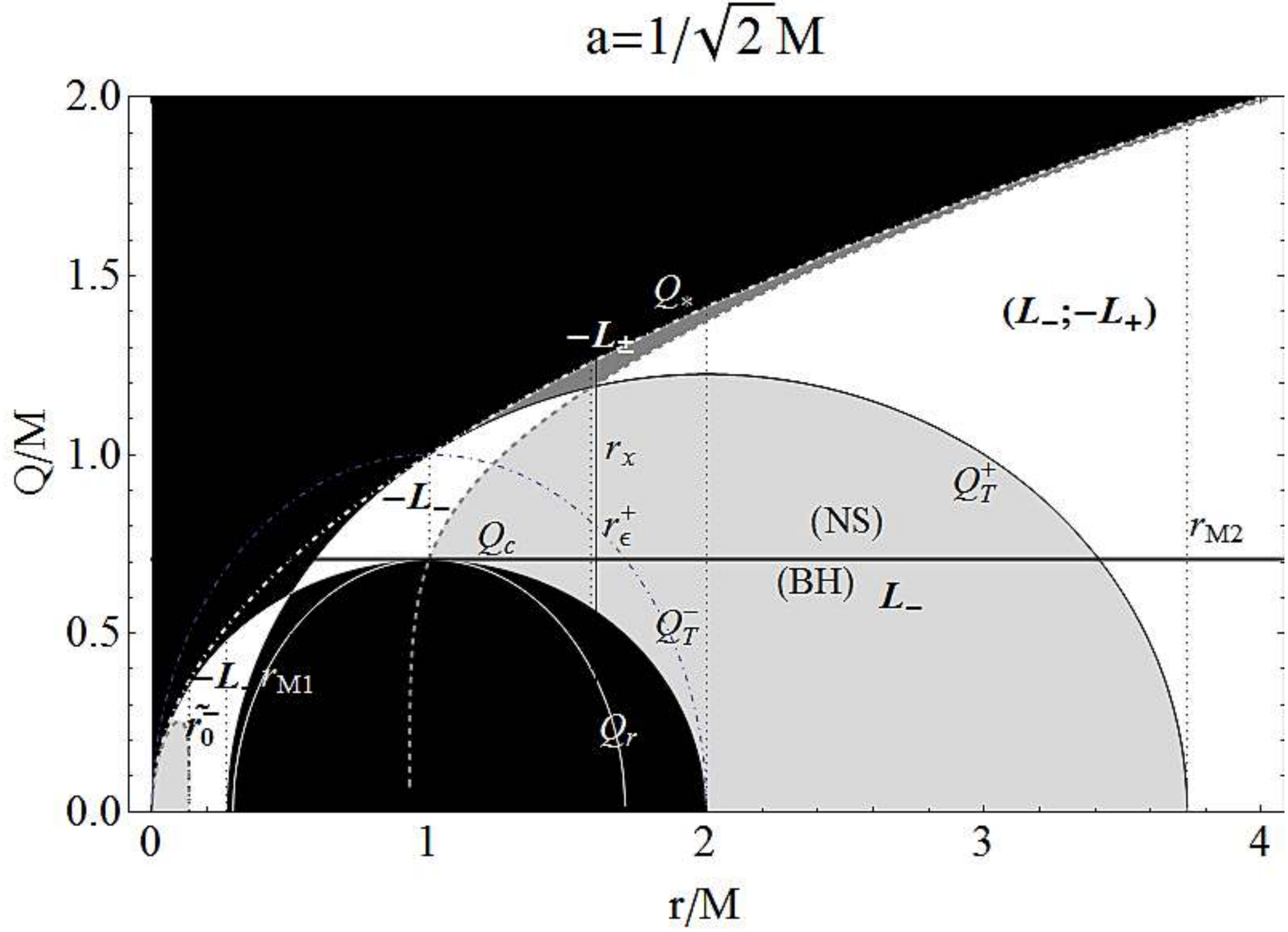}
\includegraphics[width=0.3\hsize,clip]{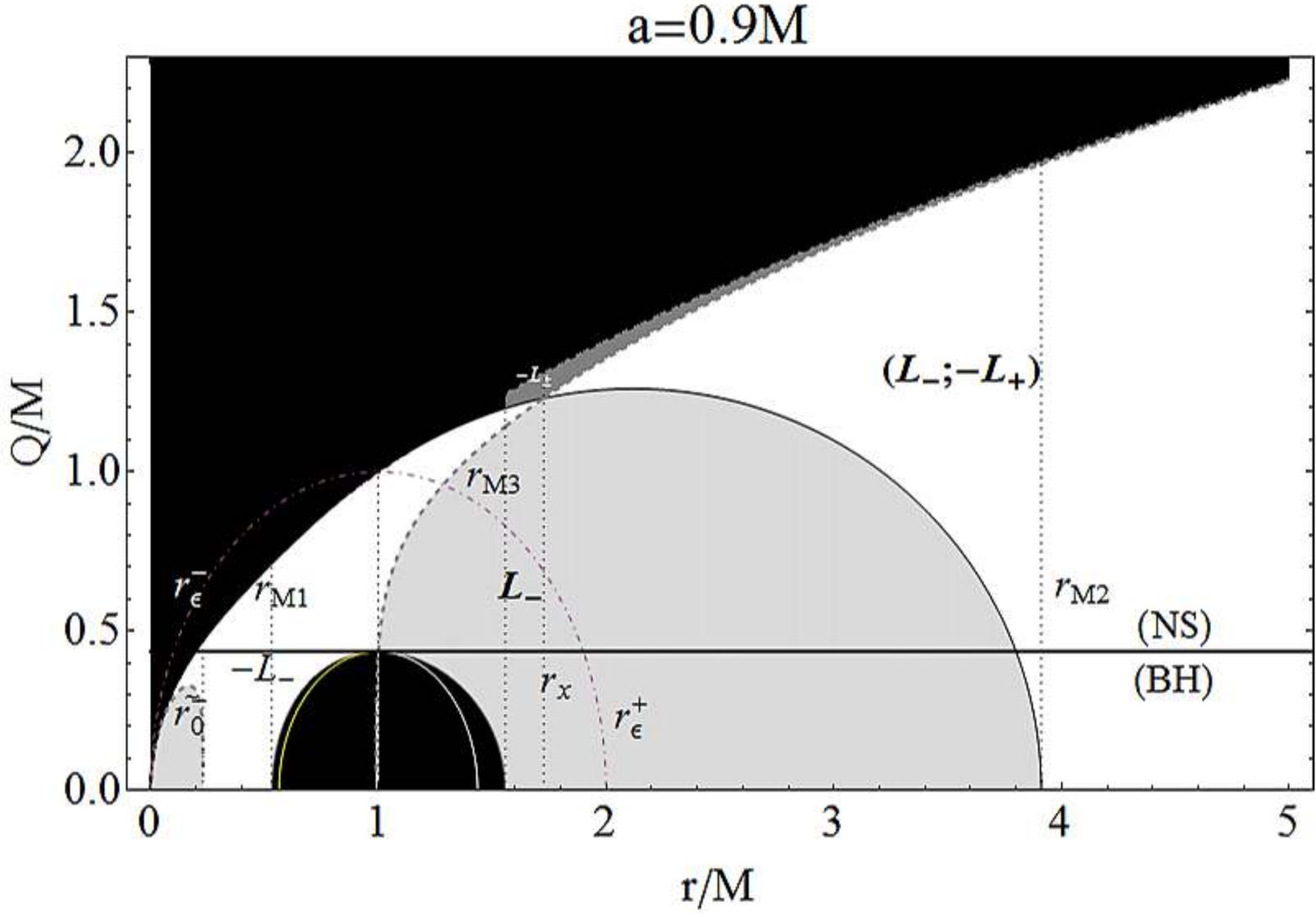}\\
\includegraphics[width=0.3\hsize,clip]{PLOT/Allei}
\includegraphics[width=0.3\hsize,clip]{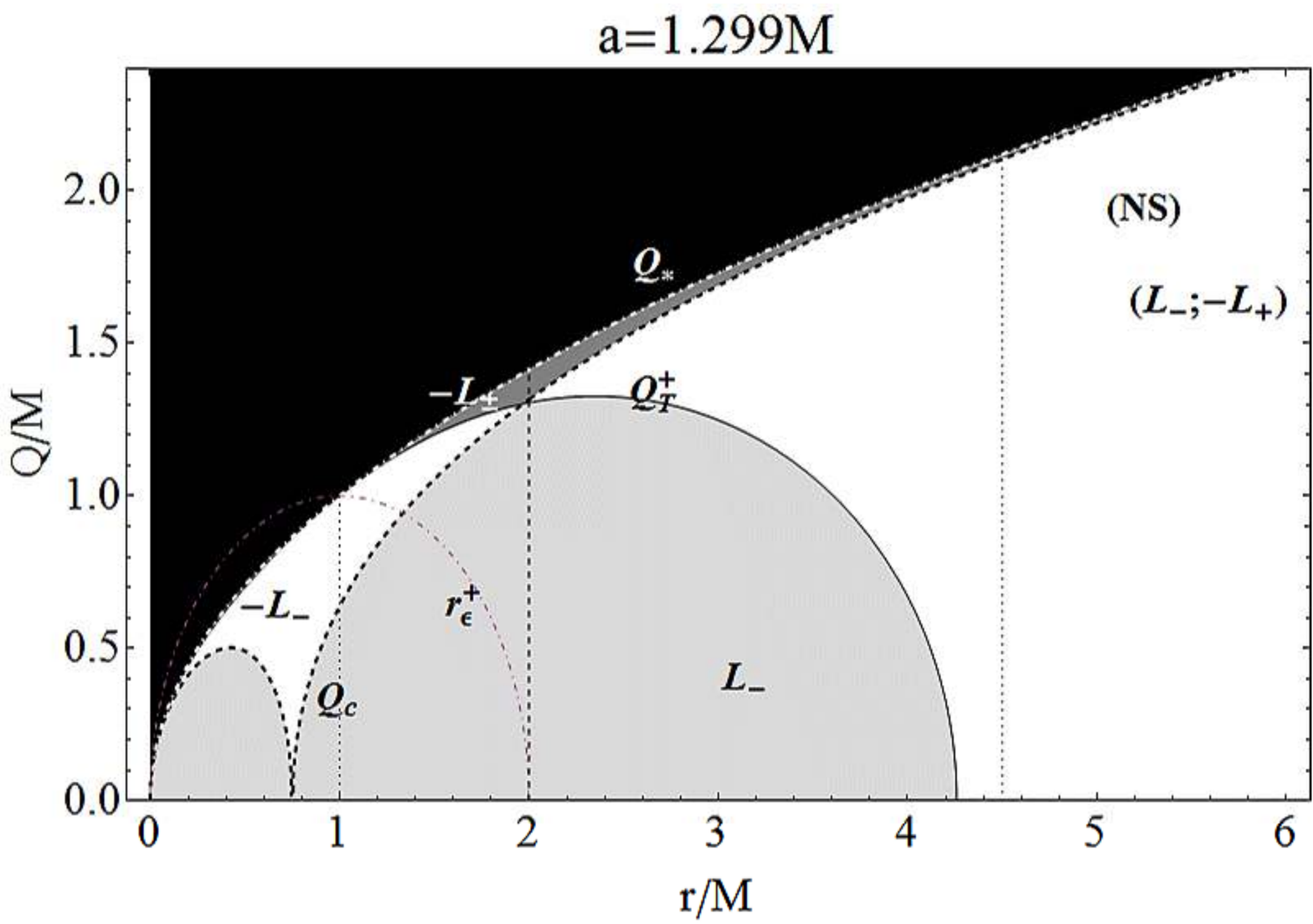}
\includegraphics[width=0.3\hsize,clip]{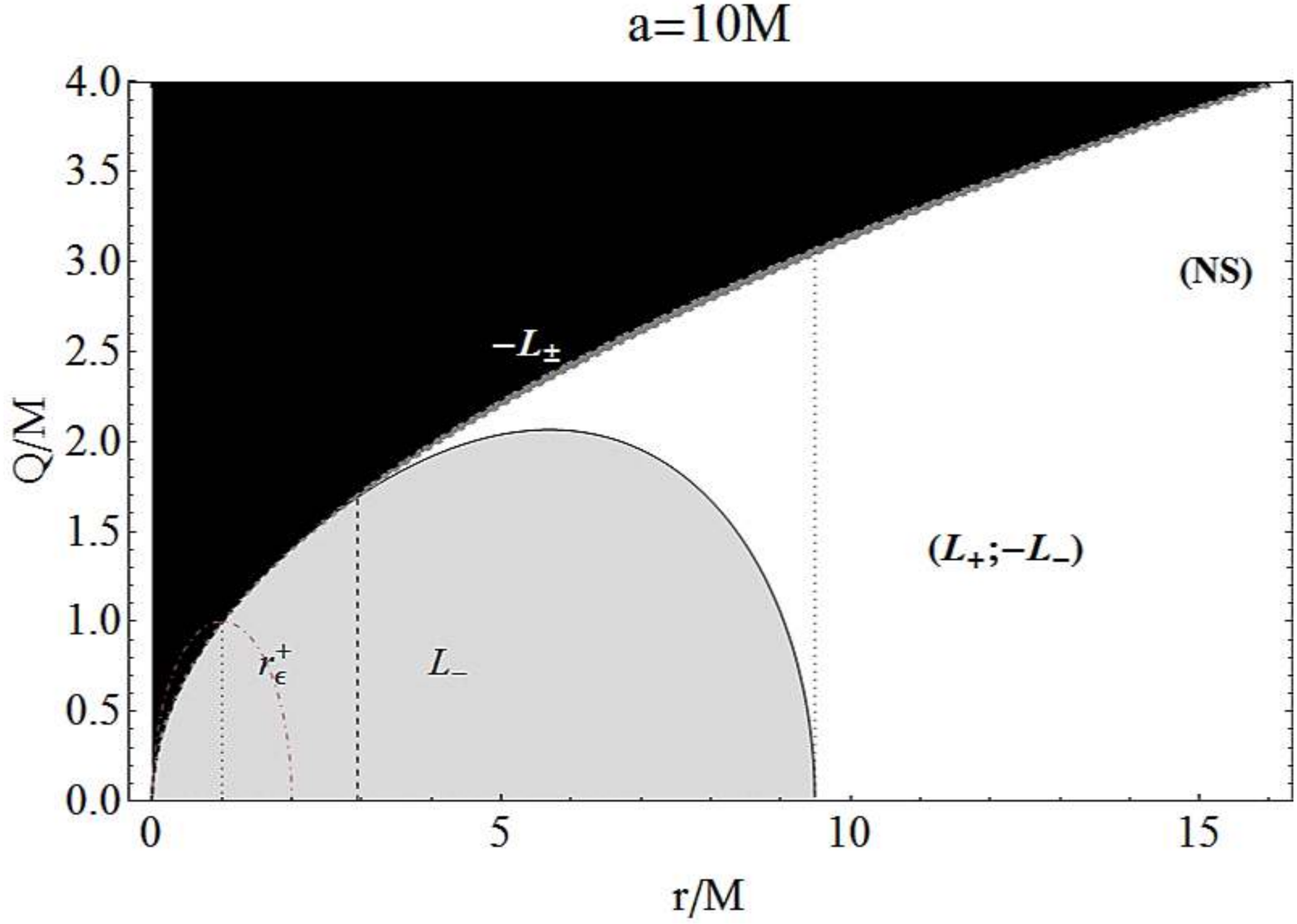}
\end{tabular}
\caption[font={footnotesize,it}]{\footnotesize{
The charge-to-mass ratio as a function of the radial coordinate for increasing values of the source angular momentum. The ratios   $Q_{\ti{T}}^{\pm}/M$, $Q_*/M$ and $Q_{c_2}$ are also plotted.  The angular momentum of the orbiting  test particle is shown in each region. Black regions are forbidden; no circular motion is possible there. Dotted lines represent the radii $\tilde{r}_0$,  $r_*$, and $r_x$.    We denoted with $\tilde{r}^-_0$  the lowest value of $\tilde{r}(Q=0)$. The horizontal thick black line represents $Q=Q_s$. A naked singularity occurs for $a>M$  and $Q>Q_s$, i.e., $Q_s=0.994987M$ for $a=0.1M$, $Q_s=0.707107M$ for $a=1/\sqrt{2}M$,  $Q_s=0.43589M$ for $a=0.9M$ and $Q_s=0$ for $a=M$. The black hole \textbf{(BH)} and naked singularity \textbf{(NS)} regions are denoted explicitly. The dotted-dashed curve represents  the  ergosphere boundaries $r_{\epsilon}^{\pm}$.} }\label{bicchic1}
\end{figure}

\begin{figure}[h!]
\begin{tabular}{ccc}
\includegraphics[width=0.3\hsize,clip]{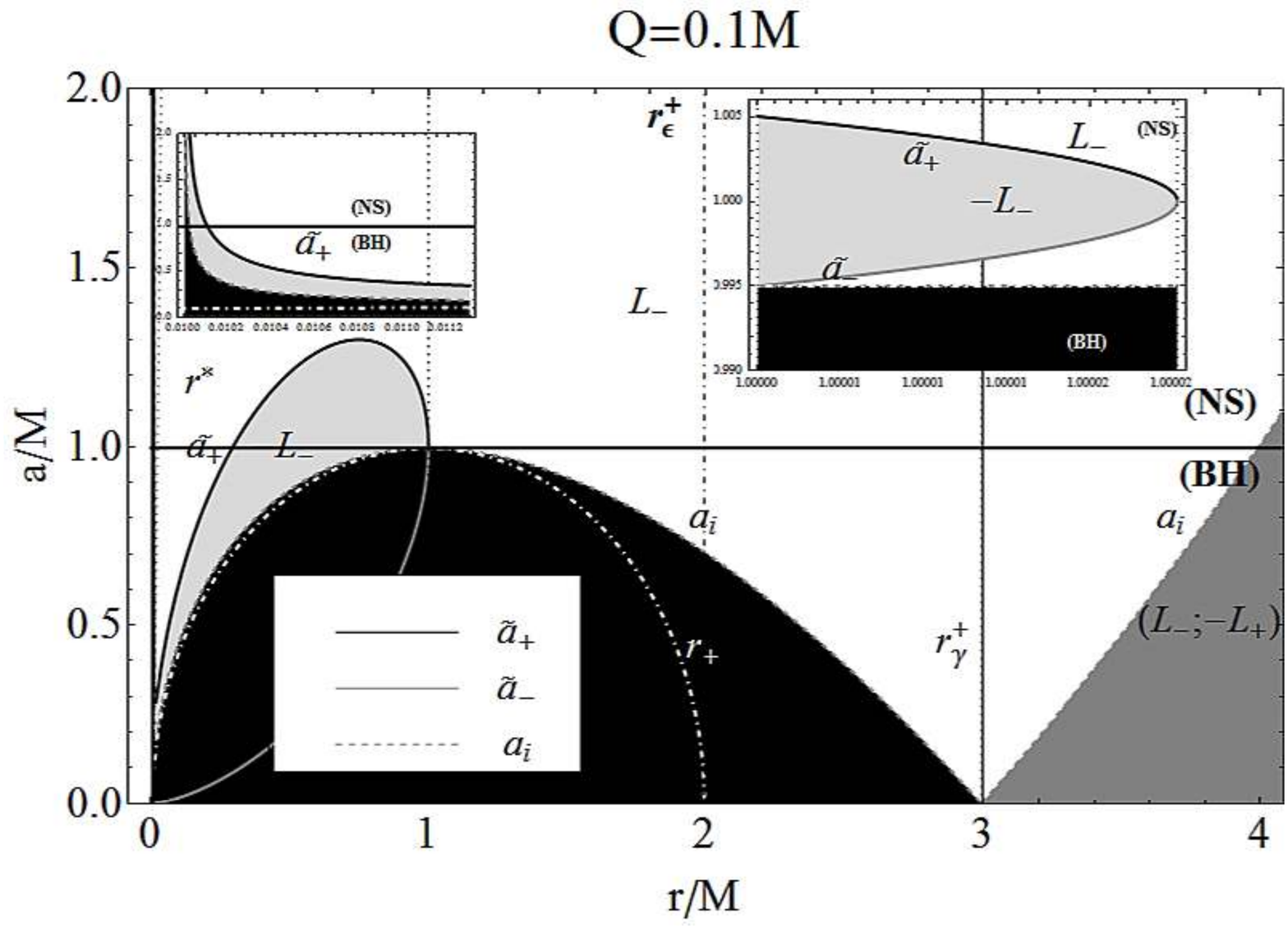}
\includegraphics[width=0.3\hsize,clip]{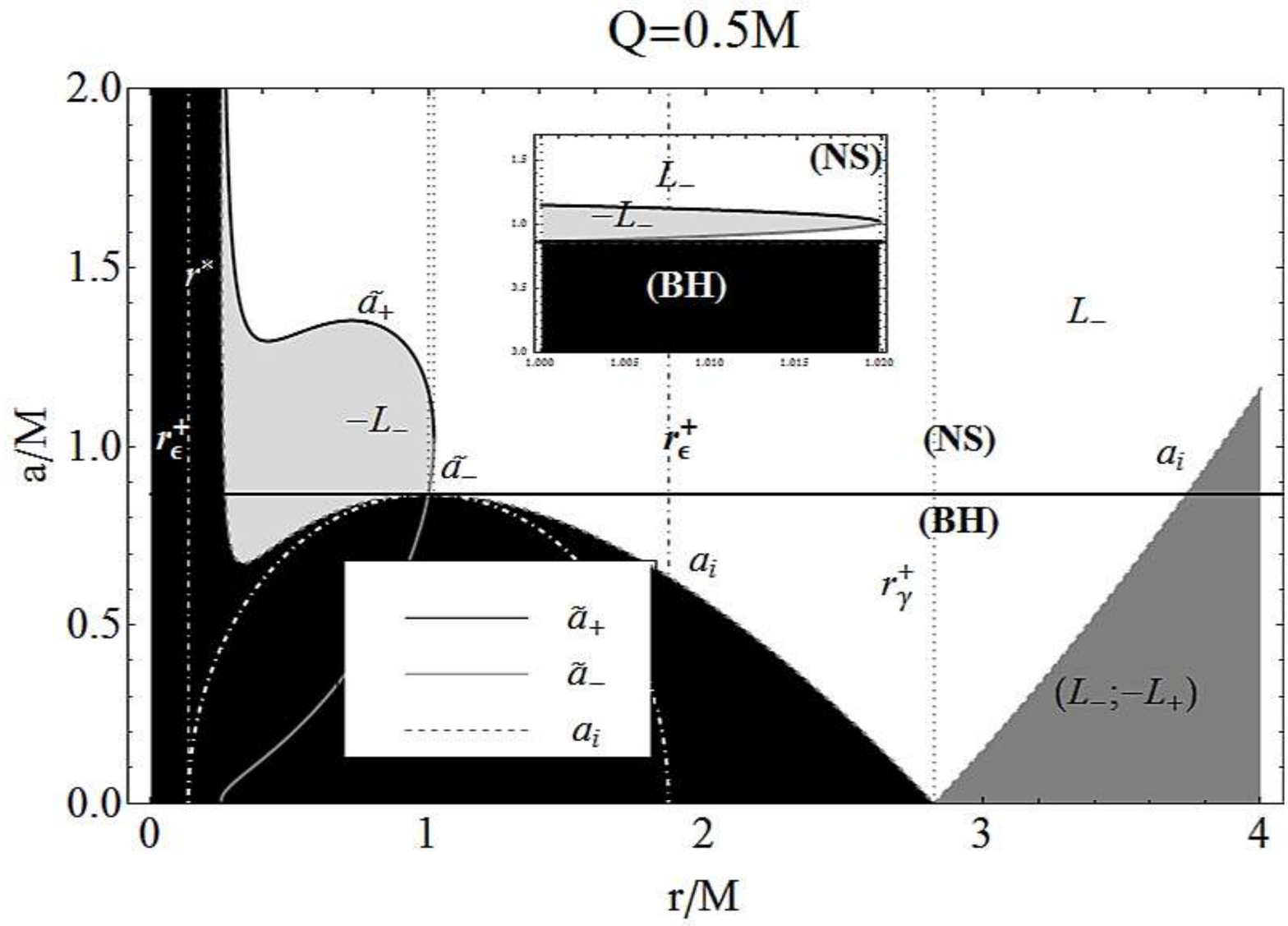}
\includegraphics[width=0.3\hsize,clip]{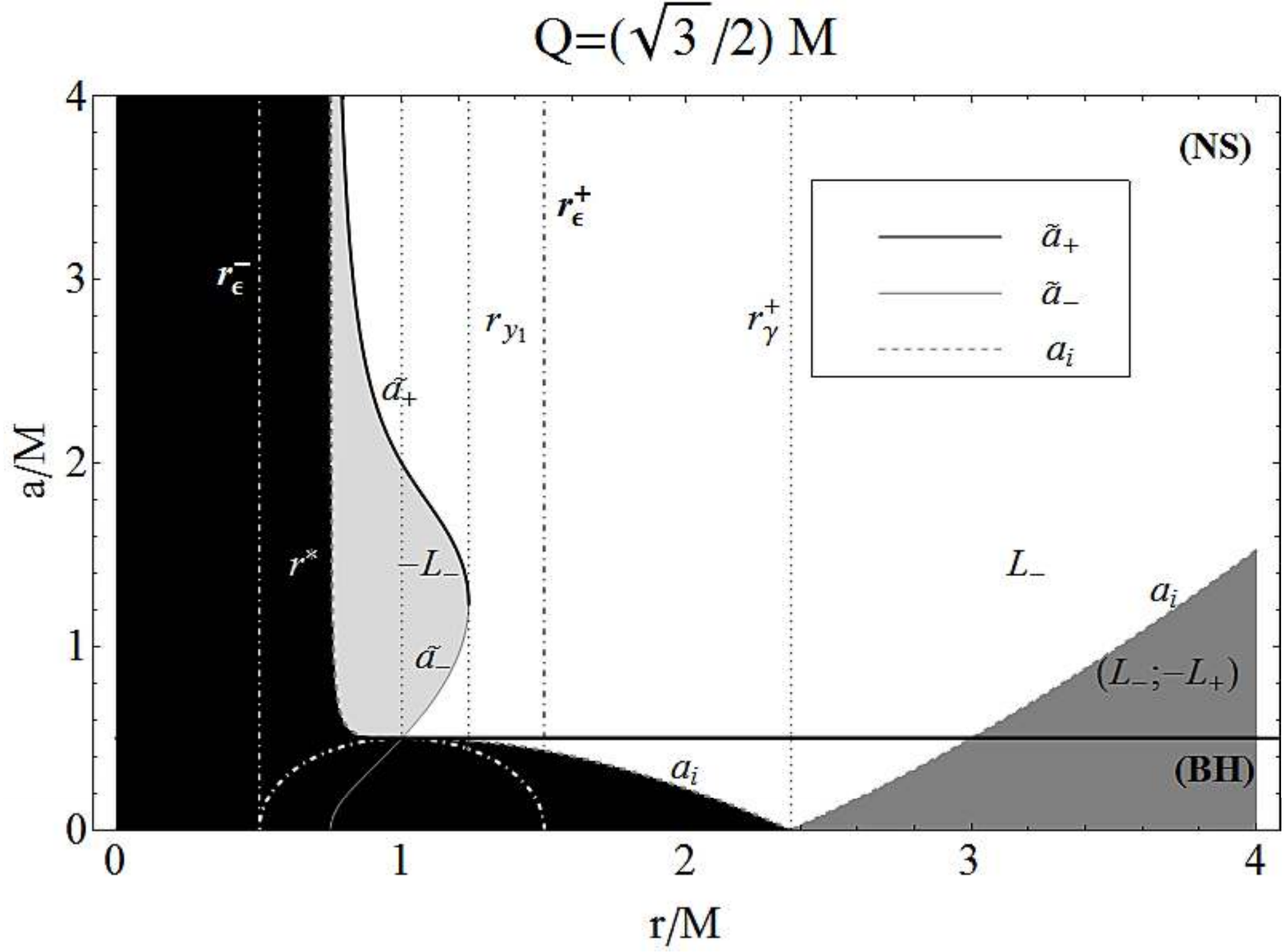}\\
\includegraphics[width=0.3\hsize,clip]{PLOT/concence}
\includegraphics[width=0.3\hsize,clip]{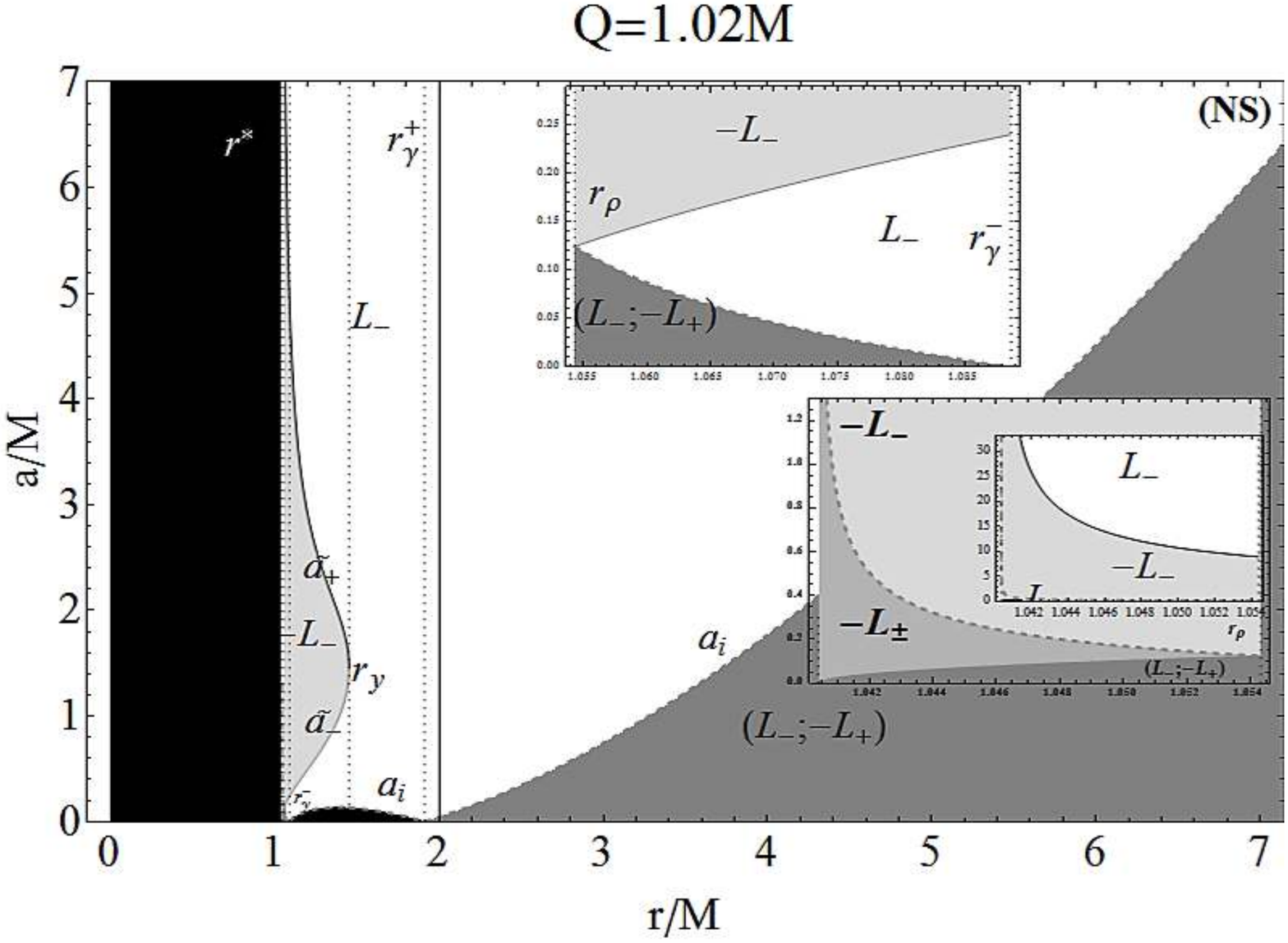}
\includegraphics[width=0.3\hsize,clip]{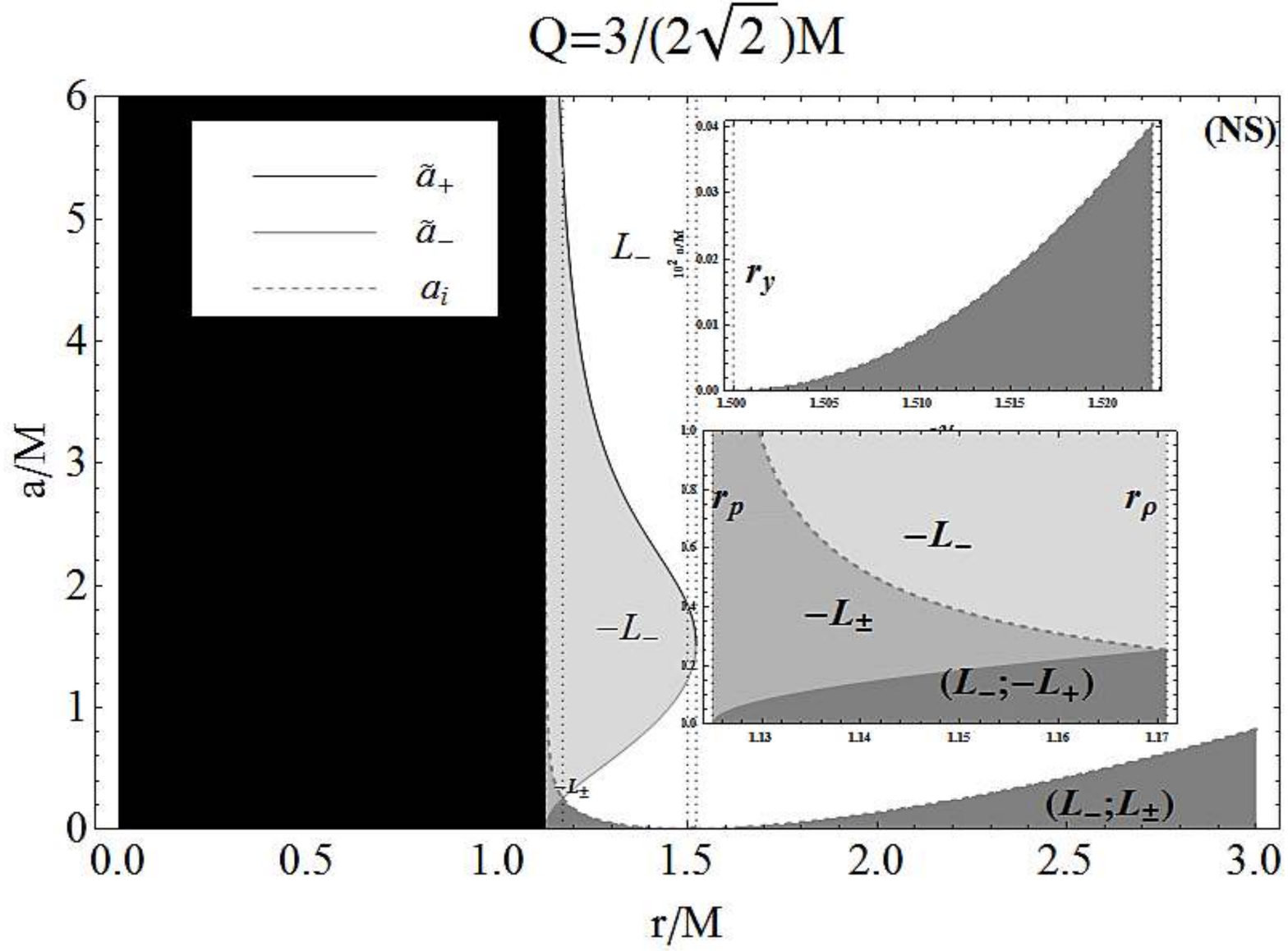}\\
\includegraphics[width=0.3\hsize,clip]{PLOT/sovie}
\includegraphics[width=0.3\hsize,clip]{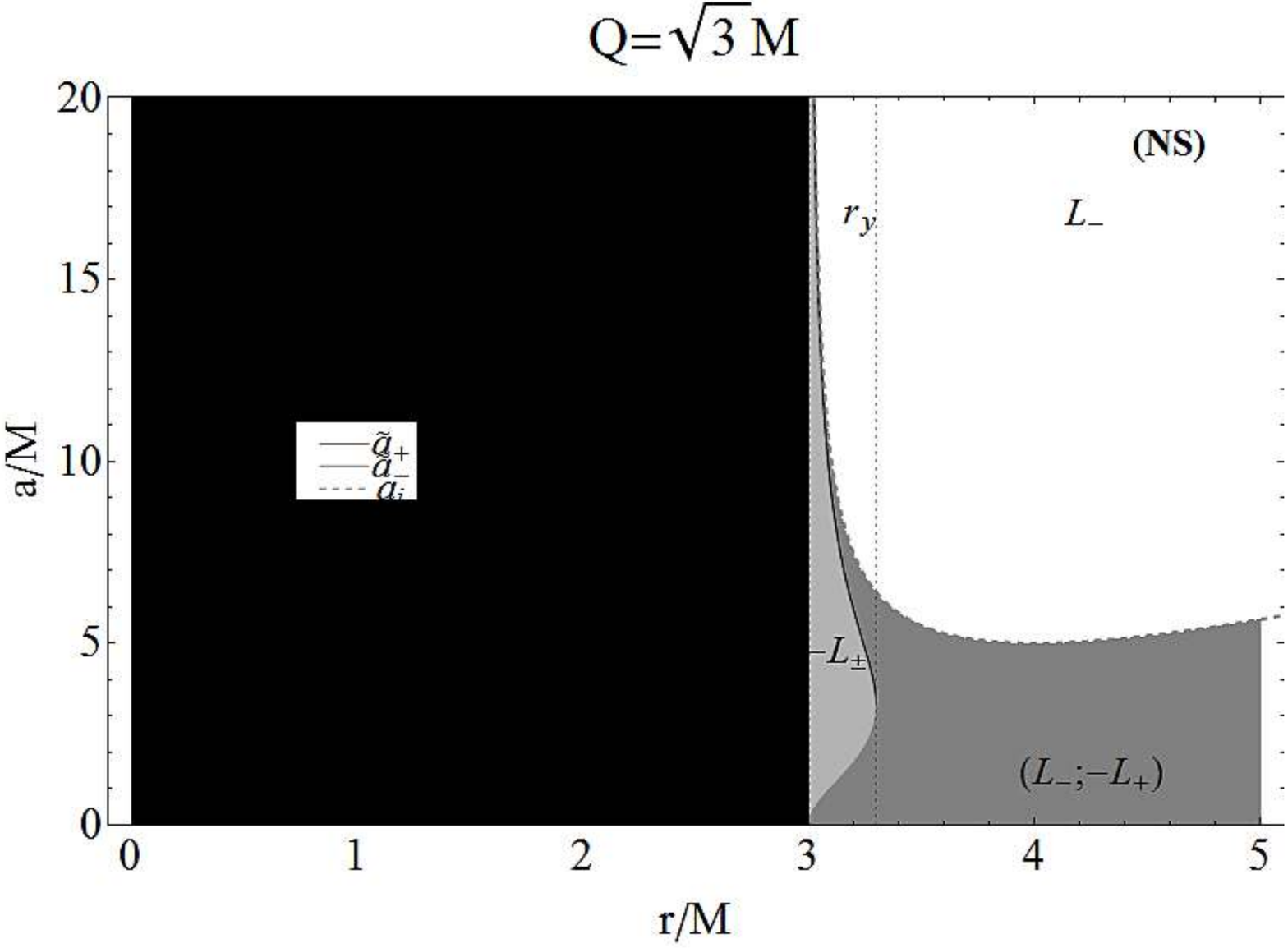}
\includegraphics[width=0.3\hsize,clip]{PLOT/Sphera}
\end{tabular}
\caption[font={footnotesize,it}]{
The intrinsic KN angular momentum $\tilde{a}_-/M$ (gray curve),  $\tilde{a}_+/M$ (black curve)  and  $a_i/M$ (dashed gray  curve)  as functions of $r/M$, for increasing values of the charge-to-mass ratio $Q/M$. The angular momentum of the orbiting test particle is shown in each region. Black regions are forbidden; no circular motion is possible there. Dotted lines represent the radii  $r_*$, $r_{\gamma}^{\pm}$ and $r_{z}$.  The dotted--dashed white curve is the outer horizon $r_+$. {The dotted-dashed black curve represents  the  ergosphere boundaries $r_{\epsilon}^{\pm}$.}
  The thick black horizontal line is $a=a_s$;   a naked singularity occurs for $Q<M$ and $a>a_s$.
  The black hole \textbf{(BH)} and the naked singularity \textbf{(NS)} regions are explicitly denoted. In this case,
   $a_s=  0.994987M$ for $Q=0.1M$, $a_s= 0.866025M$ for $Q=0.5M$, and $a_s=0.5$ for $Q=\sqrt{3}/2 M$.}\label{bicchic1due}
\end{figure}

We now summarize the results of comparing the properties of circular motion in the entire KN spacetime. It is convenient to confront our results with those obtained for the limiting cases $a=0$ and $Q=0$, which were presented in   \cite{Pu:Neutral,Pu:Kerr,Pu:Charged}, and are here schematically represented in  Figs.\il\ref{Fig:xyzns}.
It was found that black holes and naked singularities are characterized  by connected  and disconnected regions of stability, respectively. The analysis of the generalization that includes the spin and charge of the source simultaneously is
shown in Figs.\il\ref{bicchi}--\ref{bicchic1} for a varying charge-to-mass ratio for selected and increasing values of the spin $a/M$. Figures\il\ref{bicchic1due}  show the behavior of the circular orbits in terms of the spin $a/M $, for selected and increasing values of the charge $ Q/M $.

We first consider the analysis of Figs.\il\ref{bicchi}--\ref{bicchi1io}. The regions of existence of circular orbits with their orbital moments are marked in white and gray, while the black regions are those in which the motion is not possible.
In Fig.\il\ref{bicchi},  we consider the particular case $a=0.1M$ and explore the circular orbit properties  for both the black hole and naked singularity fields.
Notice that from a physical point of view, in the case of black holes,  we are interested in the motion outside the outer
horizon only,  $ r> r_+ $,  which in the plots corresponds to the  range $ [r_+, + \infty [\cup [0, a_s] \cup [0, Q_s] $.
Nevertheless, to understand better the behavior of the relevant radii and for the sake of completeness, we include some plots
of spatial regions located very closed to the origin.

In Figs.\il\ref{bicchi9} and \il\ref{bicchi1io}, we consider the cases $a=0.8M$ and $a=M$, respectively. We see that the location of the regions where  circular motion is allowed depends drastically on the value of the ratios $a/M$ and $Q/M$.

We summarize the analysis in Fig.\il\ref{bicchic1}, where we show the behavior of the orbital radius for increasing values of the source angular momentum.
As pointed out in Sec.\il\ref{Sec:NS_Case}, it is possible to distinguish   four different regions where  circular motion is allowed with different orbital angular momenta: The region  \textbf{I} with $L=L_-$,  \textbf{II} with $L\in (L_-,-L_+)$,  \textbf{III} associated to $L=-L_{\pm}$  and,  finally, the region  \textbf{IV} with $L=-L_-$.
The region $r<r_*$ (or equivalently $Q>Q_*$) is forbidden for both black holes and naked singularities.
A second forbidden region appears for configurations with $a<M$ and $Q<Q_{\ti{T}}^-$. This region becomes smaller as $a/M$ approaches the value $a=M$, where it disappears. We note the peculiar value   $a/M=1/\sqrt{2}$  which determines the interval $1/\sqrt{2}<a/M<1$, where the two forbidden regions are completely disconnected and separated on the left by a region with $L=-L_-$ and on the right by a  region with $L=L_-$. In other words, for $a\lesssim1.299M$ there exist two closed  regions with $L=L_-$ which are
disconnected, becoming connected in a particular region for larger values of $a/M$.
The region with $L=-L_{\pm}$ exist within the interval  $Q_{c}<Q<Q_*$.
The open range with $L=-L_+$ and $L=L_-$ begins in the region $r>r_x$. Furthermore, the region with $L=-L_-$ becomes smaller as $a/M$ increases to become practically nil as the area of the region $L=-L_{\pm}$. The region with orbits $L_-$ becomes increasingly wider as the intrinsic angular momentum and the charge of the source increase.
We conclude that the structure of the orbital regions around a KN source is very complex in the intervals $a<M$ and $r<r_{M_2}$.
%

As seen in Sec.\il\ref{Sec:BH_KN},
an alternative way to study the neutral test particle circular  motion in a KN spacetime  is to explore the orbital regions, for different values of the ratio $a/M$, as functions of the radial coordinate. This  is particularly suitable in order to compare the results with those of the Kerr spacetime depicted in the left plot of Fig.\il\ref{Fig:xyzns}.
In Fig.\il\ref{bicchic1due}, we show the behavior of the circular orbits for different values of the charge-to-mass ratio of the source.
In this case, fixing the source charge and  varying its
intrinsic angular momentum, we plot the  regions where circular
motion occurs  and point out the value of the orbital angular momentum
associated with each region.
The boundaries of the regions are defined by the intrinsic spin $\tilde{a}_{\pm}$ and $a_i$. For $Q <M$ there are  essentially  three types of orbital regions
characterized respectively by  $L=-L_-$, $L=L_-$, and $L=(L_-,-L_+)$.
We can see that there are two forbidden regions, the first determined by the conditions $r<r_*$ and $a<a_i$ for $Q^2/M^2<9/8$, and the second one by   $r<r_*$ for $Q^2/M^2\geq9/8$. The region within the interval $(L_-,-L_+)$ increases as $Q/M$ increases, while the regions with $L=-L_-$ and $L=L_-$ decrease as $Q/M$ increases and, finally, they coincide at $r=r_*$. Interestingly, the region with $L=-L_-$ disappears for $Q^2/M^2>3$.

One can see that the orbital region $(L=L_-)$ diminishes as the parameters
 $Q/M$ and $a/M$ decrease, and that it  extends  from a rather small distance from the
central source (light gray region).
For $Q/M$ small enough, say $Q\ll M/2$, this region disappears and, as the charge increases,
it moves towards areas with higher spin and smaller radii.
For $Q>M$  a new orbital region appears,
characterized by two different types of counter-rotating orbits with  $L=-L_ {\pm}$; this region becomes increasingly larger as the ratio $Q>M$ increases.
Finally, we note that the regions with $L=-L_{\pm}$ and $L=-L_-$ are bounded and extend up to a
maximum radius and a maximum angular momentum of the source.

The curve $a=a_i$  divides the plot into two
regions, the left black one, in which no circular motion is allowed, and the right one, in which orbital motion occurs for $a>a_i$ with $L=L_-$ and for  $a<a_i$  with  $L=(L_-,-L_ +)$.
As the source spin increases, the left forbidden region first grows to a maximum value of the orbital radius and then begins to decrease.
On the other hand, the region with  $L=(L_-,-L_ +)$  (right gray region) covers larger orbital radii as the spin increases, and moves towards the central source as the charge-to-mass ratio decreases.
\subsection{Discussion of the stability problem}
\begin{figure}[h!]
\begin{tabular}{ccc}
\includegraphics[width=0.3\hsize,clip]{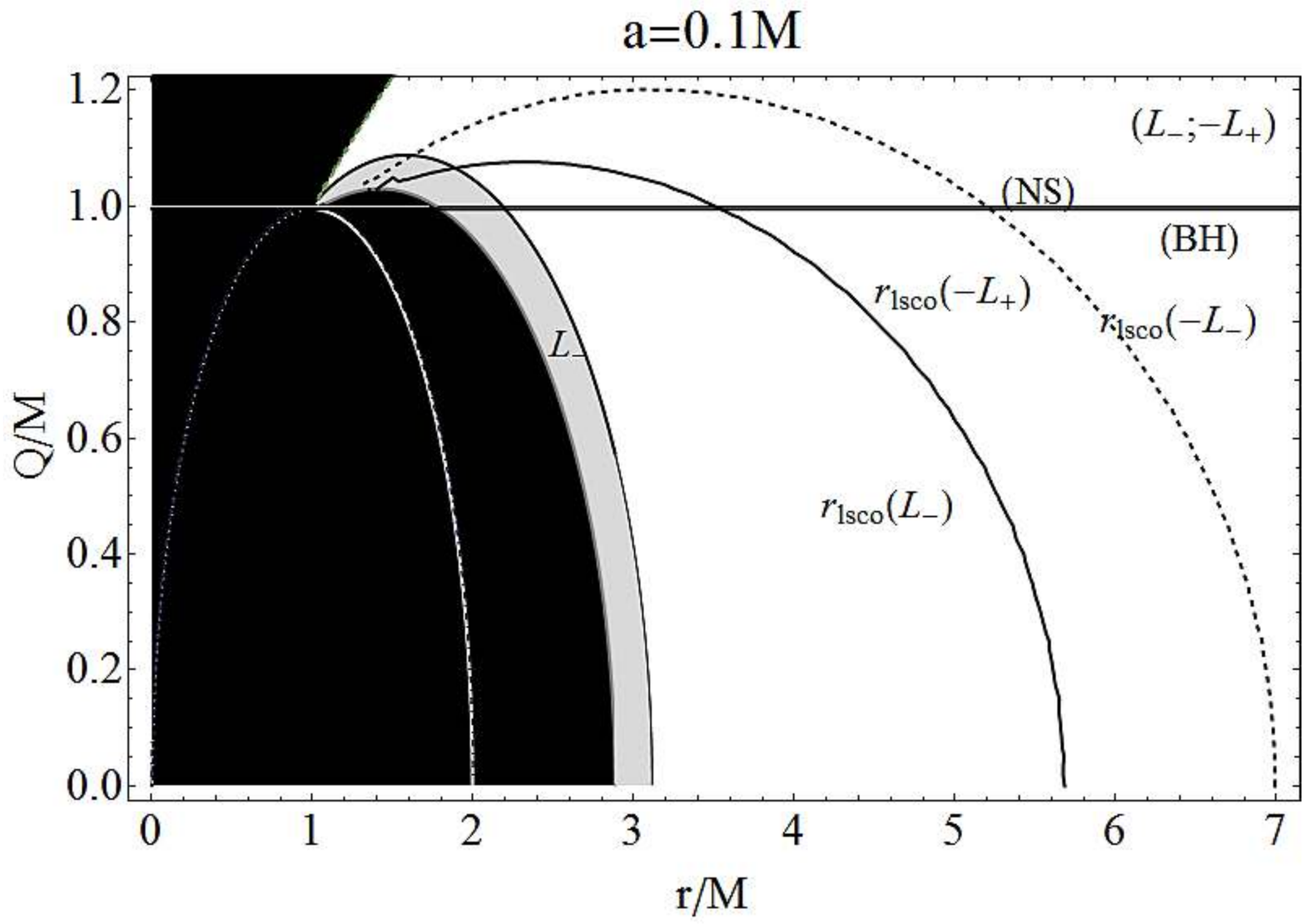}&
\includegraphics[width=0.3\hsize,clip]{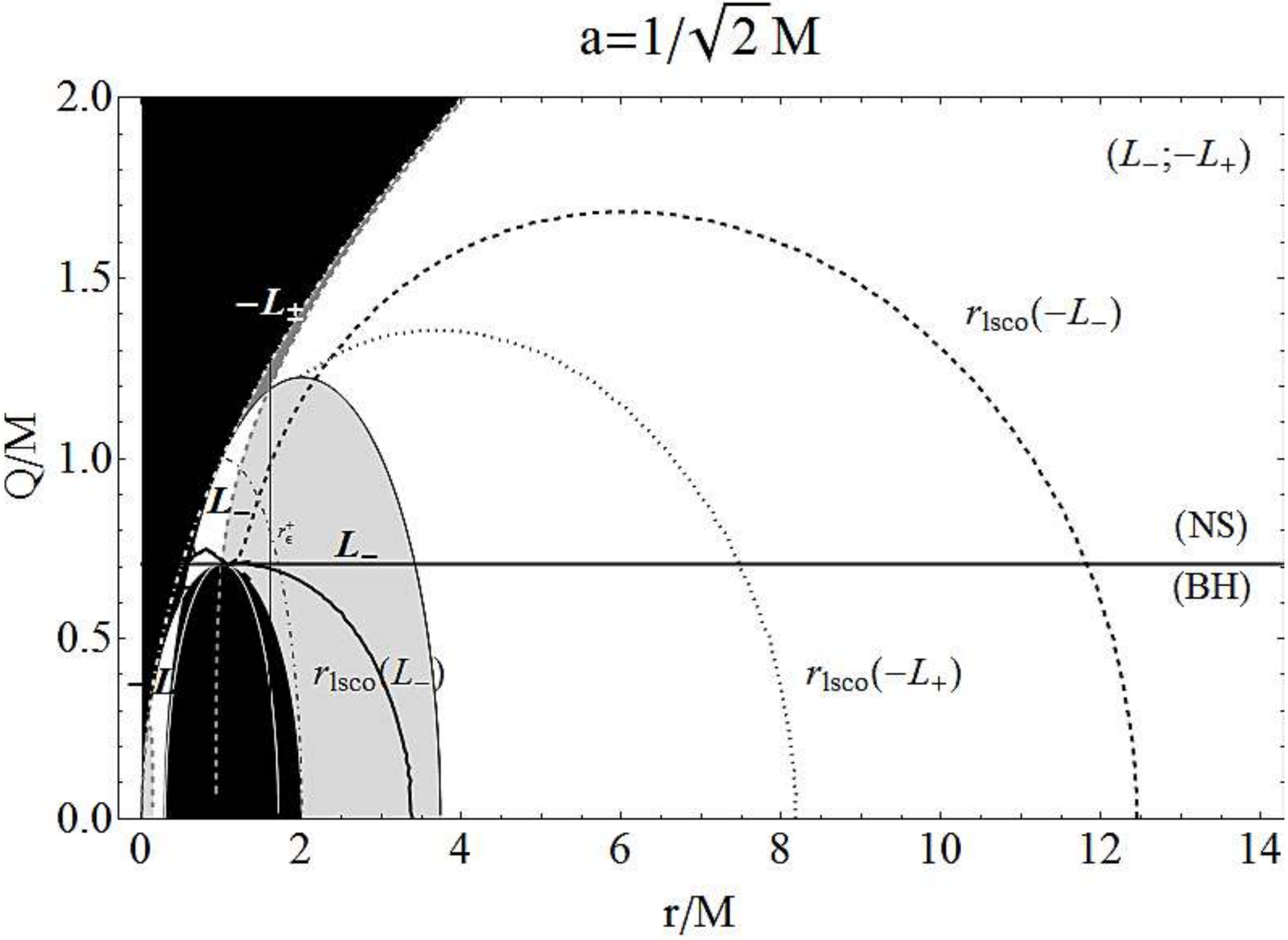}&
\includegraphics[width=0.3\hsize,clip]{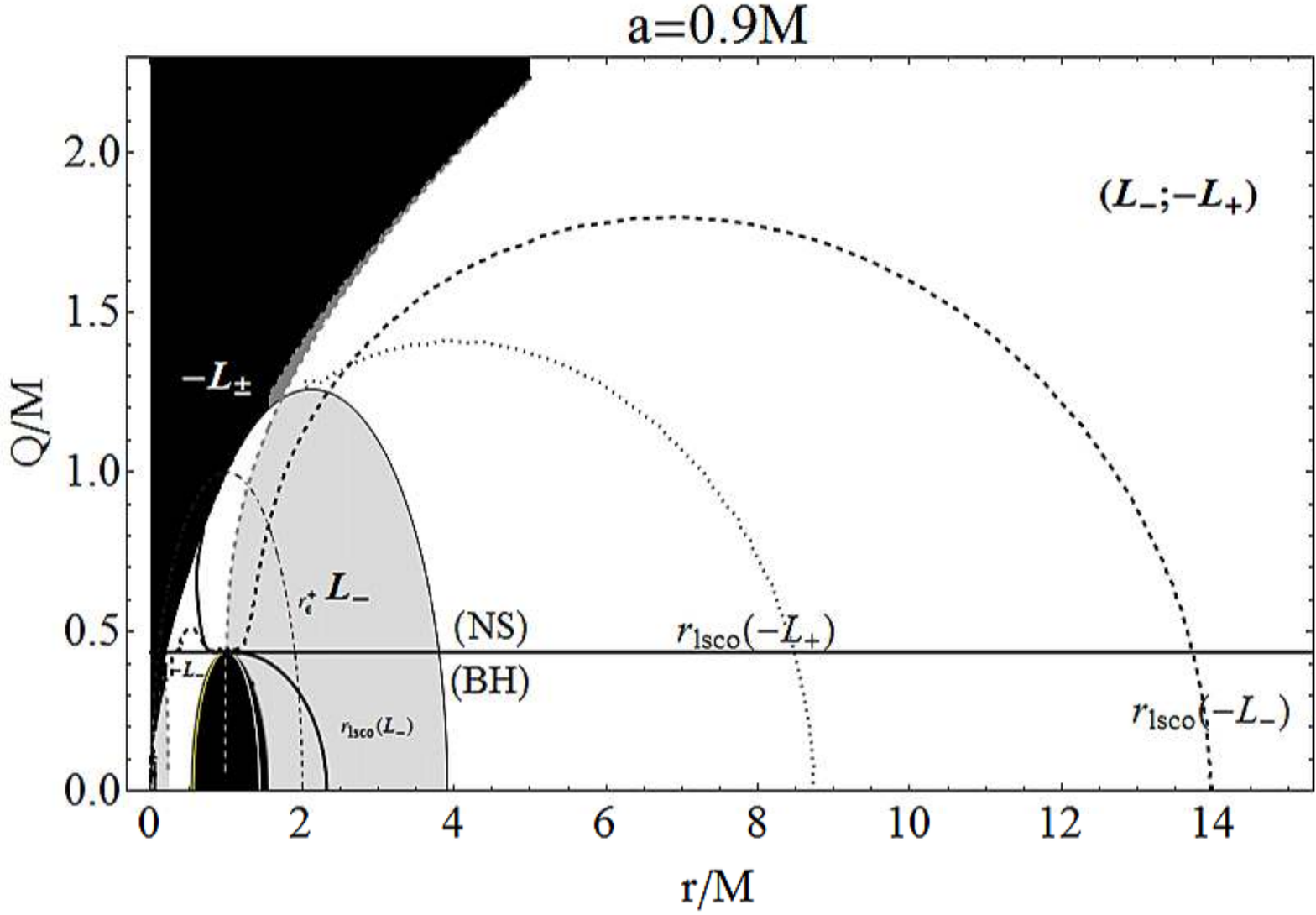}
\\
\includegraphics[width=0.3\hsize,clip]{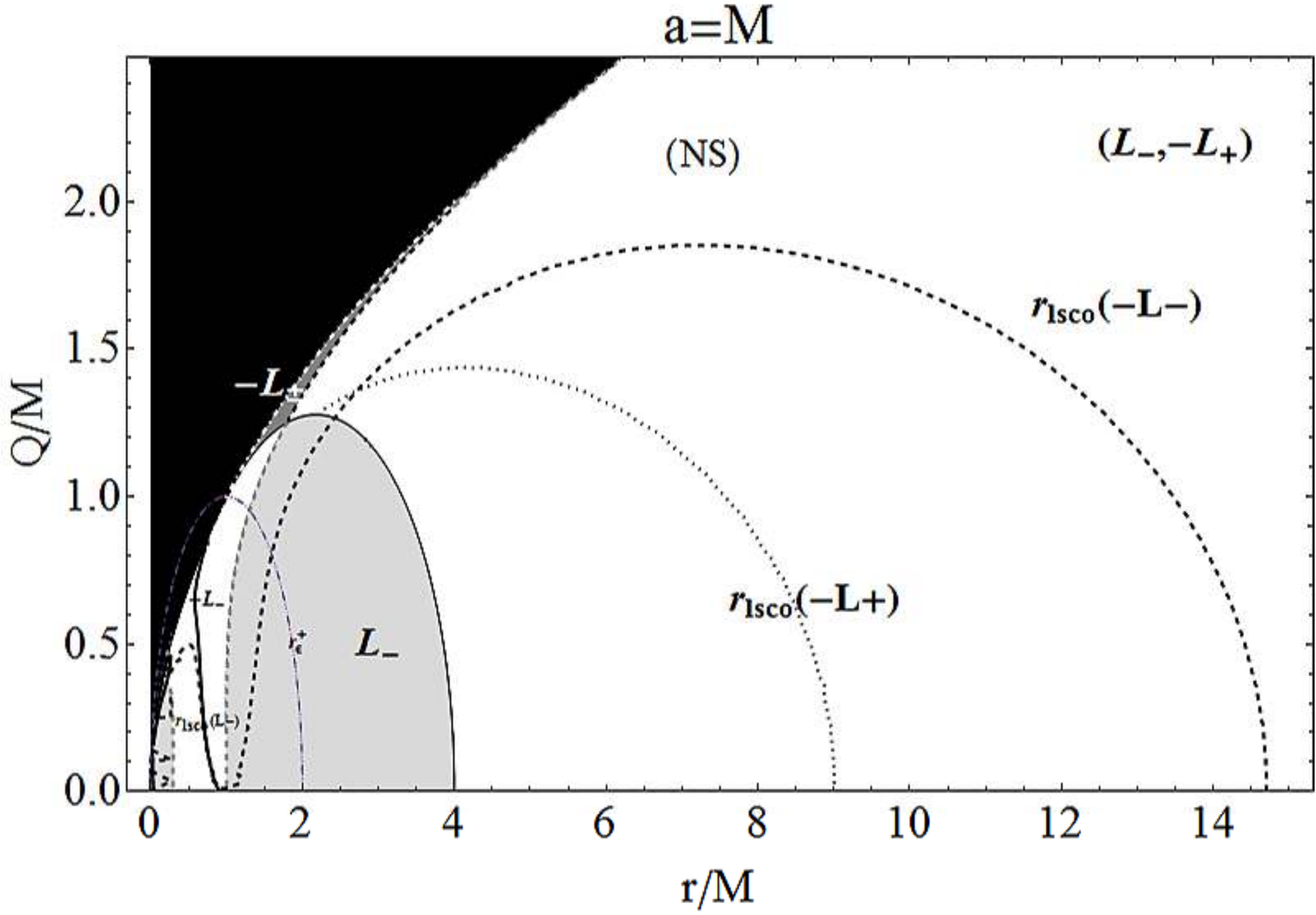}&
\includegraphics[width=0.3\hsize,clip]{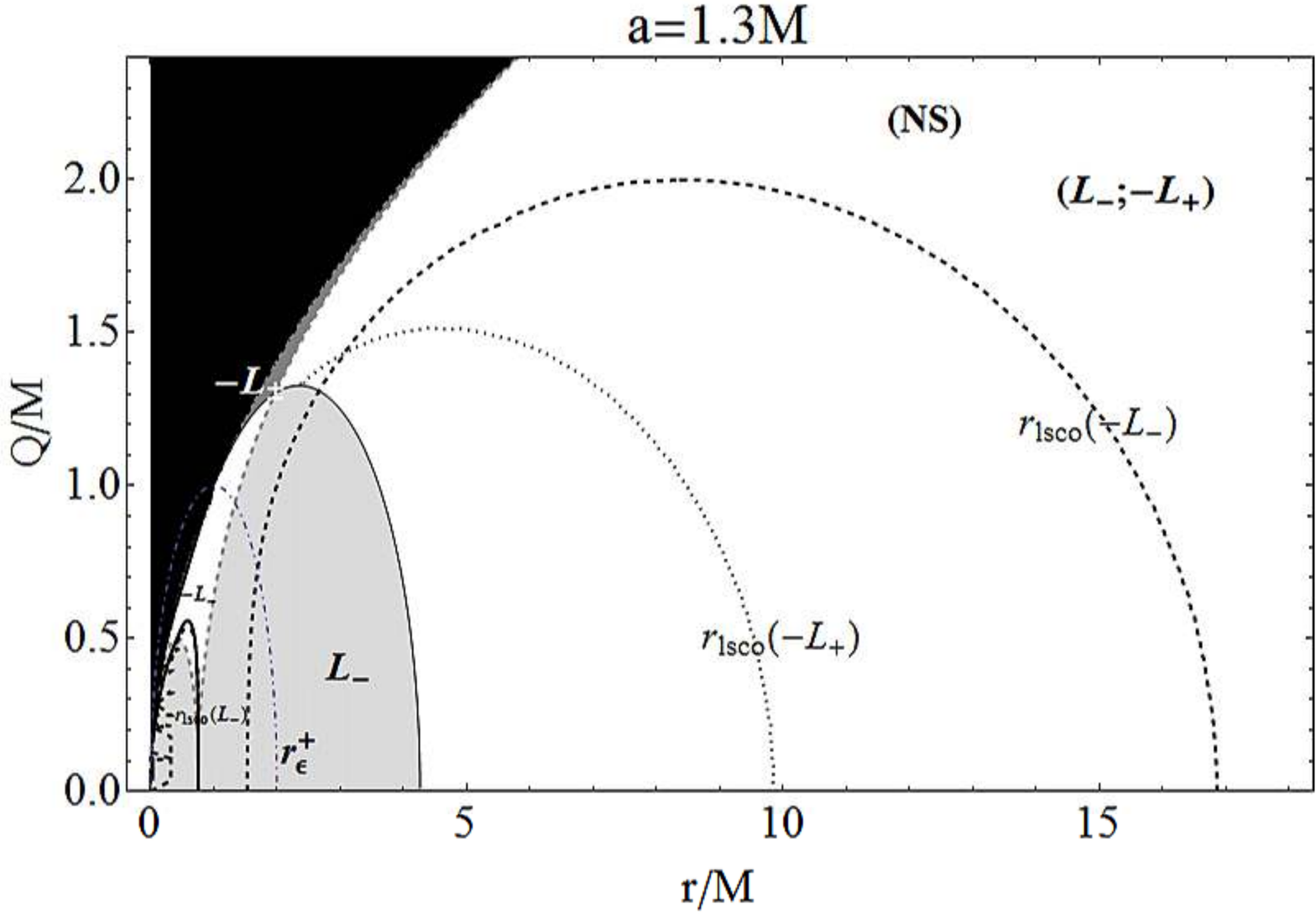}&
\includegraphics[width=0.3\hsize,clip]{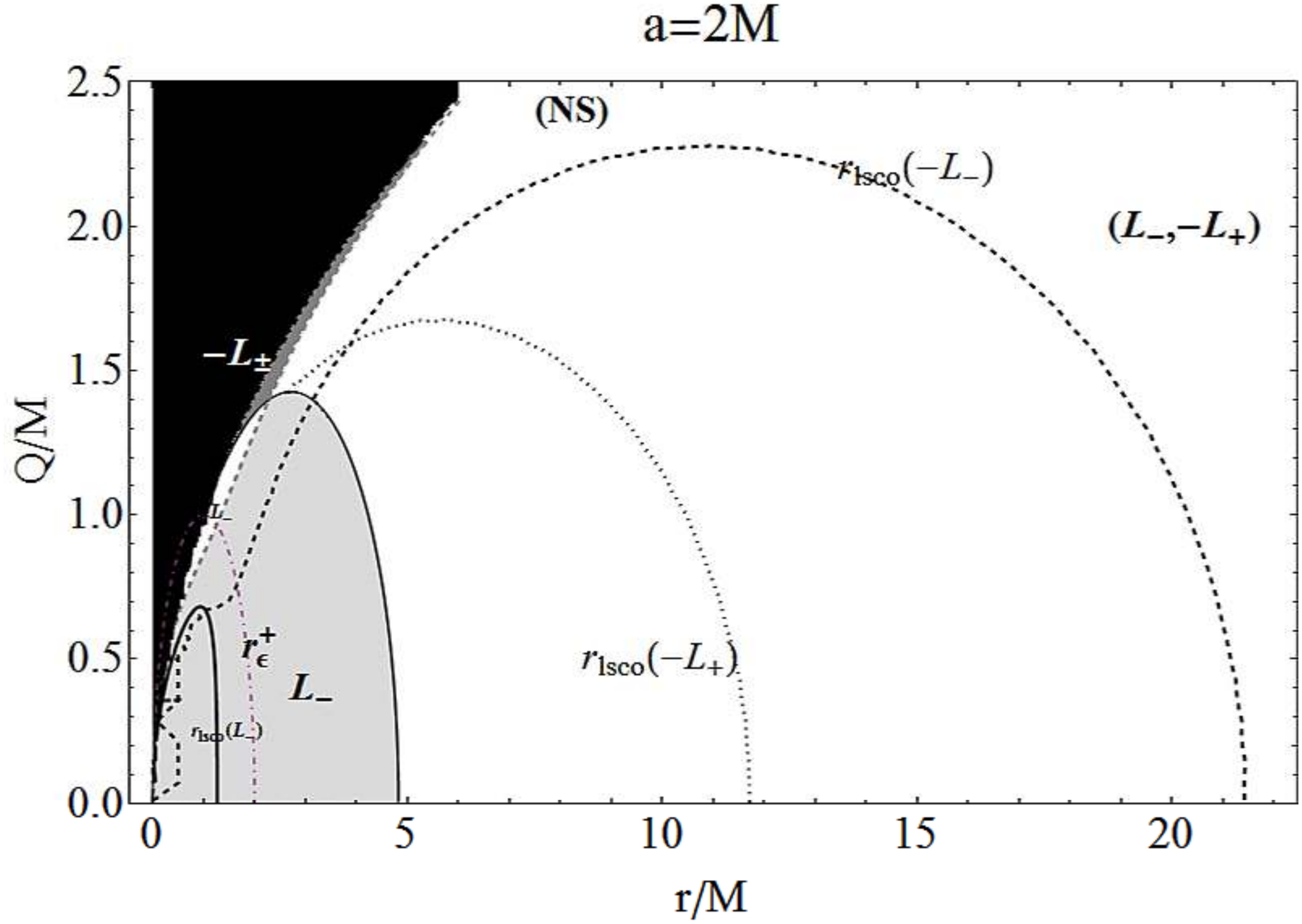}\\
\hspace{6cm}
&\includegraphics[width=0.3\hsize,clip]{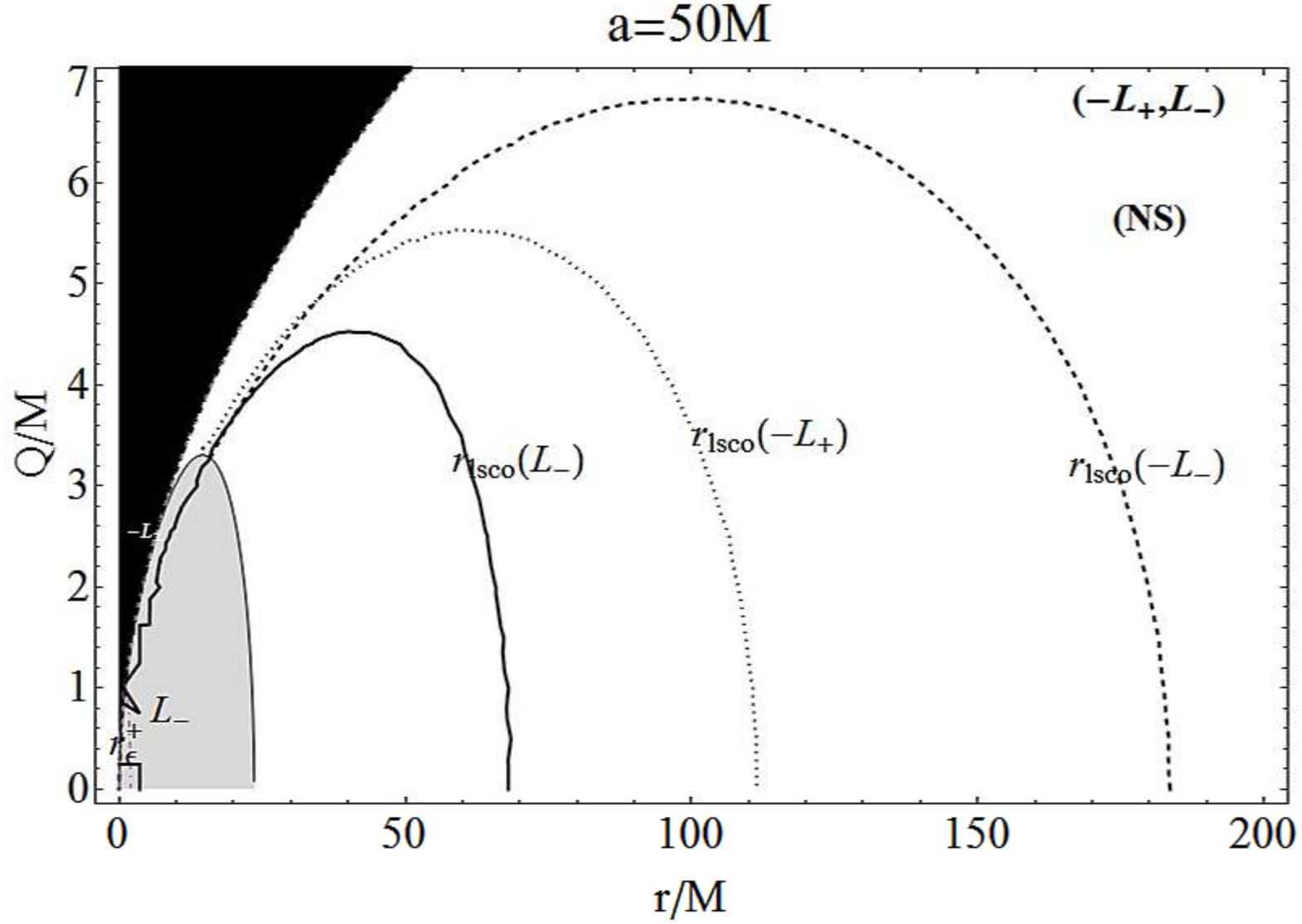}
\end{tabular}
\caption[font={footnotesize,it}]{\footnotesize{
The charge-to-mass ratios   $Q_{\ti{T}}^{\pm}/M$, $Q_*/M$ and $Q_{c_2}$ as  functions of $r/M$, for increasing values of the source spin. The angular momentum of the orbiting test particles is shown  in each region. Black regions are forbidden. The radius of the last stable circular motion $r_{lsco}$ is plotted: $r_{lsco}(L_-)$ (thick black curve),   $r_{lsco}(-L_-)$ (thick black dashed curve) and  $r_{lsco}(-L_+)$ (thick black  dotted curve). In the region $r<r_{lsco}$ ($r>r_{lsco}$), circular orbits are unstable (stable).  The horizontal thick black line represents $Q=Q_s$. A naked singularity occurs for $a>M$  and $Q>Q_s$, i.e., $Q_s=0.994987M$ for $a=0.1M$, $Q_s=0.707107M$ for $a=1/\sqrt{2}M$, $Q_s=0.43589M$ for $a=0.9M$ and $Q_s=0$ for $a=M$. The black hole \textbf{(BH)} and naked singularity \textbf{(NS)} regions are denoted explicitly. {The dotted-dashed curve  represents  the ergosphere boundaries $r_{\epsilon}^{\pm}$.}}} \label{Fig:enolim}
\end{figure}

In this last section, we will study the influence of the intrinsic angular momentum and charge of the source on the stability properties of test particles, moving along circular orbits around the central source.
In Sec.\il\ref{SubSec:BH-sta} and Sec.\il\ref{SubSec:NS-stab}, we studied the case of a black hole and a naked singularity, respectively. In  Fig.\il\ref{Fig:enolim}, we show a review of the results when both sources are analyzed
simultaneously.
First, we  note the presence of last stable circular orbit radii $r_{lsco}$ for all possible orbits with angular momentum $-L_+$, $-L_-$ and $L_-$.
The location of $r_{lsco}$ depends on the value of the intrinsic angular moment. The larger is the rotation, the further from the origin is $r_{lsco}$. This is physically reasonable since the rotation is known to generate its own gravitational field in general relativity. On the other hand, for a fixed value of $a/M$, as the charge-to-mass ratio increases, the value of $r_{lsco}$ first increases until it reaches a maximum value and then it decreases.
We have, in general, that  $r_{lsco}(L_-)>r_{lsco}(-L_+)>r_{lsco}(-L_-)$, but this situation becomes significantly more complex with increasing  spin and  decreasing source charge.

To formulate the main results in a plausible manner, let us imagine an idealized accretion disk made of test particles which are moving along circular orbits on the equatorial plane of a KN spacetime. It then follows that the structure of the accretion disk will be determined  by the locations of the radii of the last stable circular orbits. Consider, for instance, in Fig.\ref{Fig:enolim} the case $a=0.9M$.
The radius $r_{lsco}(L_-)$ determines the minimum radius of the idealized accretion disk which in the case of a black hole could be extended, in principle, to infinity, because the stability zone has no upper bound. The situation is different in the case of a naked singularity.
The disconnected structure  of the stability regions is especially clear for $a>M$. However, for $a=0.9M$, for example, the radius $r_{lsco}(-L_+)$ clearly goes into the NS zone creating a stable (left) orbital region with $L=-L_+$, an unstable region limited by $r_{lsco}(-L_+)$, and a stable region (right) for $r>r_{lsco}(-L_+)$.  This situation does not occur in the black hole case.
Notice a second unstable region of type $L_-$ separated by a stable zone (right)  from a $-L_-$ region.


%% file: Appendix_spin_radius13.tex
\appendix
\section{Limiting cases}\label{Sec:appendix_spin_radius}
To study circular motion in the gravitational field described by  the KN spacetime, in Sec. \il\ref{Sec:circular_motion_BH+NS} we introduced the spin parameters $a_i$, $a_{\ti{T}}$ and $ \tilde{a}_{\pm}$ which determine the boundaries of the regions where circular motion is allowed for both black holes and naked singularities. When these parameters coincide, the extended regions allow only a limited number of orbits
which depend on the  charge-to-mass ratio  of the central source. In this Appendix, we present the characteristics of these limiting orbits.

\begin{figure}[h!]
\begin{tabular}{cc}
\includegraphics[width=0.4\hsize,clip]{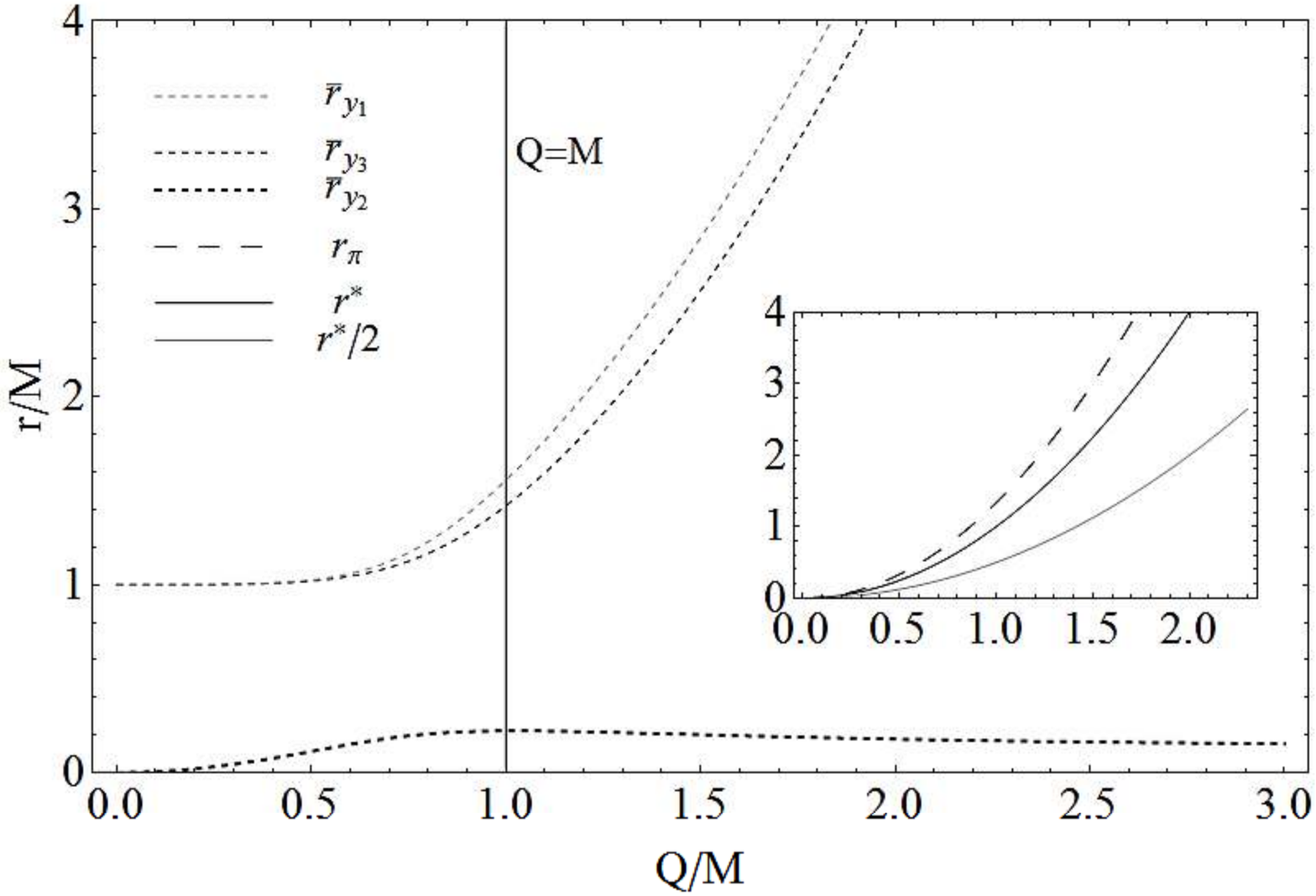}&
\includegraphics[width=0.4\hsize,clip]{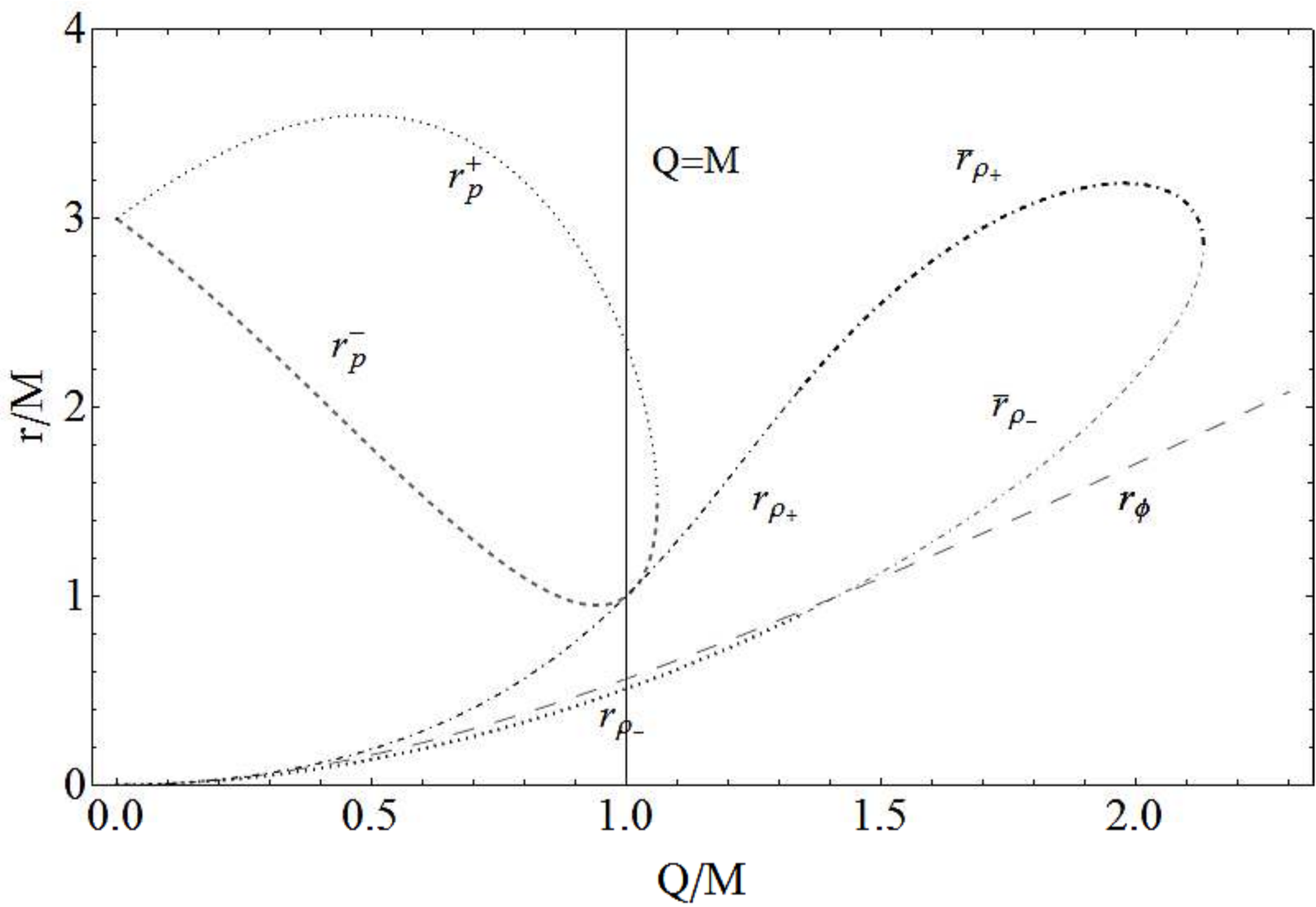}
\end{tabular}
\caption[font={footnotesize,it}]{\footnotesize{Left: Plot of $r_*/2$, $r_*$, $r_{\pi}$, $\bar{r}_{y_1}$, $\bar{r}_{y_2}$ and $\bar{r}_{y_3}$ as functions of $Q/M$. Right: Plot of $r_{\rho_\pm}$, $\bar{r}_{\rho_\pm}$, $r_{\phi}$ and $r_{p}^{\pm}$ as functions of $Q/M$.  }} \label{Fig:Two}
\end{figure}

First, orbits with  $\tilde{a}_{+}=\tilde{a}_{-}$ are characterized by the radii
\bea
r_{\epsilon}&\equiv&Q^2/2M=r_{*}/2,
\quad
r_{y_1}\equiv\frac{1}{3M} \left(M^2+Q^2+2 \bar{\bar{y}} \cos\left[\frac{1}{3} \arccos\left(\frac{\hat{\hat{y}}}{8  \bar{\bar{y}}^{3}}\right)\right]\right),
\\
r_{y_2}&\equiv&\frac{1}{3M} \left[M^2+Q^2-2 \bar{\bar{y}} \sin\left(\frac{1}{6} \left[\pi +2 \arccos\left(\frac{\hat{\hat{y}}}{8 \bar{\bar{y}}^{3}}\right)\right]\right)\right],
\\
r_{y_3}&\equiv&\frac{1}{3M} \left(M^2+Q^2-2\bar{\bar{y}} \sin\left[\frac{1}{3} \arcsin\left(\frac{\hat{\hat{y}}}{8 \bar{\bar{y}}^{3}}\right)\right]\right),
\\
\bar{r}_{{y}_1}&\equiv&\frac{1}{3M} \left(M^2+Q^2+\hat{\hat{y}}^{1/3}\right),
\quad
\bar{r}_{y_2}\equiv\frac{1}{6M} \left(2M^2+2 Q^2-\hat{\hat{y}}^{1/3}\right),
\\
\bar{r}_{y_3}&\equiv&\frac{1}{6M} \left(2M^2+2 Q^2+\left(\hat{\hat{y}}-3 \sqrt{3} \sqrt{Q^6 \left(8-13 Q^2+16 Q^4\right)}\right)^{1/3}\right.\\
&&\left.+\left(\hat{\hat{y}}+3 \sqrt{3} \sqrt{Q^6 \left(8-13 Q^2+16 Q^4\right)}\right)^{1/3}\right),
\eea
with $\bar{\bar{y}}\equiv \sqrt{M^4-M^2Q^2+Q^4}$, and $\hat{\hat{y}}\equiv 8M^6-12M^4 Q^2+15M^2 Q^4+8 Q^6$. The behavior of these radii is illustrated in Fig.\il\ref{Fig:Two}.

Moreover, for the limiting case $a_i=a_{\ti{T}}$ we obtain the following radii
\be
r_{\pi}=\frac{4 Q^2}{3M^2},\quad r^{\pm}_{p}\equiv\frac{1}{3M} \left(9M^2-4 Q^2\pm2 Q\sqrt{9M^2 -8 Q^2}\right)\ .
\ee
Finally, the condition $a_i=\tilde{a}_-$ leads to the orbits characterized by the radii
\bea
r_{\phi}&\equiv&\frac{1}{2} \left(\sqrt{9M^2+8 Q^2}-3M\right),
\\
r_{\rho_{\pm}}/M&\equiv&\frac{1}{6} \left(9-\sqrt{3} \upsilon \pm\frac{\sqrt{3} }{2M} \sqrt{-4\left(\upsilon^2M^2+45M^2+12 Q^2\right)-\frac{96 \sqrt{3} \left(5 Q^2-9M^2\right)}{\upsilon }}\right),
\\
\bar{r}_{\rho_{\pm}}/M&\equiv&\frac{1}{6} \left(9+\sqrt{3} \upsilon \pm\frac{\sqrt{3}}{2M}  \sqrt{-4\left(\upsilon^2M^2+45M^2+12 Q^2\right)+\frac{96 \sqrt{3} \left(5 Q^2-9M^2\right)}{\upsilon }}\right),
\\
\eea
where
\bea\nonumber
\zeta \equiv\left(9261M^6-12096M^4 Q^2+5814M^2 Q^4-568 Q^6+\right.
\\
\left.6 \sqrt{3}Q^2 \sqrt{5292M^8+65340M^6 Q^2-91881M^4 Q^4+45512M^2 Q^{6}-6272 Q^{8}}\right)^{1/3},
\eea
 and \(\upsilon \equiv\sqrt{\frac{441M^6+100M^2 Q^4-15M^3 \zeta +\zeta^2-4M Q^2 (96M^3+\zeta )}{M^3\zeta }}\).

In Fig.\il\ref{Fig:Two}, we show the behavior of the orbits in terms of the ratio $Q/M$.
In Fig.\il\ref{Fig:Two}-left, we consider  the first set of orbits $\{r_*/2,r_*,r_{\pi},\bar{r}_{y_1},\bar{r}_{y_2},\bar{r}_{y_3}\}$. These
radii exist for  all values of $Q/M$; they decrease when $Q/M$ decreases and diverge for as
 $Q/M$ increases. This means that the ranges where these radii have boundaries become larger as $Q/M$ increases.
The situation is completely different for  the radii of the second set, $ \{r_{\rho_\pm}, ,\bar{r}_{\rho_\pm},r_{\phi},r_{p}^{\pm}\}$,  plotted in   Fig.\il\ref{Fig:Two}(b). These orbits determine a closed region of the space $r\times Q$. The
radii $r^{\pm}_p$, where $a_i=a_T$,  and $(r_{\rho_{\pm}},\bar{r}_{\rho_{\pm}})$, where $a_i=\tilde{a}_-$, are defined for a closed range of values of $Q/M$ and are bounded. The distance   between these two radii can become null and  has a maximum as a function of the charge.
\section{Naked singularities with $Q<M$ }\label{App:QmM}
\begin{figure}[h!]
\centering
\begin{tabular}{c}
\includegraphics[width=0.5\hsize,clip]{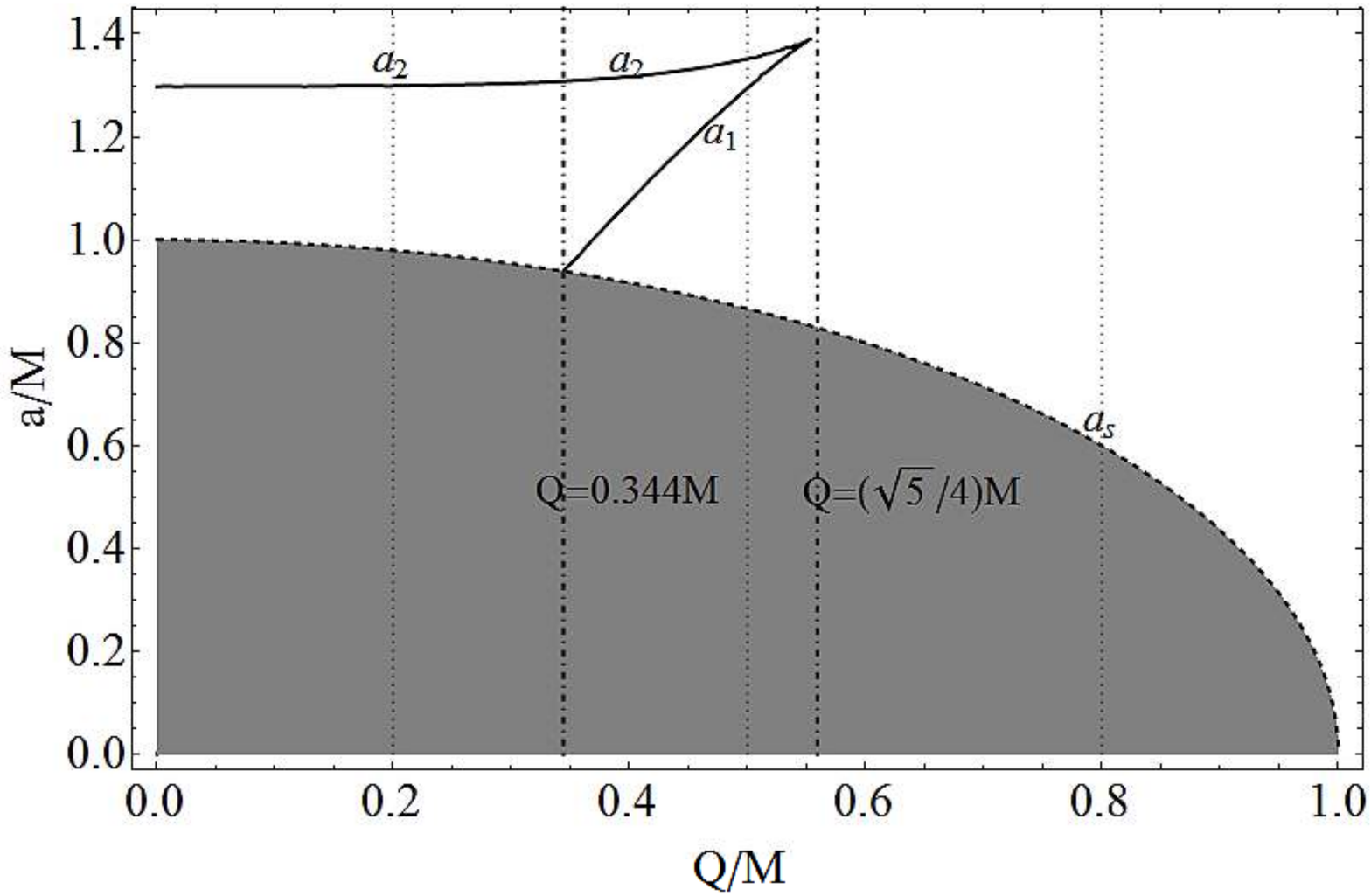}
\end{tabular}
\caption[font={footnotesize,it}]{\footnotesize{Circular
orbits in Kerr--Newman naked singularity  for $Q<M$
are explored. We
plot the parameters $a_1, a_2$ and $a_0$ as functions of the charge
$Q/M$.}} \label{Fig:AAA3}
\end{figure}

In Sec.\il\ref{Sec:NS_L0}, we use  the expressions for the spin $a_1\equiv a_-$ and $a_2\equiv a_+$ with
\be
a_{\pm}\equiv\sqrt{\left(\frac{27}{64}+2 Q^2-\frac{3 Q^4}{4}+\frac{1}{2} \sqrt{\frac{\iota_3}{768\ 2^{1/3}}+\iota_4}\pm\frac{\sqrt{\iota_5}}{32 \sqrt{6}}\right)},
\ee
in the study of solutions of the equations of motion with $L=0$,
where
\bea\nonumber
\iota_5&\equiv&3 \left(27+128 Q^2-48 Q^4\right)^2-192 Q^2 \left(144-81 Q^2+92 Q^4+32 Q^6\right)-
\\ \nonumber
&&\frac{2592\ 2^{1/3} Q^4 \left(9216-10368 Q^2+15529 Q^4+2496 Q^6+5232 Q^8+2560 Q^{10}\right)}{\iota_3}-2^{2/3} \iota_3+
\\
&&\frac{3 \left(19683-93312 Q^2-337392 Q^4+2522240 Q^6-3866496 Q^8+780288 Q^{10}+4096 Q^{12}\right)}{\sqrt{\frac{2}{3} 2^{2/3} \iota_3+1024 \iota_4}};
\eea
\bea\nonumber
\iota_4&\equiv&\frac{1}{1024 \iota_3}\left[1728\ 2^{1/3} Q^4 \left(9216-10368 Q^2+15529 Q^4+2496 Q^6+5232 Q^8+2560 Q^{10}\right)+\right.
\\
&&\left.\left(729+32 Q^2 \left(-72+593 Q^2-568 Q^4+8 Q^6\right)\right) \iota_3\right];
\eea
\be
\iota_3\equiv\left(\iota_2+\sqrt{ \left(\iota_1\right)}\right){}^{1/3}
\ee
\bea\nonumber
\iota_2&\equiv&93312 Q^6 \left(884736-1492992 Q^2+2656080 Q^4-819693 Q^6+1183756 Q^8+426790 Q^{10}+\right.
\\&&+\left.924864 Q^{12}+33280 Q^{14}+65536 Q^{16}\right);
\eea
\bea
\iota_1&\equiv&34828517376 Q^{20} \left(10-37 Q^2+16 Q^4\right)^3 \left(8-13 Q^2+16 Q^4\right) \left(783+311 Q^2+128 Q^4\right)^2\ .
\eea
To facilitate the presentation we use in the above formulas the normalized quantities $Q\rightarrow Q/M$ and $r\rightarrow r/M$.
The behavior of the spin parameters is depicted in Fig.\il\ref{Fig:AAA3}.
These parameters were used in Table\il\ref{Tab:di_gno} to determine the existence regions for solutions with $L=0$.
As  discussed in Sec.\il\ref{Sec:NS_L0}, it is important to consider in particular  the ranges
$Q/M\in]0,0.344263]$  with $(a_1,a_s)$,  $Q/M\in]0.344263,\sqrt{5}/4[$ with $(a_1,a_2,a_s)$,  and  $Q/M\in[\sqrt{5}/4,1]$ with $a_s$ only.

%% file: KN13.bbl
\begin{thebibliography}{99}

\bibitem{BiHoe85}
J. Bi\v{c}ak, C. Hoenselaers,
Phys. Rev. D \textbf{31},  10, 2476–2479 (1985).


\bibitem{Saa:2011wq}
  A.~Saa and R.~Santarelli,
  Phys.\ Rev.\ D {\bf 84}  027501 (2011).


\bibitem{Lopez83}
C. A. L\'opez,
Nuovo Cimento B,  11,  76 B, 9-27 (1983).

\bibitem{Bambi:2013eb}
  C.~Bambi and G.~Lukes-Gerakopoulos,
  arXiv:1302.0565 [gr-qc].

 \bibitem{Wunsch:2013s}
  A.~Wunsch, T.~Muller, D.~Weiskopf and G.~Wunner,
  Phys.\ Rev.\ D {\bf 87} 024007 (2013).

\bibitem{Pu:Neutral}
  D.~Pugliese, H.~Quevedo and R.~Ruffini,
  Phys.\ Rev.\  D {\bf 83}, 024021 (2011).


\bibitem{Pu:Kerr}
  D.~Pugliese, H.~Quevedo and R.~Ruffini,
  Phys.\ Rev.\ D {\bf 84} 044030 (2011).


\bibitem{Pu:Charged}
  D.~Pugliese, H.~Quevedo and R.~Ruffini,
  Phys.\ Rev.\ D {\bf 83} 104052 (2011).


\bibitem{MTW}
C.~W. Misner, K.~S. Thorne, \& J.~A. Wheeler,
\newblock {\em Gravitation},
\newblock W. H. Freeman, 1973.



\bibitem{Stu80}
Z. Stuchlik,
Astronomical Institutes of Czechoslovakia, Bulletin,  \textbf{31},  3,  129-144, (1980).

\bibitem{BiStuBa89}J. Bièák,  Z. Stuchlík, and V. Balek,
Bull. Astronom. Inst.
Czechoslovakia, \textbf{40}(2),  65–92, (1989).

\bibitem{BaBiStu89}
V. Balek, J.Bicak,   Z. Stuchlik,
 Astronomical Institutes of Czechoslovakia, Bulletin (ISSN 0004-6248),  \textbf{40},  3,   133-165, (1989).

\bibitem{Kovar:2010ty}
  J.~Kovar, O.~Kopacek, V.~Karas and Z.~Stuchlik,
  Class.\ Quant.\ Grav.\  {\bf 27} 135006 (2010).


\bibitem{Kovar:2008ta}
  J.~Kovar, Z.~Stuchlik and V.~Karas,
  Class.\ Quant.\ Grav.\  {\bf 25}  095011 (2008).


\bibitem{StuHle2000}
Z. Stuchlík and S. Hledík,
Class. Quantum Grav. \textbf{17}, 21, 4541, (2000).

\bibitem{Cin89}
J. L. Cindra,
  Class. Quantum Grav. \textbf{6}, 857, (1989).

\bibitem{Stuc81}
Z. Stuchlik,
 Astronomical Institutes of Czechoslovakia, Bulletin,  \textbf{32},  6,  366-373, (1981).

\bibitem{BekDenTre74}
R.
 Bekgamini, G. Denaedo, A. Treves
Lettere al Nuovo Cimento,  \textbf{11},  3,  183-186, (1974).


\bibitem{RaVi96}
N. K.  Rajesh,  C. V.  Vishveshwara,
  Class. Quantum Grav. \textbf{13} (7), 1783-1795, (1996).

\bibitem{StuHleJu00}
Z. Stuchlík, S. Hledík and J. Jurán,
Class. Quantum Grav. \textbf{17}, 14 2691, (2000).


\bibitem{Kovar:2013pf}
  J.~Kovar, O.~Kopacek, V.~Karas and Y.~Kojima,
  Class.\ Quant.\ Grav.\  {\bf 30}  025010 (2013).


\bibitem{AbraMillStuc93}
M. A. Abramowicz, J. C. Miller, Z. Stuchlík,
Phys. Rev. D \textbf{47}, 1440–1447 (1993).

\bibitem{Burinskii:2005mm}
  A.~Burinskii,
  Grav.\ Cosmol.\  {\bf 14}  109 (2008).



 \bibitem{Arcos:2002ip}
  H.~I.~Arcos and J.~G.~Pereira,
  Gen.\ Rel.\ Grav.\  {\bf 36} 2441 (2004).



\bibitem{deFelice74}
F. de Felice,   	
Astronomy and Astrophysics,  \textbf{34},  15 (1974).

\bibitem{Babichev:2008dy}
  E.~Babichev, S.~Chernov, V.~Dokuchaev and Y.~.Eroshenko,
  Phys.\ Rev.\ D {\bf 78} 104027 (2008).

\bibitem{StuBicBal01}
Z. Stuchlik, J. Bicak, V. Balek,
Gen. Rel.  Grav.,  \textbf{31},  1,  53-71 (1999).




\bibitem{StuJu96}
Z.  Stuchlík,   J. Jurán,
Physical Processe in Interacting Binaries, proceedings of the 19th Stellar Conference of the Slovak and Czech Astronomical Institutes, 7th-9th November 1996, Tatranska Lomnica, Slovakia. Edited by D. Chochol, A. Skopal, T. Pribulla. Astronomical Institute of the Slovak Academy of Sciences, 115, (1996).


\bibitem{StuHle98}
Z. Stuchlik,  S. Hledik,
Acta Phys. Slovaca,  \textbf{48},  5, 549-562, (1998).






\bibitem{KoStuKara0607}
J. Kovár, Z. Stuchlík, V. Karas,
Proceedings of RAGtime 8/9  2006/2007, Hradec nad Moravicí, Opava, Czech Republic, S. Hledík and Z. Stuchlík, editors, Silesian University in Opava, 125-138, (2007).

\bibitem{EsteMarr90}
E. P. Esteban, N. Marrero.
Il Nuovo Cimento B,  \textbf{11},
105,  6,  647-656, (1990).

\bibitem{JaAbraPac90}
 M. Jaroszynski,  M. A. Abramowicz, B. Paczynski,
 Acta Astronomica,  \textbf{30},  1, 1-34, (1980).



\bibitem{Kovacs:2010xm}
  Z.~Kovacs and T.~Harko,
  Phys.\ Rev.\ D {\bf 82} 124047  (2010).

  \bibitem{Pat2010} M. Patil, P. S. Joshi,
Phys. Rev. D \textbf{82} 104049 (2010).

\bibitem{Joshi:2012mk}
  P.~S.~Joshi and D.~Malafarina,
  Int.\ J.\ Mod.\ Phys.\ D {\bf 20}  2641 (2011).


\bibitem{Patil:2011uf}
  M.~Patil, P.~S.~Joshi, M.~Kimura and K.~-i.~Nakao,
  Phys.\ Rev.\ D {\bf 86}  084023 (2012).


 \bibitem{Gautrou} J. M. Cohen,
R. Gautreau, Phys.\   Rev.\ D \textbf{19},  8, (1979).

\bibitem{DiCriscienzo:2010gh}
  R.~Di Criscienzo, L.~Vanzo and S.~Zerbini,
  JHEP {\bf 1005}  092 (2010).

  \bibitem{Liang74}
E.P.T. Liang,
 Phys. Rev. D, \textbf{9}, 12, (1974).


  \bibitem{Dotti:2008ta}
  G.~Dotti and R.~J.~Gleiser,
  Class.\ Quant.\ Grav.\  {\bf 26} 215002 (2009).

\bibitem{Casadio:2003iv}
  R.~Casadio, S.~Fabi and B.~Harms,
  Phys.\ Rev.\ D {\bf 70}  044026 (2004).

\bibitem{Toth:2011ab}
  G.~Z.~Toth,
  Gen.\ Rel.\ Grav.\  {\bf 44}2019 (2012).

\bibitem{Virbhadra:2007kw}
  K.~S.~Virbhadra and C.~R.~Keeton,
  Phys.\ Rev.\  D {\bf 77} 124014 (2008).



\bibitem{Virbhadra:2002ju}
  K.~S.~Virbhadra and G.~F.~R.~Ellis,
  Phys.\ Rev.\  D {\bf 65} 103004 (2002).




 \bibitem{dotti}
  G.~Dotti, R.~J.~Gleiser, I.~F.~Ranea-Sandoval and H.~Vucetich,
  Class.\ Quant.\ Grav.\  {\bf 25} 245012 (2008).

  \bibitem{pani} 
  P.~Pani, E.~Barausse, E.~Berti and V.~Cardoso,
  Phys.\ Rev.\ D {\bf 82}, 044009 (2010).



\bibitem{quev90} H. Quevedo, Fort. Phys. {\bf 38}, 733 (1990).

\bibitem{weyl} H. Weyl, Ann. Phys. {\bf 54}, 117 (1917).

\bibitem{chaz} J. Chazy, Bull. Soc. Math. France {\bf 52} 17 (1924).

\bibitem{cur} H. Curzon, Proc. London Math. Soc. {\bf 23}, 477 (1924).

\bibitem{erro59} G. Erez and N. Rosen, Bull. Res. Counc. of Israel {\bf 8F}, 47 (1959).

\bibitem{gm} Ts. I. Gutsunaev and V. S. Manko, Gen. Rel. Grav. {\bf 17}, 1025 (1985).

\bibitem{quev87} H. Quevedo, Gen. Rel. Grav. {\bf 19}, 1013 (1987).

\bibitem{solutions}
H. Stephani, D. Kramer, M. A. H. MacCallum, C. Hoenselaers, and E. Herlt,
{\it Exact Solutions of Einstein's Field Equations} (Cambridge University Press, Cambridge, UK, 2003).



\bibitem{zip66} D. M. Zipoy,
J. Math. Phys. {\bf 7}, 1137 (1966).

\bibitem{voor70} B. Voorhees,
Phys. Rev. D {\bf 2}, 2119 (1970).

\bibitem{quev11} H. Quevedo,
Int. J. Mod. Phys. D {\bf 20} 1779 (2011).

\bibitem{Pani:2013ija}
  P.~Pani, E.~Berti and L.~Gualtieri,
  {\it Gravito-Electromagnetic Perturbations of Kerr-Newman Black Holes: Stability and Isospectrality in the Slow-Rotation Limit,}
  arXiv:1304.1160 [gr-qc].
 \end{thebibliography}
